\pdfoutput=1

\documentclass[12pt,a4paper]{article}

\usepackage{ifthen} 
\newboolean{pdflatex}
\setboolean{pdflatex}{true} 

\newboolean{articletitles}
\setboolean{articletitles}{true} 

\newboolean{uprightparticles}
\setboolean{uprightparticles}{false} 

\newboolean{inbibliography}
\setboolean{inbibliography}{false} 

\usepackage[top=1in, bottom=1.25in, left=1in, right=1in]{geometry}

\usepackage{microtype}
\usepackage{lineno}  
\usepackage{xspace} 
\usepackage{caption} 

\usepackage{graphicx}  
\usepackage{color}
\usepackage{colortbl}
\graphicspath{{./figs/}} 

\usepackage{amsmath} 
\usepackage{amssymb}
\usepackage{amsfonts}
\usepackage{upgreek} 

\newcommand*\patchAmsMathEnvironmentForLineno[1]{%
\expandafter\let\csname old#1\expandafter\endcsname\csname #1\endcsname
\expandafter\let\csname oldend#1\expandafter\endcsname\csname
end#1\endcsname
 \renewenvironment{#1}%
   {\linenomath\csname old#1\endcsname}%
   {\csname oldend#1\endcsname\endlinenomath}%
}
\newcommand*\patchBothAmsMathEnvironmentsForLineno[1]{%
  \patchAmsMathEnvironmentForLineno{#1}%
  \patchAmsMathEnvironmentForLineno{#1*}%
}
\AtBeginDocument{%
\patchBothAmsMathEnvironmentsForLineno{equation}%
\patchBothAmsMathEnvironmentsForLineno{align}%
\patchBothAmsMathEnvironmentsForLineno{flalign}%
\patchBothAmsMathEnvironmentsForLineno{alignat}%
\patchBothAmsMathEnvironmentsForLineno{gather}%
\patchBothAmsMathEnvironmentsForLineno{multline}%
\patchBothAmsMathEnvironmentsForLineno{eqnarray}%
}

\usepackage{hyperref}    
\usepackage[all]{hypcap} 

\usepackage{xspace} 
\usepackage{upgreek}

\def\lhcb {\mbox{LHCb}\xspace}

\def\MagUp {\mbox{\em Mag\kern -0.05em Up}\xspace}

\ifthenelse{\boolean{uprightparticles}}%
{

 \def\Ppi         {\ensuremath{\uppi}\xspace}

 \def\Ppsi        {\ensuremath{\uppsi}\xspace}

 \def\PDelta      {\ensuremath{\Delta}\xspace}                 
 \def\PXi      {\ensuremath{\Xi}\xspace}                 
 \def\PLambda      {\ensuremath{\Lambda}\xspace}                 
 \def\PSigma      {\ensuremath{\Sigma}\xspace}                 
 \def\POmega      {\ensuremath{\Omega}\xspace}                 
 \def\PUpsilon      {\ensuremath{\Upsilon}\xspace}                 
 
 \def\PB      {\ensuremath{\mathrm{B}}\xspace}                 
                  
 \def\PD      {\ensuremath{\mathrm{D}}\xspace}

 \def\PJ      {\ensuremath{\mathrm{J}}\xspace}                 
 \def\PK      {\ensuremath{\mathrm{K}}\xspace}

 \def\Pb      {\ensuremath{\mathrm{b}}\xspace}                 
 \def\Pc      {\ensuremath{\mathrm{c}}\xspace}                 
 \def\Pd      {\ensuremath{\mathrm{d}}\xspace}

 \def\Pi      {\ensuremath{\mathrm{i}}\xspace}

 \def\Ps      {\ensuremath{\mathrm{s}}\xspace}                 
                  
 \def\Pu      {\ensuremath{\mathrm{u}}\xspace}

}
{

 \def\Ppi         {\ensuremath{\pi}\xspace}

 \def\Ppsi        {\ensuremath{\psi}\xspace}                 
                  
 \mathchardef\PDelta="7101
 \mathchardef\PXi="7104
 \mathchardef\PLambda="7103
 \mathchardef\PSigma="7106
 \mathchardef\POmega="710A
 \mathchardef\PUpsilon="7107
                  
 \def\PB      {\ensuremath{B}\xspace}                 
                  
 \def\PD      {\ensuremath{D}\xspace}

 \def\PJ      {\ensuremath{J}\xspace}                 
 \def\PK      {\ensuremath{K}\xspace}

 \def\Pb      {\ensuremath{b}\xspace}                 
 \def\Pc      {\ensuremath{c}\xspace}                 
 \def\Pd      {\ensuremath{d}\xspace}

 \def\Pi      {\ensuremath{i}\xspace}

 \def\Ps      {\ensuremath{s}\xspace}                 
                  
 \def\Pu      {\ensuremath{u}\xspace}

}

\makeatletter
\ifcase \@ptsize \relax
  \newcommand{\miniscule}{\@setfontsize\miniscule{4}{5}}
\or
  \newcommand{\miniscule}{\@setfontsize\miniscule{5}{6}}
\or
  \newcommand{\miniscule}{\@setfontsize\miniscule{5}{6}}
\fi
\makeatother

\DeclareRobustCommand{\optbar}[1]{\shortstack{{\miniscule (\rule[.5ex]{1.25em}{.18mm})}
  \\ [-.7ex] $#1$}}

\def\g      {{\ensuremath{\Pgamma}}\xspace}

\def\uquark    {{\ensuremath{\Pu}}\xspace}

\def\dquark    {{\ensuremath{\Pd}}\xspace}

\def\squark    {{\ensuremath{\Ps}}\xspace}

\def\cquark    {{\ensuremath{\Pc}}\xspace}

\def\bquark    {{\ensuremath{\Pb}}\xspace}
\def\bquarkbar {{\ensuremath{\overline \bquark}}\xspace}

\def\pion   {{\ensuremath{\Ppi}}\xspace}
\def\piz    {{\ensuremath{\pion^0}}\xspace}

\def\pip    {{\ensuremath{\pion^+}}\xspace}
\def\pim    {{\ensuremath{\pion^-}}\xspace}
\def\pipm   {{\ensuremath{\pion^\pm}}\xspace}

\def\kaon    {{\ensuremath{\PK}}\xspace}
  \def\Kbar    {{\kern 0.2em\overline{\kern -0.2em \PK}{}}\xspace}

\def\KorKbar    {\kern 0.18em\optbar{\kern -0.18em K}{}\xspace}
\def\Kz      {{\ensuremath{\kaon^0}}\xspace}

\def\Kp      {{\ensuremath{\kaon^+}}\xspace}
\def\Km      {{\ensuremath{\kaon^-}}\xspace}

\def\Kmp     {{\ensuremath{\kaon^\mp}}\xspace}
\def\KS      {{\ensuremath{\kaon^0_{\mathrm{ \scriptscriptstyle S}}}}\xspace}

\def\Kstarz  {{\ensuremath{\kaon^{*0}}}\xspace}

\def\Kstar   {{\ensuremath{\kaon^*}}\xspace}

\def\Kstarp  {{\ensuremath{\kaon^{*+}}}\xspace}

  \def\Dbar    {{\kern 0.2em\overline{\kern -0.2em \PD}{}}\xspace}
\def\D       {{\ensuremath{\PD}}\xspace}

\def\DorDbar    {\kern 0.18em\optbar{\kern -0.18em D}{}\xspace}
\def\Dz      {{\ensuremath{\D^0}}\xspace}
\def\Dzb     {{\ensuremath{\Dbar{}^0}}\xspace}

\def\Ds      {{\ensuremath{\D^+_\squark}}\xspace}
\def\Dsp     {{\ensuremath{\D^+_\squark}}\xspace}
\def\Dsm     {{\ensuremath{\D^-_\squark}}\xspace}

\def\B       {{\ensuremath{\PB}}\xspace}
\def\Bbar    {{\ensuremath{\kern 0.18em\overline{\kern -0.18em \PB}{}}}\xspace}

\def\BorBbar    {\kern 0.18em\optbar{\kern -0.18em B}{}\xspace}
\def\Bz      {{\ensuremath{\B^0}}\xspace}

\def\Bu      {{\ensuremath{\B^+}}\xspace}
\def\Bub     {{\ensuremath{\B^-}}\xspace}
\def\Bp      {{\ensuremath{\Bu}}\xspace}
\def\Bm      {{\ensuremath{\Bub}}\xspace}

\def\Bd      {{\ensuremath{\B^0}}\xspace}
\def\Bs      {{\ensuremath{\B^0_\squark}}\xspace}
\def\Bsb     {{\ensuremath{\Bbar{}^0_\squark}}\xspace}

\def\jpsi     {{\ensuremath{{\PJ\mskip -3mu/\mskip -2mu\Ppsi\mskip 2mu}}}\xspace}

  \def\Y#1S{\ensuremath{\PUpsilon{(#1S)}}\xspace}

\def\Lbar        {{\ensuremath{\kern 0.1em\overline{\kern -0.1em\PLambda}}}\xspace}
\def\LorLbar    {\kern 0.18em\optbar{\kern -0.18em \PLambda}{}\xspace}

\newcommand{\decay}[2]{\ensuremath{#1\!\to #2}\xspace}         

\def\to                 {\ensuremath{\rightarrow}\xspace}

\def\CP                {{\ensuremath{C\!P}}\xspace}

\def\Vud  {{\ensuremath{V_{\uquark\dquark}}}\xspace}
\def\Vcd  {{\ensuremath{V_{\cquark\dquark}}}\xspace}

\def\Vus  {{\ensuremath{V_{\uquark\squark}}}\xspace}
\def\Vcs  {{\ensuremath{V_{\cquark\squark}}}\xspace}

\newcommand{\dms}{{\ensuremath{\Delta m_{\squark}}}\xspace}

\newcommand{\phis}{{\ensuremath{\phi_{\squark}}}\xspace}

\def\AT#1     {\ensuremath{A_{\mathrm{T}}^{#1}}\xspace}           

\def\C#1      {\ensuremath{\mathcal{C}_{#1}}\xspace}                       
\def\Cp#1     {\ensuremath{\mathcal{C}_{#1}^{'}}\xspace}                    
\def\Ceff#1   {\ensuremath{\mathcal{C}_{#1}^{\mathrm{(eff)}}}\xspace}        
\def\Cpeff#1  {\ensuremath{\mathcal{C}_{#1}^{'\mathrm{(eff)}}}\xspace}       
\def\Ope#1    {\ensuremath{\mathcal{O}_{#1}}\xspace}                       
\def\Opep#1   {\ensuremath{\mathcal{O}_{#1}^{'}}\xspace}                    

\newcommand{\tev}{\ifthenelse{\boolean{inbibliography}}{\ensuremath{~T\kern -0.05em eV}\xspace}{\ensuremath{\mathrm{\,Te\kern -0.1em V}}}\xspace}
\newcommand{\gev}{\ensuremath{\mathrm{\,Ge\kern -0.1em V}}\xspace}
\newcommand{\mev}{\ensuremath{\mathrm{\,Me\kern -0.1em V}}\xspace}
\newcommand{\kev}{\ensuremath{\mathrm{\,ke\kern -0.1em V}}\xspace}
\newcommand{\ev}{\ensuremath{\mathrm{\,e\kern -0.1em V}}\xspace}
\newcommand{\gevc}{\ensuremath{{\mathrm{\,Ge\kern -0.1em V\!/}c}}\xspace}
\newcommand{\mevc}{\ensuremath{{\mathrm{\,Me\kern -0.1em V\!/}c}}\xspace}
\newcommand{\gevcc}{\ensuremath{{\mathrm{\,Ge\kern -0.1em V\!/}c^2}}\xspace}
\newcommand{\gevgevcccc}{\ensuremath{{\mathrm{\,Ge\kern -0.1em V^2\!/}c^4}}\xspace}
\newcommand{\mevcc}{\ensuremath{{\mathrm{\,Me\kern -0.1em V\!/}c^2}}\xspace}

\def\fb   {\ensuremath{\mbox{\,fb}}\xspace}
\def\invfb   {\ensuremath{\mbox{\,fb}^{-1}}\xspace}

\newcommand{\chisq}{\ensuremath{\chi^2}\xspace}

\def\gsim{{~\raise.15em\hbox{$>$}\kern-.85em
          \lower.35em\hbox{$\sim$}~}\xspace}
\def\lsim{{~\raise.15em\hbox{$<$}\kern-.85em
          \lower.35em\hbox{$\sim$}~}\xspace}

\def\sqs   {\ensuremath{\protect\sqrt{s}}\xspace}

\def\degrees{\ensuremath{^{\circ}}\xspace}

\def\rad{\ensuremath{\mathrm{ \,rad}}\xspace}

\def\tell1  {TELL1\xspace}
\def\ukl1   {UKL1\xspace}

\newcommand{\eg}{\mbox{\itshape e.g.}\xspace}

\newcommand{\etc}{\mbox{\itshape etc.}\xspace}

\newcommand{\vs}{\mbox{\itshape vs.}\xspace}

\usepackage{pdflscape}

\usepackage{cite} 
\usepackage{mciteplus}

\usepackage{multirow}

\usepackage{rotating}

\allowdisplaybreaks

\newcommand{\fbar}	{\ensuremath{\bar{f}}\xspace}

\newcommand{\deltams}   {\ensuremath{\Delta m_s}\xspace}
\renewcommand{\dms}     {\deltams}
\newcommand{\gs}        {\ensuremath{\Gamma_s}\xspace}
\newcommand{\dgs}       {\ensuremath{\Delta\Gamma_s}\xspace}
\newcommand{\f}         {\ensuremath{f}\xspace} 
\renewcommand{\fb}      {\fbar} 
\newcommand{\Af}        {\ensuremath{A_{\f}}\xspace} 

\newcommand{\Abf}       {\ensuremath{\bar{A}_{f}}\xspace}

\newcommand{\lf}        {\ensuremath{\lambda_{\f}}\xspace}

\newcommand{\omcl}{\ensuremath{1-{\rm CL}}\xspace}

\newcommand{\plugin}{{\sc Plugin}\xspace}

\newcommand{\Dh}{\ensuremath{Dh}\xspace}
\newcommand{\DK}{\ensuremath{DK}\xspace}
\newcommand{\robust}{\DK}

\renewcommand{\Kz}     {\texorpdfstring{\ensuremath{K^0}}{K0}\xspace}

\newcommand{\hp} {\texorpdfstring{\ensuremath{h^{+}}}{h+}\xspace}
\newcommand{\hm} {\texorpdfstring{\ensuremath{h^{-}}}{h-}\xspace}

\newcommand{\BuDpi}   {\texorpdfstring{\ensuremath{\Bu\to D \pip}}		{B+ -> Dpi+}\xspace}
\newcommand{\BuDK}    {\texorpdfstring{\ensuremath{\Bu\to D \Kp}}		{B+ -> DK+}\xspace}
\newcommand{\BuDh}    {\texorpdfstring{\ensuremath{\Bu\to D h^+}}		{B+ -> Dh+}\xspace}

\newcommand{\BuDKpipi}{\texorpdfstring{\ensuremath{\Bu\to D \Kp\pip\pim}}		{B+ -> DK+pi+pi-}\xspace}
\newcommand{\BuDhpipi}{\texorpdfstring{\ensuremath{\Bu\to D h^+\pim\pip}}		{B+ -> Dh+pi+pi-}\xspace}
\newcommand{\BuDpipipi}{\texorpdfstring{\ensuremath{\Bu\to D \pip\pim\pip}}		{B+ -> Dpi+pi+pi-}\xspace}

\newcommand{\BdDzKpi} {\texorpdfstring{\decay{\Bz}{\Dz K\pi}}		{}}
\newcommand{\BdDKstz} {\texorpdfstring{\decay{\Bz}{D K^{*0}}}		{}}

\newcommand{\BdDKpi}  {\texorpdfstring{\decay{\Bd}{\D\Kp\pim}} {B0 -> DK+pi-}}
\newcommand{\BdDzKstz}{\texorpdfstring{\ensuremath{\Bd\to D K^{*0}}}		{B0 -> DK*0}\xspace}

\newcommand{\BsDsK}	{\texorpdfstring{\ensuremath{\Bs\to D_s^\mp K^\pm}}	{Bs -> DsK}\xspace}

\newcommand{\Dshhh}    {\decay{\Ds}{\hp\hm\pip}}

\newcommand{\DKpipipi} {\texorpdfstring{\ensuremath{D\to K^{\pm}\pi^{\mp}\pip\pim}}{D -> K3pi}\xspace}

\newcommand{\DKpipiz}  {\texorpdfstring{\ensuremath{D\to K^{\pm}\pi^{\mp}\piz}}{D -> K2pi}\xspace}

\newcommand{\DKpi}     {\texorpdfstring{\ensuremath{D\to K^+\pi^-}}{D -> Kpi}\xspace}
\newcommand{\DKSKpi}   {\texorpdfstring{\ensuremath{D\to \KS K^-\pi^+}}{D -> KSKpi}\xspace}

\newcommand{\DzKpi}    {\texorpdfstring{\ensuremath{\Dz\to K^\pm\pi^\mp}}{D0 -> Kpi}\xspace}

\newcommand{\DzKpiFav} {\texorpdfstring{\ensuremath{\Dz\to\Km\pip}}{D0 -> K-pi+}\xspace}
\newcommand{\DzKpiSup} {\texorpdfstring{\ensuremath{\Dz\to\pim\Kp}}{D0 -> pi-K+}\xspace}

\newcommand{\DzKpipipi}{\texorpdfstring{\ensuremath{\Dz\to K^\pm\pi^\mp\pi^+\pi^-}}{D -> K3pi}\xspace}
\newcommand{\Dzhpipipi}{\texorpdfstring{\ensuremath{\Dz\to h^\pm\pi^\mp\pi^+\pi^-}}{D -> h3pi}\xspace}
\newcommand{\Dzhhpiz}  {\texorpdfstring{\ensuremath{\Dz \to h^\pm h^\mp \piz}}{D -> hhpi0}\xspace}
\newcommand{\DzKpipiz} {\texorpdfstring{\ensuremath{\Dz \to K^\pm\pi^\mp \piz}}{D -> Kpipi0}\xspace}

\newcommand{\Dhh}      {\texorpdfstring{\ensuremath{D\to h^+h^-}}{D -> hh}\xspace}

\newcommand{\Dhpipipi} {\texorpdfstring{\ensuremath{D\to h^+\pi^-\pi^+\pi^-}}{D -> hpipipi}\xspace}
\newcommand{\Dpipipipi} {\texorpdfstring{\ensuremath{D\to \pi^+\pi^-\pi^+\pi^-}}{D -> hpipipi}\xspace}

\newcommand{\Dhhpiz}   {\texorpdfstring{\ensuremath{D\to h^+h^-\piz}}{D -> hhpi0}\xspace}
\newcommand{\Dzhh}     {\texorpdfstring{\ensuremath{\Dz\to h^+h^-}}{D -> hh}\xspace}
\newcommand{\DzKK}     {\texorpdfstring{\ensuremath{\Dz\to K^+K^-}}{D -> KK}\xspace}
\newcommand{\Dzpipi}   {\texorpdfstring{\ensuremath{\Dz\to\pi^+\pi^-}}{D -> pipi}\xspace}
\newcommand{\DzKShh}   {\texorpdfstring{\ensuremath{\Dz\to\KS h^+h^-}}{D0 -> KShh}\xspace}
\newcommand{\DKSpipi}  {\texorpdfstring{\ensuremath{D\to\KS\pi^+\pi^-}}{D -> KSpipi}\xspace}
\newcommand{\DKSKK}    {\texorpdfstring{\ensuremath{D\to\KS K^+K^-}}{D -> KSKK}\xspace}
\newcommand{\DKShh}    {\texorpdfstring{\ensuremath{D\to\KS h^+h^-}}{D -> KShh}\xspace}

\newcommand{\Dpipipiz} {\texorpdfstring{\ensuremath{\D\to\pi^+\pi^-\piz}}{D -> pipipi0}\xspace}
\newcommand{\DKKpiz}   {\texorpdfstring{\ensuremath{\D\to K^+ K^-\piz}}{D -> KKpi0}\xspace}

\renewcommand{\g}{\texorpdfstring{\ensuremath{\gamma}}{gamma}\xspace}

\newcommand{\rbh}  {\texorpdfstring{\ensuremath{r_B^h}}{rBh}\xspace}

\newcommand{\rD}{\texorpdfstring{\ensuremath{r_{D}}}{rD}\xspace}
\newcommand{\dD}{\texorpdfstring{\ensuremath{\delta_{D}}}{dD}\xspace}

\newcommand{\Fp}{\texorpdfstring{\ensuremath{F_{+}}}{F+}\xspace}

\newcommand{\rb}  {\texorpdfstring{\ensuremath{r_B^{DK}}}{rBDK}\xspace}
\newcommand{\rbsq}{\texorpdfstring{\ensuremath{(r_B^{DK})^2}}{rBDK2}\xspace}
\newcommand{\db}  {\ensuremath{\delta_B^{DK}}\xspace}

\newcommand{\rbpi}  {\texorpdfstring{\ensuremath{r_B^{D\pi}}}{rBDpi}\xspace}
\newcommand{\rbpisq}{\texorpdfstring{\ensuremath{(r_B^{D\pi})^2}}{rBDpi2}\xspace}
\newcommand{\dbpi}  {\ensuremath{\delta_B^{D\pi}}\xspace}

\newcommand{\rbDkpp}{\ensuremath{r_B^{DK\pi\pi}}\xspace}
\newcommand{\rbDkppsq}{\ensuremath{(\rbDkpp)^2}\xspace}
\newcommand{\dbDkpp}{\ensuremath{\delta_B^{DK\pi\pi}}\xspace}
\newcommand{\kbDkpp}{\ensuremath{\kappa_B^{DK\pi\pi}}\xspace}
\newcommand{\rbDppp}{\ensuremath{r_B^{D\pi\pi\pi}}\xspace}
\newcommand{\rbDpppsq}{\ensuremath{(\rbDppp)^2}\xspace}
\newcommand{\dbDppp}{\ensuremath{\delta_B^{D\pi\pi\pi}}\xspace}
\newcommand{\kbDppp}{\ensuremath{\kappa_B^{D\pi\pi\pi}}\xspace}

\newcommand{\rbDKstz}{\ensuremath{r_B^{DK^{*0}}}\xspace}
\newcommand{\rbbarDKstz}{\ensuremath{\bar{r}_B^{DK^{*0}}}\xspace}

\newcommand{\dbDKstz}{\ensuremath{\delta_B^{DK^{*0}}}\xspace}

\newcommand{\dbbarDKstz}{\ensuremath{\bar{\delta}_B^{DK^{*0}}}\xspace}
\newcommand{\kbDKstz}{\ensuremath{\kappa_B^{DK^{*0}}}\xspace}
\newcommand{\RbDKstz}{\ensuremath{\bar{R}_{B}^{DK^{*0}}}\xspace}
\newcommand{\DbDKstz}{\ensuremath{\Delta\bar{\delta}_{B}^{DK^{*0}}}\xspace}
\newcommand{\RBDKstz}{\ensuremath{\bar{R}_{B}^{DK^{*0}}}\xspace}
\newcommand{\DBDKstz}{\ensuremath{\Delta\bar{\delta}_{B}^{DK^{*0}}}\xspace}

\newcommand{\rdsk}{\texorpdfstring{\ensuremath{r_B^{D_s K}}}{rDsK}\xspace}
\newcommand{\rdsksq}{\texorpdfstring{\ensuremath{(r_B^{D_s K})^2}}{rDsK^2}\xspace}
\newcommand{\ddsk}{\texorpdfstring{\ensuremath{\delta_B^{D_s K}}}{dDsK}\xspace}
\renewcommand{\phis}{\texorpdfstring{\ensuremath{\phi_s}}{phis}\xspace}

\newcommand{\rdKpi}  {\ensuremath{r_D^{K\pi}}\xspace}
\newcommand{\rdKpisq}{\ensuremath{(r_D^{K\pi})^2}\xspace}
\newcommand{\ddKpi}  {\ensuremath{\delta_D^{K\pi}}\xspace}
\newcommand{\RdKpi}  {\ensuremath{R_D^{K\pi}}\xspace}

\newcommand{\rdKpp}{\ensuremath{r_{D}^{K2\pi}}\xspace}

\newcommand{\ddKpp}{\ensuremath{\delta_{D}^{K2\pi}}\xspace}
\newcommand{\kdKpp}{\ensuremath{\kappa_{D}^{K2\pi}}\xspace}
\newcommand{\Fppp}{\ensuremath{F_{\pi\pi\piz}}\xspace}
\newcommand{\FKKp}{\ensuremath{F_{KK\piz}}\xspace}

\newcommand{\rdKskpi}{\ensuremath{r_D^{K_SK\pi}}\xspace}
\newcommand{\rdKskpisq}{\ensuremath{(r_D^{K_SK\pi})^2}\xspace}
\newcommand{\ddKskpi}{\ensuremath{\delta_D^{K_SK\pi}}\xspace}
\newcommand{\kdKskpi}{\ensuremath{\kappa_D^{K_SK\pi}}\xspace}
\newcommand{\RdKskpi}{\ensuremath{R_D^{K_SK\pi}}\xspace}

\newcommand{\rdKppp}{\texorpdfstring{\ensuremath{r_D^{K3\pi}}}{rD(K3pi)}\xspace}
\newcommand{\rdKpppsq}{\ensuremath{(r_D^{K3\pi})^2}\xspace}
\newcommand{\ddKppp}{\ensuremath{\delta_D^{K3\pi}}\xspace}
\newcommand{\kdKppp}{\ensuremath{\kappa_D^{K3\pi}}\xspace}
\newcommand{\Fpppp}{\ensuremath{F_{\pi\pi\pi\pi}}\xspace}

\newcommand{\DAcpKK}  {\texorpdfstring{\ensuremath{A_{KK}^{\rm dir}}\xspace}{AcpDir(KK)}}
\newcommand{\DAcpPipi}{\texorpdfstring{\ensuremath{A_{\pi\pi}^{\rm dir}}\xspace}{AcpDir(pipi}}

\newcommand{\xd}{\texorpdfstring{\ensuremath{x_D}\xspace}{xD}}
\newcommand{\yd}{\texorpdfstring{\ensuremath{y_D}\xspace}{yD}}

\newcommand{\xm} {\ensuremath{x_-}\xspace}
\newcommand{\ym} {\ensuremath{y_-}\xspace}
\newcommand{\xpm}{\ensuremath{x_\pm}\xspace}
\newcommand{\xp} {\ensuremath{x_+}\xspace}
\newcommand{\yp} {\ensuremath{y_+}\xspace}
\newcommand{\ypm}{\ensuremath{y_\pm}\xspace}
\newcommand{\xmdk} {\ensuremath{x_-^{DK}}\xspace}
\newcommand{\ymdk} {\ensuremath{y_-^{DK}}\xspace}

\newcommand{\xpdk} {\ensuremath{x_+^{DK}}\xspace}
\newcommand{\ypdk} {\ensuremath{y_+^{DK}}\xspace}

\newcommand{\xmdkst} {\ensuremath{\bar{x}_-^{D\Kstarz}}\xspace}
\newcommand{\ymdkst} {\ensuremath{\bar{y}_-^{D\Kstarz}}\xspace}

\newcommand{\xpdkst} {\ensuremath{\bar{x}_+^{D\Kstarz}}\xspace}
\newcommand{\ypdkst} {\ensuremath{\bar{y}_+^{D\Kstarz}}\xspace}

\newcommand{\xmdkpi} {\ensuremath{x_-^{D\Kstarz}}\xspace}
\newcommand{\ymdkpi} {\ensuremath{y_-^{D\Kstarz}}\xspace}
\newcommand{\xpmdkpi}{\ensuremath{x_\pm^{D\Kstarz}}\xspace}
\newcommand{\xpdkpi} {\ensuremath{x_+^{D\Kstarz}}\xspace}
\newcommand{\ypdkpi} {\ensuremath{y_+^{D\Kstarz}}\xspace}
\newcommand{\ypmdkpi}{\ensuremath{y_\pm^{D\Kstarz}}\xspace}

\newcommand{\AcpDkKK}{\ensuremath{A_{\CP}^{DK, KK}}\xspace}
\newcommand{\AcpDkPipi}{\ensuremath{A_{\CP}^{DK, \pi\pi}}\xspace}
\newcommand{\AcpDpiKK}{\ensuremath{A_{\CP}^{D\pi, KK}}\xspace}
\newcommand{\AcpDpiPipi}{\ensuremath{A_{\CP}^{D\pi, \pi\pi}}\xspace}
\newcommand{\AfavDkKpi}{\ensuremath{A_{\rm fav}^{DK, K\pi}}\xspace}

\newcommand{\RkpKpi}{\ensuremath{R_{K/\pi}^{K\pi}}\xspace}
\newcommand{\RkpKK}{\ensuremath{R_{K/\pi}^{KK}}\xspace}
\newcommand{\RkpPipi}{\ensuremath{R_{K/\pi}^{\pi\pi}}\xspace}
\newcommand{\RCPKK}{\ensuremath{R_{CP}^{KK}}\xspace}
\newcommand{\RCPPiPi}{\ensuremath{R_{CP}^{\pi\pi}}\xspace}
\newcommand{\RadsDK}{\ensuremath{R_{\rm{ADS}}^{DK, \pi K}}\xspace}
\newcommand{\RadsDPi}{\ensuremath{R_{\rm{ADS}}^{D\pi, \pi K}}\xspace}
\newcommand{\AadsDK}{\ensuremath{A_{\rm{ADS}}^{DK, \pi K}}\xspace}
\newcommand{\AadsDPi}{\ensuremath{A_{\rm{ADS}}^{D\pi, \pi K}}\xspace}

\newcommand{\AfavDkKppp}{\ensuremath{A_{\rm fav}^{DK, \pi K \pi\pi}}\xspace}
\newcommand{\AcpDKpppp}{\ensuremath{A_{\CP}^{DK, \pi\pi\pi\pi}}\xspace}
\newcommand{\AcpDpipppp}{\ensuremath{A_{\CP}^{D\pi, \pi\pi\pi\pi}}\xspace}

\newcommand{\RkpKppp}{\ensuremath{R_{K/\pi}^{K\pi\pi\pi}}\xspace}
\newcommand{\Rkppppp}{\ensuremath{R_{K/\pi}^{4\pi}}\xspace}

\newcommand{\RCPpppp}{\ensuremath{R_{CP}^{\pi\pi\pi\pi}}\xspace}
\newcommand{\RadsDKKppp}{\ensuremath{R_{\rm{ADS}}^{DK, \pi K\pi\pi}}\xspace}
\newcommand{\RadsDPiKppp}{\ensuremath{R_{\rm{ADS}}^{D\pi, \pi K\pi\pi}}\xspace}
\newcommand{\AadsDKKppp}{\ensuremath{A_{\rm{ADS}}^{DK, \pi K\pi\pi}}\xspace}
\newcommand{\AadsDPiKppp}{\ensuremath{A_{\rm{ADS}}^{D\pi, \pi K\pi\pi}}\xspace}

\newcommand{\AcpDkKKPiz}{\ensuremath{A_{\CP}^{DK, KK\piz}}\xspace}
\newcommand{\AcpDkPiPiPiz}{\ensuremath{A_{\CP}^{DK, \pi\pi\piz}}\xspace}
\newcommand{\AcpDpiKKPiz}{\ensuremath{A_{\CP}^{D\pi, KK\piz}}\xspace}
\newcommand{\AcpDpiPiPiPiz}{\ensuremath{A_{\CP}^{D\pi, \pi\pi\piz}}\xspace}
\newcommand{\AfavDkKPiPiz}{\ensuremath{A_{\rm fav}^{DK, K\pi\piz}}\xspace}

\newcommand{\RCPKKPiz}{\ensuremath{R_{CP}^{KK\piz}}\xspace}
\newcommand{\RCPPiPiPiz}{\ensuremath{R_{CP}^{\pi\pi\piz}}\xspace}
\newcommand{\RadsDKkpp}{\ensuremath{R_{\rm{ADS}}^{DK, \pi K\piz}}\xspace}
\newcommand{\RadsDPikpp}{\ensuremath{R_{\rm{ADS}}^{D\pi, \pi K\piz}}\xspace}
\newcommand{\AadsDKkpp}{\ensuremath{A_{\rm{ADS}}^{DK, \pi K\piz}}\xspace}
\newcommand{\AadsDPikpp}{\ensuremath{A_{\rm{ADS}}^{D\pi, \pi K\piz}}\xspace}
\newcommand{\RkpKpp}{\ensuremath{R_{K/\pi}^{K\pi\piz}}\xspace}
\newcommand{\RkpKKp}{\ensuremath{R_{K/\pi}^{KK\piz}}\xspace}
\newcommand{\Rkpppp}{\ensuremath{R_{K/\pi}^{\pi\pi\piz}}\xspace}

\newcommand{\AfavDkKskpi}{\ensuremath{A_{\rm fav}^{DK, K_SK\pi}}\xspace}
\newcommand{\AsupDkKskpi}{\ensuremath{A_{\rm ADS}^{DK, K_SK\pi}}\xspace}

\newcommand{\RfavsupDkKskpi}{\ensuremath{R_{\rm ADS}^{DK,K_SK\pi}}\xspace}

\newcommand{\AfavDkstKpi}{\ensuremath{\bar{A}_{\rm fav}^{D\Kstarz,\,K\pi}}\xspace}

\newcommand{\RpDkstKpi}{\ensuremath{\bar{R}_+^{D\Kstarz,\,K\pi}}\xspace}
\newcommand{\RmDkstKpi}{\ensuremath{\bar{R}_-^{D\Kstarz,\,K\pi}}\xspace}

\newcommand{\RcpDkpipi}{\ensuremath{R_{\CP}^{DK\pi\pi}}\xspace}
\newcommand{\AfavDkpipiKpi}{\ensuremath{A_{\rm fav}^{DK\pi\pi,\,K\pi}}\xspace}
\newcommand{\AfavDpipipiKpi}{\ensuremath{A_{\rm fav}^{D\pi\pi\pi,\,K\pi}}\xspace}
\newcommand{\AcpDkpipiKK}{\ensuremath{A_{\CP}^{DK\pi\pi,\,KK}}\xspace}
\newcommand{\AcpDkpipiPipi}{\ensuremath{A_{\CP}^{DK\pi\pi,\,\pi\pi}}\xspace}
\newcommand{\AcpDpipipiKK}{\ensuremath{A_{\CP}^{D\pi\pi\pi,\,KK}}\xspace}
\newcommand{\AcpDpipipiPipi}{\ensuremath{A_{\CP}^{D\pi\pi\pi,\,\pi\pi}}\xspace}
\newcommand{\RpDkpipiKpi}{\ensuremath{R_+^{DK\pi\pi,\,K\pi}}\xspace}
\newcommand{\RmDkpipiKpi}{\ensuremath{R_-^{DK\pi\pi,\,K\pi}}\xspace}
\newcommand{\RpDpipipiKpi}{\ensuremath{R_+^{D\pi\pi\pi,\,K\pi}}\xspace}
\newcommand{\RmDpipipiKpi}{\ensuremath{R_-^{D\pi\pi\pi,\,K\pi}}\xspace}

\newcommand{\Cpar}	{\ensuremath{C}\xspace}
\newcommand{\Dpar}	{\ensuremath{D_f}\xspace}
\newcommand{\Dbpar}	{\ensuremath{D_{\bar{f}}}\xspace}
\newcommand{\Spar}	{\ensuremath{S_f}\xspace}
\newcommand{\Sbpar}	{\ensuremath{S_{\bar{f}}}\xspace}
\newcommand{\Cbpar}     {\ensuremath{C_{\fb}}\xspace}
\renewcommand{\Cpar}      {\ensuremath{C_{\f}}\xspace}
\renewcommand{\Sbpar}     {\ensuremath{S_{\fb}}\xspace}
\renewcommand{\Spar}      {\ensuremath{S_{\f}}\xspace}
\renewcommand{\Dbpar}     {\ensuremath{{A_{\fb}^{\Delta\Gamma}}}\xspace}
\renewcommand{\Dpar}      {\ensuremath{{A_{\f}^{\Delta\Gamma}}}\xspace}

\newcommand{\gQuoted}                                   {\ensuremath{72.2\,^{+6.8}_{-7.3}}\xspace}

\newcommand{\gRobustCentral}         {\ensuremath{72.2}\xspace}

\newcommand{\gRobustOnesig}          {\ensuremath{[64.9,79.0]}\xspace}
\newcommand{\gRobustTwosig}          {\ensuremath{[55.9,85.2]}\xspace}
\newcommand{\gRobustThreesig}        {\ensuremath{[43.7,90.9]}\xspace}
\newcommand{\rbRobustCentral}        {\ensuremath{0.1019}\xspace}
\newcommand{\rbRobustOnesig}         {\ensuremath{[0.0963,0.1075]}\xspace}
\newcommand{\rbRobustTwosig}         {\ensuremath{[0.0907,0.1128]}\xspace}
\newcommand{\rbRobustThreesig}       {\ensuremath{[0.0849,0.1182]}\xspace}
\newcommand{\dbRobustCentral}        {\ensuremath{142.6}\xspace}
\newcommand{\dbRobustOnesig}         {\ensuremath{[136.0,148.3]}\xspace}
\newcommand{\dbRobustTwosig}         {\ensuremath{[127.8,153.6]}\xspace}
\newcommand{\dbRobustThreesig}       {\ensuremath{[116.2,158.7]}\xspace}
\newcommand{\rbDKstzRobustCentral}   {\ensuremath{0.218}\xspace}
\newcommand{\rbDKstzRobustOnesig}    {\ensuremath{[0.171,0.263]}\xspace}
\newcommand{\rbDKstzRobustTwosig}    {\ensuremath{[0.118,0.305]}\xspace}
\newcommand{\rbDKstzRobustThreesig}  {\ensuremath{[0.000,0.348]}\xspace}
\newcommand{\dbDKstzRobustCentral}   {\ensuremath{189}\xspace}
\newcommand{\dbDKstzRobustOnesig}    {\ensuremath{[169,212]}\xspace}
\newcommand{\dbDKstzRobustTwosig}    {\ensuremath{[148,241]}\xspace}
\newcommand{\dbDKstzRobustThreesig}  {\ensuremath{[123,283]}\xspace}

\newcommand{\gRobustCentralPMProb}  {\ensuremath{\gRobustCentralPMProb\,^{+6.8}_{-8.1}}\xspace}

\newcommand{\gFullCentral} { \ensuremath{73.5}\xspace }
\newcommand{\rbFullCentral} { \ensuremath{0.1017}\xspace }
\newcommand{\dbFullCentral} { \ensuremath{141.6}\xspace }
\newcommand{\rbDKstzFullCentral} { \ensuremath{0.220}\xspace }
\newcommand{\dbDKstzFullCentral} { \ensuremath{188}\xspace }
\newcommand{\rbpiFullCentral} { \ensuremath{0.027}\xspace }
\newcommand{\dbpiFullCentral} { \ensuremath{348.3}\xspace }

\newcommand{\gFullOnesig}    { \ensuremath{[70.5,76.8]}\xspace }
\newcommand{\gFullTwosig}    { \ensuremath{[56.7,83.4]}\xspace }
\newcommand{\gFullThreesig}  { \ensuremath{[40.1,90.8]}\xspace }

\newcommand{\rbFullOnesig}   { \ensuremath{[0.0970,0.1064]}\xspace }
\newcommand{\rbFullTwosig}   { \ensuremath{[0.0914,0.1110]}\xspace }
\newcommand{\rbFullThreesig} { \ensuremath{[0.0844,0.1163]}\xspace }

\newcommand{\dbFullOnesig}   { \ensuremath{[136.6,146.3]}\xspace }
\newcommand{\dbFullTwosig}   { \ensuremath{[127.2,151.1]}\xspace }
\newcommand{\dbFullThreesig} { \ensuremath{[114.6,155.7]}\xspace }

\newcommand{\rbDKstzFullOnesig} { \ensuremath{[0.173,0.264]}\xspace }
\newcommand{\rbDKstzFullTwosig} { \ensuremath{[0.121,0.307]}\xspace }
\newcommand{\rbDKstzFullThreesig} { \ensuremath{[0.000,0.355]}\xspace }

\newcommand{\dbDKstzFullOnesig} { \ensuremath{[168,211]}\xspace }
\newcommand{\dbDKstzFullTwosig} { \ensuremath{[148,239]}\xspace }
\newcommand{\dbDKstzFullThreesig} { \ensuremath{[120,280]}\xspace }

\newcommand{\rbpiFullOnesig} { \ensuremath{[0.0207,0.0318]}\xspace }
\newcommand{\rbpiFullTwosig} { \ensuremath{[0.0020,0.0365]}\xspace }
\newcommand{\rbpiFullThreesig} { \ensuremath{[0.0008,0.0425]}\xspace }

\newcommand{\dbpiFullOnesig} { \ensuremath{[343.2,352.9]}\xspace }
\newcommand{\dbpiFullTwosig} { \ensuremath{[220.5,356.4]}\xspace }
\newcommand{\dbpiFullThreesig} { \ensuremath{[192.9,359.8]}\xspace }

\newcommand{\gRobustCentralBayes}         {\ensuremath{70.3}\xspace}

\newcommand{\gRobustOnesigBayes}          {\ensuremath{[62.4,77.4]}\xspace}
\newcommand{\gRobustTwosigBayes}          {\ensuremath{[52.6,83.5]}\xspace}
\newcommand{\gRobustThreesigBayes}        {\ensuremath{[42.1,88.4]}\xspace}
\newcommand{\rbRobustCentralBayes}        {\ensuremath{0.1012}\xspace}
\newcommand{\rbRobustOnesigBayes}         {\ensuremath{[0.0954,0.1064]}\xspace}
\newcommand{\rbRobustTwosigBayes}         {\ensuremath{[0.0900,0.1120]}\xspace}
\newcommand{\rbRobustThreesigBayes}       {\ensuremath{[0.0846,0.1171]}\xspace}
\newcommand{\dbRobustCentralBayes}        {\ensuremath{142.2}\xspace}
\newcommand{\dbRobustOnesigBayes}         {\ensuremath{[134.7,148.1]}\xspace}
\newcommand{\dbRobustTwosigBayes}         {\ensuremath{[125.3,153.7]}\xspace}
\newcommand{\dbRobustThreesigBayes}       {\ensuremath{[113.2,157.9]}\xspace}
\newcommand{\rbDKstzRobustCentralBayes}   {\ensuremath{0.204}\xspace}
\newcommand{\rbDKstzRobustOnesigBayes}    {\ensuremath{[0.149,0.253]}\xspace}
\newcommand{\rbDKstzRobustTwosigBayes}    {\ensuremath{[0.073,0.299]}\xspace}
\newcommand{\rbDKstzRobustThreesigBayes}  {\ensuremath{[0.000,0.322]}\xspace}
\newcommand{\dbDKstzRobustCentralBayes}   {\ensuremath{190.3}\xspace}
\newcommand{\dbDKstzRobustOnesigBayes}    {\ensuremath{[165.8,218.4]}\xspace}
\newcommand{\dbDKstzRobustTwosigBayes}    {\ensuremath{[139.5,263.4]}\xspace}
\newcommand{\dbDKstzRobustThreesigBayes}  {\ensuremath{[117.8,292.4]}\xspace}

\newcommand{\gFullCentralBayes}        {\ensuremath{72.4}\xspace}

\newcommand{\gFullOnesigBayes}         {\ensuremath{[63.9,79.0]}\xspace}
\newcommand{\gFullTwosigBayes}         {\ensuremath{[52.1,84.6]}\xspace}
\newcommand{\gFullThreesigBayes}       {\ensuremath{[40.1,89.5]}\xspace}
\newcommand{\rbFullCentralBayes}       {\ensuremath{0.1003}\xspace}
\newcommand{\rbFullOnesigBayes}        {\ensuremath{[0.0948,0.1057]}\xspace}
\newcommand{\rbFullTwosigBayes}        {\ensuremath{[0.0893,0.1109]}\xspace}
\newcommand{\rbFullThreesigBayes}      {\ensuremath{[0.0838,0.1159]}\xspace}
\newcommand{\dbFullCentralBayes}       {\ensuremath{141.0}\xspace}
\newcommand{\dbFullOnesigBayes}        {\ensuremath{[133.3,147.5]}\xspace}
\newcommand{\dbFullTwosigBayes}        {\ensuremath{[122.1,153.1]}\xspace}
\newcommand{\dbFullThreesigBayes}      {\ensuremath{[108.6,157.5]}\xspace}
\newcommand{\rbDKstzFullCentralBayes}   {\ensuremath{0.2072}\xspace}
\newcommand{\rbDKstzFullOnesigBayes}    {\ensuremath{[0.1514,0.2555]}\xspace}
\newcommand{\rbDKstzFullTwosigBayes}    {\ensuremath{[0.0788,0.3007]}\xspace}
\newcommand{\rbDKstzFullThreesigBayes}  {\ensuremath{[0.0031,0.3291]}\xspace}
\newcommand{\dbDKstzFullCentralBayes}   {\ensuremath{189.8}\xspace}
\newcommand{\dbDKstzFullOnesigBayes}    {\ensuremath{[166.3,216.5]}\xspace}
\newcommand{\dbDKstzFullTwosigBayes}    {\ensuremath{[143.9,255.2]}\xspace}
\newcommand{\dbDKstzFullThreesigBayes}  {\ensuremath{[120.2,286.0]}\xspace}
\newcommand{\rbpiFullCentralBayes}     {\ensuremath{0.0043}\xspace}
\newcommand{\rbpiFullOnesigBayes}      {\ensuremath{[0.0027,0.0063]}\xspace}
\newcommand{\rbpiFullTwosigBayes}      {\ensuremath{[0.0011,0.0281]}\xspace}
\newcommand{\rbpiFullThreesigBayes}    {\ensuremath{[0.0008,0.0329]}\xspace}
\newcommand{\dbpiFullCentralBayes}     {\ensuremath{303.7}\xspace}
\newcommand{\dbpiFullOnesigBayes}      {\ensuremath{[264.7,332.7]}\xspace}
\newcommand{\dbpiFullTwosigBayes}      {\ensuremath{[231.5,355.2]}\xspace}
\newcommand{\dbpiFullThreesigBayes}    {\ensuremath{[202.7,359.0]}\xspace}

\usepackage{longtable} 

\begin{document}

\renewcommand{\thefootnote}{\fnsymbol{footnote}}
\setcounter{footnote}{1}


\begin{titlepage}
\pagenumbering{roman}

\vspace*{-1.5cm}
\centerline{\large EUROPEAN ORGANIZATION FOR NUCLEAR RESEARCH (CERN)}
\vspace*{1.5cm}
\noindent
\begin{tabular*}{\linewidth}{lc@{\extracolsep{\fill}}r@{\extracolsep{0pt}}}
\ifthenelse{\boolean{pdflatex}}
{\vspace*{-2.7cm}\mbox{\!\!\!\includegraphics[width=.14\textwidth]{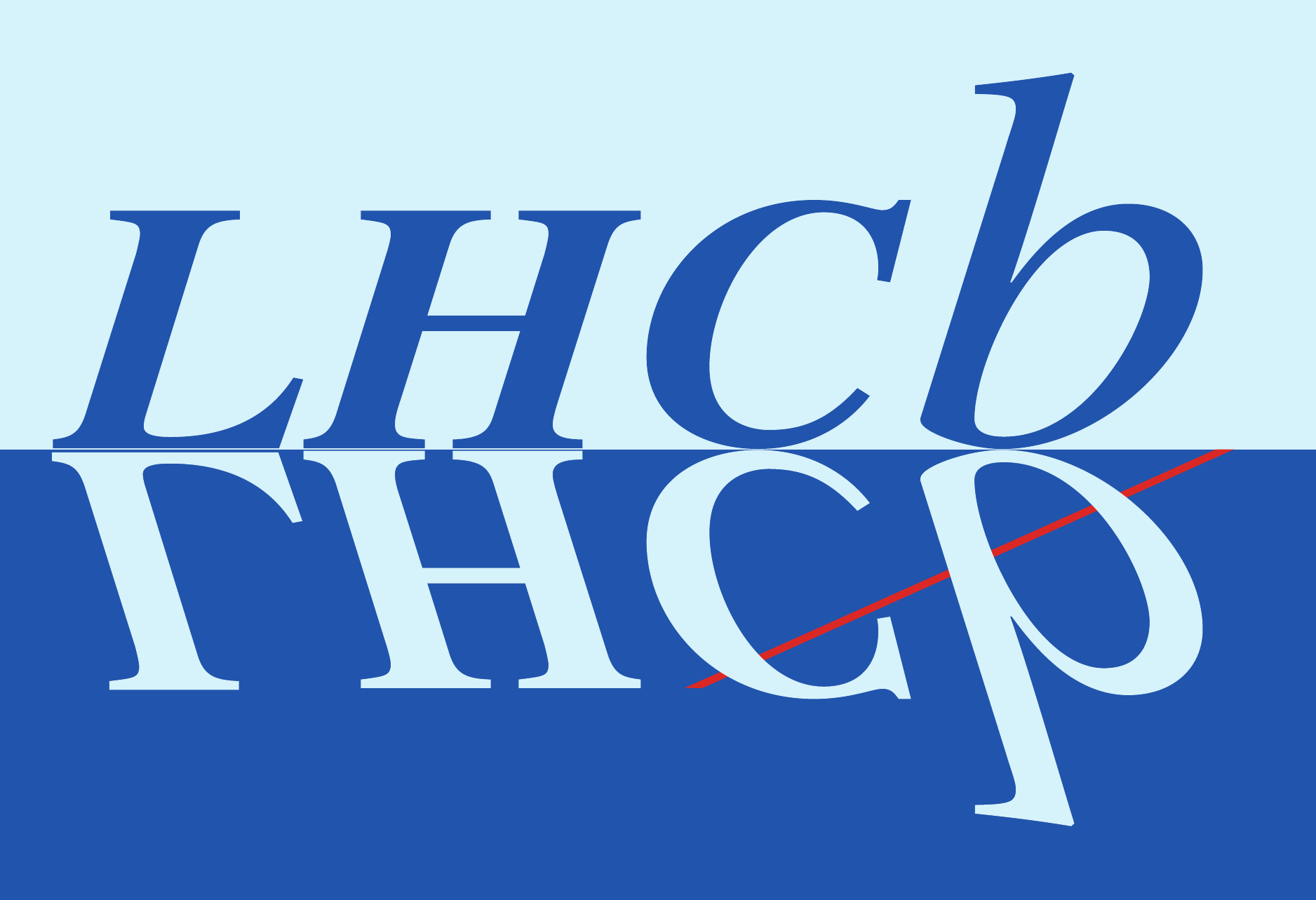}} & &}%
{\vspace*{-1.2cm}\mbox{\!\!\!\includegraphics[width=.12\textwidth]{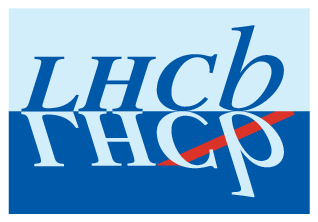}} & &}%
\\
& & CERN-EP-2016-270 \\  
& & LHCb-PAPER-2016-032 \\  
& & November 9, 2016 \\ 
\end{tabular*}

\vspace*{2.0cm}

{\normalfont\bfseries\boldmath\huge
\begin{center}
Measurement of the CKM angle $\gamma$ from a combination of LHCb results
\end{center}}

\vspace*{2.0cm}

\begin{center}
The LHCb collaboration\footnote{Authors are listed at the end of this paper.}
\end{center}

\vspace{\fill}

\begin{abstract}
\noindent
A combination of measurements sensitive to the CKM angle $\gamma$ from \lhcb is performed.
The inputs are from analyses of time-integrated \BuDK, \BdDKstz, \BdDKpi and \BuDKpipi tree-level decays.
In addition, results from a time-dependent analysis of \BsDsK decays are included.
The combination yields $\g = (\gQuoted)^\circ$, 
where the uncertainty includes systematic effects.
The 95.5\%~confidence level interval is determined to be $\g \in \gRobustTwosig^\circ$.
A second combination is investigated, also including measurements from \BuDpi and \BuDpipipi decays,
which yields compatible results.

\end{abstract}

\vspace*{1.0cm}

\begin{center}
  Published in JHEP 12 (2016) 087
\end{center}

\vspace{\fill}

{\footnotesize
\centerline{\copyright~CERN on behalf of the \lhcb collaboration, licence \href{http://creativecommons.org/licenses/by/4.0/}{CC-BY-4.0}.}}
\vspace*{2mm}

\end{titlepage}


\newpage
\setcounter{page}{2}
\mbox{~}
%
%
%
%

\cleardoublepage


\renewcommand{\thefootnote}{\arabic{footnote}}
\setcounter{footnote}{0}

\tableofcontents
\cleardoublepage


\pagestyle{plain} 
\setcounter{page}{1}
\pagenumbering{arabic}



\section{Introduction}
\label{sec:Introduction}

Understanding the origin of the baryon asymmetry of the Universe is one of the key issues of modern physics.
Sakharov showed that such an asymmetry can arise if three conditions are fulfilled~\cite{Sakharov:1991},
one of which is the requirement that both charge ($C$) and charge-parity (\CP) symmetries are broken.
The latter phenomenon arises in the Standard Model (SM)
of particle physics through the complex phase of
the Cabibbo-Kobayashi-Maskawa (CKM) quark mixing matrix~\cite{Cabibbo:1963yz,Kobayashi:1973fv},
although the effect in the SM is not large enough to account for the observed baryon asymmetry in the Universe~\cite{Gavela:1993ts}.
Violation of \CP symmetry can be studied by measuring the angles
of the CKM unitarity triangle~\cite{Wolfenstein:1983yz,Chau:1984fp,Buras:1994ec}.
The least precisely known of these angles, $\gamma\equiv\arg[-V^{}_{ud}V^*_{ub}/V^{}_{cd}V^*_{cb}]$,
can be measured using only tree-level processes~\cite{Gronau:1991dp,Gronau:1990ra,Atwood:1996ci,Atwood:2000ck}; a method that, assuming
new physics is not present in tree-level decays~\cite{Brod:2014bfa}, has negligible theoretical uncertainty~\cite{Brod:2013sga}.
Disagreement between such direct measurements of $\gamma$ and the value inferred from global CKM fits, assuming the validity of the SM,
would indicate new physics beyond the SM.

The value of \g can be determined by exploiting the interference between favoured $b\to cW$ ($V^{}_{cb}$) and suppressed $b \to uW$ ($V^{}_{ub}$) transition amplitudes using decay
channels such as \BuDh, \BdDKstz, \BdDKpi, \BuDhpipi and \BsDsK~\cite{Gronau:1991dp,Gronau:1990ra,Atwood:1996ci,Atwood:2000ck,Bondar:2001vn,Giri:2003ty,Grossman:2002aq,Fleischer:2003yb,Aleksan:1991nh,Dunietz:1987bv,Gershon:2008pe,Gershon:2009qc}, where $h$ is a kaon or pion and $D$
refers to a neutral charm meson that is a mixture of the \Dz and \Dzb flavour eigenstates.
The inclusion of charge conjugate processes is implied throughout, unless otherwise stated.
The most precise way to determine \g is through a combination of measurements from analyses of many decay modes.
Hadronic parameters such as those that describe the ratio ($r^X_B$) or strong phase difference ($\delta^X_B$) between the $V^{}_{cb}$ and $V^{}_{ub}$ transition amplitudes
and where $X$ is a specific final state of a $B$ meson decay, are also simultaneously determined. The ratio of the suppressed to favoured \B decay amplitudes is related
to \g and the hadronic parameters by $\mathcal{A}_{\rm sup}/\mathcal{A}_{\rm fav} = r^X_B e^{i(\delta^X_B\pm\g)}$, where the $+$ ($-$)
sign refers to the decay of a meson containing a \bquarkbar (\bquark).
The statistical uncertainty with which \g can be measured is approximately inversely proportional to the value of $r^X_B$, which is around 0.1 for \BuDK decays~\cite{Amhis:2014hma}.
In the \BuDpi channel, \rbpi is expected to be of order 0.005~\cite{Tuning:2016} because the favoured amplitude is enhanced by $|\Vud|/|\Vus|$ while the suppressed amplitude is further reduced by $|\Vcd|/|\Vcs|$ with respect to \BuDK decays.
Consequently, the expected sensitivity to \g in \BuDpi decays is considerably lower than for \BuDK decays, although the signal yields are higher.
For \BdDKstz (and also \BsDsK) decays a higher value is expected~\cite{DeBruyn:2012jp}, $\rbDKstz\sim\rdsk\sim 0.3$,
which compensates for the lower branching fraction~\cite{PDG2014}, whilst the expected value
for \rbDkpp is similar to \rb.
The current world average, using only direct measurements of $B\to DK$-like decays, is $\g = (73.2\,^{+6.3}_{-7.0})^{\circ}$~\cite{CKMfitter} (or, using different inputs with an alternative statistical approach, $\g = (68.3\pm7.5)^{\circ}$~\cite{UTfit}).
The previous LHCb combination found $\g = (73\,^{+9}_{-10})^{\circ}$~\cite{LHCb-CONF-2014-004}.

This paper presents the latest combination of LHCb measurements of tree-level decays that are sensitive to \g.
The results supersede those previously reported in Refs.~\cite{LHCb-CONF-2013-006,Aaij:2013zfa,LHCb-CONF-2014-004,LHCb-CONF-2016-001},
including more decay channels and updating selected channels to the full Run 1 dataset of $pp$ collisions at $\sqs=7$ and $8$~TeV, corresponding to an integrated luminosity of 3\invfb.
Two combinations are performed, one including all inputs from $B\to DK$-like modes (referred to as \DK)
and one additionally including inputs from \BuDpi and \BuDpipipi decays (referred to as \Dh).
The \DK combination includes 71 observables depending on 32 parameters, whilst the \Dh combination has 89 observables and 38 parameters.

The analyses included in the combinations use a variety of methods to measure \g, which are reviewed in Ref.~\cite{Gershon:2016fda}.
The observables are briefly summarised below; their dependence on \g and various hadronic parameters is given in Appendix~\ref{sec:relations}.
The Gronau-London-Wyler (GLW) method~\cite{Gronau:1991dp,Gronau:1990ra} considers the
decays of $D$ mesons to \CP eigenstates, for example the \CP-even decays $D\to\Kp\Km$ and $D\to\pip\pim$.
The Atwood-Dunietz-Soni (ADS) approach~\cite{Atwood:1996ci,Atwood:2000ck} extends this to include final states that are not \CP eigenstates,
for example $\Dz\to\pim\Kp$, where the interference between the Cabibbo-allowed and doubly Cabibbo-suppressed decay modes in both the \B and
$D$ decays gives rise to large charge asymmetries. This introduces an additional dependence on the $D$ decay dynamics through the ratio of suppressed and favoured
$D$ decay amplitudes, \rD, and their phase difference, \dD. The GLW/ADS formalism is easily extended to multibody $D$ decays~\cite{Atwood:1996ci,Atwood:2000ck,Nayak:2014tea}
although the multiple interfering amplitudes dilute the sensitivity to \g. For multibody ADS modes this dilution is parameterised
in terms of a coherence factor, $\kappa_{D}$, and for the GLW modes it is parametrised by \Fp, which describes the
fraction of \CP-even content in a multibody decay.
For multibody $D$ decays these parameters are measured independently and used as external constraints in the combination as discussed
in Sec.~\ref{sec:inputsauxilary}. The GLW/ADS observables are constructed from decay-rate ratios, double ratios and charge asymmetries as outlined in the following.

For GLW analyses the observables are the charge-averaged rate and the partial-rate asymmetry. The former is defined as
\begin{align}
  \label{eq:rcp}
  R_{\CP} = 2\, \frac{\Gamma(\Bm \to D_{\CP}K^-) + \Gamma(\Bp \to D_{\CP}K^+)}{\Gamma(\Bm \to \Dz K^-) + \Gamma(\Bp \to \Dzb K^+)}\,,
\end{align}
where $D_{\CP}$ refers to the final state of a \D meson decay into a \CP eigenstate.
Experimentally it is convenient to measure $R_{\CP}$, for a given final state $f$,
by forming a double ratio that is normalised using the rate for a Cabibbo-favoured decay (\eg $\Dz\to\Km\pip$), and the equivalent
quantities from the relevant $\Bp\to \D\pim$ decay mode.
Defining the ratio of the favoured $\Bp\to\Dzb\Kp$ and $\Bp\to\Dzb\pip$ partial widths, for a given final state $f$, as
\begin{align}
  \label{eq:rkp}
  R^{f}_{K/\pi} = \frac{\Gamma(\Bm \to D[\to\f]K^-) + \Gamma(\Bp \to D[\to\fbar]K^{+})}{\Gamma(\Bm \to D[\to\f]\pi^-) + \Gamma(\Bp \to D[\to\fbar]\pi^{+})}\,,
\end{align}
the double ratios are constructed as
\begin{align}
  \label{eq:doubleRs}
  \RCPKK \approx \frac{ R^{KK}_{K/\pi} }{ R^{K\pi}_{K/\pi}}, \ \;\;\;  \RCPPiPi \approx \frac{ R^{\pi\pi}_{K/\pi} }{ R^{K\pi}_{K/\pi}}, \ \;\;\; \RCPKKPiz \approx \frac{ R^{KK\piz}_{K/\pi} }{ R^{K\pi\piz}_{K/\pi}}, \ \;\;\;  \RCPPiPiPiz \approx \frac{ R^{\pi\pi\piz}_{K/\pi} }{ R^{K\pi\piz}_{K/\pi}}, \ \;\;\; \etc
\end{align}
These relations are exact when the suppressed \BuDpi decay amplitude ($\bquark\to\uquark$) vanishes and the flavour specific
rates, given in the denominator of Eq.~(\ref{eq:rcp}), are measured using the appropriate flavour-specific $D$ decay channel.
The GLW partial-rate asymmetry, for a given
\D meson decay into a \CP eigenstate $f$, is defined as
\begin{align}
  \label{eq:acp}
  A_{\CP}^{Dh,f} &= \frac{\Gamma(\Bm\to D_{\CP}h^-) - \Gamma(\Bp\to D_{\CP}h^+)}{\Gamma(\Bm\to D_{\CP}h^-) + \Gamma(\Bp\to D_{\CP}h^+)}\,.
\end{align}
Similarly, observables associated to the ADS modes, for a suppressed $\D\to f$ decay, are the charge-averaged rate and the partial-rate asymmetry.
For the charge-averaged rate, it is adequate to use a single ratio (normalised to the favoured $\D\to\fbar$ decay) because the detection asymmetries
cancel out. The charge-averaged rate is defined as
\begin{align}
  \label{eq:rads}
  R_{\text{ADS}}^{Dh,\fbar} = \frac{\Gamma(\Bm \to D[\to \fbar]h^-) + \Gamma(\Bp \to D[\to \f]h^+)}{\Gamma(\Bm \to D[\to \f]h^-) + \Gamma(\Bp \to D[\to \fbar]h^+ )}\,,
\end{align}
whilst the partial-rate asymmetry is defined as
\begin{align}
  \label{eq:aads}
  A_{\text{ADS}}^{Dh,\fbar} = \frac{\Gamma(\Bm \to D[\to \fbar]h^-) - \Gamma(\Bp \to D[\to \f]h^+)}{\Gamma(\Bm \to D[\to \fbar]h^-) + \Gamma(\Bp \to D[\to \f]h^+ )}\,.
\end{align}
The equivalent charge asymmetry for favoured ADS modes is defined as
\begin{align}
  \label{eq:af}
  A_{{\rm fav}}^{Dh,f} = \frac{\Gamma(\Bm\to D[\to f]h^-) - \Gamma(\Bp\to D[\to \fbar]h^+)}{\Gamma(\Bm\to D[\to f]h^-) + \Gamma(\Bp\to D[\to \fbar]h^+)}\,.
\end{align}
Some of the input analyses determined two statistically independent observables instead of those in Eqs.~(\ref{eq:rads}) and~(\ref{eq:aads}), namely the ratio of partial widths for the suppressed and favoured decays of each initial $B$ flavour,
\begin{align}
  R_{+}^{Dh,\fbar}  = \frac{\Gamma(\Bp\to D[\to f]\hp)}{\Gamma(\Bp\to D[\to \fbar]\hp)}\,, \label{eq:rp} \\
  R_{-}^{Dh,\fbar}  = \frac{\Gamma(\Bm\to D[\to \fbar]\hm)}{\Gamma(\Bm\to D[\to f]\hm)}\,.
  \label{eq:rm}
\end{align}
It should be noted that Eqs.~(\ref{eq:rads}) and~(\ref{eq:aads}) are related to Eqs.~(\ref{eq:rp}) and~(\ref{eq:rm}) by
\begin{align}
  R_{\text{ADS}} = \frac{R_{+}+R_{-}}{2} \,, \,\,\, A_{\text{ADS}} = \frac{ R_{-} - R_{+} }{ R_{-} + R_{+} } \,,
\end{align}
if the rates of the Cabibbo-favoured decays for \Bm and \Bp are identical.

Similar to the ADS approach is the Grossman-Ligeti-Soffer (GLS) method~\cite{Grossman:2002aq} that exploits singly Cabibbo-suppressed decays such as \DKSKpi.
The GLS observables are defined in analogy to Eqs.~(\ref{eq:rads}--\ref{eq:af}).
Note that in the GLS method the favoured decay has sensitivity to \g because the ratio between the suppressed and favoured amplitudes is much larger than in the ADS approach.
It is therefore worthwhile to include the favoured GLS decays in the combinations, which is not the case for the favoured ADS channels alone.

The Giri-Grossman-Soffer-Zupan (GGSZ) method~\cite{Bondar:2001vn,Giri:2003ty} uses self-conjugate multibody
$D$ meson decay modes like $\KS\pip\pim$. Sensitivity to \g is obtained by comparing the distributions of decays in the $D\to f$ Dalitz plot
for opposite-flavour initial-state \B and \Bbar mesons. The population of candidates in the Dalitz plot depends on
four variables, referred to as Cartesian variables which, for a given \B decay final state $X$, are defined as
\begin{align}
  \xpm^{X} = r^X_B \cos(\delta^X_B \pm \g), \label{eq:xpm}\\
  \ypm^{X} = r^X_B \sin(\delta^X_B \pm \g). \label{eq:ypm}
\end{align}
These are the preferred observables for GGSZ analyses.
The GLW/ADS and GGSZ formalisms can also be extended to multibody $B$ decays by including a coherence factor, $\kappa_{B}$,
that accounts for dilution from interference between competing amplitudes.
This inclusive approach is used for all multibody and quasi-two-body \B decays, with the exception of the GLW-Dalitz analysis of \BdDKpi
decays where an amplitude analysis is performed to determine $\xpm^{X}$ and $\ypm^{X}$. Here the term quasi-two-body decays refer to a two body
resonant decay that contributes to a three body final state (\eg $\Bd\to\D\Kstar(892)^{0}$ decays in the \BdDKpi final state).

Time-dependent (TD) analyses of \BsDsK are
also sensitive to \g~\cite{Fleischer:2003yb,Aleksan:1991nh,Dunietz:1987bv}.
Due to the interference between the mixing and decay amplitudes,
the \CP-sensitive observables, which are the coefficients of the time evolution
of \BsDsK decays, have a dependence on $(\g-2\beta_{s})$, where $\beta_{s}\equiv \arg(-V_{ts}^{}V_{tb}^{*}/V_{cs}^{}V_{cb}^{*})$.
In the SM, to a good approximation, $-2\beta_{s}$ is equal to the phase \phis determined from $\Bs\to\jpsi\phi$ and similar decays,
and therefore an external constraint on the value of \phis
provides
sensitivity to \g.
The time-dependent decay rates for the initially pure \Bs and \Bsb flavour eigenstates
are given by
\begin{align}
\frac{{\rm d}\Gamma_{\Bs\to\f}(t)}{{\rm d}t} = \,&\frac{1}{2} |\Af|^2 (1+|\lf|^2) e^{-\gs t} \left[
         \cosh\left(\frac{\dgs t}{2}\right)
  + \Dpar\sinh\left(\frac{\dgs t}{2}\right) \right. \nonumber\\
& + \Cpar\cos\left(\dms t\right)
  - \Spar\sin\left(\dms t\right)
\bigg],
\label{eq:decay_rates_1}\\
\frac{{\rm d}\Gamma_{\Bsb\to\f}(t)}{{\rm d}t} = \,&\frac{1}{2} |\Af|^2 \left|\frac{p}{q}\right|^2 (1+|\lf|^2) e^{-\gs t} \left[
         \cosh\left(\frac{\dgs t}{2}\right)
  + \Dpar\sinh\left(\frac{\dgs t}{2}\right) \right. \nonumber\\
& - \Cpar\cos\left(\dms t\right)
  + \Spar\sin\left(\dms t\right)
\bigg],
\label{eq:decay_rates_2}
\end{align}
where \mbox{$\lf \equiv (q/p)\cdot(\Abf/\Af)$} and \Af (\Abf)
is the decay amplitude of a
\Bs (\Bsb) to a final state \f.
In the convention used, \f (\fb) is the $\Dsm \Kp$ ($\Dsp \Km$) final state.
The parameter \dms is the oscillation frequency for \Bs mesons, \gs is the average \Bs decay width, and \dgs is the
decay-width difference between the heavy and light mass eigenstates in the \Bs system, which is known to be
positive~\cite{Aaij:2012eq} as expected in the SM.
The observables sensitive to \g are \Dpar, \Cpar and \Spar.
The complex coefficients $p$ and $q$ relate the \Bs meson mass
eigenstates, $|B_{L,H}\rangle$, to the flavour eigenstates,
$|\Bs\rangle$ and $|\Bsb\rangle$, as $|B_L\rangle = p|\Bs\rangle+q|\Bsb\rangle$ and $|B_H\rangle = p|\Bs\rangle-q|\Bsb\rangle$
with $|p|^2+|q|^2=1$.
Similar equations can be written for the \CP-conjugate decays replacing
\Spar by \Sbpar, and
\Dpar by \Dbpar,
and, assuming no \CP violation in either the decay or mixing
amplitudes, $\Cbpar = - \Cpar$.
The relationships between the observables, \g and the
hadronic parameters are given in Appendix~\ref{sec:relations_dsk}.


The combinations are potentially sensitive to subleading effects from \Dz--\Dzb mixing~\cite{Meca:1998ee,Silva:1999bd,Rama:2013voa}.
These are
corrected for where necessary, by taking into account the \Dz decay-time acceptances of the individual measurements.
The size of the correction is inversely proportional to $r_{B}^{X}$ and so is particularly important for the $\Bp\to\D\pip(\pip\pim)$ modes.
For consistency, the correction is also applied in the corresponding $\Bp\to\D\Kp(\pip\pim)$ modes.
The correction for other decay modes would be small and is not applied.
There can also be an effect from \CP violation in \Dhh decays~\cite{Wang:2012ie,Martone:2012nj,Bhattacharya:2013vc,Bondar:2013jxa},
which is included in the relevant $\Bp\to\Dzb\hp(\pip\pim)$ analyses using the world average values~\cite{Amhis:2014hma},
although the latest measurements indicate that the effect is negligible~\cite{LHCb-PAPER-2015-055}.
Final states that include a \KS meson are potentially affected by corrections due to \CP violation and mixing in the neutral kaon system,
parametrised by the non-zero parameter $\epsilon_K$~\cite{Grossman:2013woa}.
The effect is expected to be $\mathcal{O}(\epsilon_K/\rbh)$, which is negligible for $\BuDK$ decays since $|\epsilon_K| \approx 0.002$ and $\rb\approx 0.1$~\cite{Amhis:2014hma}.
For \BuDpi decays this ratio is expected to be $\mathcal{O}(1)$ since $\rbpi$ is expected to be around $0.5\%$~\cite{Tuning:2016}. Consequently, the \BuDpi decay modes affected,
such as those with $\D\to\KS \Kmp\pipm$, are not included in the \Dh combination.

To determine \g with the best possible precision, auxiliary information on some of the hadronic parameters is used in conjunction with
observables measured in other LHCb analyses.
More information on these quantities can be found in Secs.~\ref{sec:inputs} and~\ref{sec:inputsauxilary}, with
a summary provided in Tables~\ref{tab:inputs} and~\ref{tab:inputs_aux}.
Frequentist and Bayesian treatments are both studied. Section~\ref{sec:statinterpretation} describes the frequentist treatment with results and coverage studies reported in Sec.~\ref{sec:results}. Section~\ref{sec:bayesian} describes the results of a Bayesian analysis.

\section{\boldmath Inputs from LHCb analyses sensitive to \g}
\label{sec:inputs}

The LHCb measurements used as inputs in the combinations are summarised in Table~\ref{tab:inputs} and described briefly below.
The values and uncertainties of the observables are
provided in Appendix~\ref{sec:inputs_vals}
and the correlations are given in Appendix~\ref{sec:inputs_corrs}.
The relationships between the observables and the physics parameters are listed in Appendix~\ref{sec:relations}.
All analyses use a data sample corresponding to an integrated luminosity of 3~\invfb, unless otherwise stated.

\begin{table}[h!]
  \caption{List of the LHCb measurements used in the combinations.}
  \label{tab:inputs}
  \centering
  \renewcommand{\arraystretch}{1.4}
  \begin{tabular}{l l l l p{3.5cm}}
        \hline
        $\B$ decay  & $\D$ decay & Method   & Ref. & Status since last combination~\cite{LHCb-CONF-2014-004}  \\
        \hline
        \BuDh     & \Dhh      & GLW/ADS     & \cite{LHCb-PAPER-2016-003} & Updated to 3\invfb \\
        \BuDh     & \Dhpipipi & GLW/ADS     & \cite{LHCb-PAPER-2016-003} & Updated to 3\invfb \\
        \BuDh     & \Dhhpiz   & GLW/ADS     & \cite{LHCb-PAPER-2015-014} & New                \\
        \BuDK     & \DKShh    & GGSZ        & \cite{LHCb-PAPER-2014-041} & As before          \\
        \BuDK     & \DKSKpi   & GLS         & \cite{LHCb-PAPER-2013-068} & As before          \\
        \BuDhpipi & \Dhh      & GLW/ADS     & \cite{LHCb-PAPER-2015-020} & New                \\
        \BdDzKstz & \DKpi     & ADS         & \cite{LHCb-PAPER-2014-028} & As before          \\
        \BdDKpi   & \Dhh      & GLW-Dalitz  & \cite{LHCb-PAPER-2015-059} & New                \\
        \BdDzKstz & \DKSpipi  & GGSZ        & \cite{LHCb-PAPER-2016-007} & New                \\
        \BsDsK    & \Dshhh    & TD          & \cite{LHCb-PAPER-2014-038} & As before          \\
        \hline
      \end{tabular}
\end{table}

\begin{itemize}


  \item \textbf{\boldmath\BuDh, \Dhh.}
  The GLW/ADS measurement using \BuDh, \Dzhh decays~\cite{LHCb-PAPER-2016-003}
  is an update of a previous analysis~\cite{LHCb-PAPER-2012-001}.
The observables are defined in analogy to Eqs.~(\ref{eq:doubleRs}--\ref{eq:af}).


  \item \textbf{\boldmath\BuDh, \Dhpipipi.}
The ADS measurement using the \BuDh, \DKpipipi decay mode~\cite{LHCb-PAPER-2016-003} is an update of a previous measurement~\cite{LHCb-PAPER-2012-055}.
The quasi-GLW measurement with \BuDh, \Dpipipipi decays is included in the combination for the first time.
The label ``quasi'' is used because the \Dpipipipi decay is not completely \CP-even; the fraction of \CP-even content is given by \Fpppp as described in Sec.~\ref{sec:inputsauxilary}.
The method for constraining \g using these decays is described in Ref.~\cite{Nayak:2014tea}, with
observables defined in analogy to Eqs.~(\ref{eq:doubleRs}--\ref{eq:af}).


  \item \textbf{\boldmath\BuDh, \Dhhpiz.}
Inputs from the quasi-GLW/ADS analysis of \BuDh, \Dhhpiz decays~\cite{LHCb-PAPER-2015-014} are new to this combination.
The \CP-even content of the \DKKpiz (\Dpipipiz) decay mode is given by the parameter \FKKp (\Fppp), as described in Sec.~\ref{sec:inputsauxilary}.
The observables are defined in analogy to Eqs.~(\ref{eq:doubleRs}--\ref{eq:af}).


  \item \textbf{\boldmath\BuDK, \DKShh.}
The inputs from the model-independent GGSZ analysis
of \BuDK, \DKShh decays~\cite{LHCb-PAPER-2014-041} are the same as those used in the previous combination~\cite{LHCb-CONF-2014-004}.
The variables, defined in analogy to Eqs.~(\ref{eq:xpm}--\ref{eq:ypm}), are obtained from a simultaneous fit
to the Dalitz plots of $\DKSpipi$ and $\DKSKK$ decays. Inputs from a model-dependent GGSZ analysis of the same decay~\cite{LHCb-PAPER-2014-017} using data
corresponding to 1\invfb are not included due to the overlap of the datasets.


  \item \textbf{\boldmath\BuDK, \DKSKpi.}
The inputs from the GLS analysis
of \BuDK, \DKSKpi decays~\cite{LHCb-PAPER-2013-068}
are the same as those included in the last combination~\cite{LHCb-CONF-2014-004}.
The observables are defined in analogy to Eqs.~(\ref{eq:rads}--\ref{eq:af}).
The negligible statistical and systematic correlations are not taken into account.


  \item \textbf{\boldmath\BuDhpipi, \Dhh.}
The inputs from the LHCb GLW/ADS analysis of \BuDhpipi, \Dzhh decays~\cite{LHCb-PAPER-2015-020}
are included in the combination for the first time.
The observables are defined in analogy to Eqs.~(\ref{eq:doubleRs}--\ref{eq:acp},\ref{eq:af}--\ref{eq:rm}).
The only non-negligible correlations are statistical, $\rho(\AcpDkpipiKK,\AcpDkpipiPipi)=0.20$ and $\rho(\AcpDpipipiKK,\AcpDpipipiPipi)=0.08$. 


  \item \textbf{\boldmath\BdDzKstz, \DKpi.}
The inputs from the ADS analysis
of $\Bd\to\Dz\Kstar(892)^{0}$, \DzKpi decays~\cite{LHCb-PAPER-2014-028}
are included as they were in the previous combination~\cite{LHCb-CONF-2014-004}. However, the GLW part of this
analysis (with \DzKK and \Dzpipi) has been superseded by the Dalitz plot analysis.
The ADS observables are defined in analogy to Eqs.~(\ref{eq:af}--\ref{eq:rm}).


  \item \textbf{\boldmath\BdDKpi, \Dhh.}
Information from the GLW-Dalitz analysis of \BdDKpi, \Dzhh decays~\cite{LHCb-PAPER-2015-059}
is added to the combination for the first time.
The ``Dalitz'' label indicates the method used to determine information about \CP violation in this mode.
The variables, defined in analogy to Eqs.~(\ref{eq:xpm}--\ref{eq:ypm}), are determined from a simultaneous Dalitz
plot fit to \BdDKpi with $\Dz\to\Km\pip$, $D\to\Kp\Km$ and $D\to\pip\pim$ samples, as described in Refs.~\cite{Gershon:2008pe,Gershon:2009qc}.
Note that the observables are those associated with the $\D\Kstar(892)^{0}$ amplitudes.
Constraints on hadronic parameters are also obtained in this analysis, as described in Sec.~\ref{sec:inputsauxilary}.


  \item \textbf{\boldmath\BdDzKstz, \DKSpipi.}
    Inputs from the model-dependent GGSZ analysis of $\BdDzKstz(892)$, \DKSpipi decays~\cite{LHCb-PAPER-2016-007}
are included in the combination for the first time.
The observables, defined in analogy to Eqs.~(\ref{eq:xpm}--\ref{eq:ypm}), are measured by fitting
the $\DKSpipi$ Dalitz plot using a model developed by the BaBar collaboration~\cite{delAmoSanchez:2010xz}.

A model-independent GGSZ analysis~\cite{LHCb-PAPER-2016-006} is also performed by LHCb on the same data sample. Currently, the model-dependent analysis
has the best sensitivity to the parameters $\xpm$ and $\ypm$. Therefore the model-dependent results
are used in the combination. The numerical results of the combination change insignificantly if the model-independent results are used instead.


  \item \textbf{\boldmath\BsDsK.}
The inputs used from the time-dependent analysis
of \BsDsK decays
using data corresponding to 1\invfb~\cite{LHCb-PAPER-2014-038}
are identical to those used in Ref.~\cite{LHCb-CONF-2014-004}. Note however that a different sign convention is used here, as defined in Eqs.(\ref{eq:decay_rates_1}--\ref{eq:decay_rates_2}) and Appendix~\ref{sec:relations_dsk}.

\end{itemize}

\section{Auxiliary inputs}
\label{sec:inputsauxilary}

The external inputs are briefly described below and summarised in Table~\ref{tab:inputs_aux}.
These measurements provide constraints on unknown parameters and result in better precision on \g.
The values and uncertainties of the observables are provided in Appendix~\ref{sec:inputs_aux_vals}
and the correlations are
given in Appendix~\ref{sec:inputs_aux_corrs}.

\begin{table}[b]
  \caption{List of the auxiliary inputs used in the combinations.}
  \label{tab:inputs_aux}
  \renewcommand{\arraystretch}{1.4}
  \centering
      \begin{tabular}{l l l l }
        \hline
        Decay      & Parameters                  & Source & Ref. \\
        \hline
         \Dz--\Dzb-mixing   & \xd, \yd                             & HFAG       &  \cite{Amhis:2014hma}       \\
         \DKpi              & \rdKpi, \ddKpi                       & HFAG       &  \cite{Amhis:2014hma}       \\
         \Dhh               & \DAcpKK, \DAcpPipi                   & HFAG       &  \cite{Amhis:2014hma}       \\
         \DKpipipi          & \ddKppp, \kdKppp, \rdKppp            & CLEO+LHCb  &  \cite{Evans:2016tlp}       \\
         \Dpipipipi         & \Fpppp                               & CLEO       &  \cite{Malde:2015mha}       \\
         \DKpipiz           & \ddKpp, \kdKpp, \rdKpp               & CLEO+LHCb  &  \cite{Evans:2016tlp}       \\
         \Dhhpiz            & \Fppp, \FKKp                         & CLEO       &  \cite{Malde:2015mha}       \\
         \DKSKpi            & \ddKskpi, \kdKskpi, \rdKskpi         & CLEO       &  \cite{Insler:2012pm}       \\
         \DKSKpi            & \rdKskpi                             & LHCb       &  \cite{LHCb-PAPER-2015-026} \\
         \BdDzKstz          & \kbDKstz, \RbDKstz, \DbDKstz         & LHCb       &  \cite{LHCb-PAPER-2015-059} \\
         \BsDsK             & \phis                                & LHCb       &  \cite{LHCb-PAPER-2014-059} \\
        \hline
      \end{tabular}
\end{table}

\begin{itemize}


  \item \textbf{Input from global fit to charm data.}
The GLW/ADS measurements need input to constrain the charm
system in three areas: the ratio and strong phase difference for \DzKpiFav and \DzKpiSup decays (\rdKpi, \ddKpi),
charm mixing (\xd, \yd) and direct \CP violation in \Dzhh decays
(\DAcpKK, \DAcpPipi), taken from a recent HFAG
charm fit~\cite{Amhis:2014hma}.
These do not include the latest results on $\Delta A_{\CP}$ from LHCb~\cite{LHCb-PAPER-2015-055} but
their impact has been checked and found to be negligible.
The value of \ddKpi is shifted by $180\degrees$ compared to the HFAG result in order to match the phase convention adopted in this paper.
The parameter \RdKpi is related to the amplitude ratio \rdKpi through $\RdKpi \equiv \rdKpisq$.


  \item \textbf{\boldmath Input for \DzKpipiz and \DzKpipipi decays.}
The ADS measurements with \DzKpipiz and \DzKpipipi decays require
knowledge of the hadronic parameters describing the $D$ decays. These are the ratio, strong phase difference
and coherence factors of the two decays: \rdKpp, \ddKpp, \kdKpp, \rdKppp, \ddKppp and \kdKppp.
Recently an analysis of \DzKpipipi decays has been performed by LHCb~\cite{LHCb-PAPER-2015-057} that is sensitive to \rdKppp, \ddKppp and
\kdKppp. Furthermore, an updated measurement has been performed using CLEO-c data, and the results have been combined with those from LHCb~\cite{Evans:2016tlp}
to yield constraints and correlations of the six parameters.
These are included as Gaussian constraints in this combination, in line with the treatment of the other auxiliary inputs.


  \item \textbf{\boldmath \CP content of \Dhhpiz and \Dpipipipi decays.}
For both the three-body \Dhhpiz and four-body \Dpipipipi quasi-GLW measurements the fractional \CP-even content of the decays, \FKKp, \Fppp and \Fpppp, are used as inputs.
These parameters were measured by the CLEO collaboration~\cite{Malde:2015mha}.
The uncertainty for the \CP-even content of \Dpipipipi decays is increased from $\pm 0.028$ to $\pm 0.032$ to account for the non-uniform acceptance of the LHCb detector
following the recommendation in Ref.~\cite{LHCb-PAPER-2016-003}.
For the \Dhhpiz decay the LHCb efficiency is sufficiently uniform to avoid the need to increase the \Fp uncertainty for these modes.


  \item \textbf{\boldmath Input for \DKSKpi parameters.}
The \BuDK, \DKSKpi GLS measurement needs inputs for the charm system parameters
\rdKskpi, \ddKskpi, and \kdKskpi. Constraints from Ref.~\cite{Insler:2012pm} on all three are included, along with
an additional constraint on the branching fraction ratio \RdKskpi from Ref.~\cite{LHCb-PAPER-2015-026}.
The results corresponding to
a limited region of the Dalitz plot, dominated by the $K^{*}(892)^{+}$ resonance, are used here.
The quantity \RdKskpi is related to \rdKskpi
through
\begin{equation}
\label{eq:rws_mixing_kskpi}
\RdKskpi = \frac{ \rdKskpisq - \kdKskpi\rdKskpi(\yd\cos\ddKskpi-\xd\sin\ddKskpi) } { 1 - \rdKskpi\kdKskpi(\yd\cos\ddKskpi+\xd\sin\ddKskpi)}\,.
\end{equation}
The linear correlation coefficient between \ddKskpi and \kdKskpi is extracted from the experimental likelihood
as $\rho(\ddKskpi,\kdKskpi)=-0.60$.


  \item \textbf{\boldmath Constraints on the \BdDKstz hadronic parameters.}
The quasi-two-body \BdDKstz ADS and model-dependent GGSZ measurements need input on the coherence factor
\kbDKstz and the parameters $\RbDKstz = \rbbarDKstz / \rbDKstz$ and $\DbDKstz = \dbbarDKstz - \dbDKstz$, which relate the hadronic parameters
of the quasi-two-body \BdDKstz ADS and GGSZ measurements (barred symbols)
to those of the \BdDKpi amplitude analysis (unbarred symbols).
The resulting values are taken from the LHCb GLW-Dalitz analysis described in Ref.~\cite{LHCb-PAPER-2015-059}.
These are taken to be uncorrelated with each other and with the \xpmdkpi, \ypmdkpi
parameters that are determined from the same analysis.


  \item \textbf{\boldmath Constraint on \phis.}
The time-dependent measurement of \BsDsK determines the quantity $\g-2\beta_s$.
In order to interpret this as a measurement of \g, the weak phase
$-2\beta_s\equiv \phis$ is constrained to the value measured by LHCb
in $\Bs\to\jpsi hh$ decays~\cite{LHCb-PAPER-2014-059}. It has been checked
that using the world average instead has a negligible impact on the results.

\end{itemize}

\section{Statistical treatment}
\label{sec:statinterpretation}

The baseline results of the combinations are presented using a frequentist treatment,
starting from a likelihood function built from the product of the
probability density functions (PDFs), $f_i$, of experimental observables $\vec{A}_i$,
\begin{equation}
	\label{eq:comblh}
	\mathcal{L}(\vec{\alpha}) = \prod_i f_i(\vec{A}_i^{\rm obs} | \vec{\alpha})\,,
\end{equation}
where $\vec{A}_i^{\rm obs}$ are the measured values of the observables from an input analysis $i$,
and $\vec{\alpha}$ is the set of parameters.
For each of the inputs it is assumed that the observables follow
a Gaussian distribution
\begin{equation}
  f_i(\vec{A}_{i}^{\rm obs} | \vec{\alpha}) \propto \exp\left( -\frac{1}{2} (\vec{A}_i(\vec{\alpha})-\vec{A}_{i}^{\rm obs})^T \,
	V_i^{-1} \, (\vec{A}_i(\vec{\alpha})-\vec{A}_{i}^{\rm obs}) \right)\,,
  \label{eq:gpdf}
\end{equation}
where $V_i$ is the experimental covariance matrix, which includes statistical and systematic
uncertainties and their correlations.
Correlations in the systematic uncertainties between the statistically independent input measurements are assumed to be zero.

A $\chi^2$-function is defined as
$\chi^2(\vec{\alpha}) = -2 \ln \mathcal{L}(\vec{\alpha})$.
The best-fit point is given by the global minimum of the
$\chi^2$-function, $\chi^2(\vec{\alpha}_{\min})$.
To evaluate the confidence level (CL) for a given value of a parameter,
\eg $\g=\g_0$ in the following,
the value of the $\chi^2$-function
at the new minimum is considered, $\chi^2(\vec{\alpha}'_{\min}(\g_0))$.
The associated profile likelihood function for the parameters is $\mathcal{L}(\vec{\alpha}'_{\min}(\g_0))$.
Then a test statistic is defined as
\mbox{$\Delta\chi^2 = \chi^2(\vec{\alpha}'_{\min}) - \chi^2(\vec{\alpha}_{\min})$}.
The $p$-value, or \omcl, is calculated by means of a Monte Carlo
procedure, described in Ref.~\cite{woodroofe}
and briefly recapitulated here.
For each value of $\g_0$ the test statistic $\Delta\chi^2$ is calculated,
and a set of pseudoexperiments, $\vec{A}_{j}$, is generated
according to Eq.~(\ref{eq:gpdf}) with parameters $\vec{\alpha}$ set to
the values at $\vec{\alpha}^\prime_{\min}$. A new value of the test statistic, $\Delta\chi^{2\prime}$, is
calculated for each pseudoexperiment by replacing $\vec{A}_{\rm obs} \to \vec{A}_j$ and
minimising with respect to $\vec{\alpha}$, once with \g as a free parameter, and once with \g fixed to $\gamma_0$.
The value of \omcl is then defined as the fraction of pseudoexperiments for which $\Delta\chi^2 < \Delta\chi^{2\prime}$.
This method is sometimes referred to as the ``$\hat\mu$'', or the \plugin
method. Its coverage cannot be guaranteed~\cite{woodroofe}
for the full parameter space, but can be verified for the best-fit point.
The reason is that for each value of $\gamma_0$, the nuisance parameters,
{\em i.e.} the components of $\vec{\alpha}$ other than the parameter of interest,
are set to their best-fit values for this point, as opposed to
computing an $n$-dimensional confidence region, which is computationally impractical.
The coverage of the frequentist combinations is discussed in Sec.~\ref{sec:coverage}.

\section{Results}
\label{sec:results}


Results for the \DK combination are presented in Sec.~\ref{sec:results_dk} and for the
\Dh combination in Sec.~\ref{sec:results_dh}. The coverage of the frequentist method is
discussed in Sec.~\ref{sec:coverage} whilst an interpretation of the results is provided in Sec.~\ref{sec:interpretation}.
The rate equations from which the observables are determined are invariant under the simultaneous transformation
$\g \to \g + 180 \degrees$, $\delta^X_B \to \delta^X_B + 180 \degrees$, where $\delta^X_B$ is the strong phase for each
$\B\to DX$ decay considered.
Only the solution most consistent with the
determination of \g from the global CKM fit~\cite{CKMfitter,UTfit} is shown.

\subsection{\boldmath \DK combination}
\label{sec:results_dk}

The \DK combination consists of 71 observables and 32 parameters.
The goodness of
fit computed from the \chisq value at the best fit point given the number
of degrees of freedom is $p=91.5\%$. The equivalent value calculated from the fraction of
pseudoexperiments, generated from the best fit point, which have a \chisq
larger than that found in the data is $p=(90.5\pm0.2)\%$.

\begin{figure}
  \centering
  \includegraphics[width=.48\textwidth]{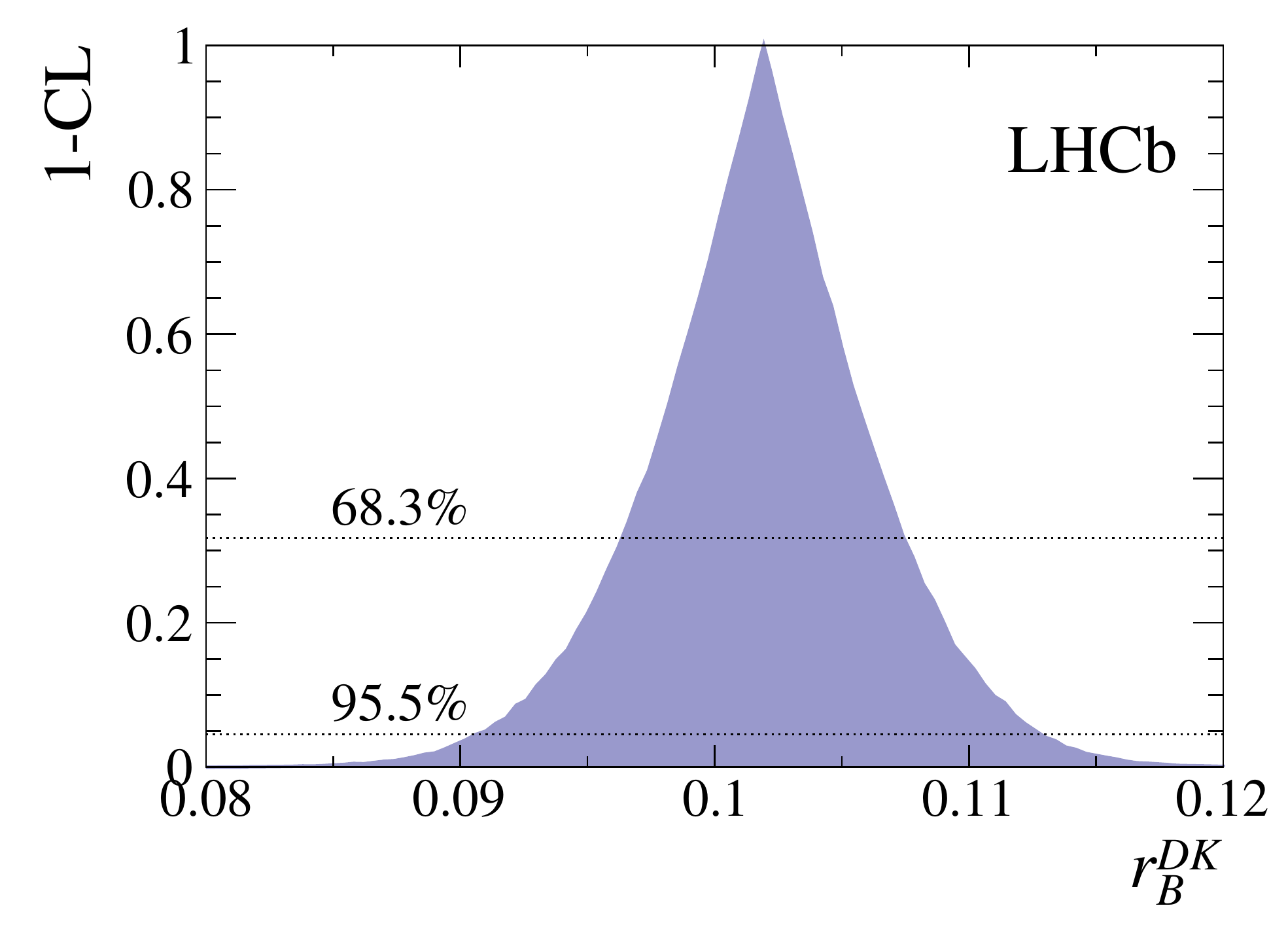}
  \includegraphics[width=.48\textwidth]{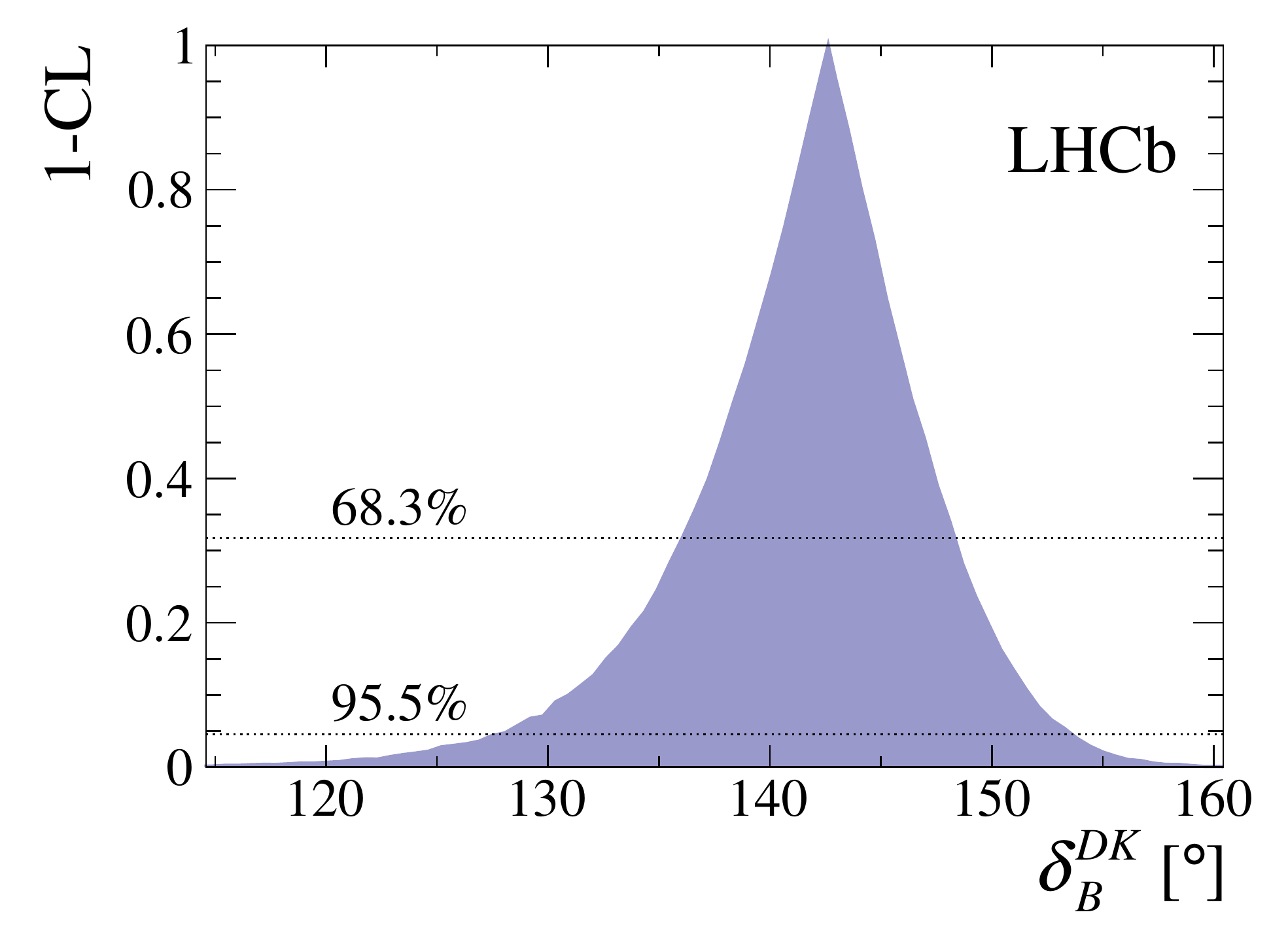}
  \includegraphics[width=.48\textwidth]{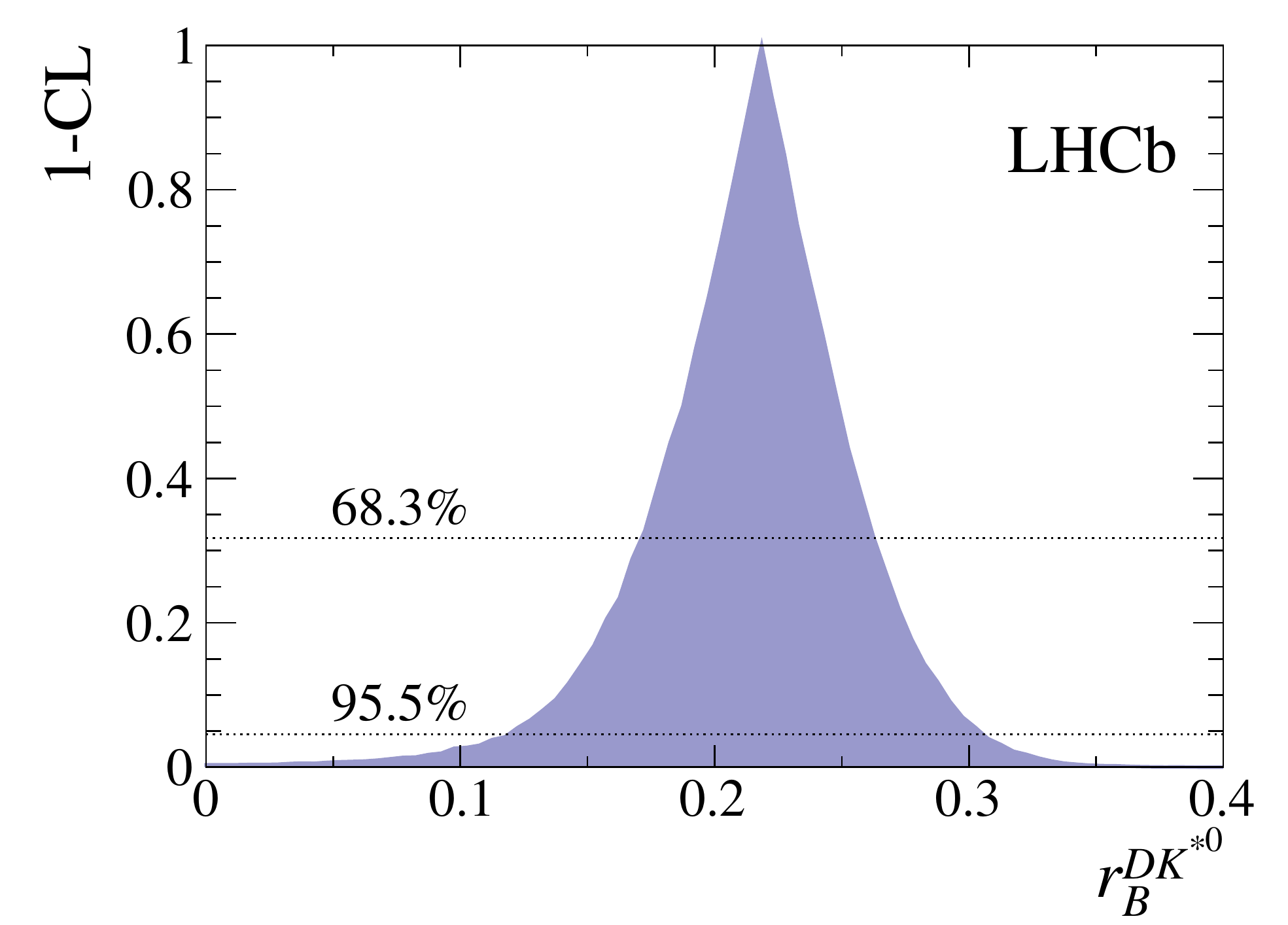}
  \includegraphics[width=.48\textwidth]{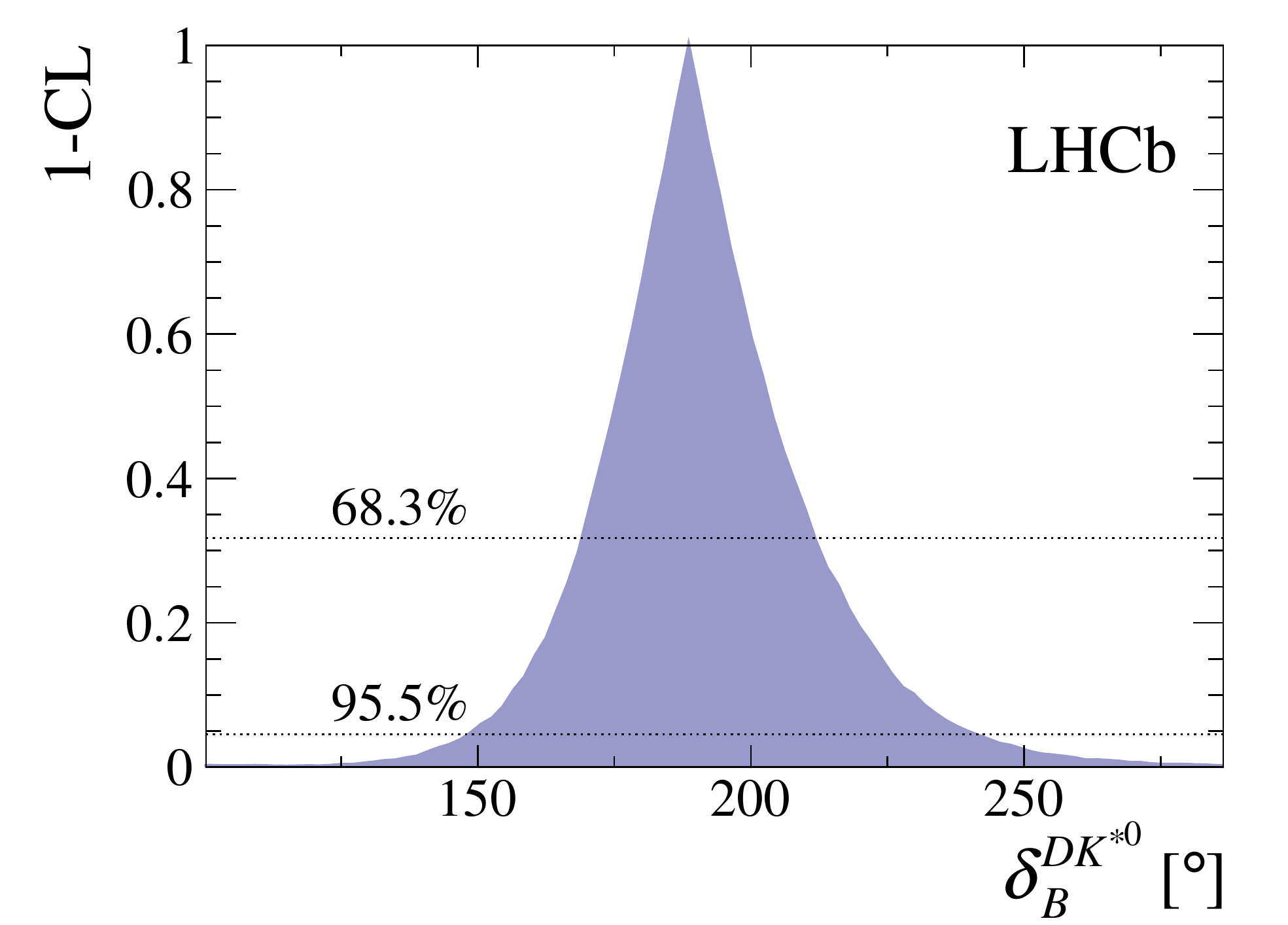}
  \includegraphics[width=.48\textwidth]{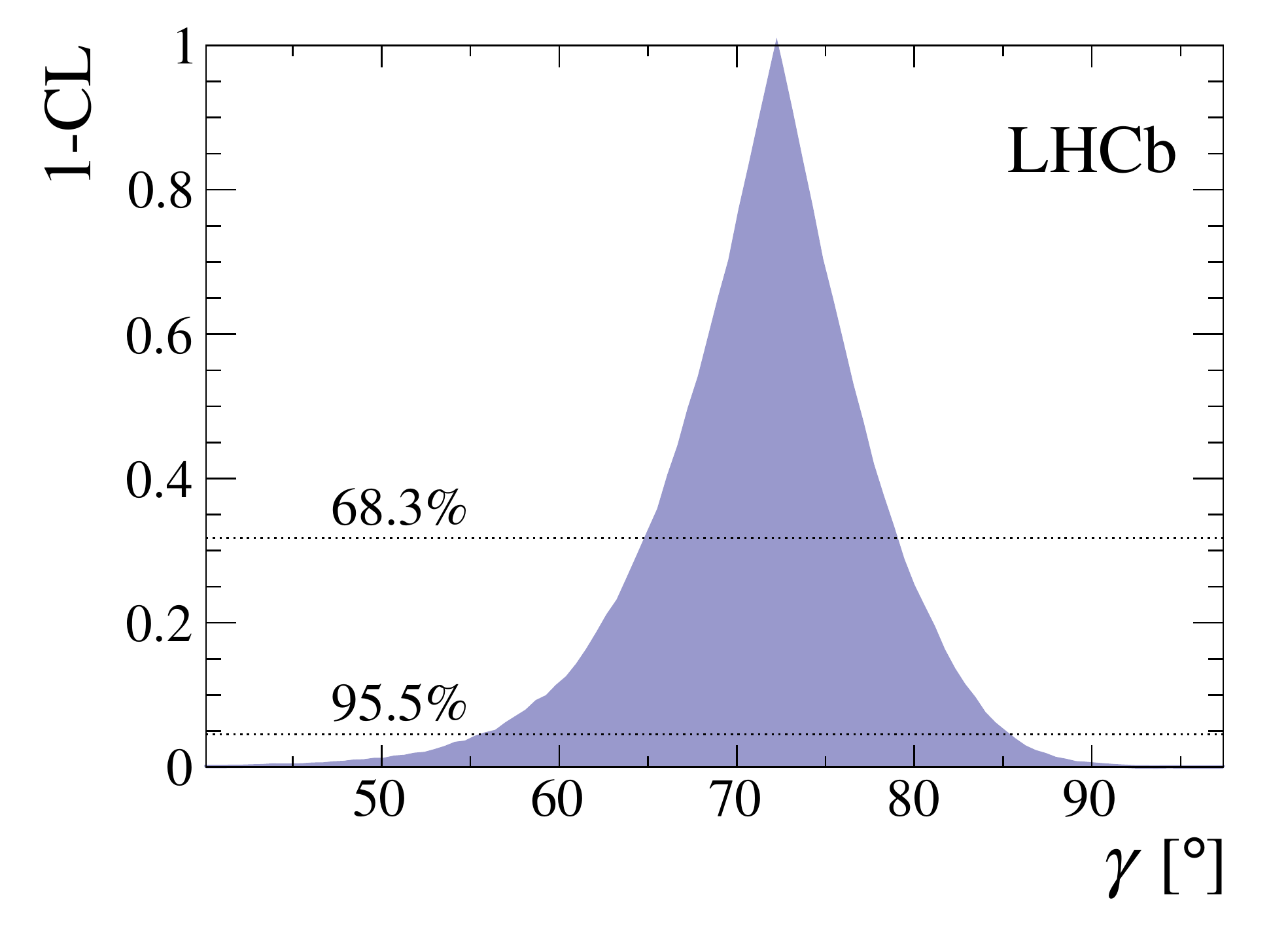}
  \caption{\omcl curves for the \DK combination obtained with the \plugin method. 
The $1\sigma$ and $2\sigma$ levels are indicated by the horizontal dotted lines.
  }
  \label{fig:resultrobust}
\end{figure}

\begin{figure}
  \centering
  \includegraphics[width=.48\textwidth]{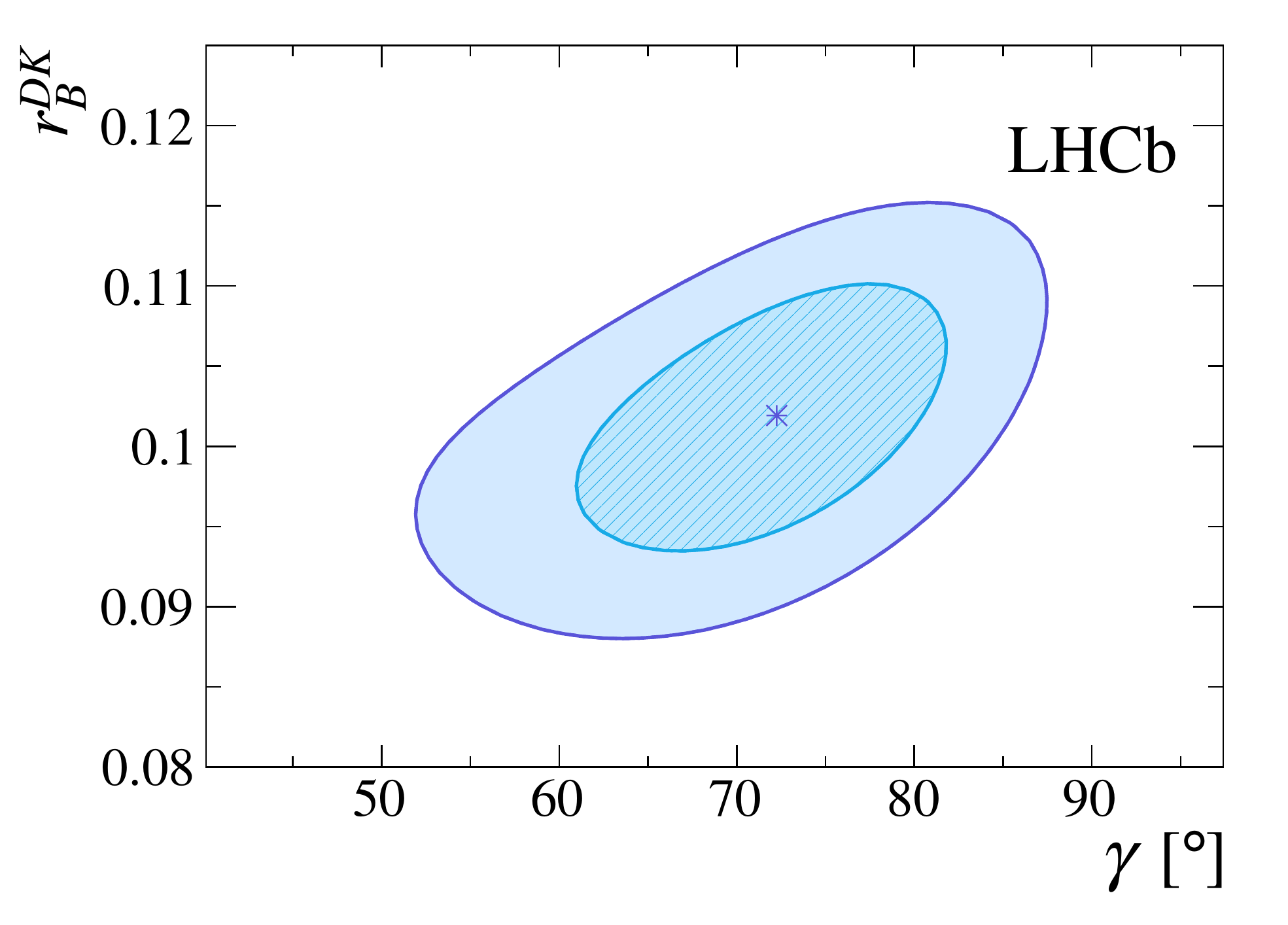}
  \includegraphics[width=.48\textwidth]{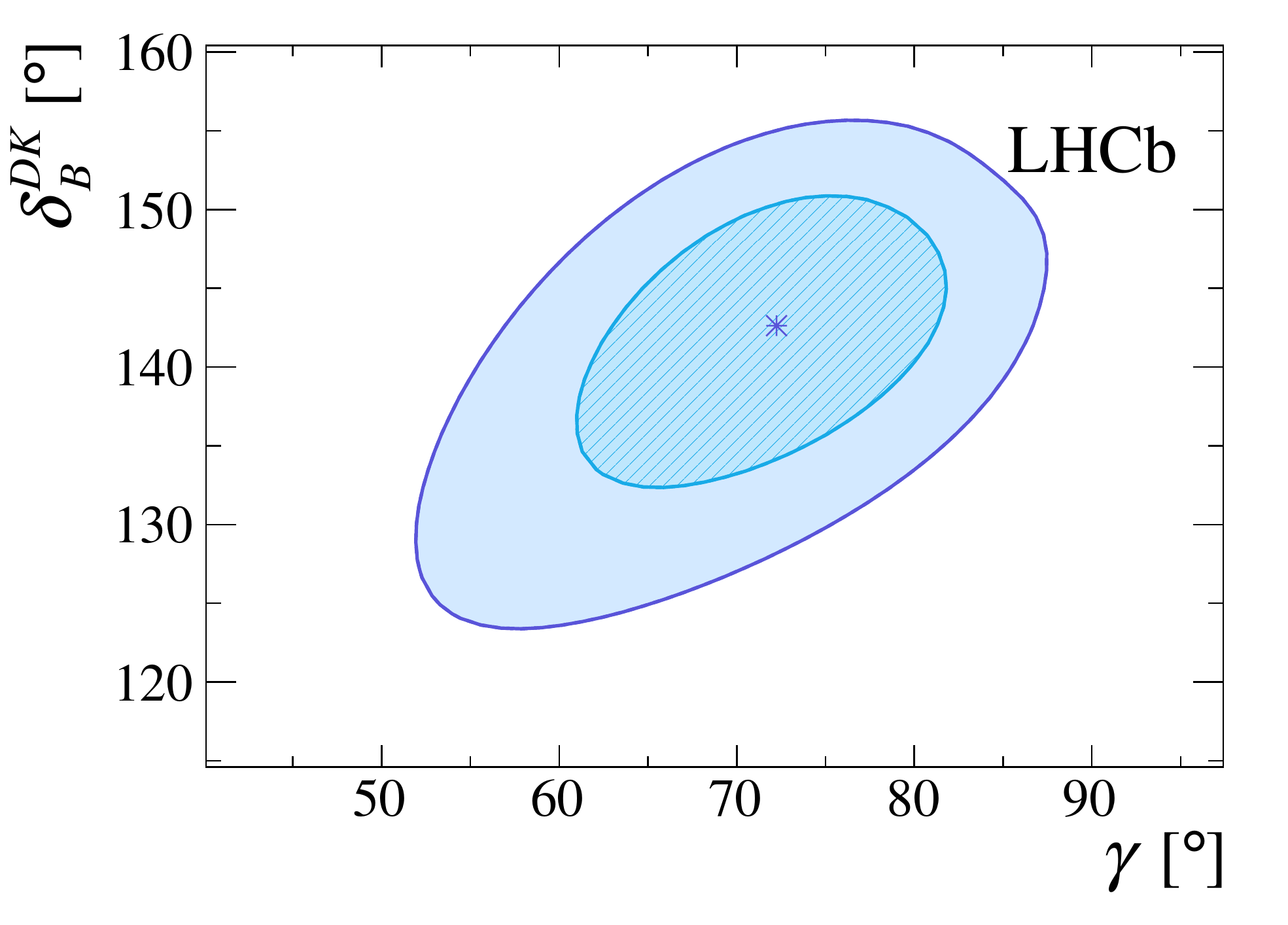}
  \includegraphics[width=.48\textwidth]{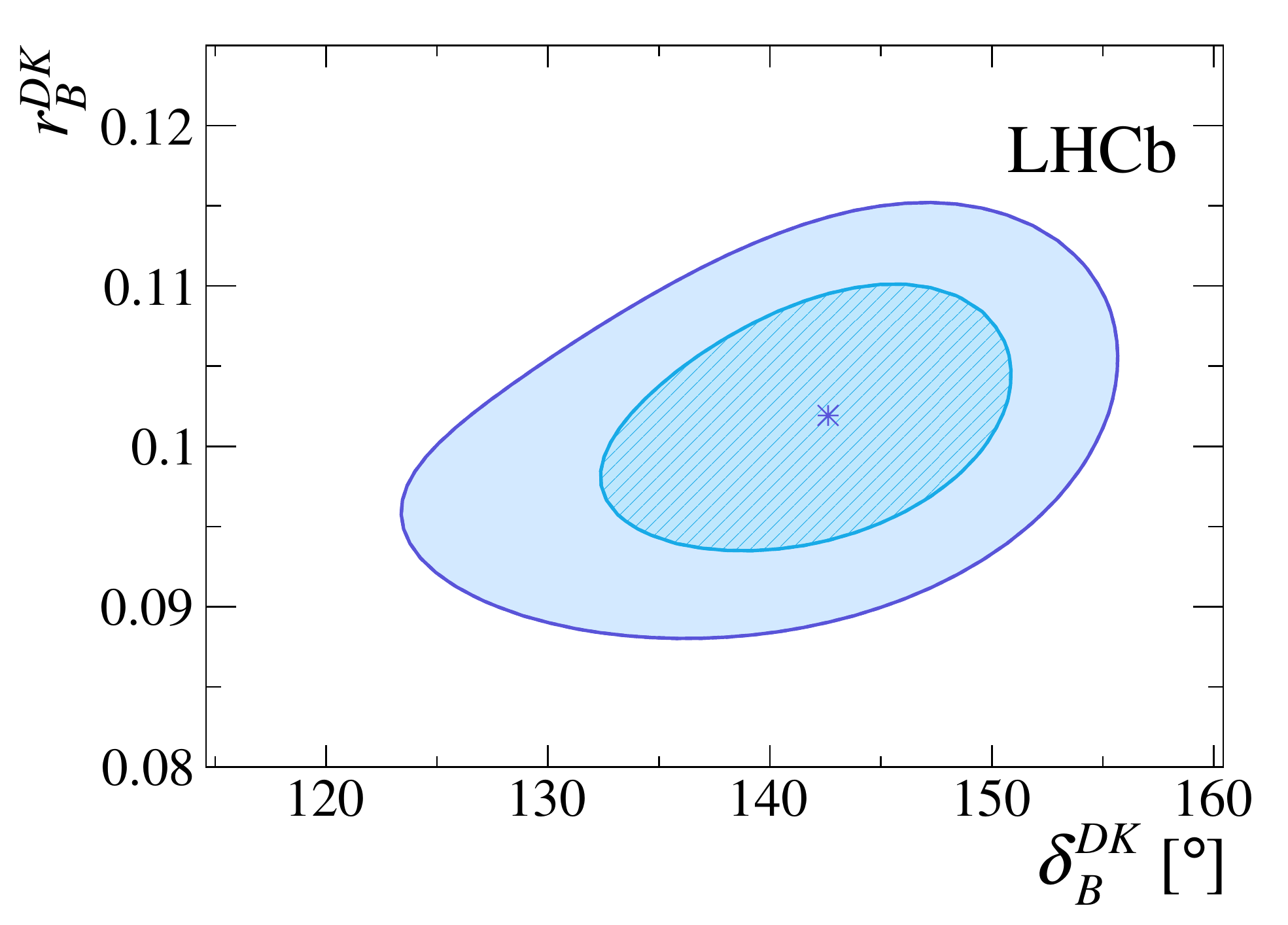}
  \includegraphics[width=.48\textwidth]{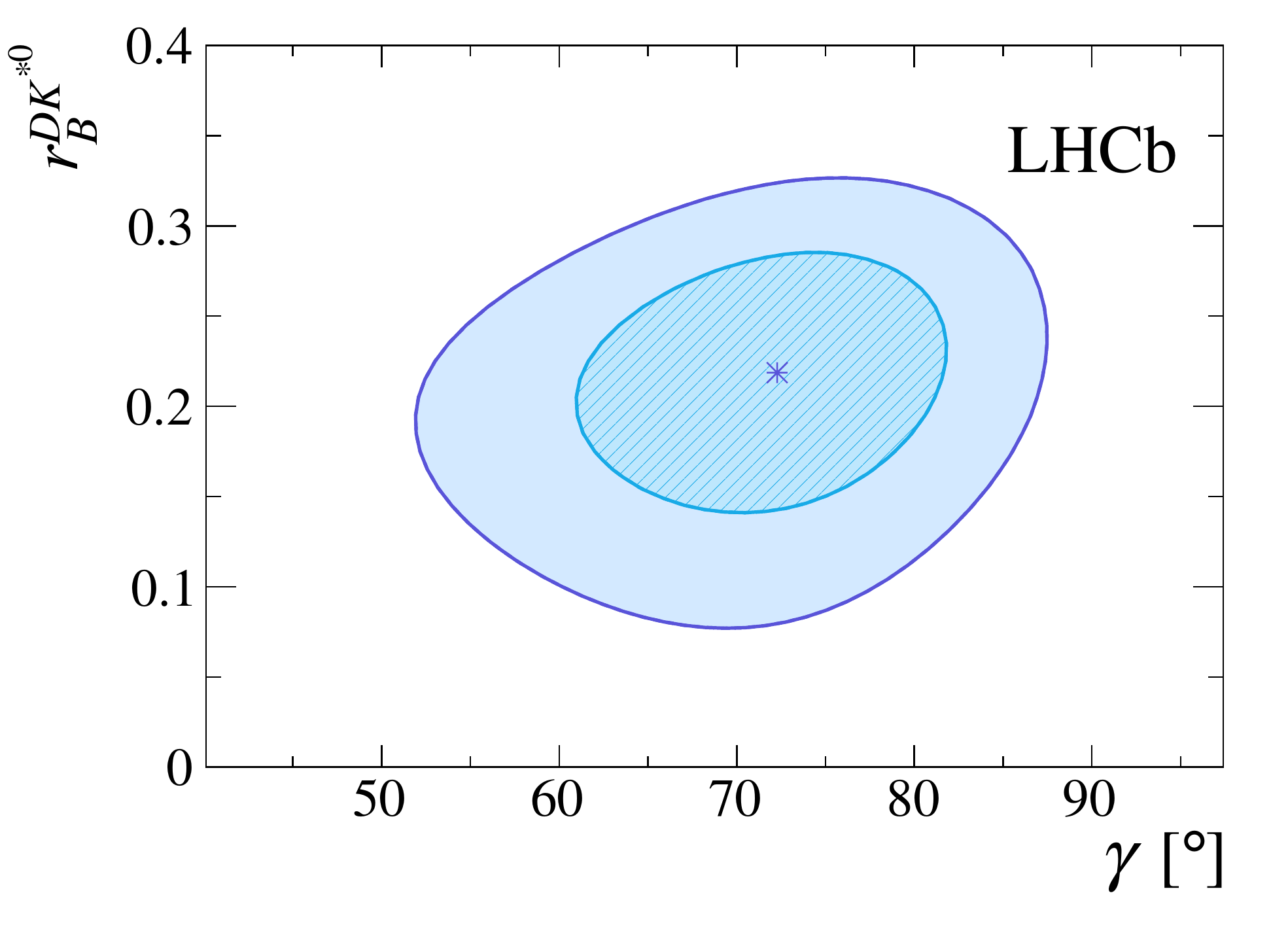}
  \includegraphics[width=.48\textwidth]{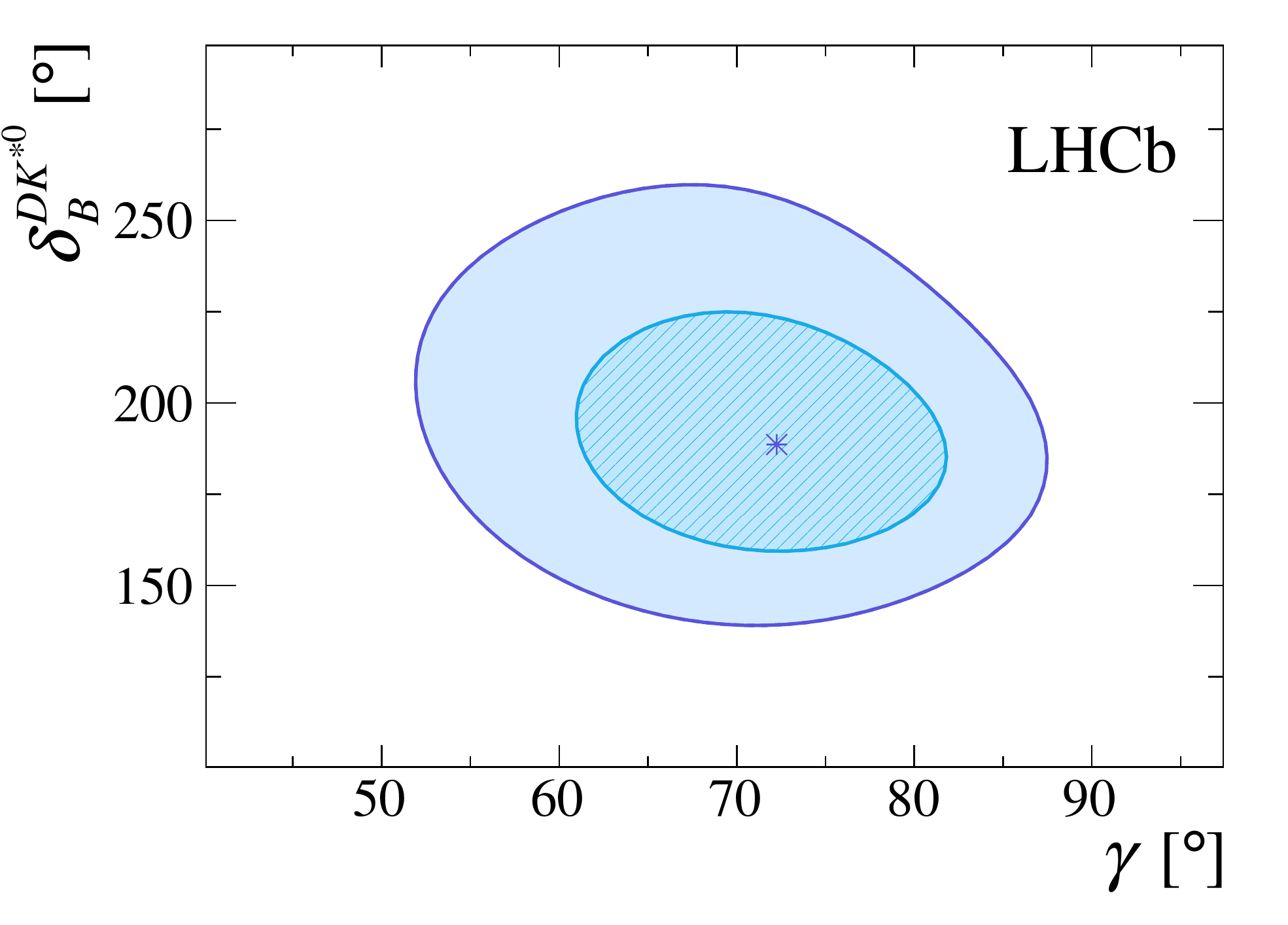}
  \includegraphics[width=.48\textwidth]{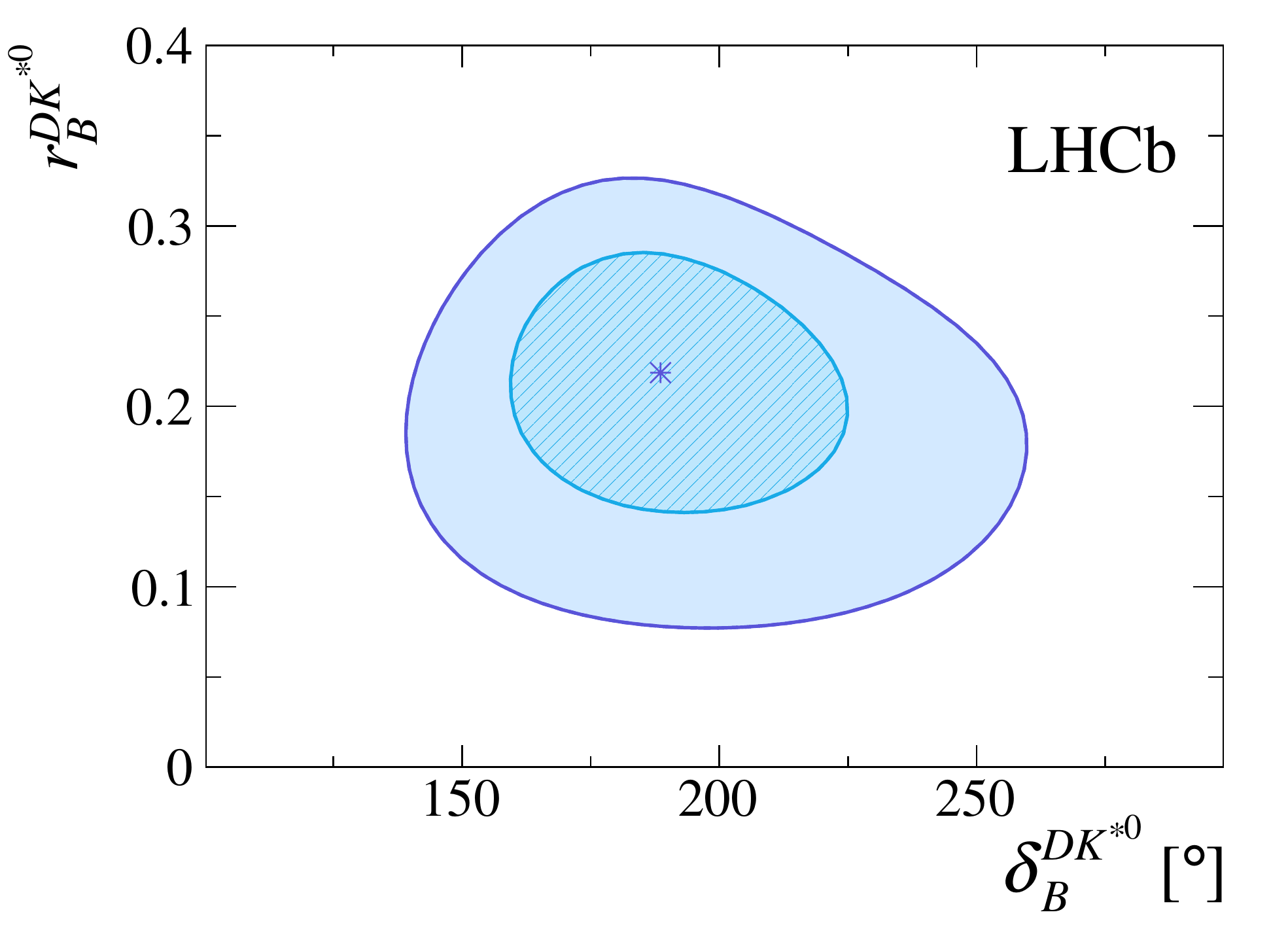}
  \caption{Profile likelihood contours from the \DK combination.
  The contours show the two-dimensional $1\sigma$ and $2\sigma$ boundaries, corresponding to $68.3\%$ and $95.5\%$~CL, respectively.}
  \label{fig:robustcombo2d_1}
\end{figure}

Table~\ref{tab:resultrobust} summarises the resulting central values
and confidence intervals that are obtained from five separate one-dimensional \plugin scans for
the parameters: \g, \rb, \db, \rbDKstz and \dbDKstz. These are shown in Fig.~\ref{fig:resultrobust}.
Due to computational constraints the two-dimensional contours, shown in Fig.~\ref{fig:robustcombo2d_1}, are obtained via the profile
likelihood method in which the value of the test statistic itself ($\Delta\chisq$) is used.
Except for the coverage, as described in Sec.~\ref{sec:coverage},
this is verified to be a good approximation of the \plugin method.
The parameter correlations obtained from the profile likelihood method are given in Appendix~\ref{sec:app_dk_corr}.


\begin{table}[!h]
\centering
\caption{Confidence intervals and central values for the
parameters of interest in the frequentist \DK combination.}
\label{tab:resultrobust}
\renewcommand{\arraystretch}{1.4}
\begin{tabular}{p{2cm}cccc}
\hline
Observable & Central value & 68.3\% Interval & 95.5\% Interval & 99.7\% Interval \\
\hline
$\gamma\, (^{\circ})$	  & \gRobustCentral	      & \gRobustOnesig	      & \gRobustTwosig        & \gRobustThreesig       \\
$\rb$			            & \rbRobustCentral	    & \rbRobustOnesig	      & \rbRobustTwosig       & \rbRobustThreesig      \\
$\db (^{\circ})$	    & \dbRobustCentral	    & \dbRobustOnesig	      & \dbRobustTwosig       & \dbRobustThreesig      \\
$\rbDKstz$			      & \rbDKstzRobustCentral	& \rbDKstzRobustOnesig	& \rbDKstzRobustTwosig  & \rbDKstzRobustThreesig \\
$\dbDKstz (^{\circ})$	& \dbDKstzRobustCentral	& \dbDKstzRobustOnesig	& \dbDKstzRobustTwosig  & \dbDKstzRobustThreesig \\
\hline
\end{tabular}
\end{table}


\subsection{\boldmath \Dh combination}
\label{sec:results_dh}

The \Dh combination includes observables measured from \mbox{\BuDpi} and \mbox{\BuDpipipi} decays, in addition to those measured in the \DK combination, for a total of
89 observables and 38 parameters. The goodness of fit calculated from the \chisq is $p=72.9\%$ and calculated from the pseudoexperiments
is $p=(71.4\pm0.3)\%$.

\begin{figure}
  \centering
  \includegraphics[width=.48\textwidth]{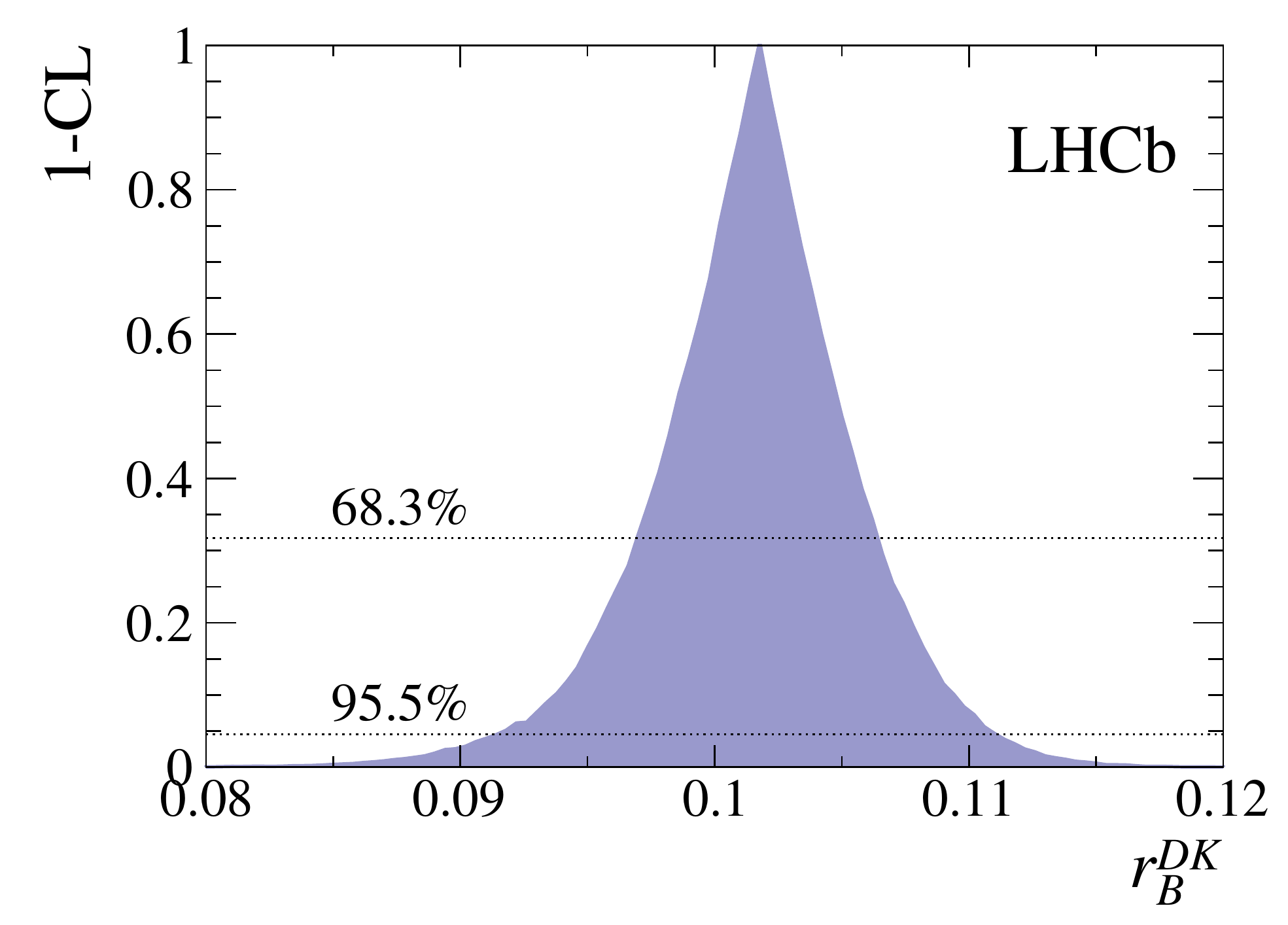}
  \includegraphics[width=.48\textwidth]{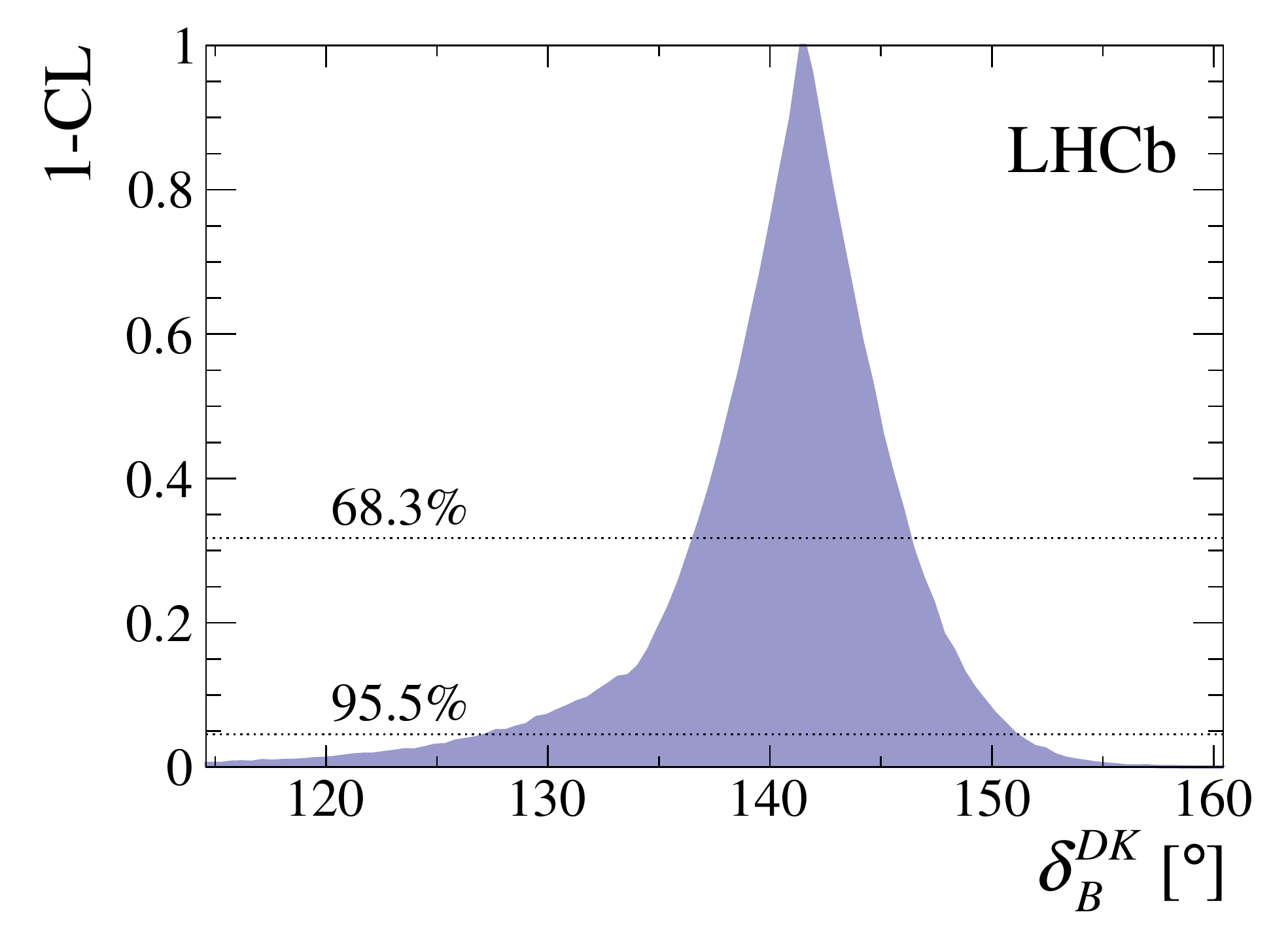}
  \includegraphics[width=.48\textwidth]{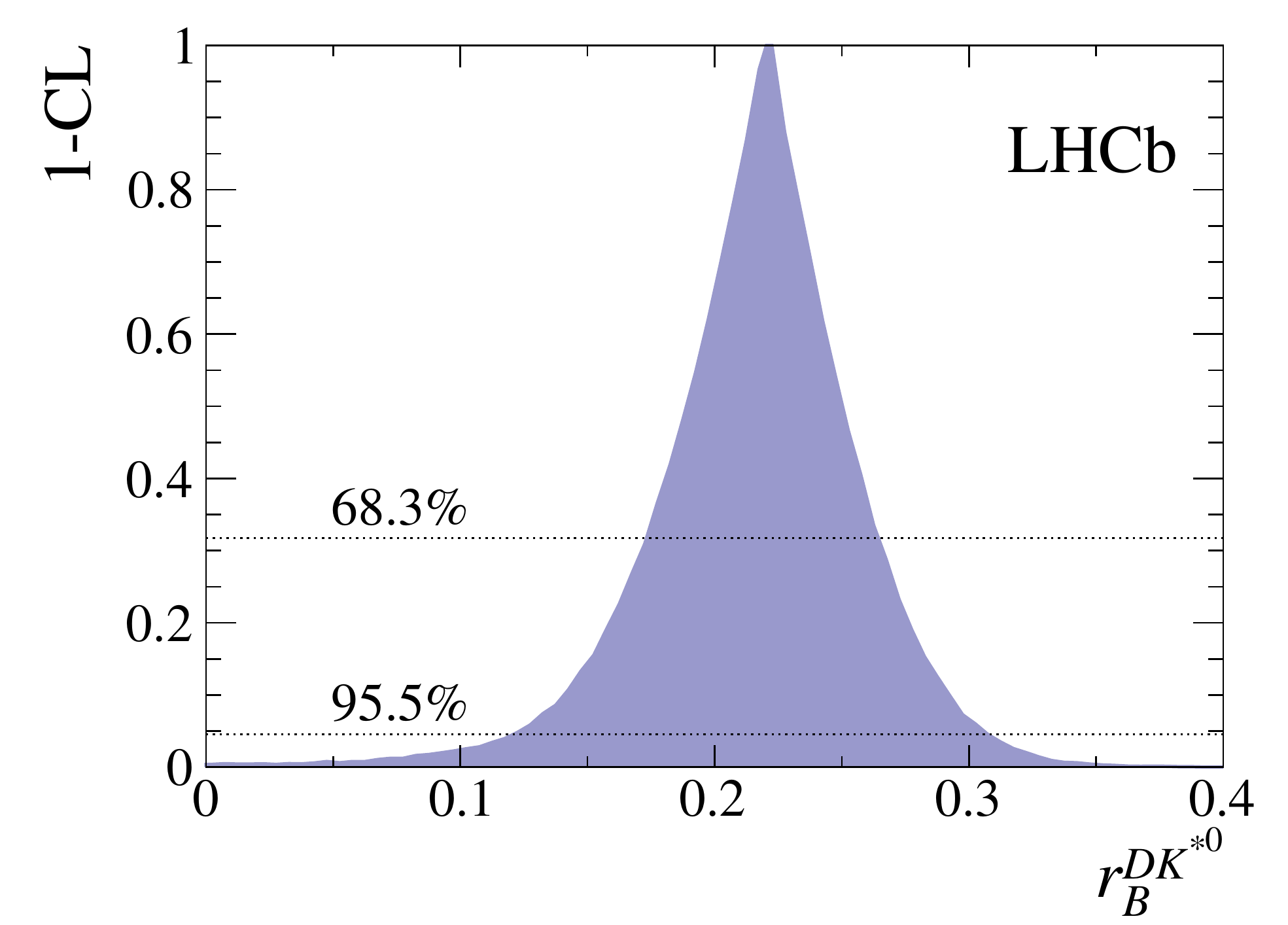}
  \includegraphics[width=.48\textwidth]{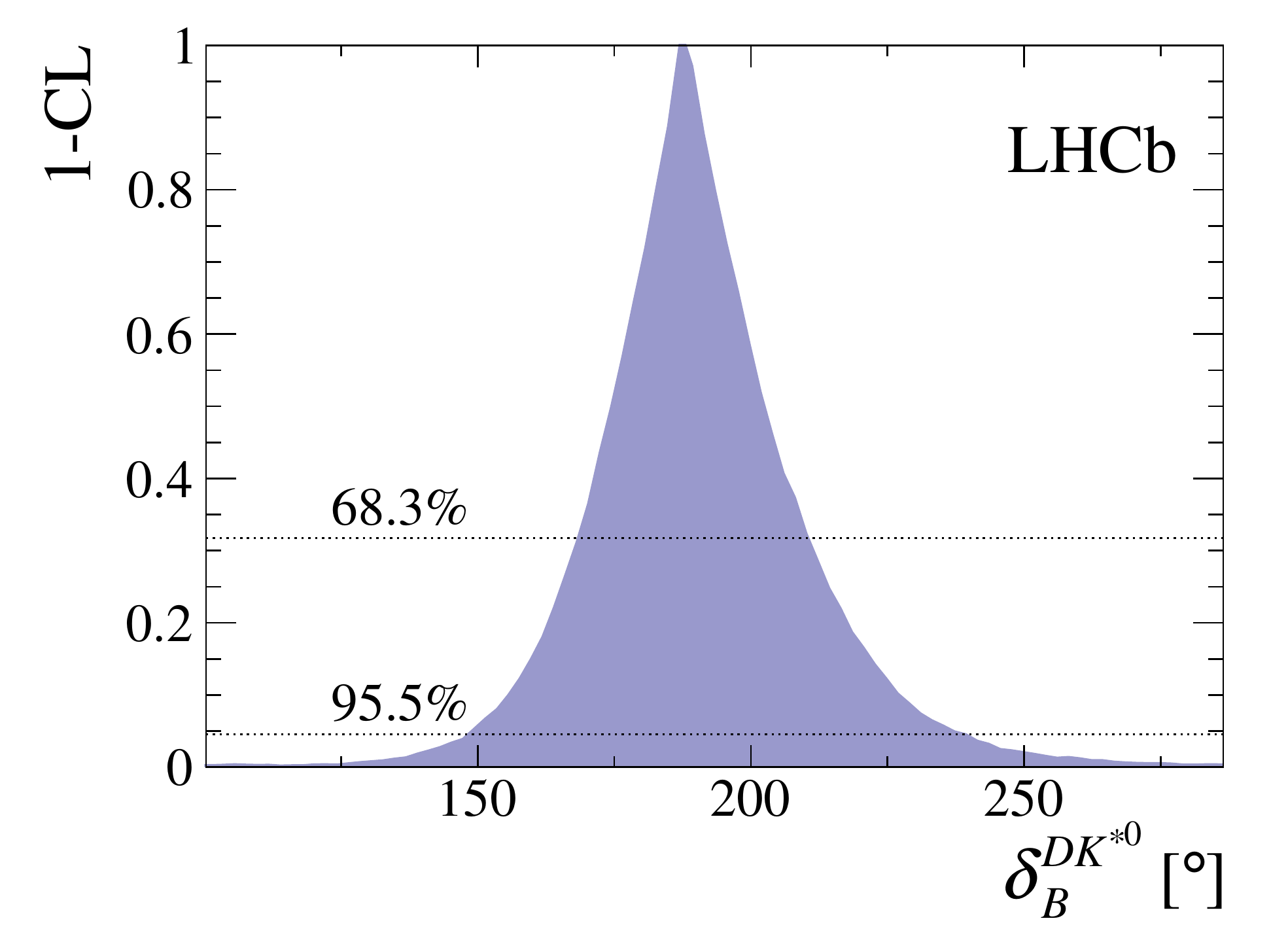}
  \includegraphics[width=.48\textwidth]{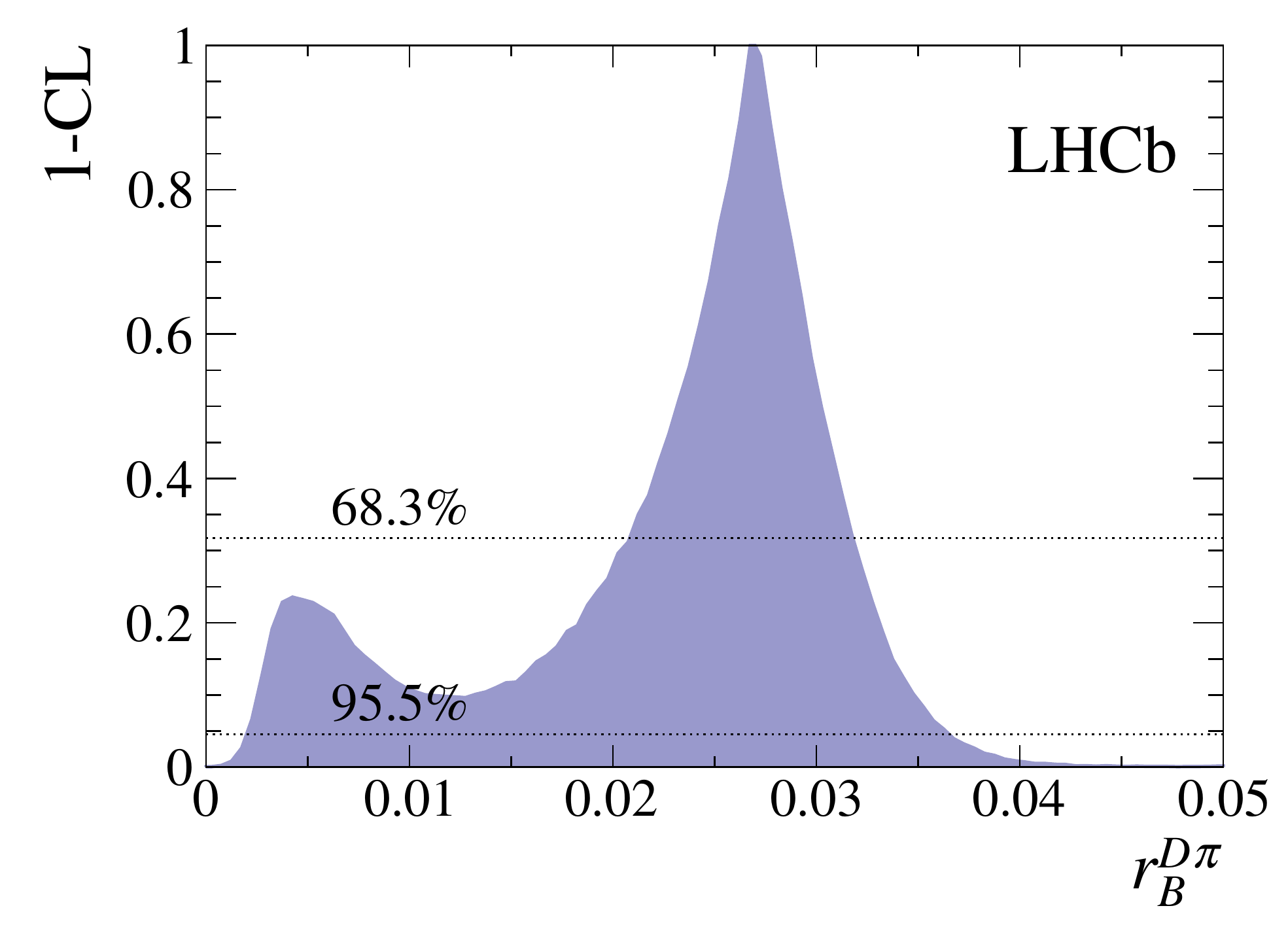}
  \includegraphics[width=.48\textwidth]{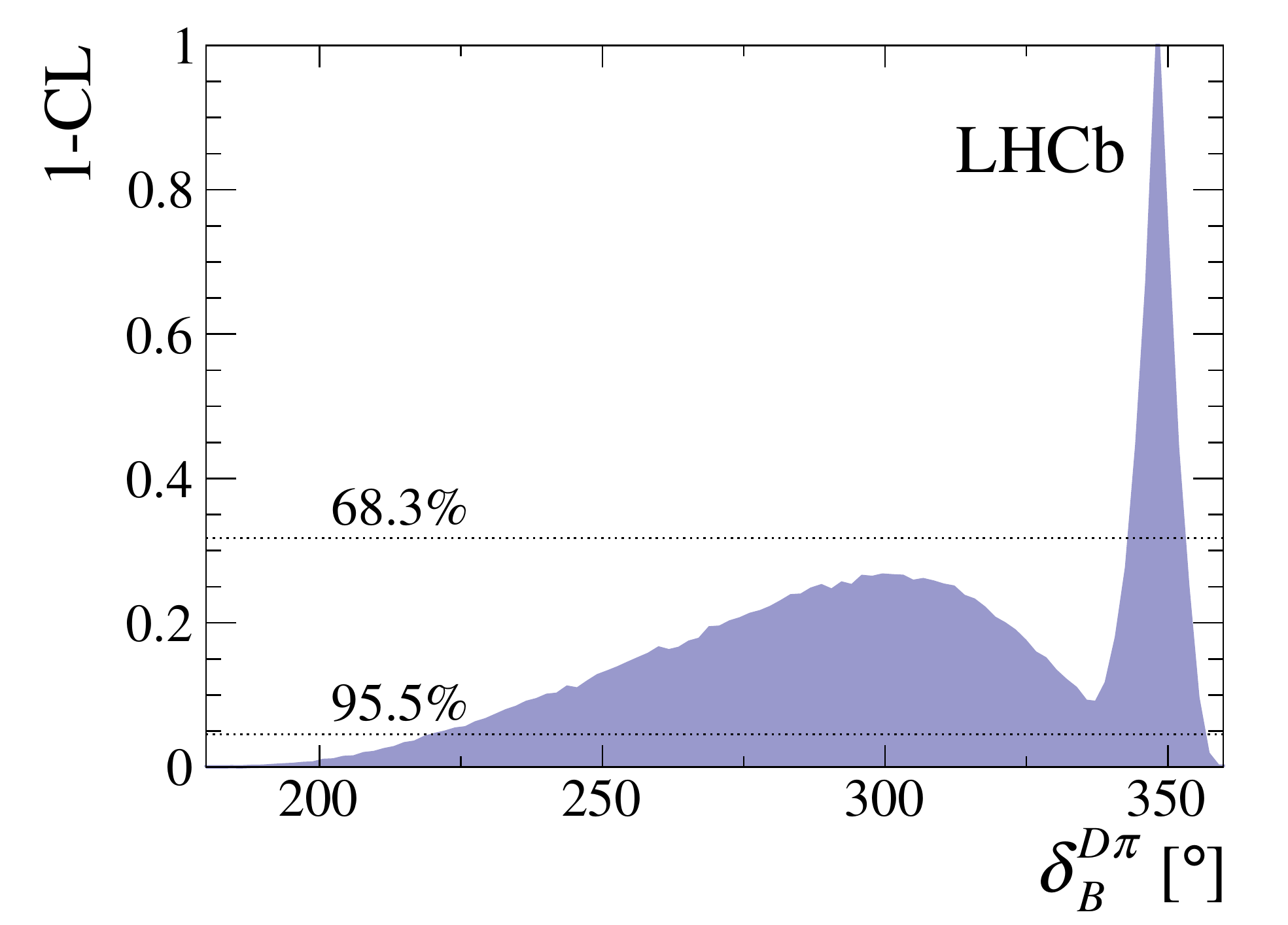}
  \includegraphics[width=.48\textwidth]{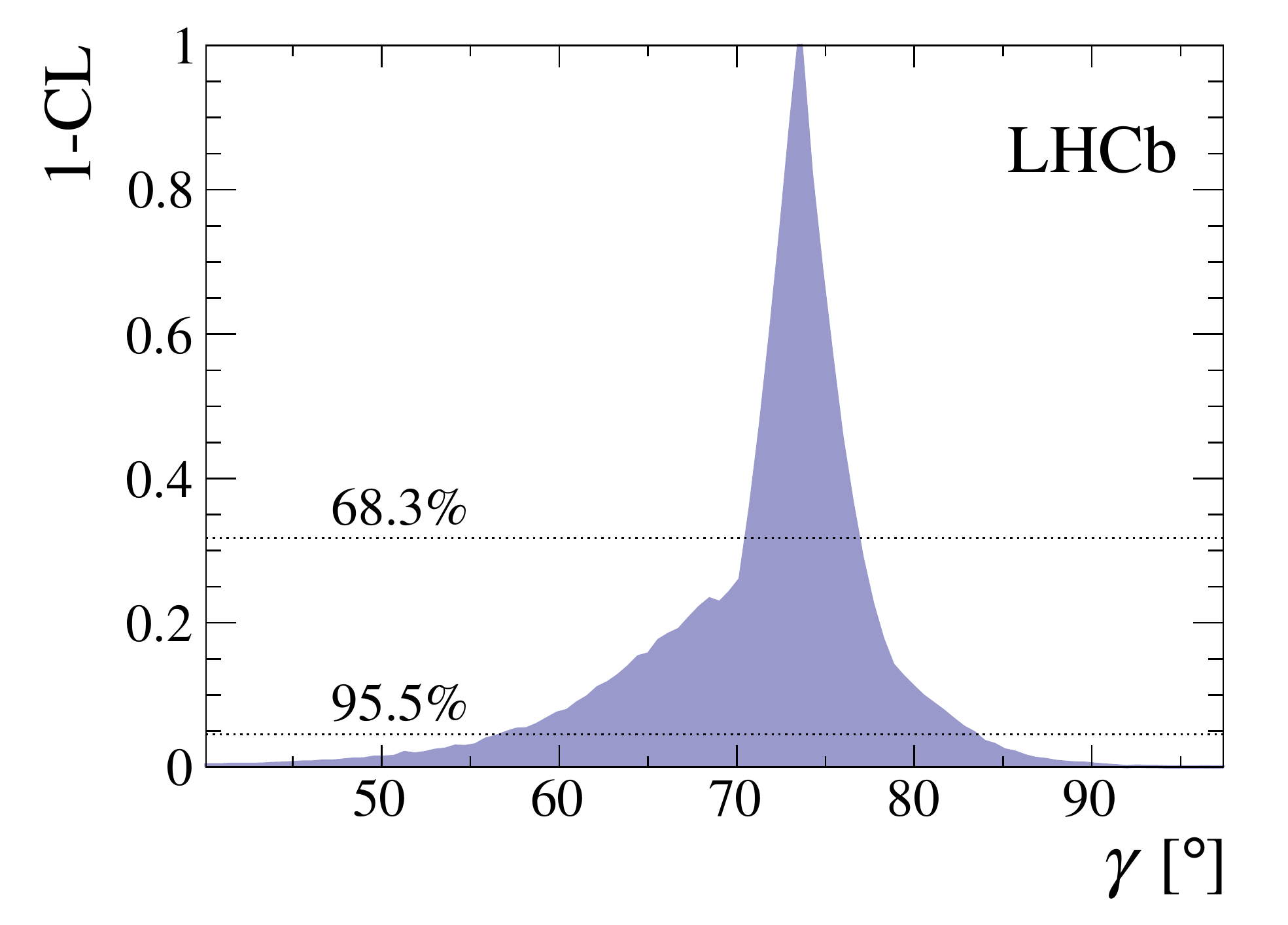}
  \caption{\omcl curves for the \Dh combination obtained with the \plugin method.
The $1\sigma$ and $2\sigma$ levels are indicated by the horizontal dotted lines.
  \vspace{2cm}
  }
  \label{fig:resultdh}
\end{figure}

\begin{figure}
  \centering
  \includegraphics[width=.40\textwidth]{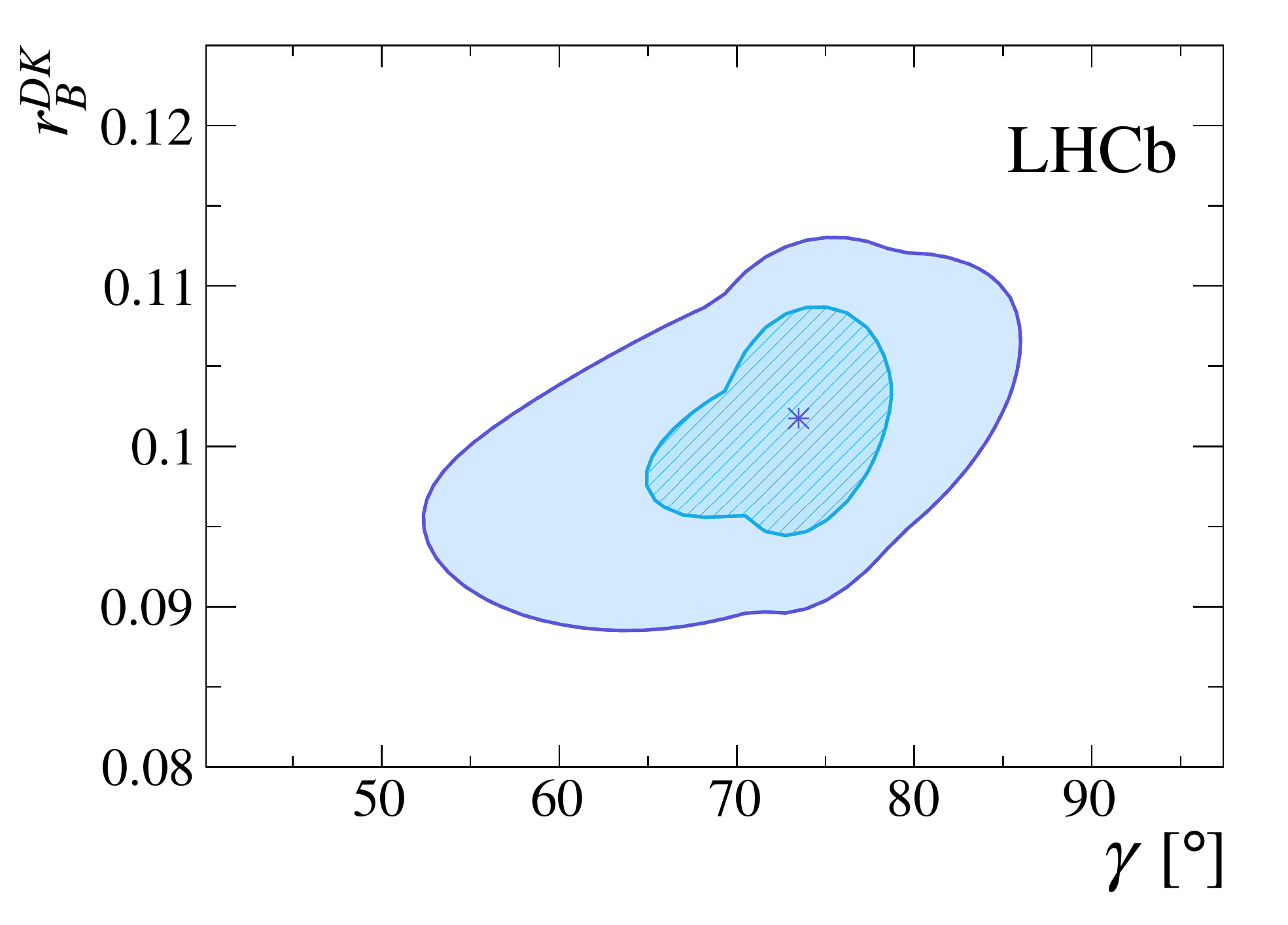}
  \includegraphics[width=.40\textwidth]{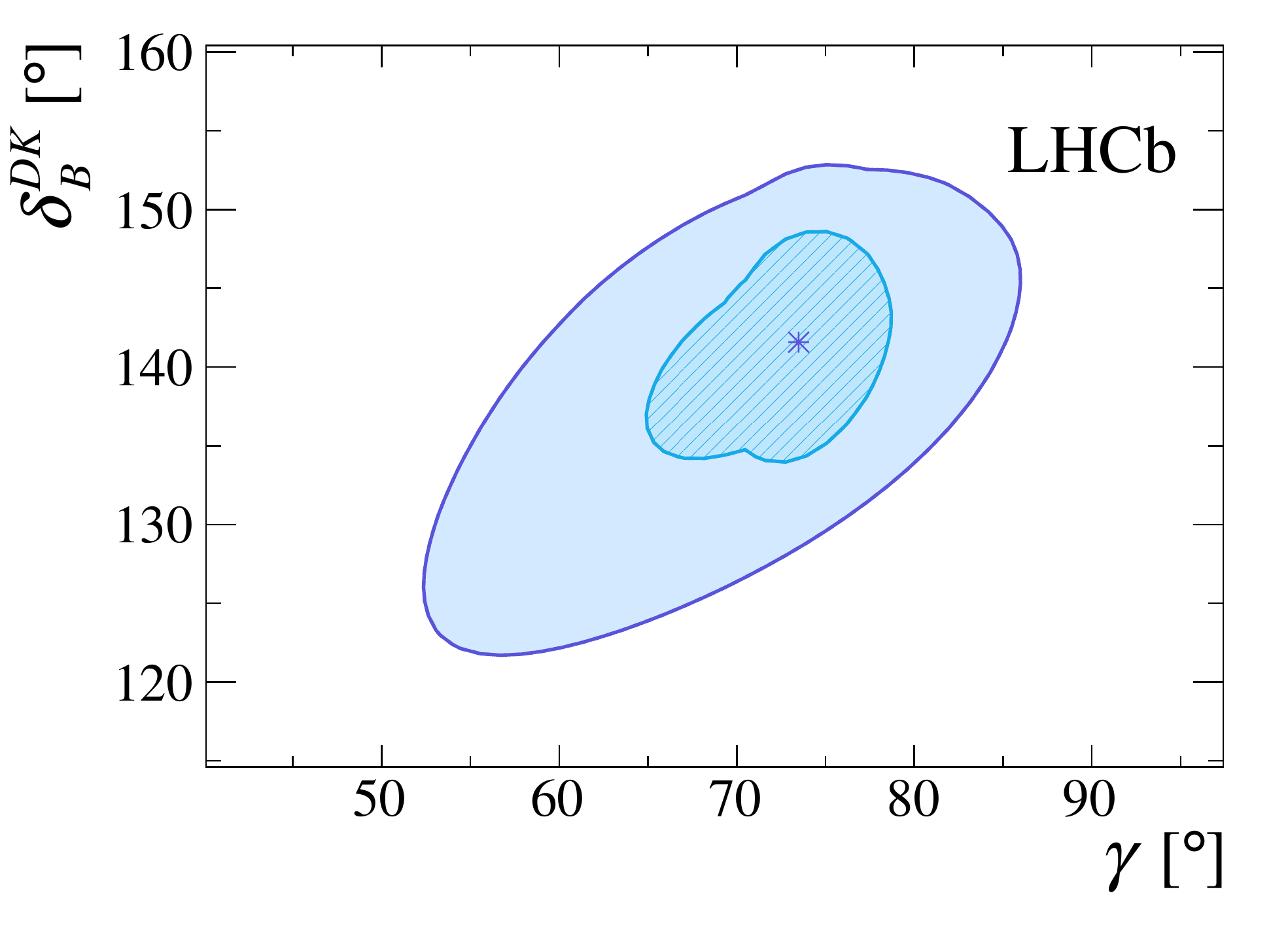}
  \includegraphics[width=.40\textwidth]{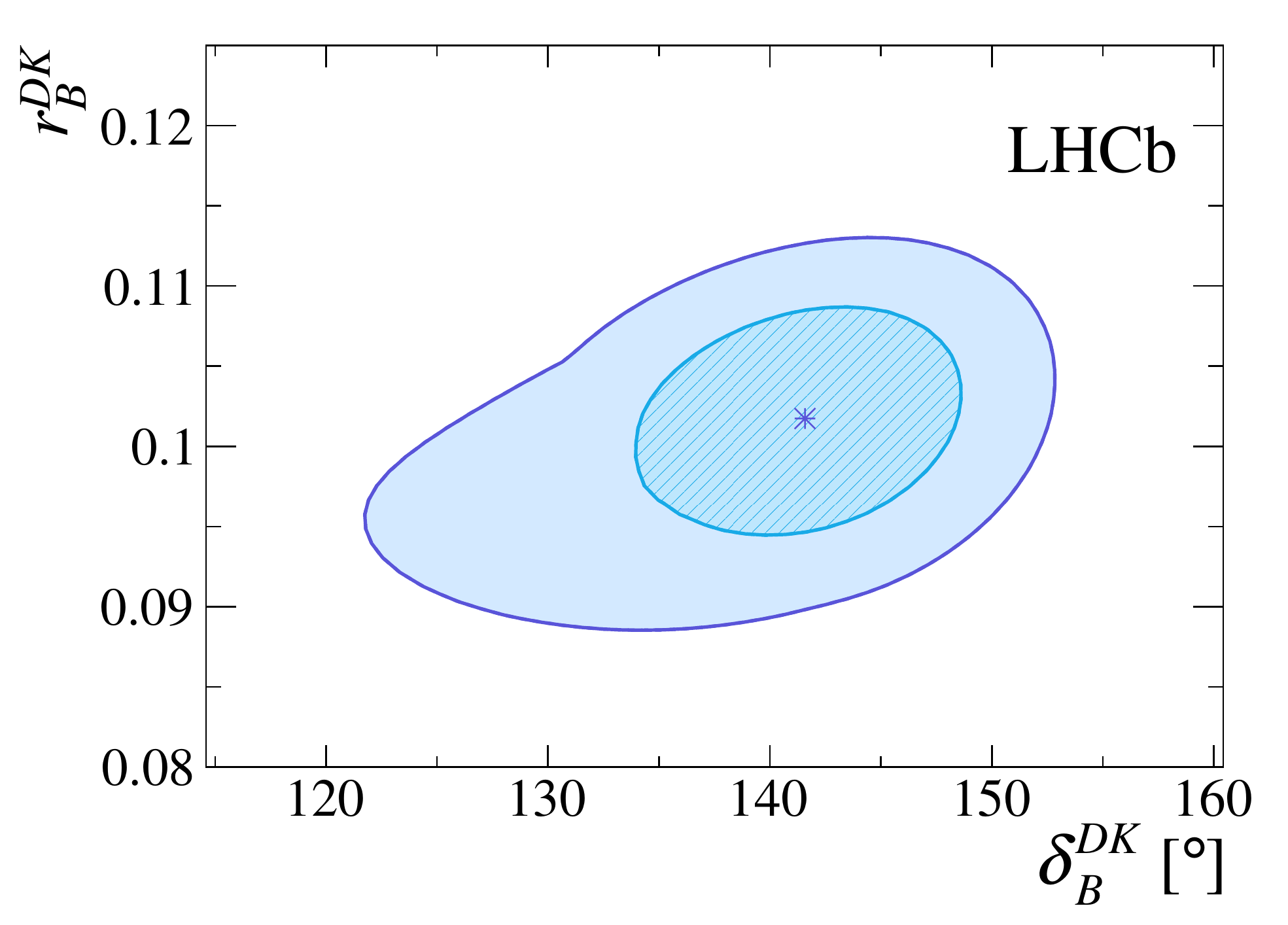}
  \includegraphics[width=.40\textwidth]{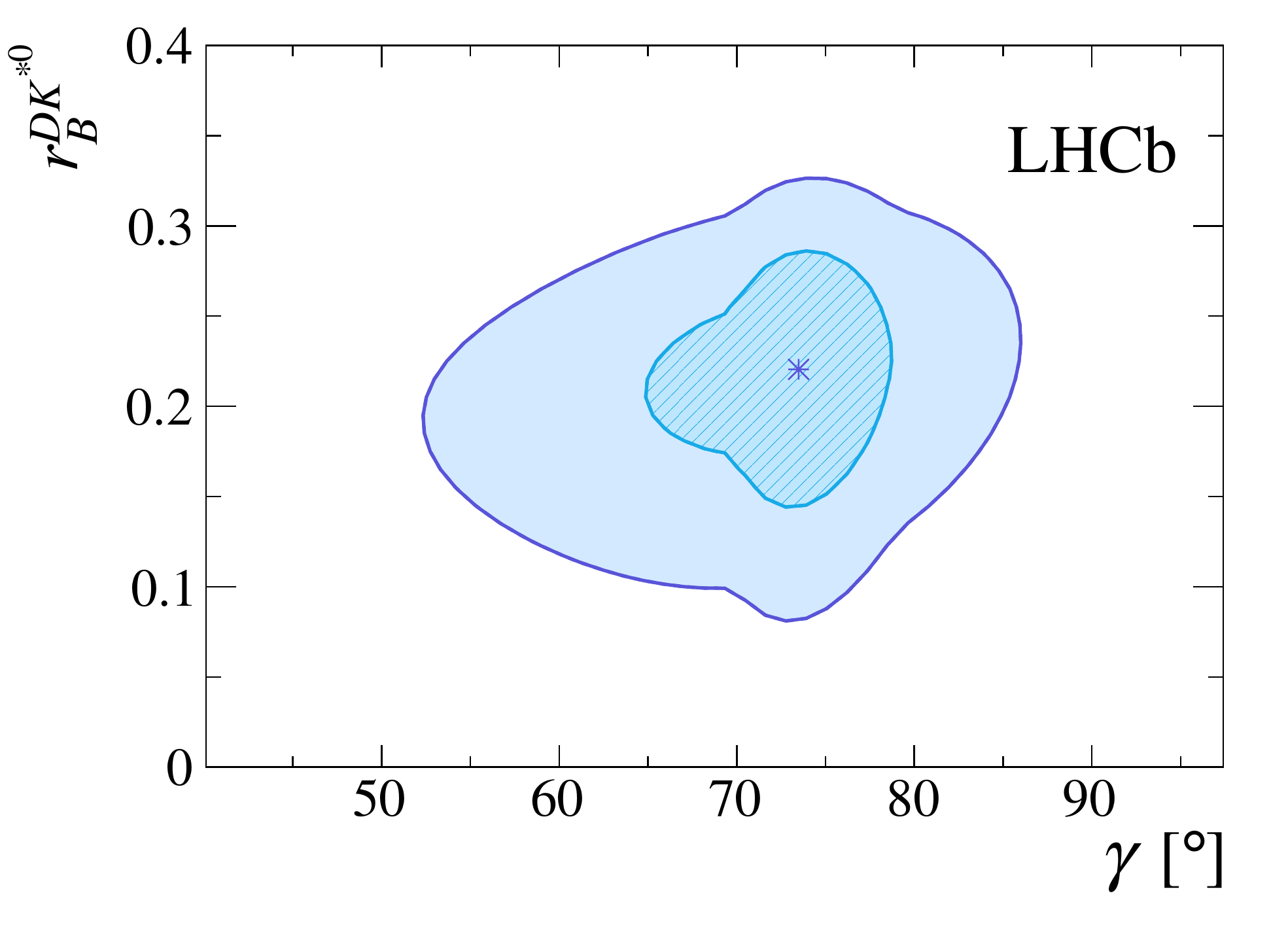}
  \includegraphics[width=.40\textwidth]{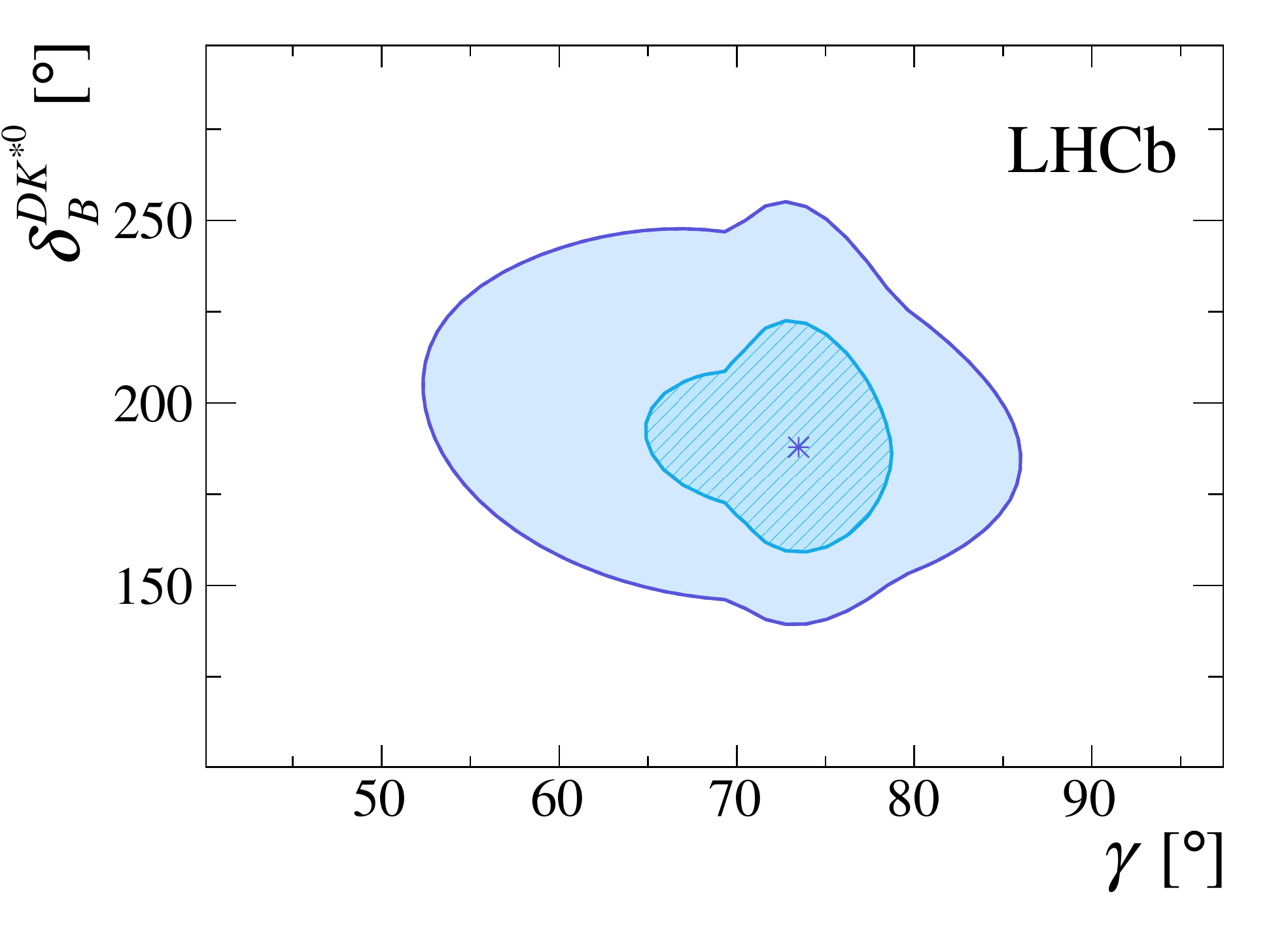}
  \includegraphics[width=.40\textwidth]{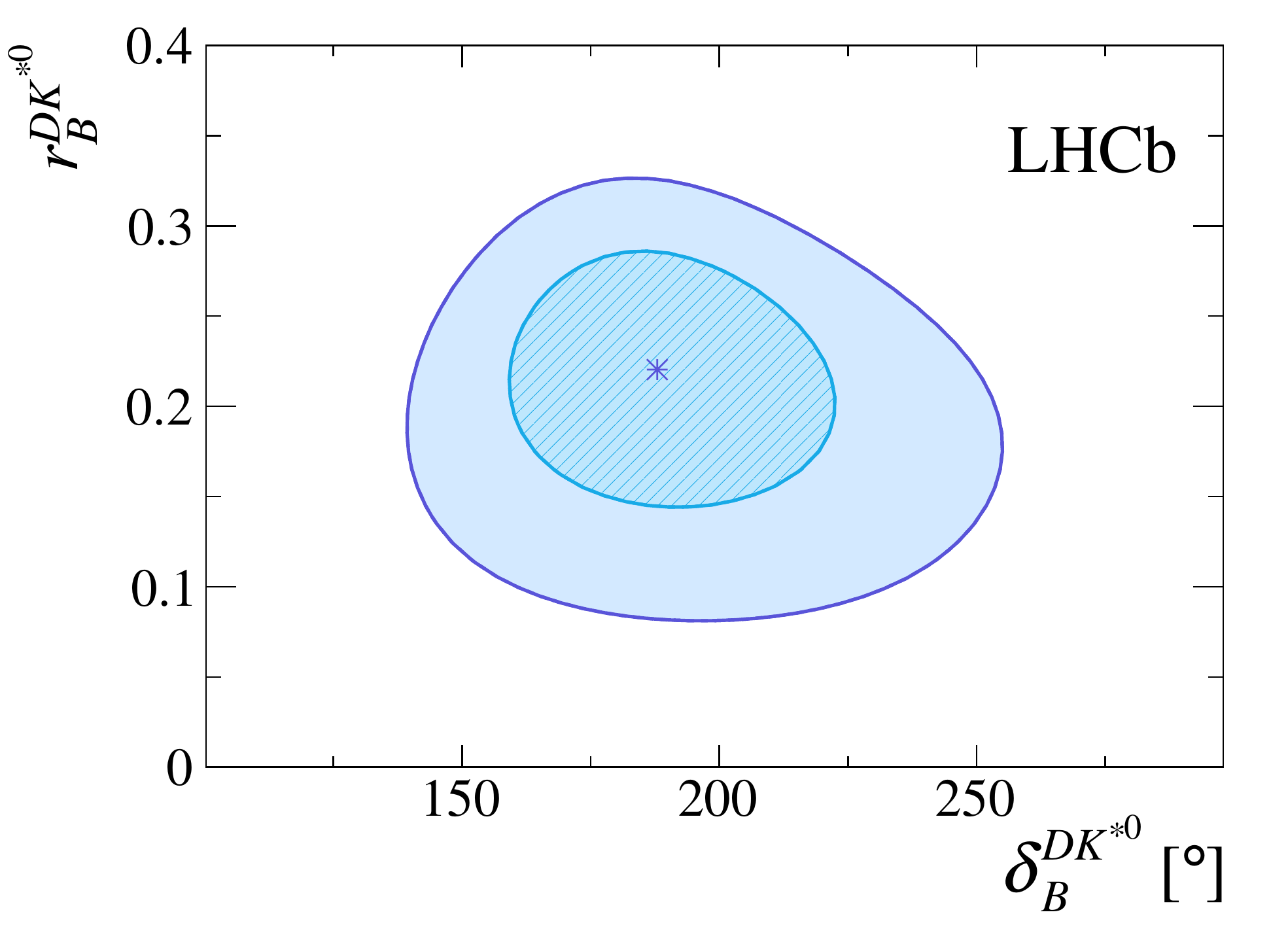}
  \includegraphics[width=.40\textwidth]{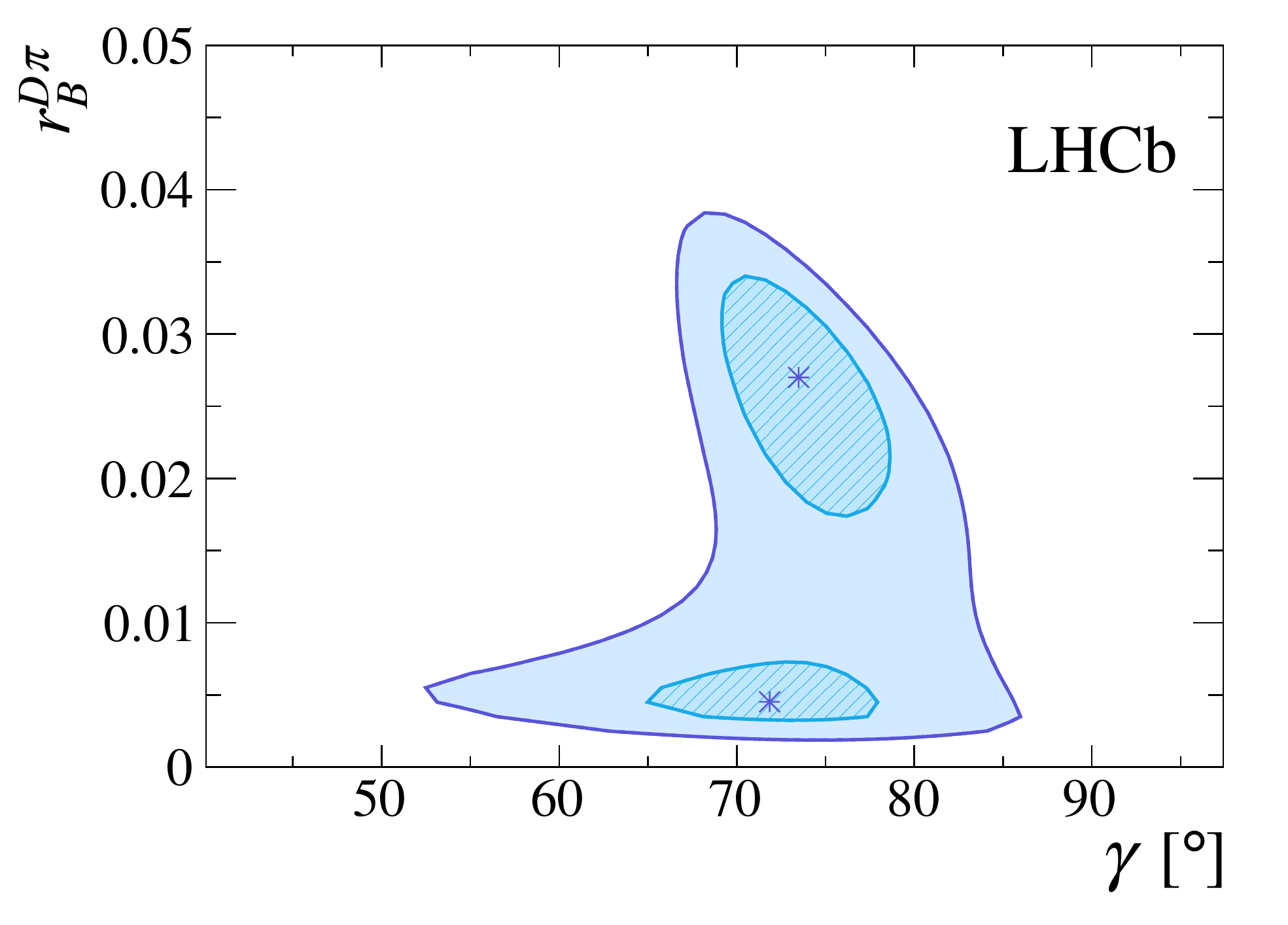}
  \includegraphics[width=.40\textwidth]{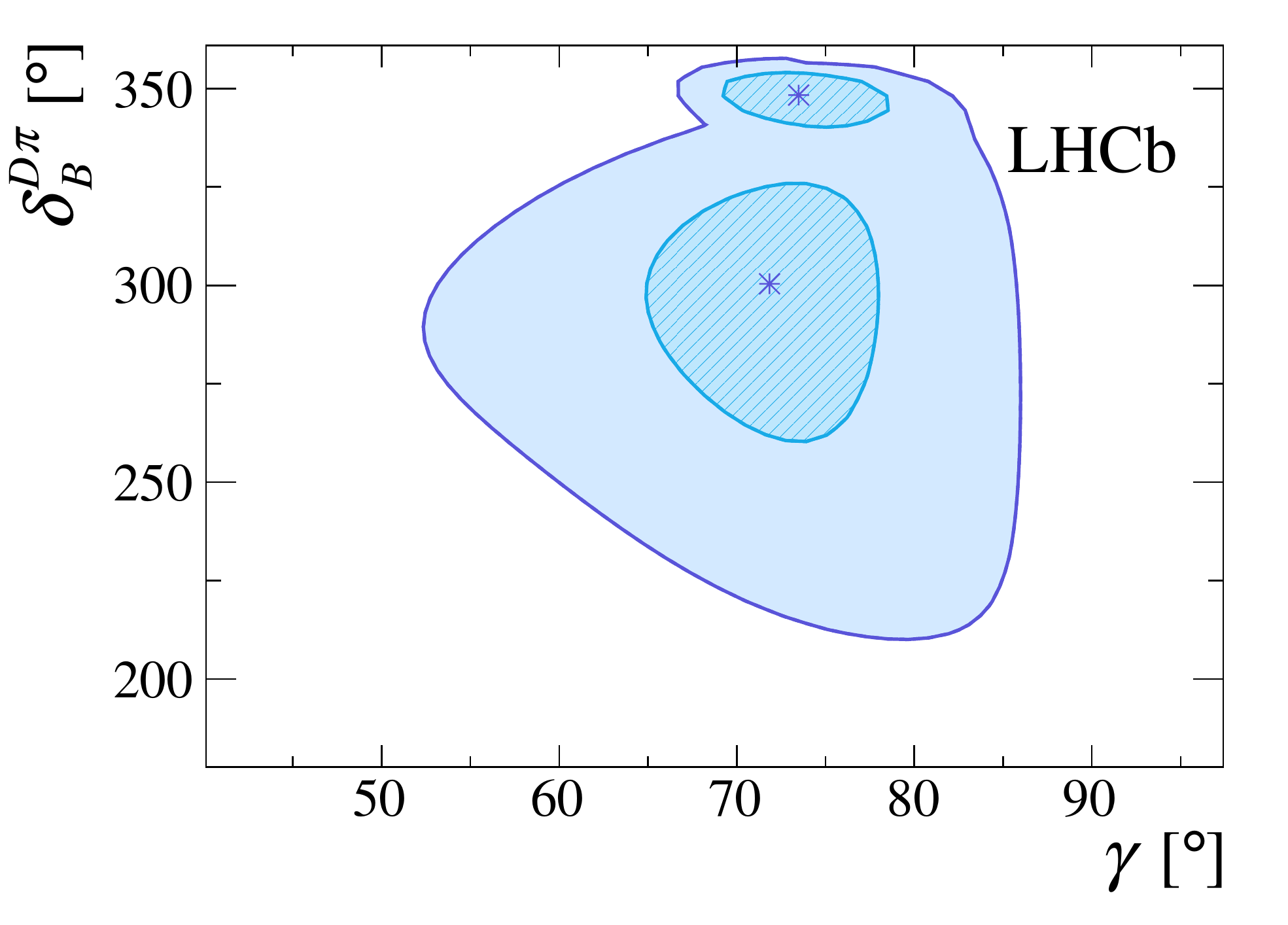}
  \includegraphics[width=.40\textwidth]{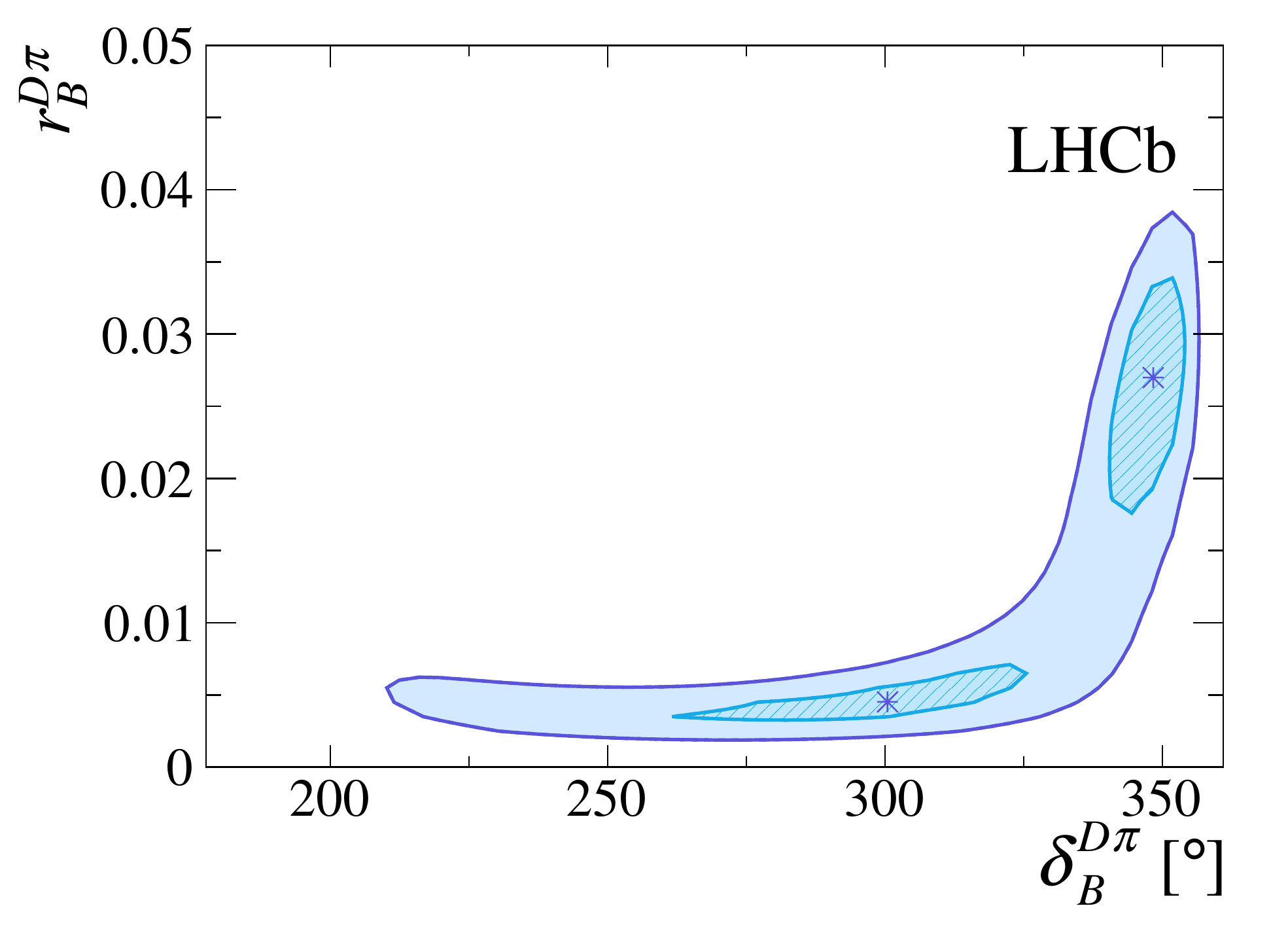}
  \caption{Profile likelihood contours from the \Dh combination.
  The contours show the two-dimensional $1\sigma$ and $2\sigma$ boundaries, corresponding to $68.3\%$ and $95.5\%$~CL, respectively.}
  \label{fig:resultdh_2d}
\end{figure}

Table~\ref{tab:resultdh} gives the results of the one-dimensional \plugin scans for \g, \rbpi, \dbpi, \rb, \db, \rbDKstz and \dbDKstz.
The scans are shown in Fig.~\ref{fig:resultdh}.
Two solutions are found, corresponding to \rbpi values of $0.027$ and $0.0045$ for the favoured and secondary solutions, respectively.
Figure~\ref{fig:resultdh} shows that the secondary solution is suppressed by slightly more than $1\sigma$.
Consequently, the $1\sigma$ interval for $\g$ is very narrow because the uncertainty scales inversely with the central value of \rbpi.
As with the \DK combination, the two-dimensional scans are performed using the profile likelihood method and are shown in Fig.~\ref{fig:resultdh_2d}.
The two solutions and the non-Gaussian contours are clearly visible.
The parameter correlations obtained from the profile likelihood method for both solutions are given in Appendix~\ref{sec:app_dh_corr}.
The coverage for the \Dh analysis is examined in Sec.~\ref{sec:coverage}, where it is found that the coverage is slightly low and then
starts to degrade when the true value of \rbpi is less than $0.01$, reaching a minimum around $0.006$, before the behaviour of the
$DK$ combination is recovered at very low values.

\begin{table}
\centering
\caption{Confidence intervals and central values for the
parameters of interest in the frequentist \Dh combination.}
\label{tab:resultdh}
\renewcommand{\arraystretch}{1.4}
\begin{tabular}{p{2cm}cccc}
\hline
Observable & Central value & 68.3\% Interval & 95.5\% Interval & 99.7\% Interval \\
\hline
$\gamma\, (^{\circ})$	  & \gFullCentral	      & \gFullOnesig	      & \gFullTwosig        & \gFullThreesig       \\
$\rb$			            & \rbFullCentral	    & \rbFullOnesig	      & \rbFullTwosig       & \rbFullThreesig      \\
$\db (^{\circ})$	    & \dbFullCentral	    & \dbFullOnesig	      & \dbFullTwosig       & \dbFullThreesig      \\
$\rbDKstz$			      & \rbDKstzFullCentral	& \rbDKstzFullOnesig	& \rbDKstzFullTwosig  & \rbDKstzFullThreesig \\
$\dbDKstz (^{\circ})$	& \dbDKstzFullCentral	& \dbDKstzFullOnesig	& \dbDKstzFullTwosig  & \dbDKstzFullThreesig \\
\rbpi			            & \rbpiFullCentral	  & \rbpiFullOnesig	    & \rbpiFullTwosig     & \rbpiFullThreesig    \\
$\dbpi (^{\circ})$	  & \dbpiFullCentral	  & \dbpiFullOnesig	    & \dbpiFullTwosig     & \dbpiFullThreesig    \\
\hline
\end{tabular}
\end{table}

Recently, attempts have been made to estimate the value of \rbpi using the known branching fractions of $\Bd\to\Dzb\Kz$ and $\Bd\to\Dzb\piz$ decays
and SU(3) symmetry~\cite{Tuning:2016}, predicting a value of $\rbpi=0.0053\pm0.0007$, consistent with the secondary solution observed in the data.
Using this as an additional external input in the \Dh combination gives $\g = (71.8 ^{+7.2}_{-8.6})\degrees$,
which shows that when \rbpi is small the uncertainties on \g are dominated by the $\B\to\D K$ inputs. This behaviour
is similarly reflected by the 95.5\% and 99.7\% confidence intervals for the \Dh combination when no external constraint on \rbpi is used.
The goodness of fit calculated from the \chisq is $p=70.5\%$ and calculated
from pseudoexperiments is $p=(69.7\pm0.6)\%$.

Given the poor expected additional sensitivity from the $B\to D\pi$-like modes, coupled with the highly non-Gaussian $p$-value distribution of the \Dh combination,
and the fact that the coverage of the \Dh combination is low near the expected value of \rbpi (see Sec.~\ref{sec:coverage}), we choose to quote as the nominal result
that of the \DK combination, namely $\g = (\gQuoted)\degrees$.


\subsection{Coverage of the frequentist method}
\label{sec:coverage}

\begin{figure}
    \centering
    \includegraphics[width=.48\textwidth]{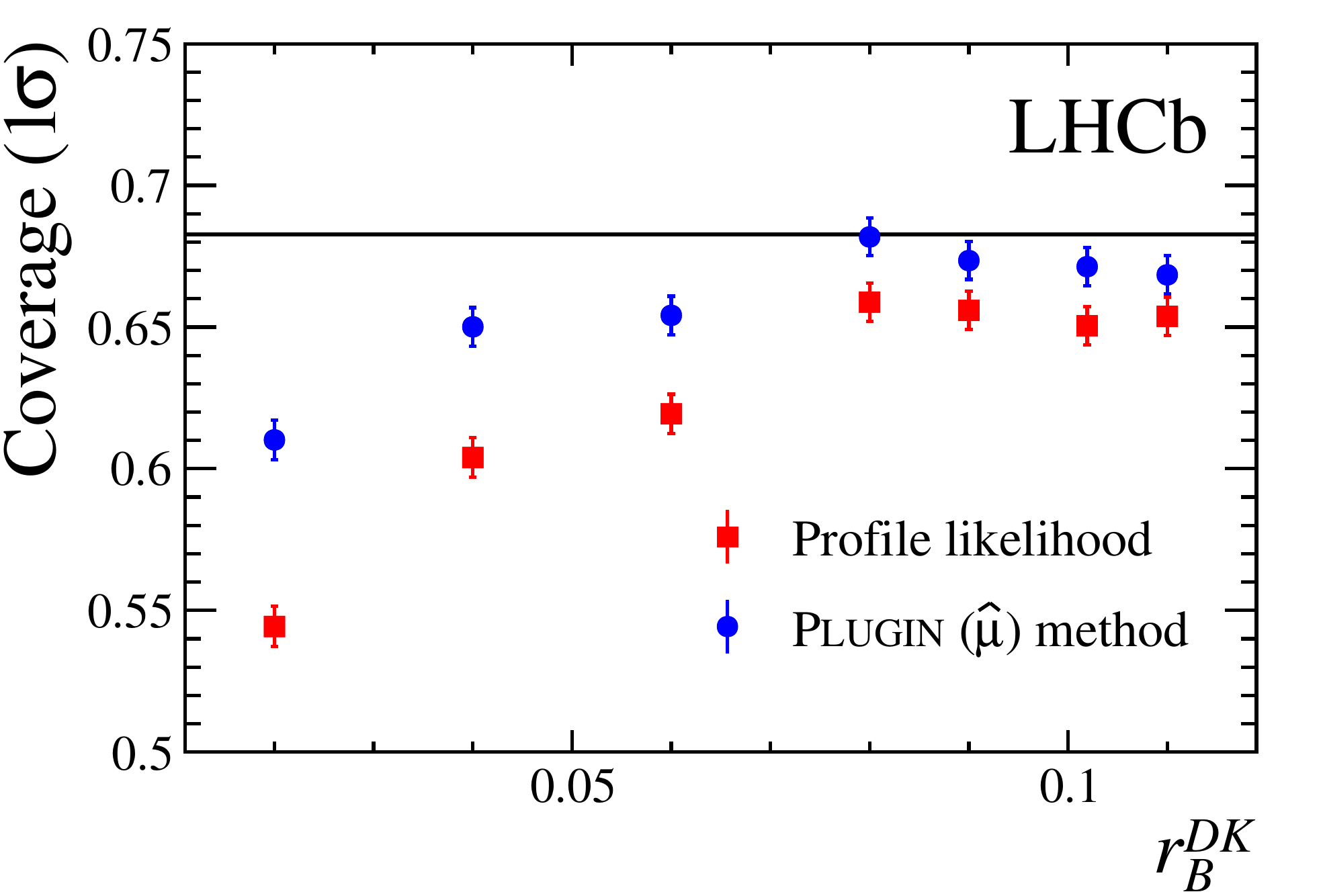}
    \includegraphics[width=.48\textwidth]{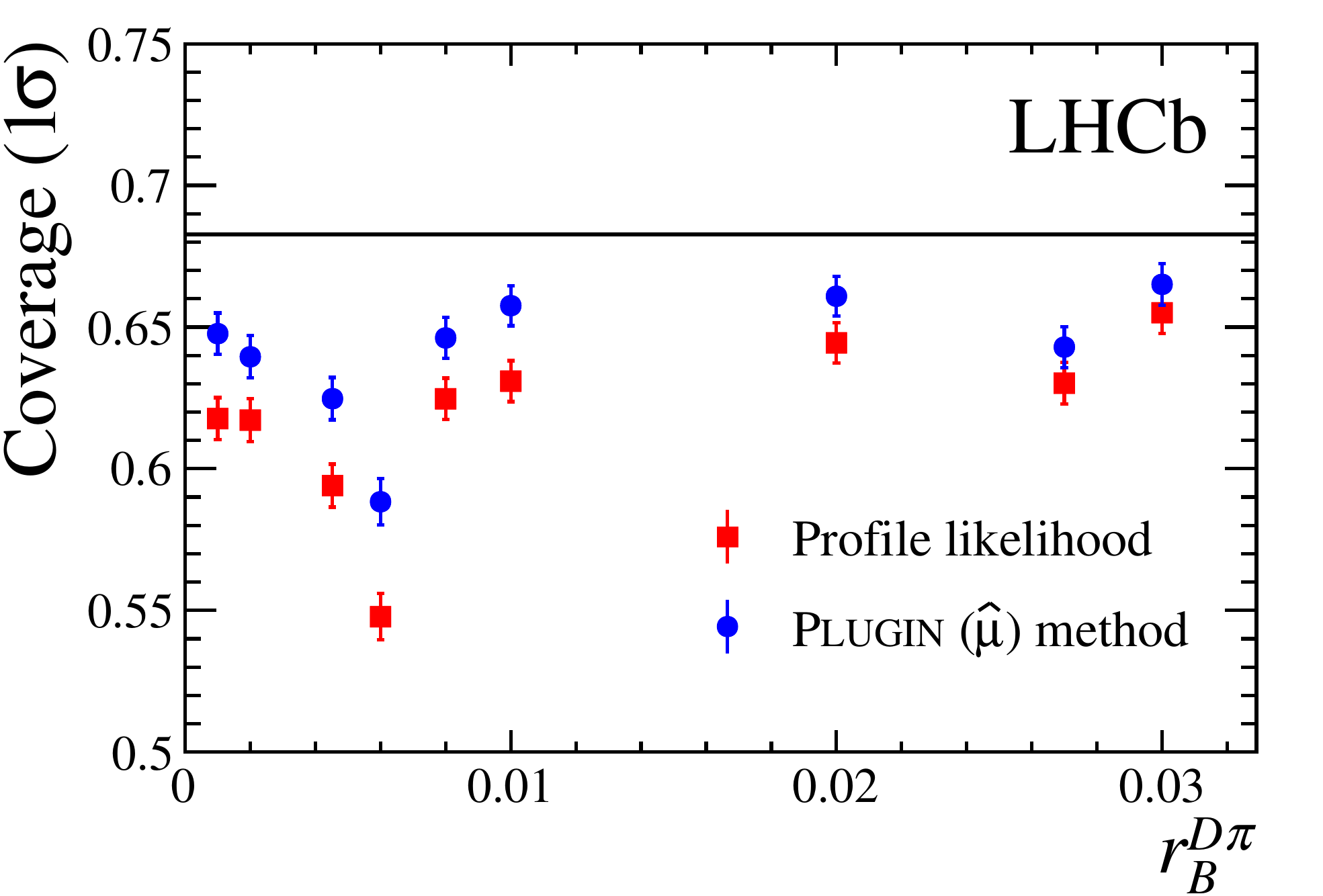}
    \caption{Dependence of the coverage for the one-dimensional \plugin method (blue circles) and the profile likelihood method (red squares) on \rb for the \DK combination (left) and on \rbpi for the \Dh combination (right). The solid horizontal line shows the nominal coverage at $1\sigma$ of 68.3\%.}
    \label{fig:coverage_vs_r}
\end{figure}

The coverage of
the \plugin (and the profile likelihood) method is tested by generating pseudoexperiments and evaluating the fraction
for which the $p$-value is less than that obtained for the data.
In general, the coverage depends on the point in parameter space.
Following the procedure described in Ref.~\cite{Aaij:2013zfa}, the coverage of the
profile likelihood and one-dimensional \plugin method intervals are tested. The
coverage is determined for each method using the same pseudoexperiments; consequently their
uncertainties are correlated.
The results for the best fit points are shown in Table~\ref{tab:coverage}.
Figure~\ref{fig:coverage_vs_r} shows the coverage of the $1\sigma$ intervals as determined from pseudoexperiments for the
\DK (\Dh) combination as a function of the value of \rb (\rbpi) used to generate the pseudoexperiments.
It can be seen that the coverage for the \DK combination
degrades as the true value of \rb gets smaller. This behaviour has previously been observed by the CKMfitter group~\cite{CKMfitter}. The fitted value
found in this combination, $\rb\approx 0.1$, is well within the regime of accurate coverage. The dependence of the coverage for the \Dh combination on \rbpi shows similar behaviour, where the coverage begins to degrade when the true value reaches $\rbpi<0.01$, worsening until
the true value of \rbpi becomes so small that the $D\pi$ modes offer no sensitivity and the behaviour seen in the \DK combination is recovered.
The fitted value of \rbpi in the $Dh$ combination ($\sim 0.03$) falls in the regime with good coverage, whilst the expected value, and indeed the value of the second minimum ($\sim0.005$), is in the regime in which the coverage starts to deteriorate. No correction for under-coverage is applied to the confidence intervals quoted in Tables~\ref{tab:resultrobust} and~\ref{tab:resultdh}.

\begin{table}
\caption{\small
Measured coverage $\alpha$ of the confidence intervals for \g, determined at
the best fit points, for both the one-dimensional \plugin and profile likelihood methods.
The nominal coverage is denoted as $\eta$.}
\label{tab:coverage}
\begin{center}
\begin{tabular}{l | c c c}
\hline
\multicolumn{1}{c}{} & $\eta$ [\%] & $\alpha$ (profile likelihood) [\%] & $\alpha$ (\plugin) [\%]  \\
\hline
\multirow{3}{*}{\DK} &
  $68.3$ & $65.1 \pm 0.7$ & $67.1 \pm 0.7$ \\
& $95.5$ & $93.5 \pm 0.4$ & $94.3 \pm 0.3$ \\
& $99.7$ & $98.7 \pm 0.2$ & $98.8 \pm 0.2$ \\
\hline
\multirow{3}{*}{$Dh$} &
  $68.3$ & $63.0 \pm 0.7$ & $64.3 \pm 0.7$ \\
& $95.5$ & $90.9 \pm 0.4$ & $91.7 \pm 0.4$ \\
& $99.7$ & $95.3 \pm 0.3$ & $95.6 \pm 0.3$ \\
\hline
\end{tabular}
\end{center}
\end{table}

\subsection{Interpretation}
\label{sec:interpretation}

Using the nominal \DK combination and the simple profile likelihood method some further interpretation of the results is presented in this section.
Performing the \DK combination with statistical uncertainties only suggests that the systematic contribution to the uncertainty on \g is approximately $3\degrees$.
Performing the combination without use of the external constraints (described in Sec.~\ref{sec:inputsauxilary}) roughly
doubles the uncertainty on \g, demonstrating the value of including this information.

\begin{figure}
  \centering
  \includegraphics[width=0.48\textwidth]{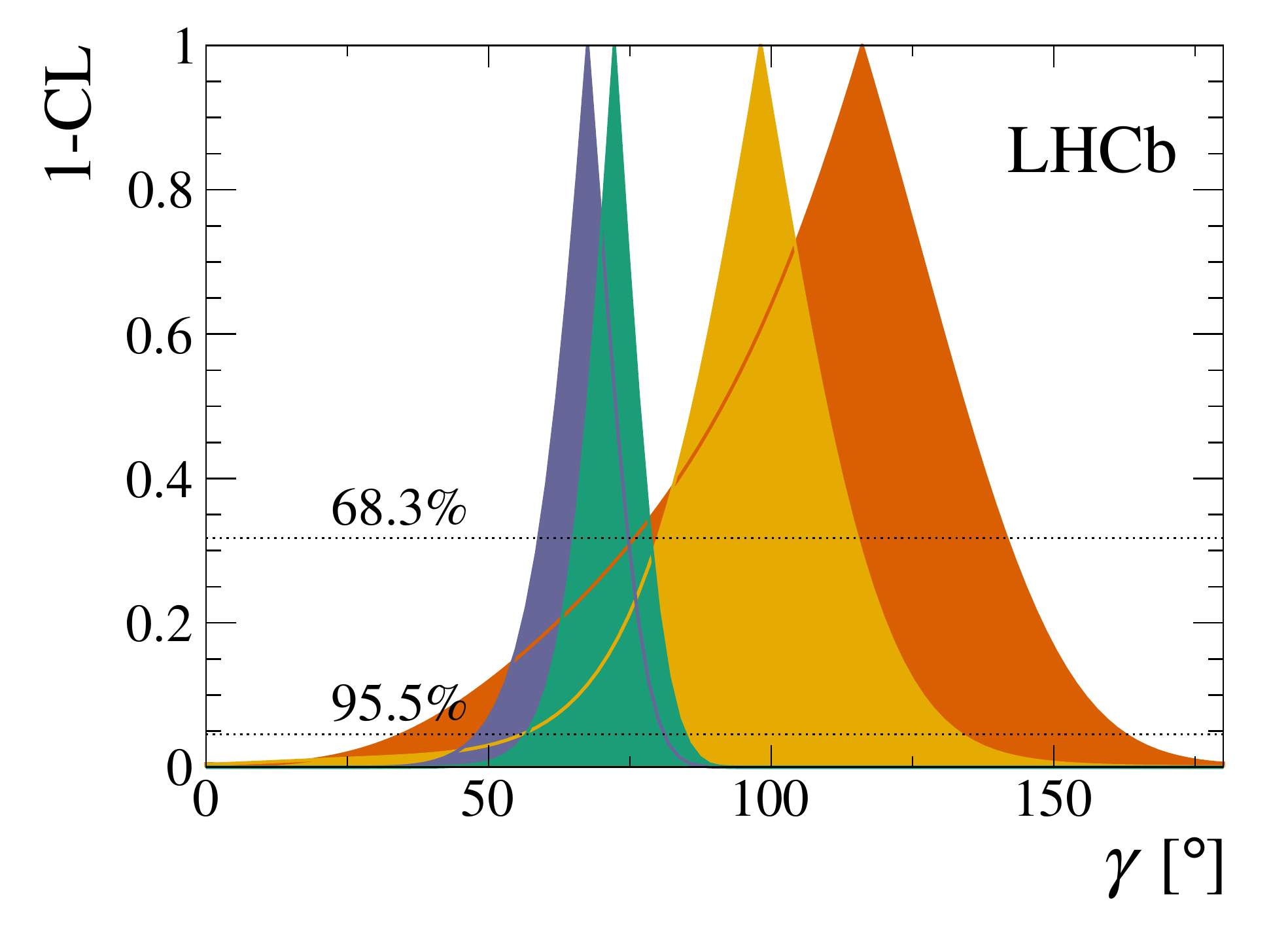}
  \includegraphics[width=0.48\textwidth]{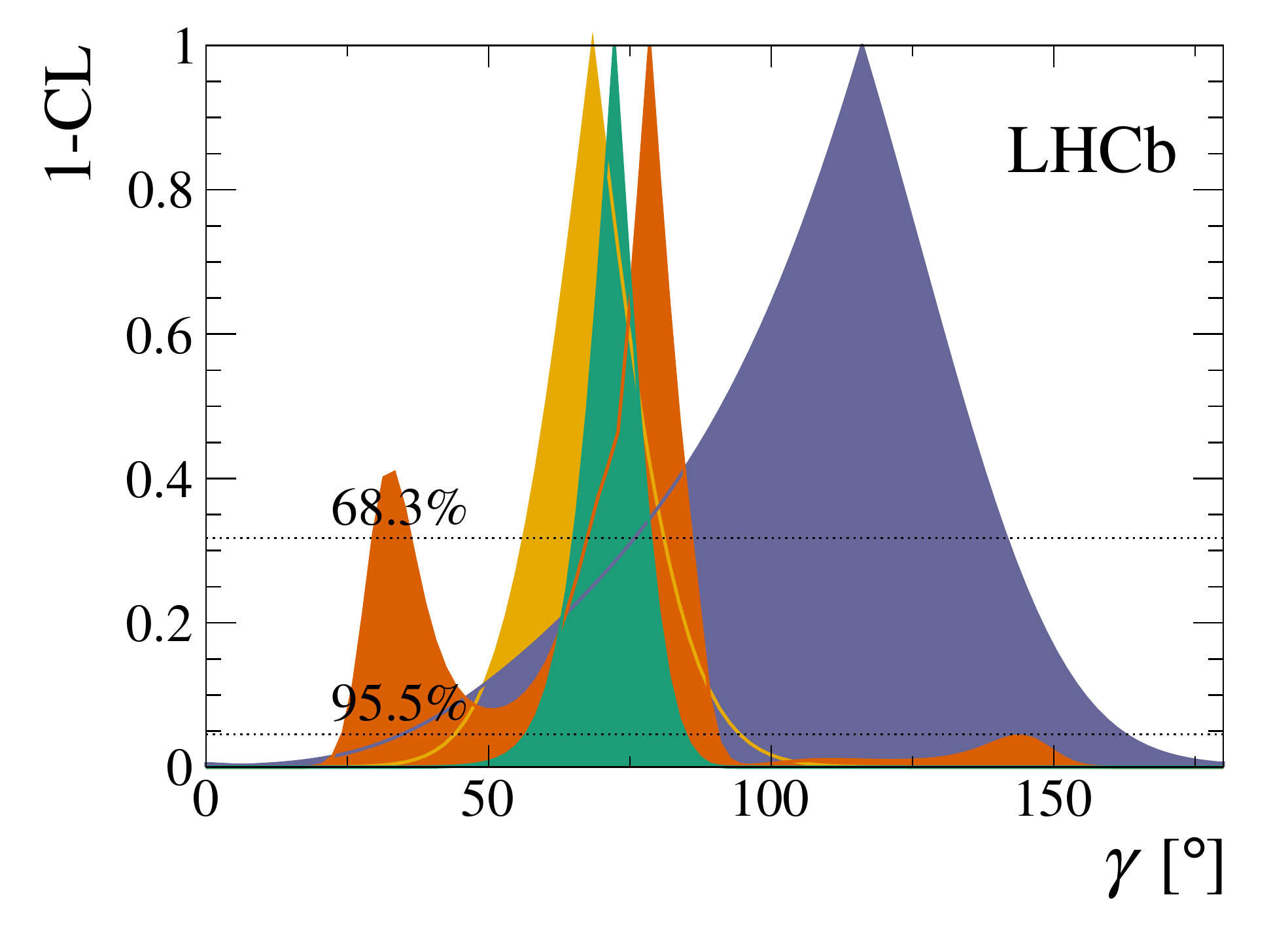} \\
  \vspace{-0.6cm}
  \includegraphics[width=0.2\textwidth]{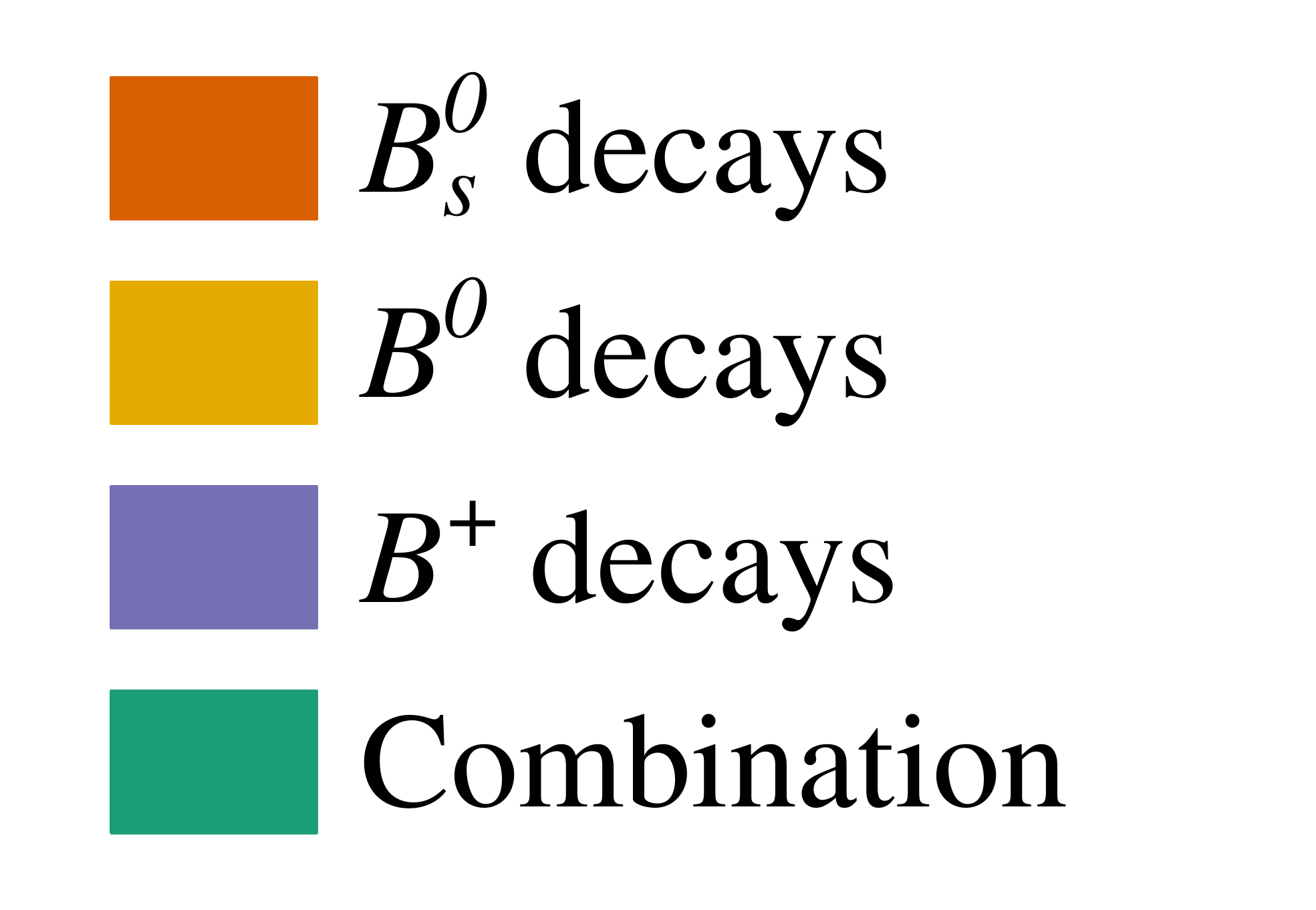}
  \hspace{4cm}
  \includegraphics[width=0.2\textwidth]{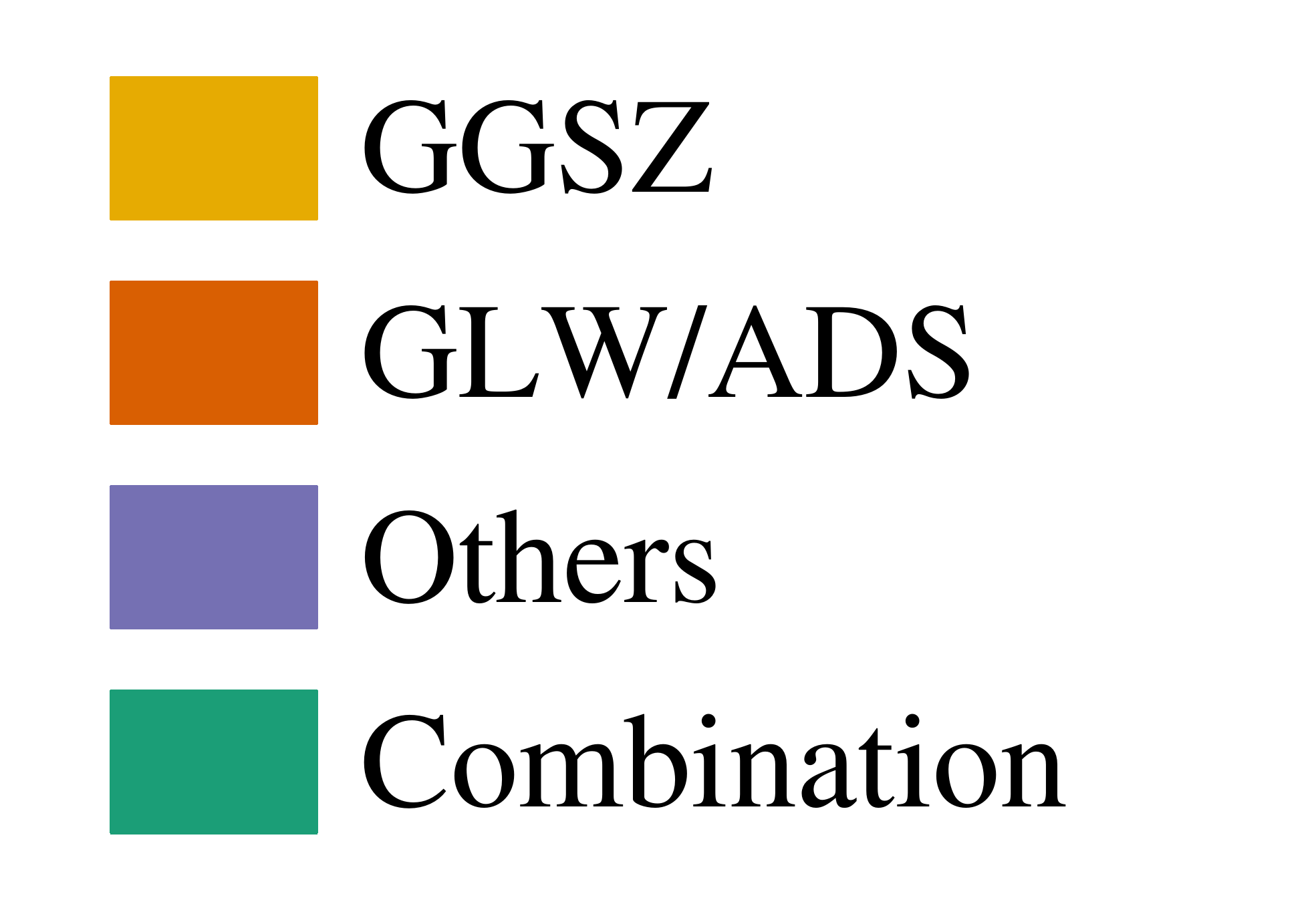}
  \caption{\omcl plots, using the profile likelihood method, for \DK combinations split by the initial $B$ meson flavour (left) and split by analysis method (right). Left: (orange) \Bs initial state, (yellow) \Bd initial states, (blue) \Bu initial states and (green) the full combination. Right: (yellow) GGSZ methods, (orange) GLW/ADS methods, (blue) other methods and (green) the full combination.}
  \label{fig:g_sensitivity}
\end{figure}

\begin{figure}
  \begin{center}
    \includegraphics[width=0.48\textwidth]{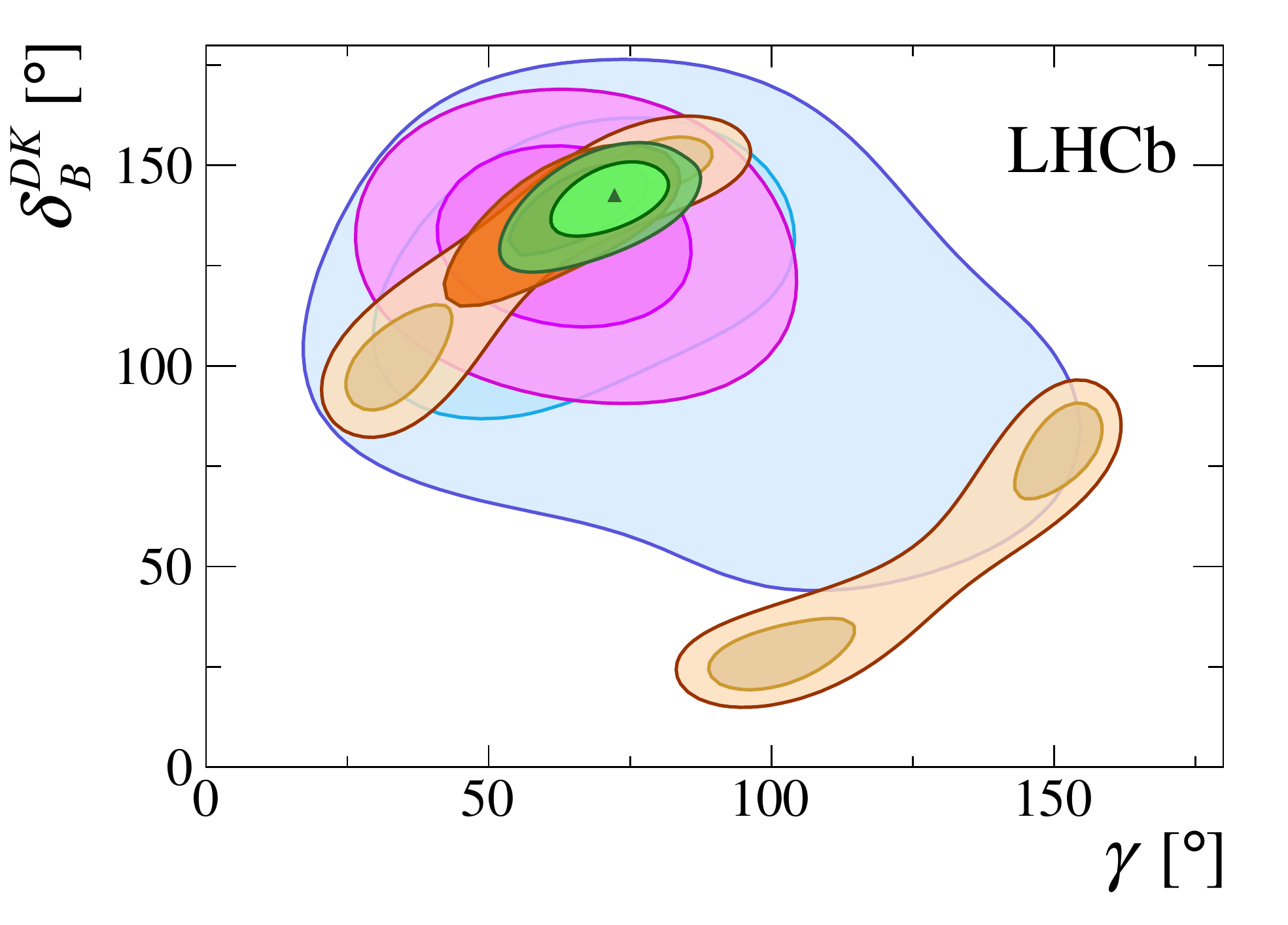}
    \includegraphics[width=0.48\textwidth]{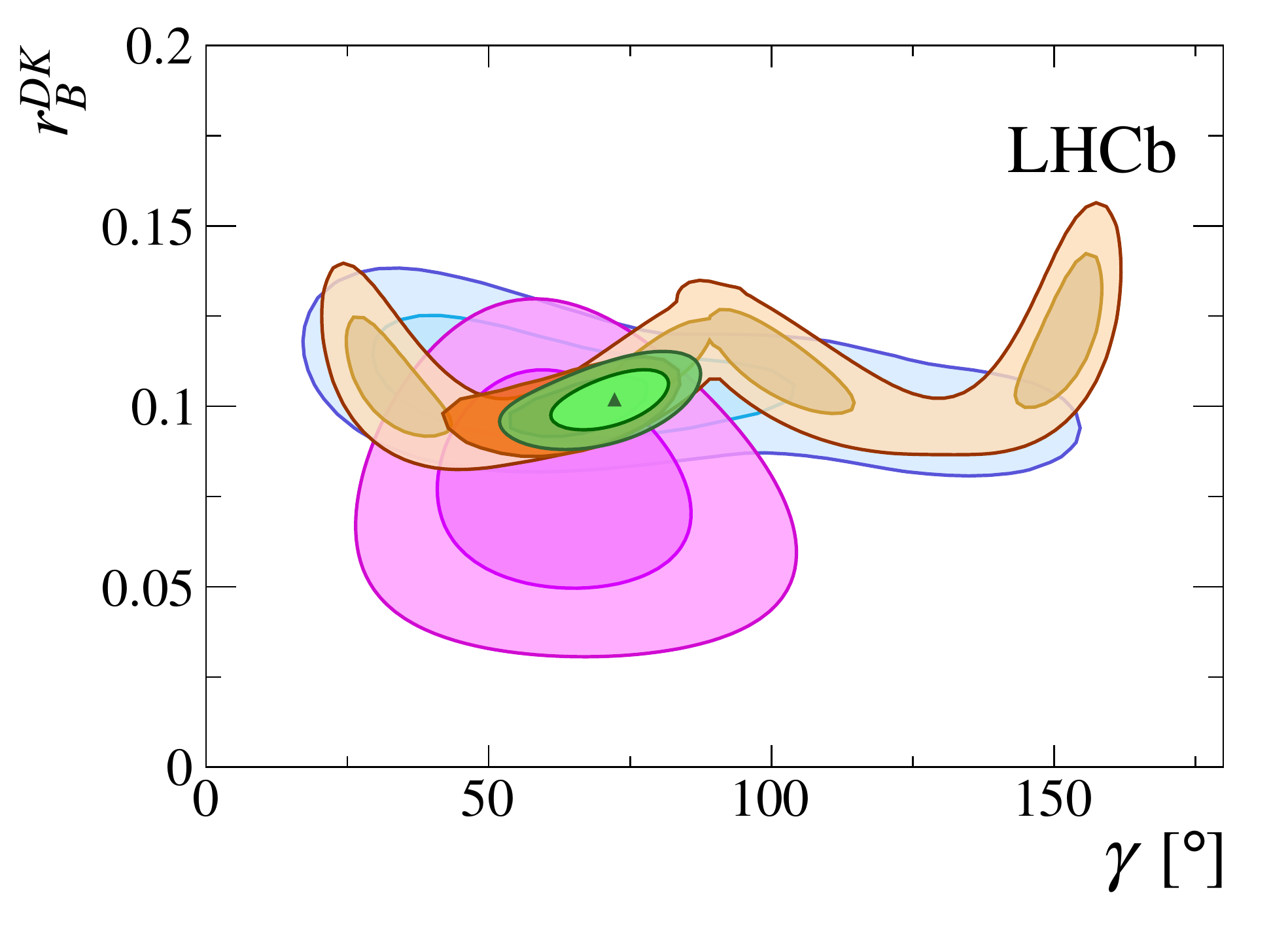} \\
    \includegraphics[width=0.3\textwidth]{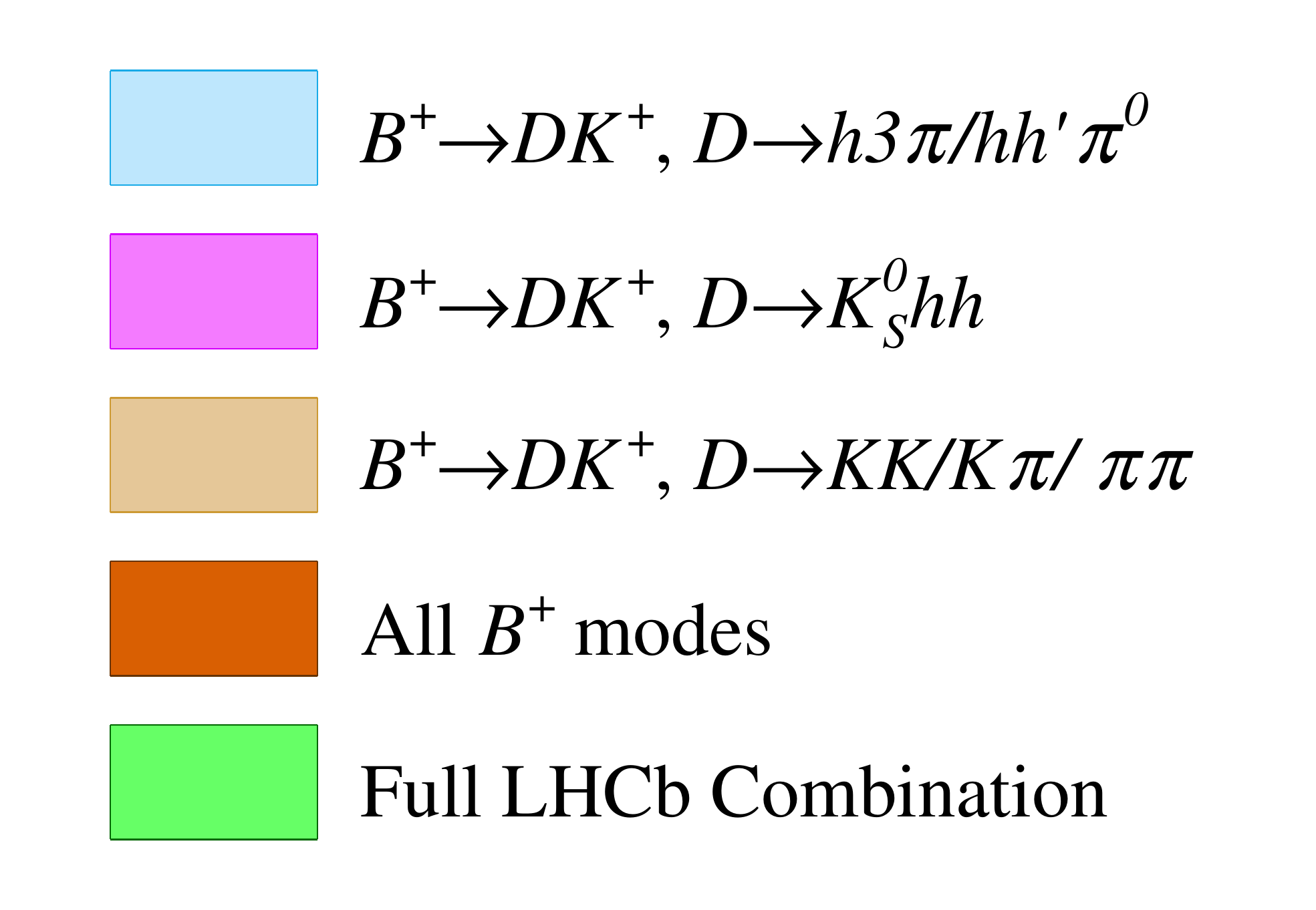}
  \end{center}
  \caption{Profile likelihood contours of \g \vs \db (left) and \g \vs \rb (right) for various \DK sub-combinations: (blue) \BuDK, $\Dz\to h\pi\pi\pi$/$hh^{\prime}\piz$, (pink) \BuDK, $\Dz\to\KS hh$, (light brown) \BuDK, $\Dz\to KK$/$K\pi$/$\pi\pi$, (orange) all \Bu modes and (green) the full combination. Dark and light regions show the intervals containing 68.3\% and 95.5\% respectively.}
  \label{fig:g_d_dk}
\end{figure}

\begin{figure}
  \begin{center}
    \includegraphics[width=0.48\textwidth]{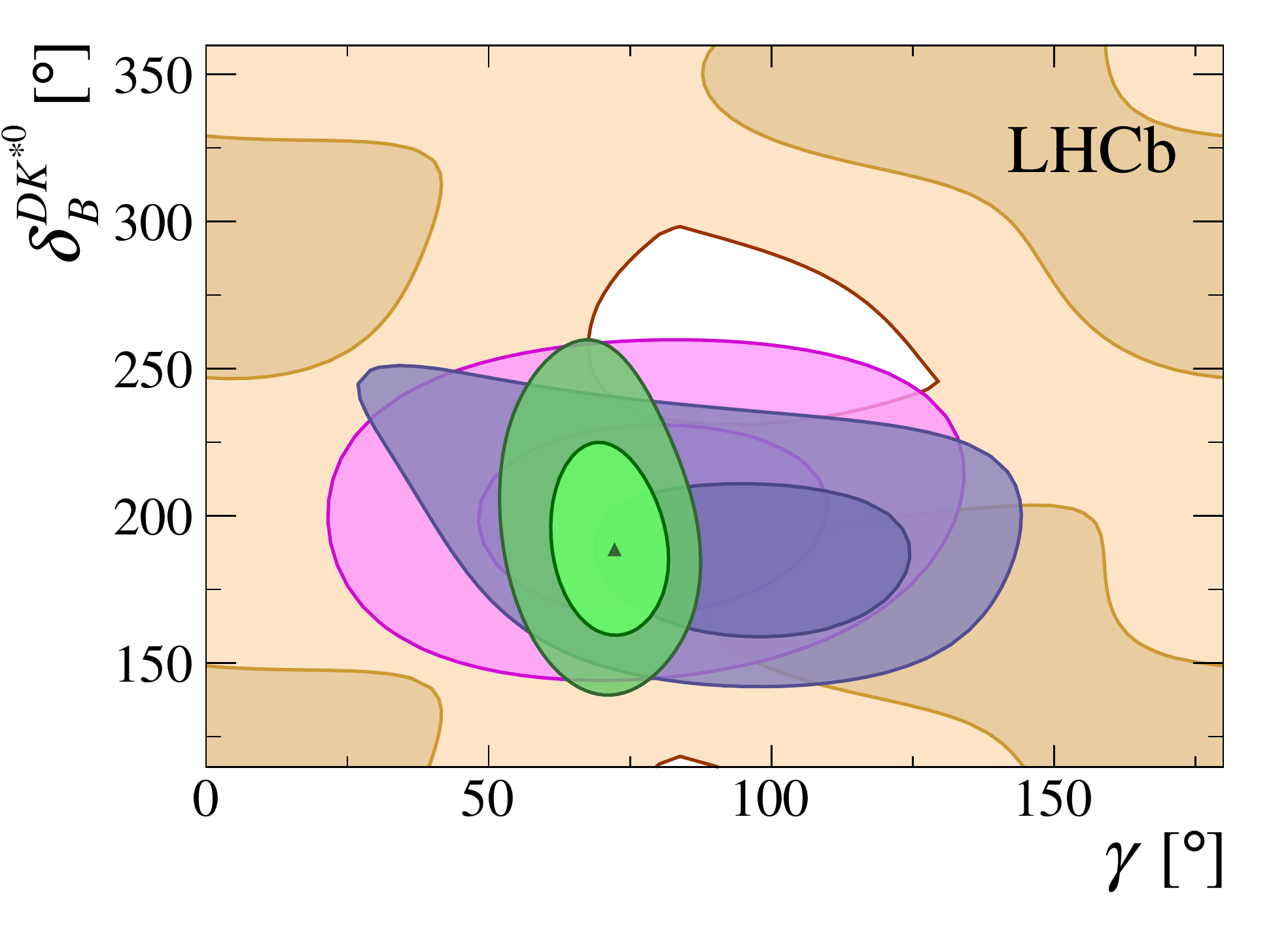}
    \includegraphics[width=0.48\textwidth]{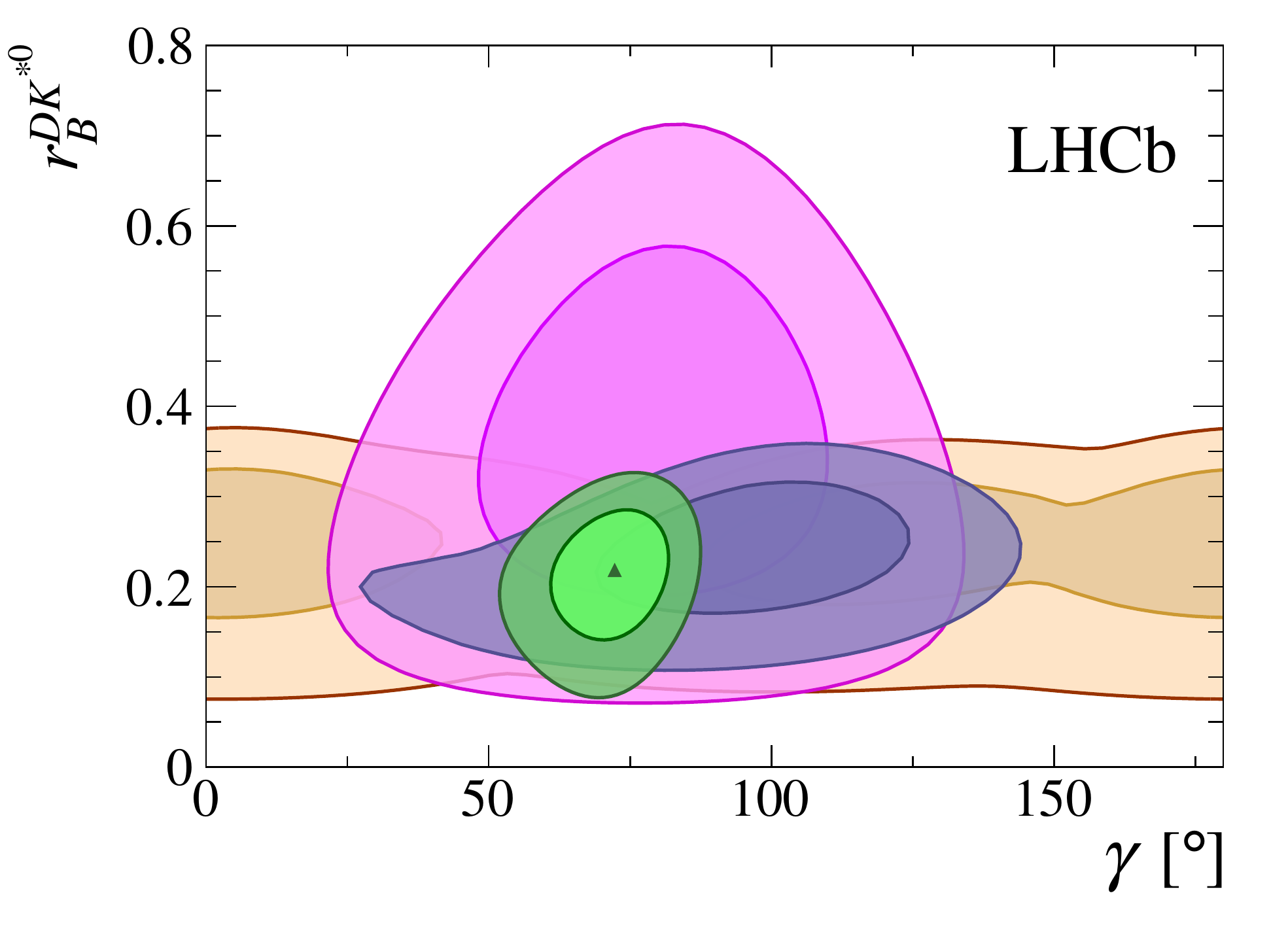} \\
    \includegraphics[width=0.3\textwidth]{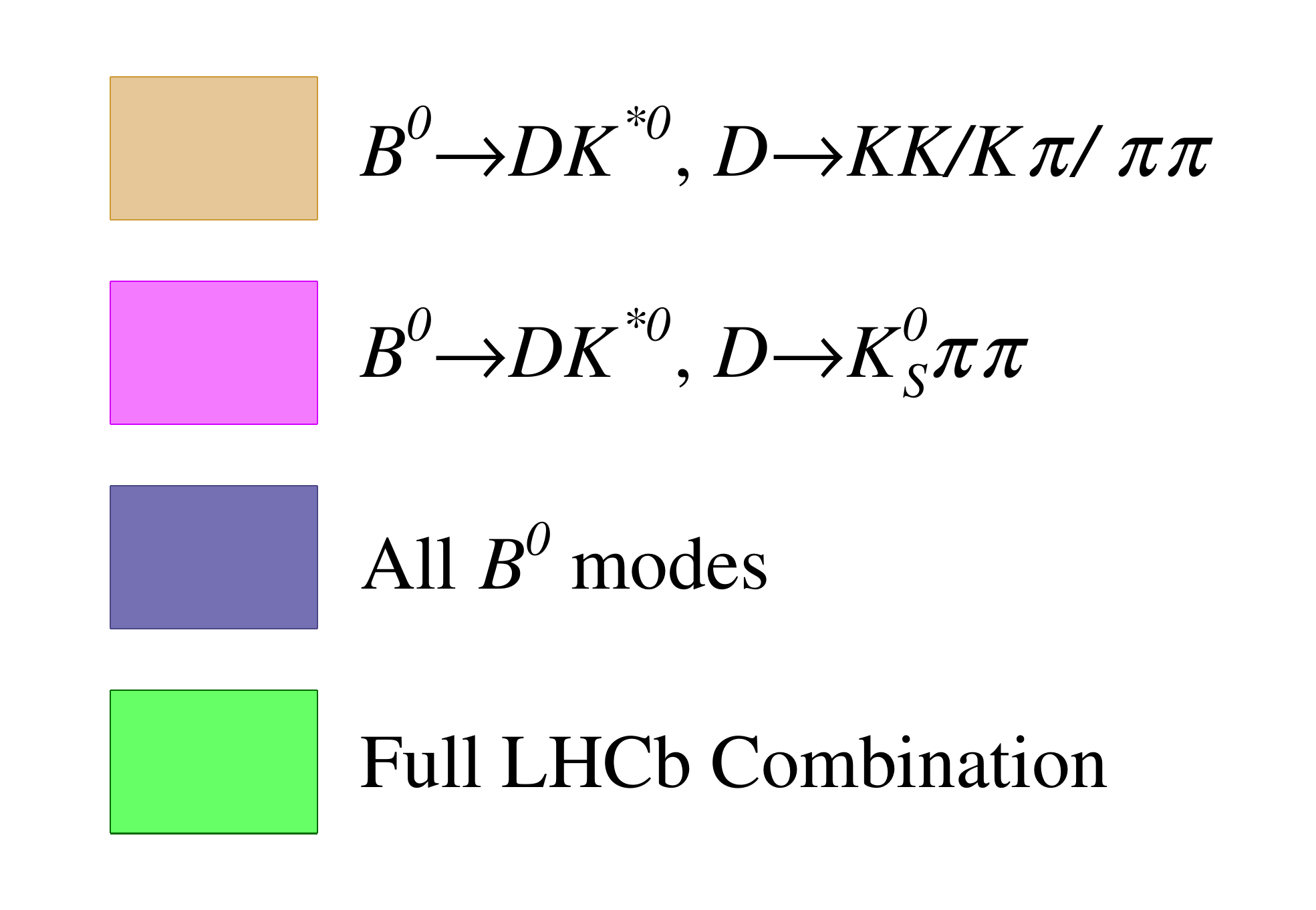}
  \end{center}
  \caption{Profile likelihood contours of \g \vs \dbDKstz (left) and \g \vs \rbDKstz (right) for various \DK sub-combinations: (brown) \BdDKstz, $\Dz\to KK$/$K\pi$/$\pi\pi$, (pink) \BdDKstz, $\Dz\to\KS \pi\pi$, (purple) all \Bd modes and (green) the full combination. Dark and light regions show the intervals containing 68.3\% and 95.5\% respectively.}
  \label{fig:g_d_dkstz}
\end{figure}

The origin of the sensitivity to \g of the various decay modes and analysis methods in the \DK combination is demonstrated in Fig.~\ref{fig:g_sensitivity}.
It can be seen that \BuDK decays offer the best sensitivity (see Fig.~\ref{fig:g_sensitivity} left) and that the GLW/ADS methods offer multiple narrow solutions
compared to the single broader solution of the GGSZ method (see Fig.~\ref{fig:g_sensitivity} right). Figures~\ref{fig:g_d_dk} and~\ref{fig:g_d_dkstz}
further demonstrate the complementarity of the input methods in the $(\g\,\vs\,\delta_{B}^{X})$ and $(\g\,\vs\,r_{B}^{X})$ planes, for the \Bp and \Bd systems respectively.

\section{Bayesian analysis}
\label{sec:bayesian}
The combinations are also performed using a Bayesian procedure.
Probability (or credible) intervals (or regions) are obtained according to a highest posterior density approach.
A highest posterior density interval (region) is defined by the property that the minimum density of any point within the interval (region)
is equal to, or larger than, the density of any point outside that interval (region).

\subsection{\boldmath \DK combination}
\label{sec:bayesian-dk}
\begin{figure}
  \centering
  \includegraphics[width=.45\textwidth]{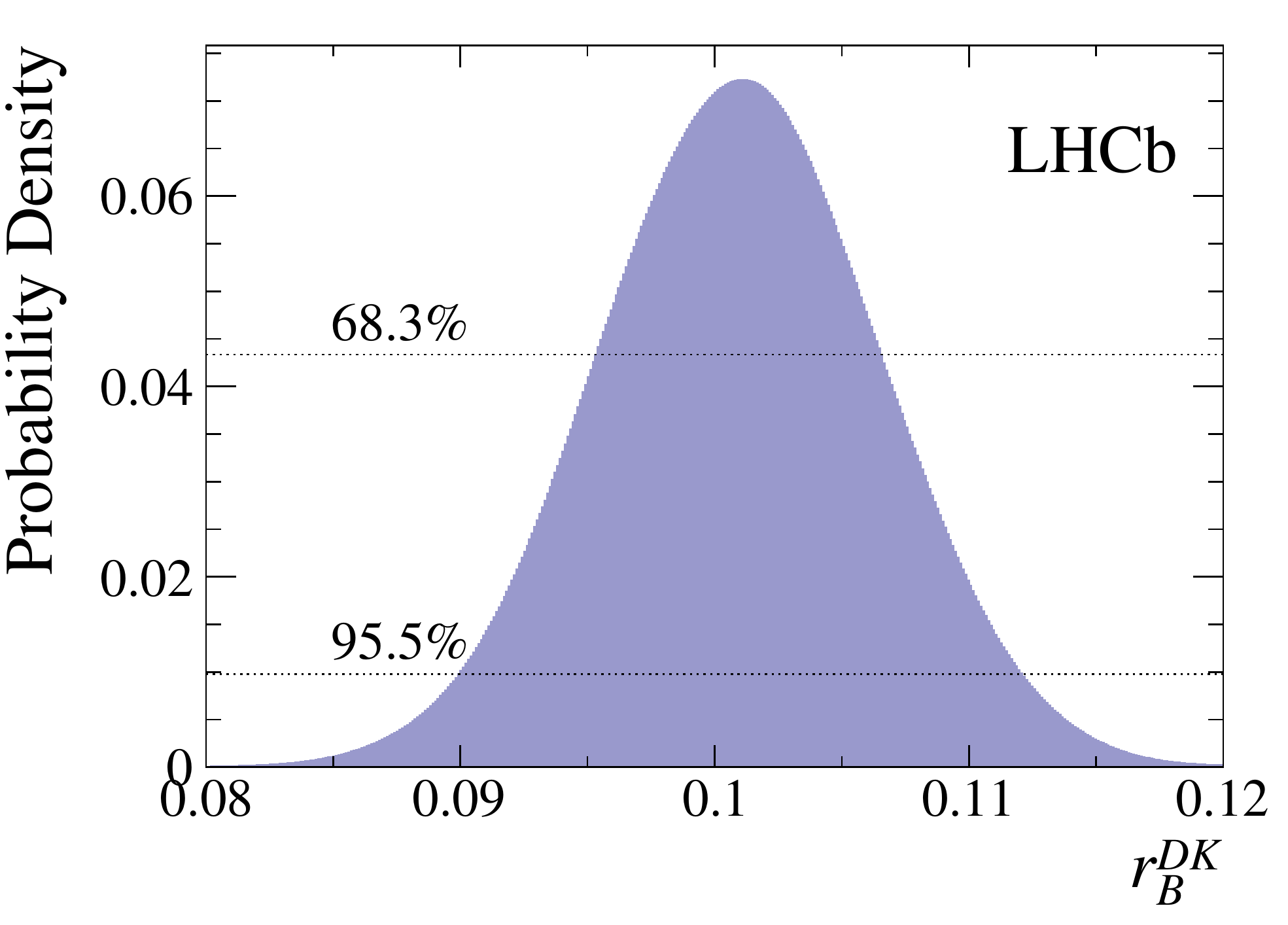}
  \includegraphics[width=.45\textwidth]{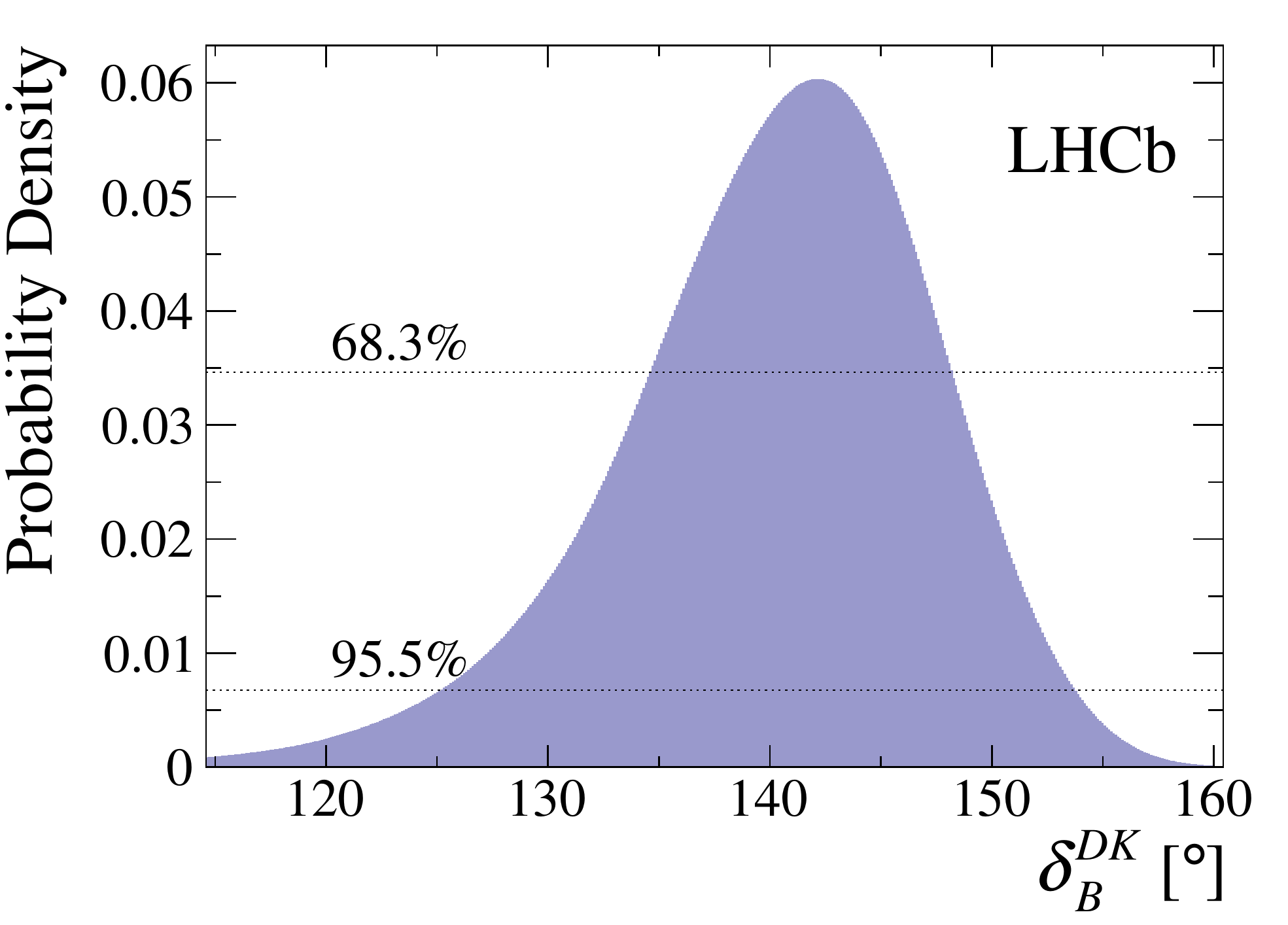}
  \includegraphics[width=.45\textwidth]{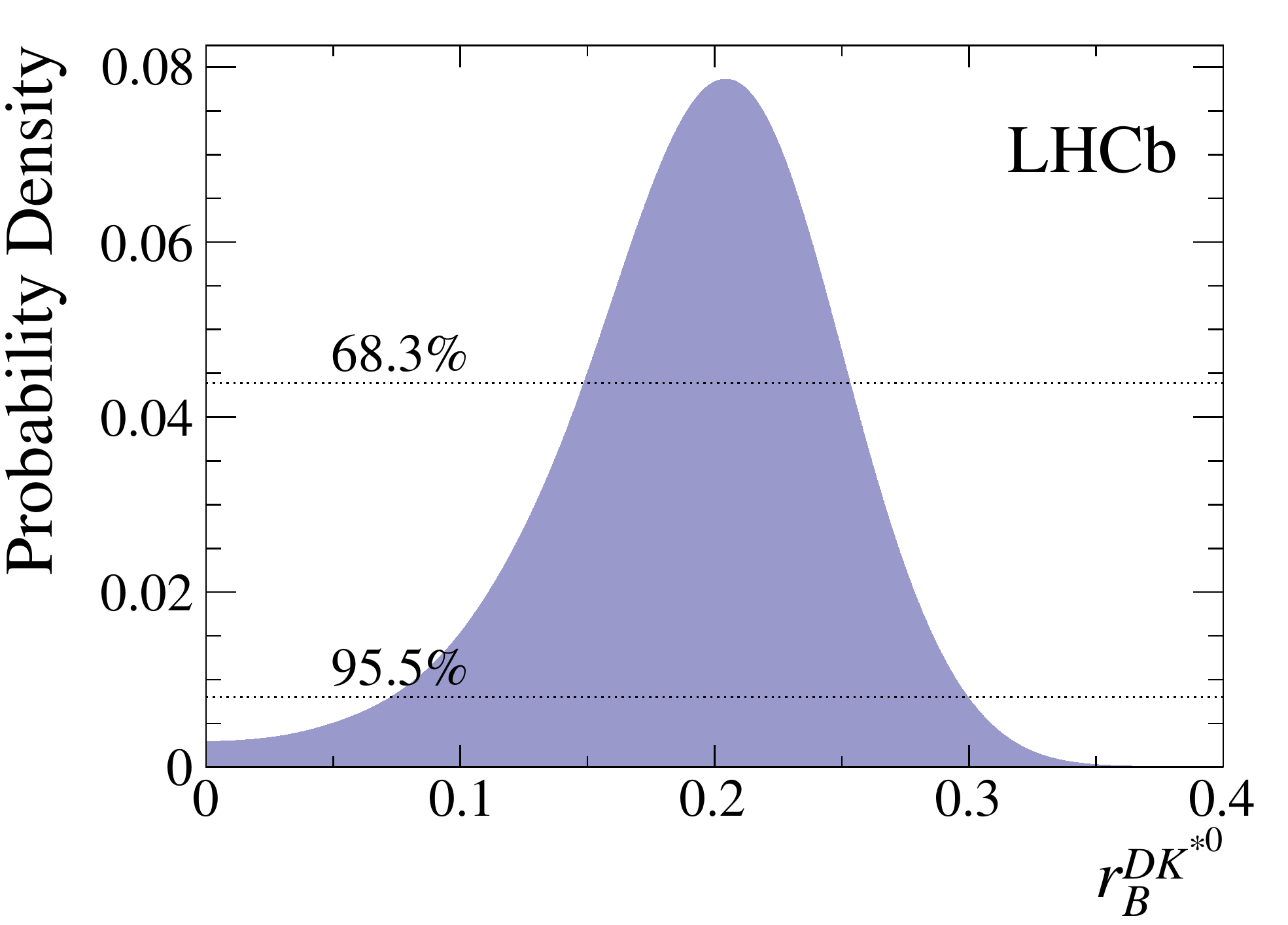}
  \includegraphics[width=.45\textwidth]{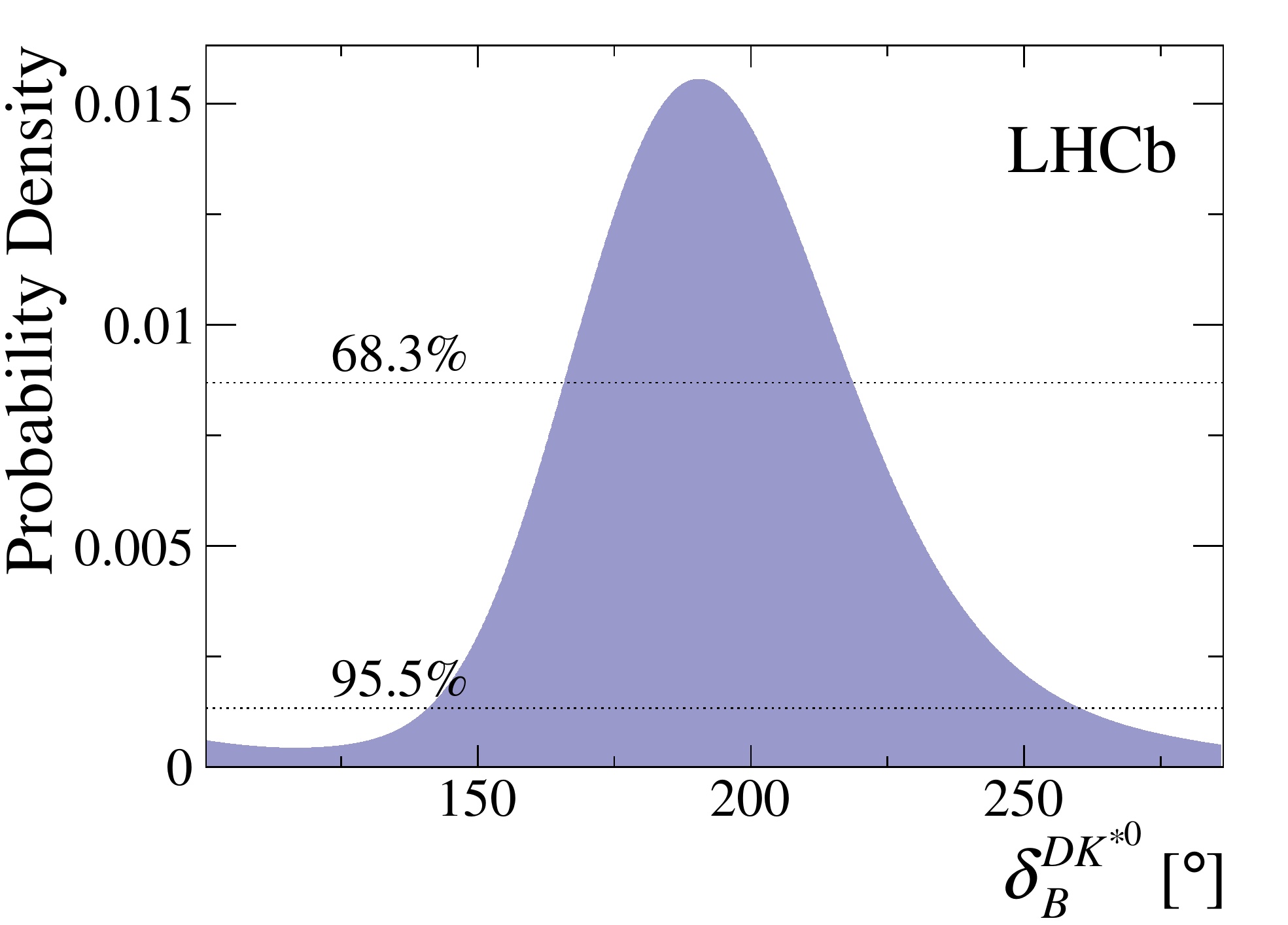}
  \includegraphics[width=.45\textwidth]{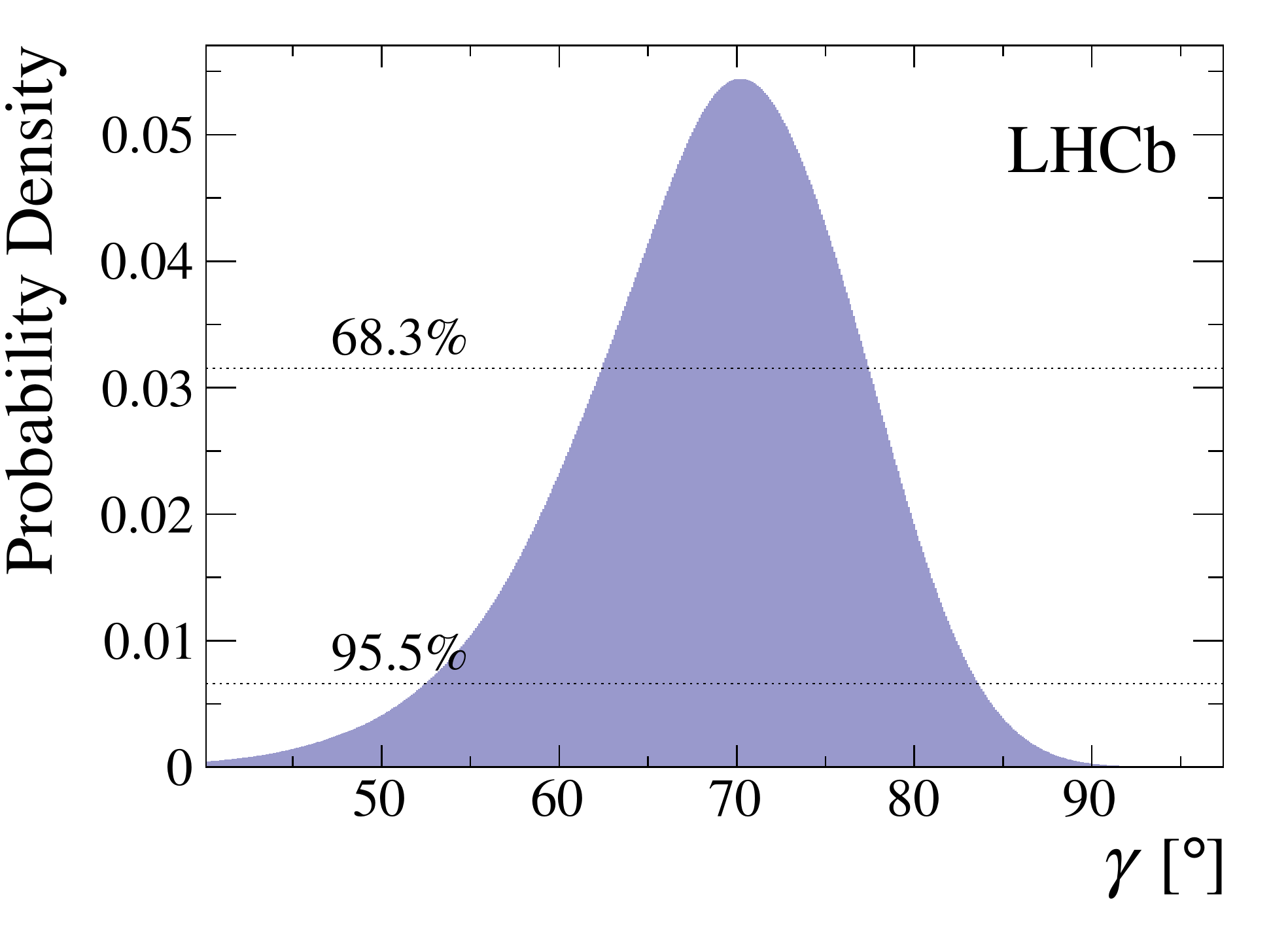}
  \caption{Posterior probability density from the Bayesian interpretation for the \DK combination.}
  \label{fig:DKcombBayes}
\end{figure}

\begin{figure}
  \centering
  \includegraphics[width=.45\textwidth]{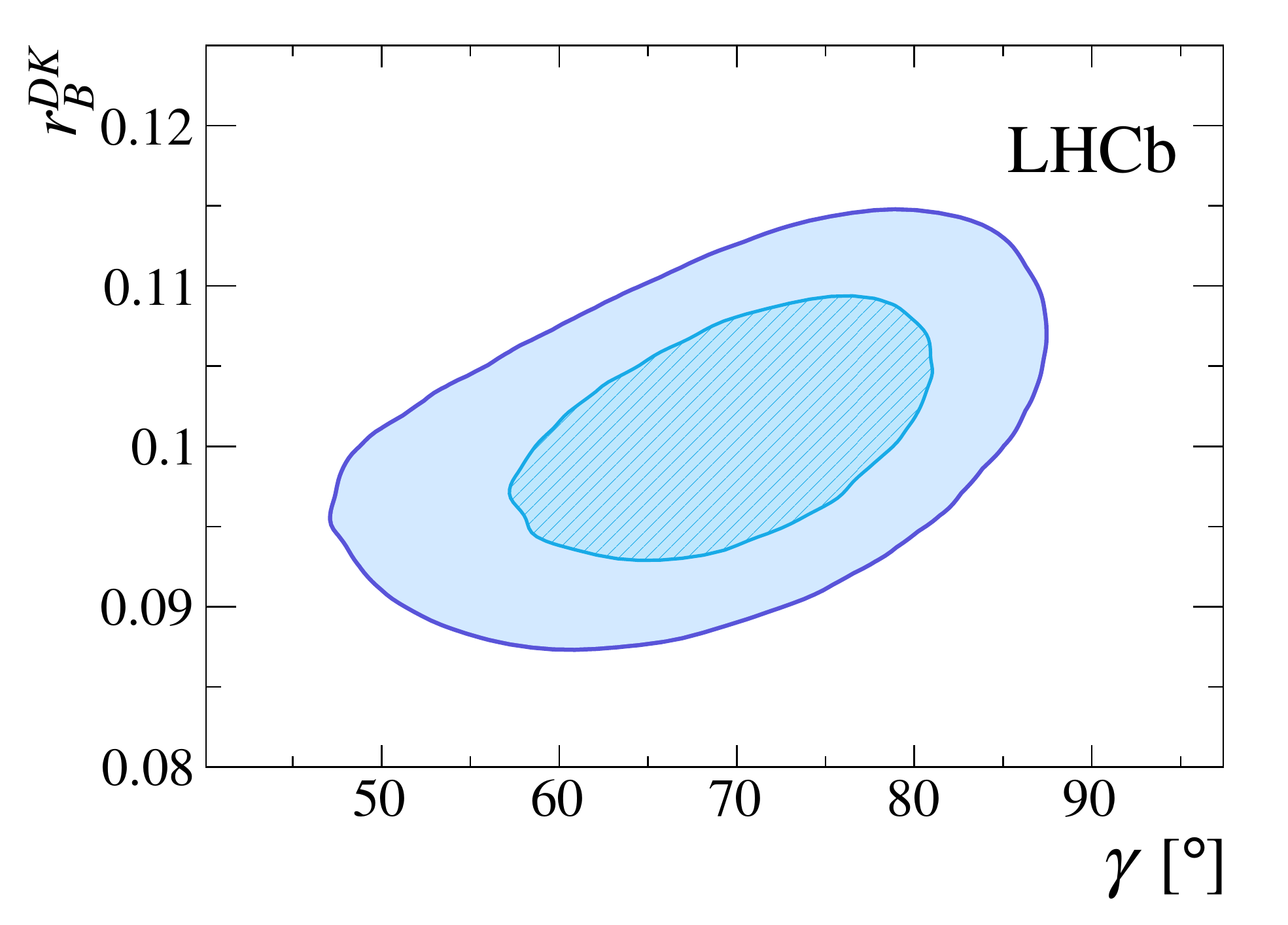}
  \includegraphics[width=.45\textwidth]{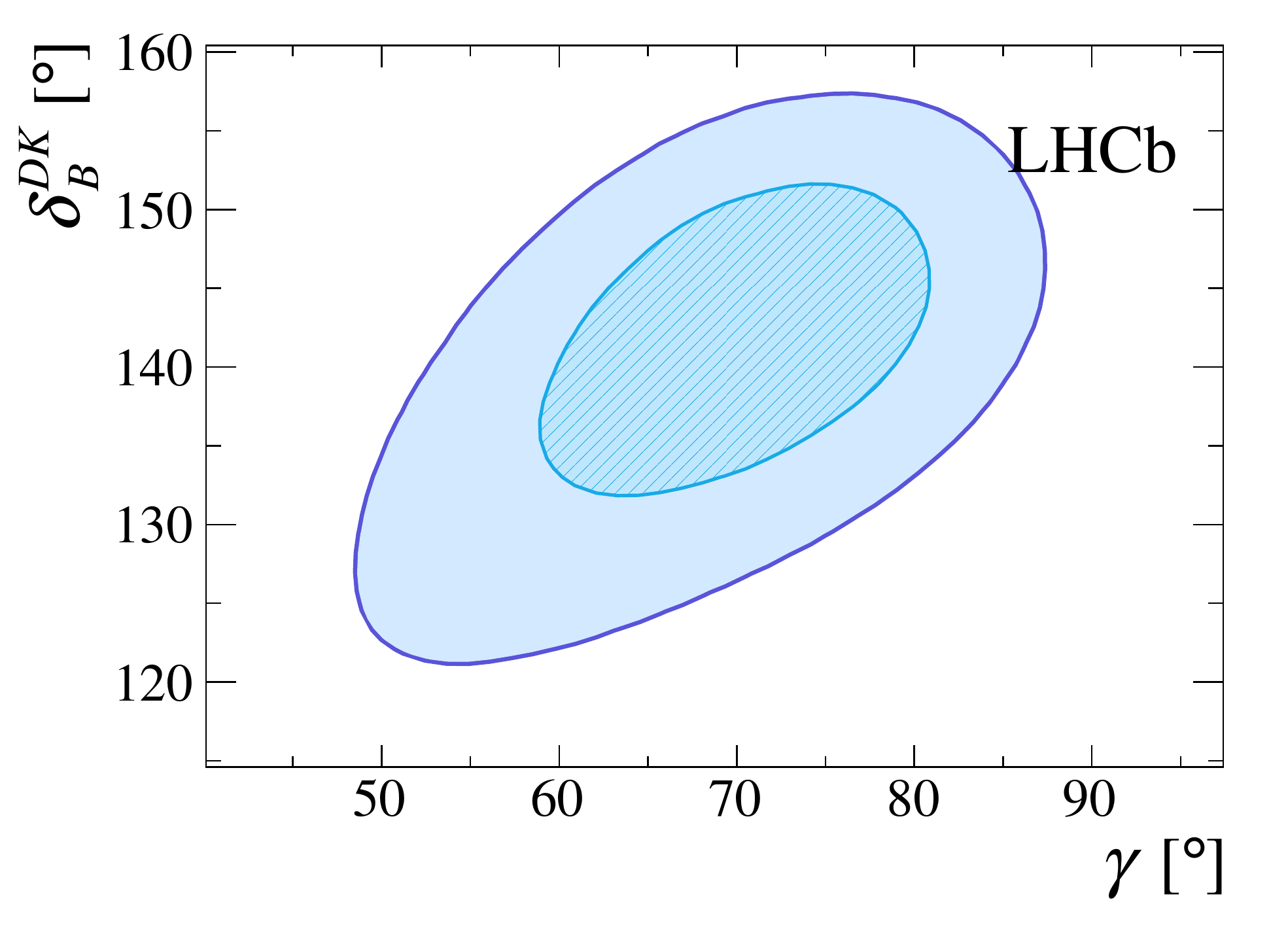}
  \includegraphics[width=.45\textwidth]{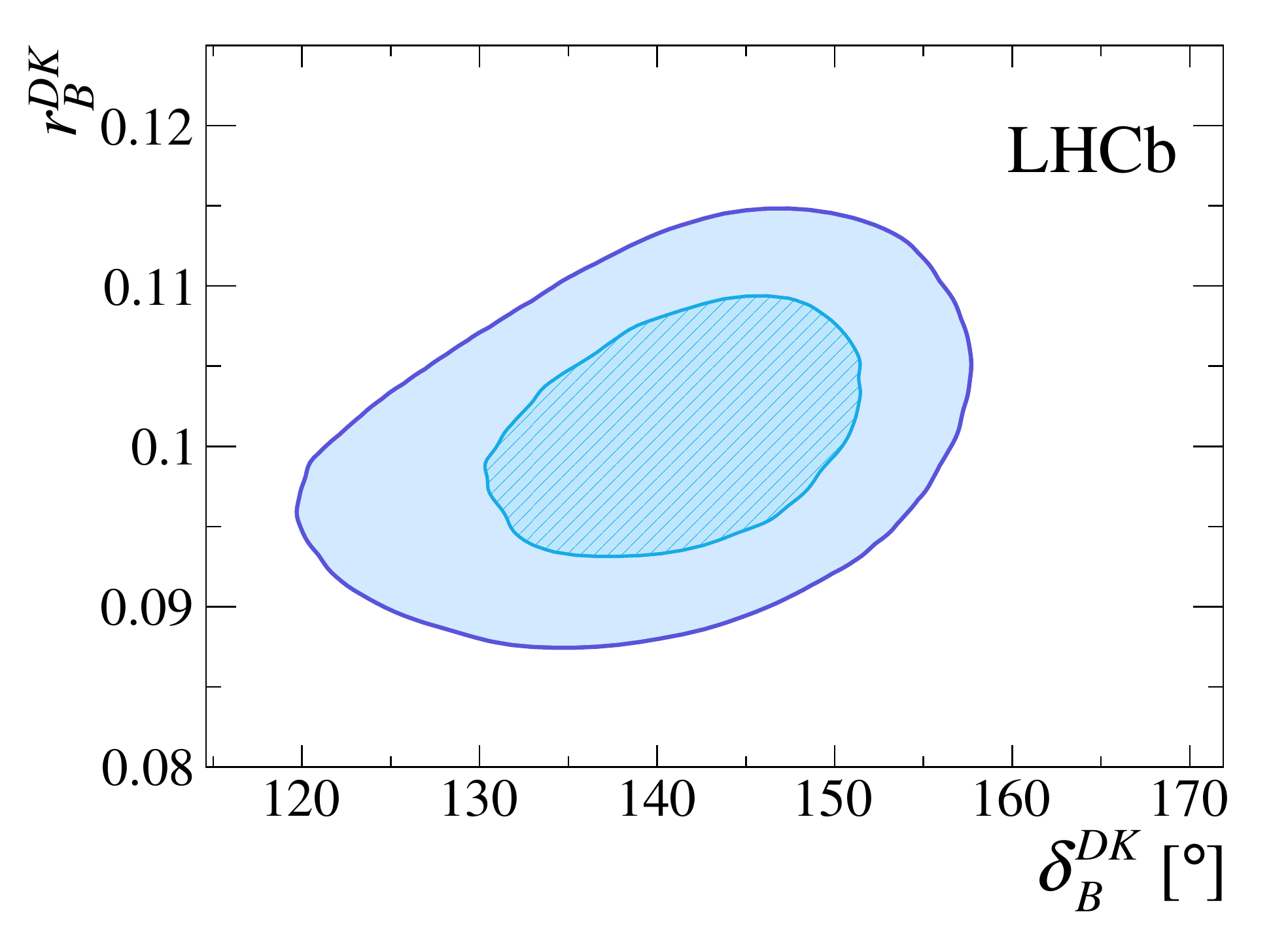}
  \includegraphics[width=.45\textwidth]{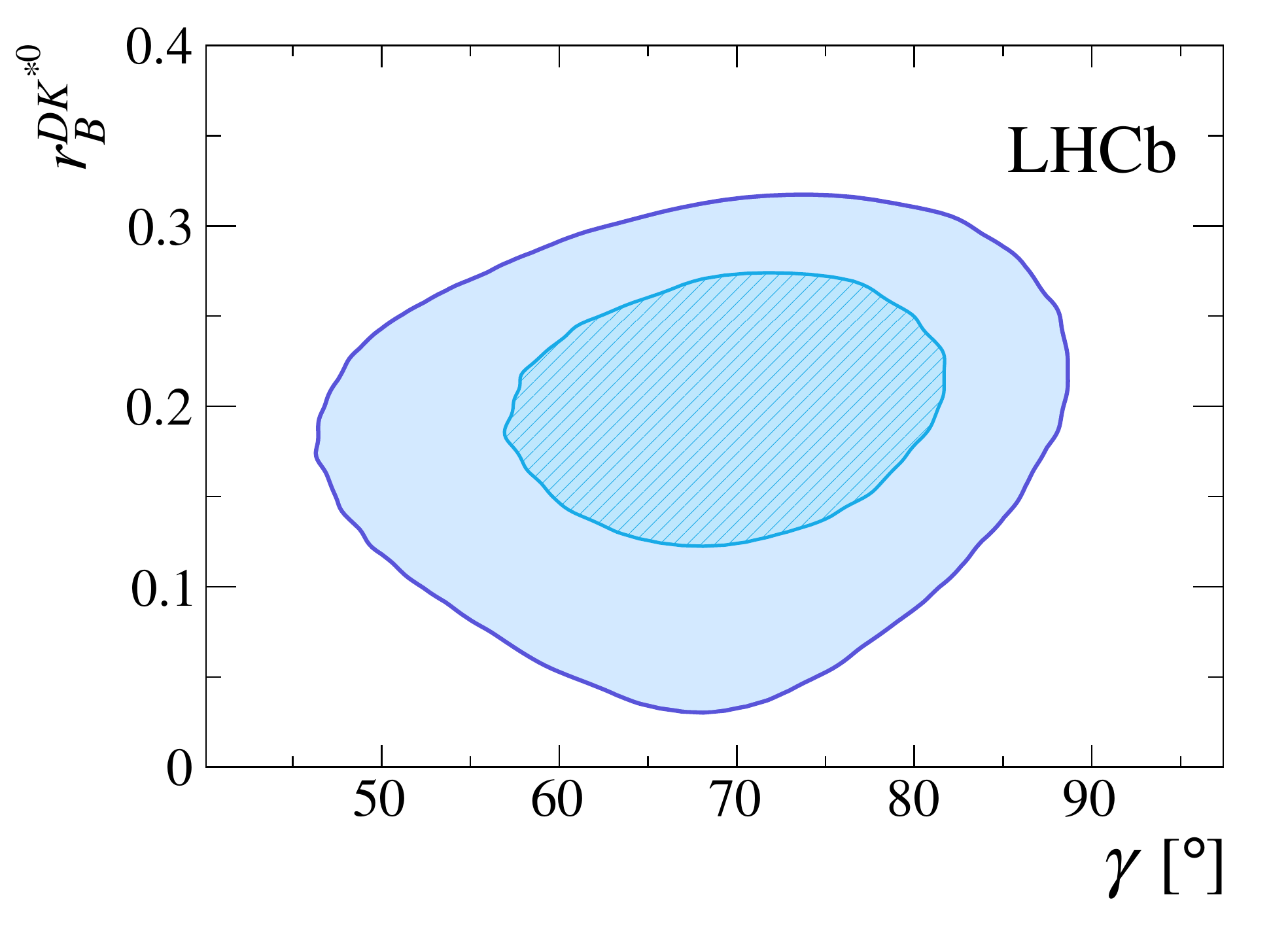}
  \includegraphics[width=.45\textwidth]{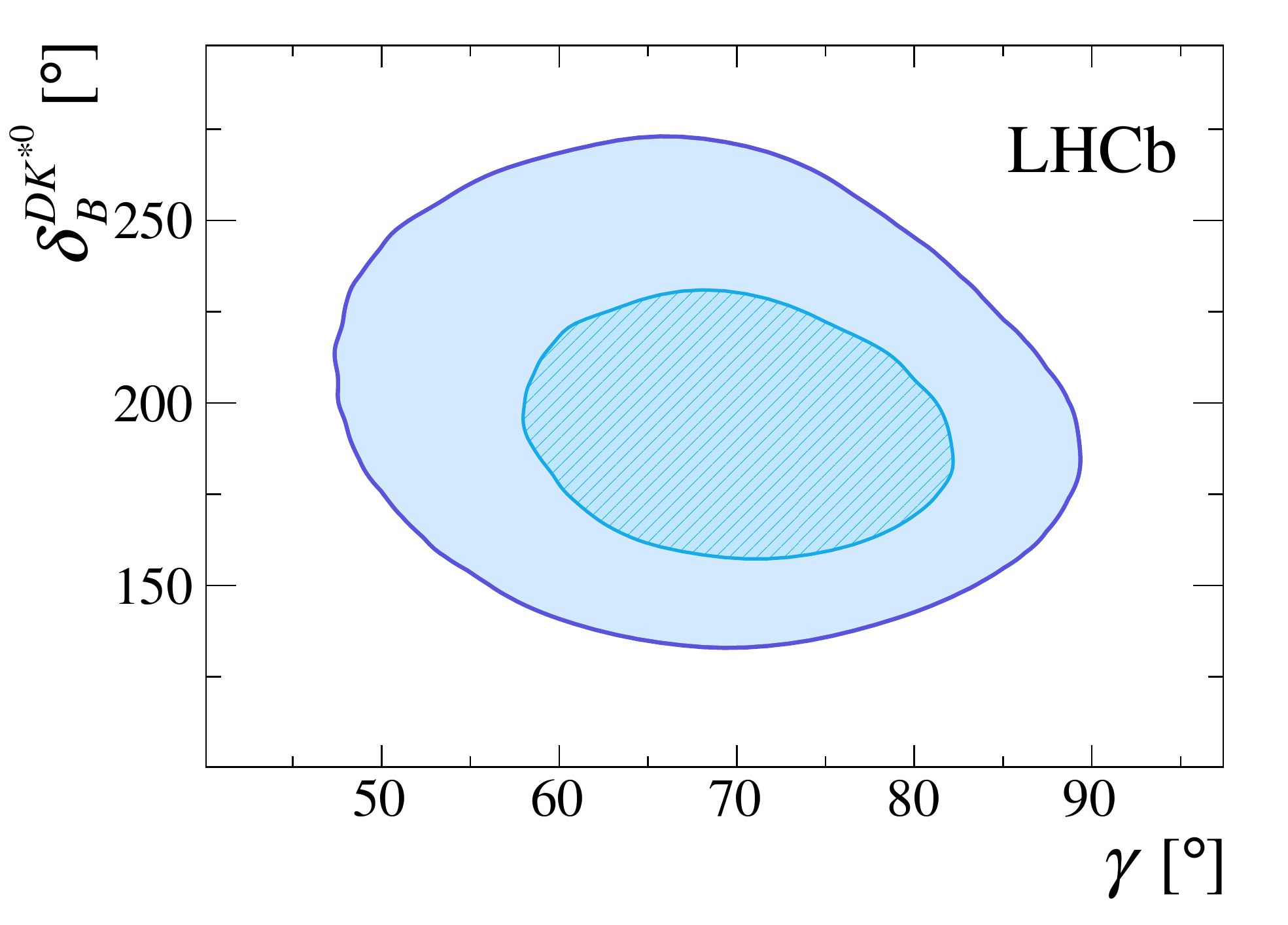}
  \includegraphics[width=.45\textwidth]{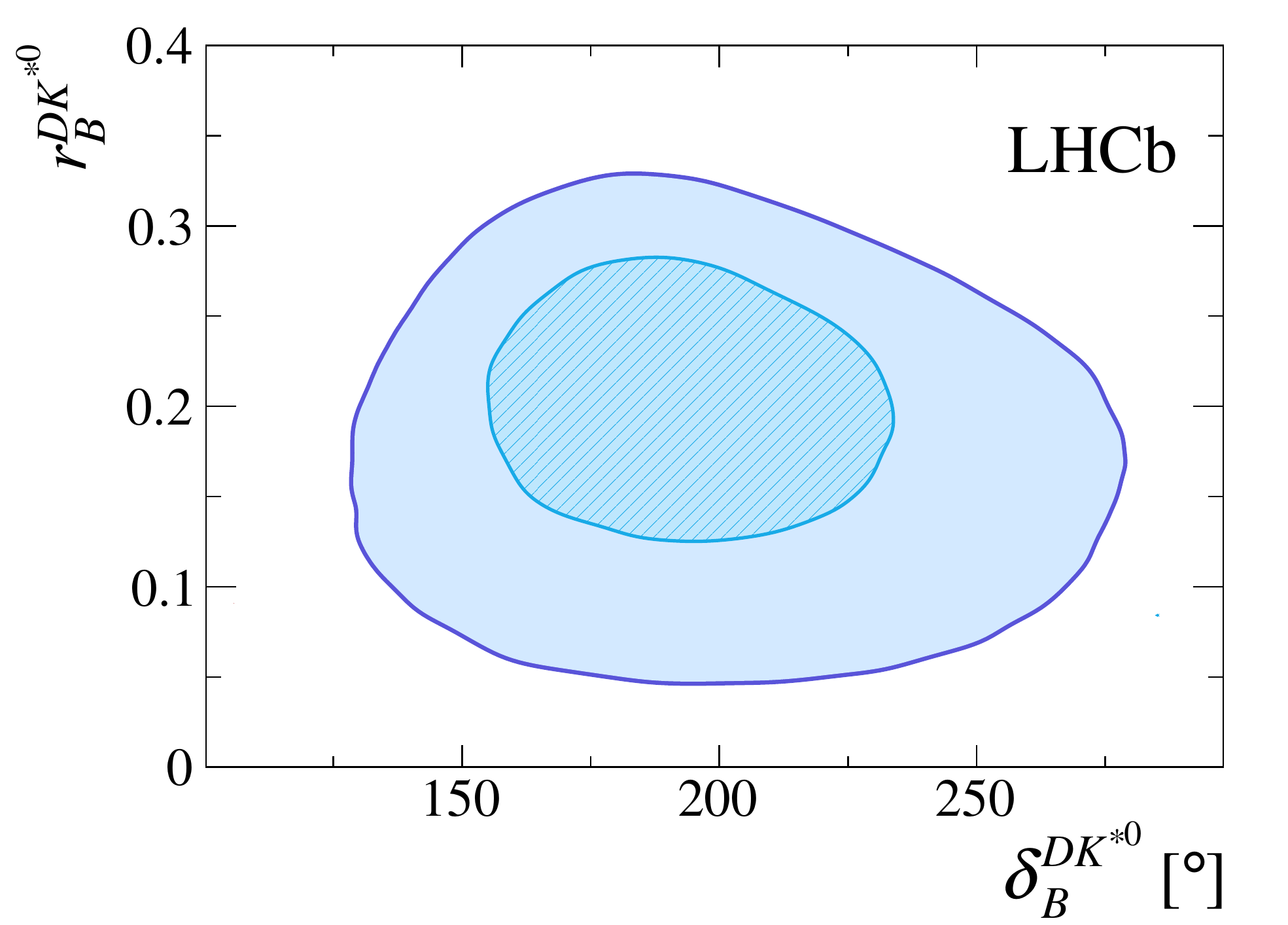}
  \caption{Two-dimensional posterior probability regions from the Bayesian interpretation for the \DK combination. Light and dark regions show the 68.3\% and 95.5\% credible intervals respectively.}
  \label{fig:DKcombBayes2D}
\end{figure}

Uniform prior probability distributions (hereafter referred to as priors) are used for \g and the $B$-meson hadronic parameters in the \DK combination, allowing them to vary inside the following ranges:
$\gamma\in[0^{\circ}, 180^{\circ}]$,
$\delta_B^{DK}\in[-180^{\circ}, 180^{\circ}]$,
$r_B^{DK}\in[0.06, 0.14]$. The priors for \dbDkpp and \ddsk are identical to that for \db; the range for \dbDKstz is $[0^{\circ}, 360^{\circ}]$.
The allowed ranges for \rbDKstz, \rbDkpp and \rdsk are [0, 0.45], \mbox{[0, 0.16]} and [0, 0.2].
The remaining auxiliary parameters are constrained with Gaussian priors
according to the externally measured values and their uncertainties.
A range of alternative prior distributions have been found to have negligible impact on the results for \g.
The results are shown in Table~\ref{tab:resBayes} and in
Figs.~\ref{fig:DKcombBayes} and~\ref{fig:DKcombBayes2D}.
The Bayesian credible intervals are found to be in good agreement with the frequentist confidence intervals.

\begin{table}[!htb]
\caption{\label{tab:resBayes}Credible intervals and most probable values for
the hadronic parameters determined from the \DK Bayesian combination.
}
\renewcommand{\arraystretch}{1.4}
\begin{center}
\begin{tabular}{p{2cm}cccc}
\hline
Observable & Central value & 68.3\% Interval & 95.5\% Interval & 99.7\% Interval \\
\hline
$\gamma\, (^{\circ})$	  & \gRobustCentralBayes	      & \gRobustOnesigBayes  	    & \gRobustTwosigBayes       & \gRobustThreesigBayes        \\
$\rb$		              & \rbRobustCentralBayes	      & \rbRobustOnesigBayes	    & \rbRobustTwosigBayes      & \rbRobustThreesigBayes       \\
$\db (^{\circ})$      & \dbRobustCentralBayes	      & \dbRobustOnesigBayes	    & \dbRobustTwosigBayes      & \dbRobustThreesigBayes       \\
$\rbDKstz$			      & \rbDKstzRobustCentralBayes	& \rbDKstzRobustOnesigBayes	& \rbDKstzRobustTwosigBayes	& \rbDKstzRobustThreesigBayes  \\
$\dbDKstz (^{\circ})$	& \dbDKstzRobustCentralBayes	& \dbDKstzRobustOnesigBayes	& \dbDKstzRobustTwosigBayes	& \dbDKstzRobustThreesigBayes  \\
\hline
\end{tabular}
\end{center}
\end{table}

\subsection{\boldmath \Dh combination}
\label{sec:bayesian-dh}
For the $Dh$ combination additional uniform priors are introduced: $\rbpi \in [0,0.06]$, $\dbpi \in [180^{\circ},360^{\circ}]$,
$\rbDppp \in [0,0.13]$ and $\dbDppp \in [0\degrees, 360\degrees]$.
All other priors are as described above for the \DK combination.

\begin{figure}
  \centering
  \includegraphics[width=.45\textwidth]{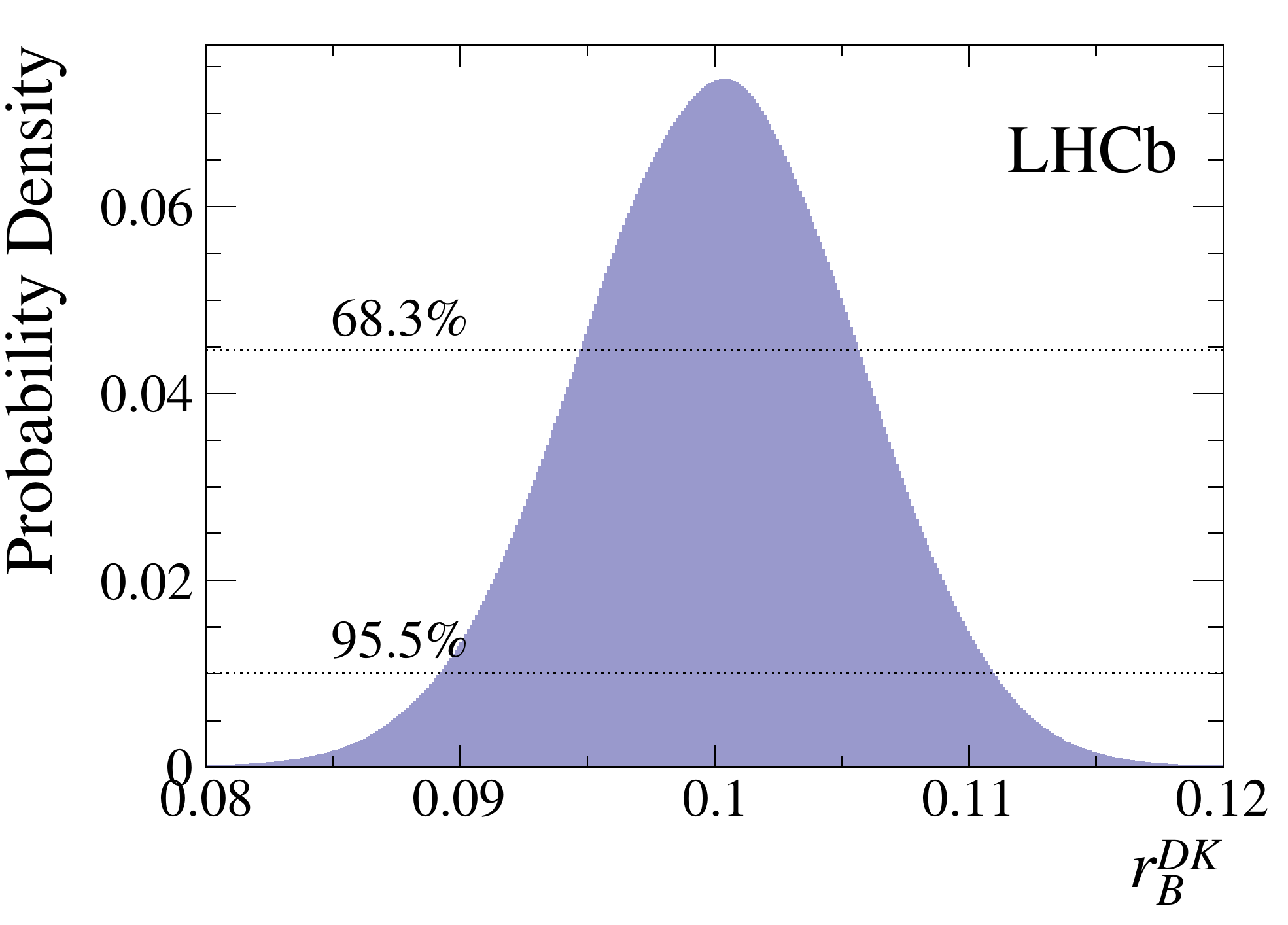}
  \includegraphics[width=.45\textwidth]{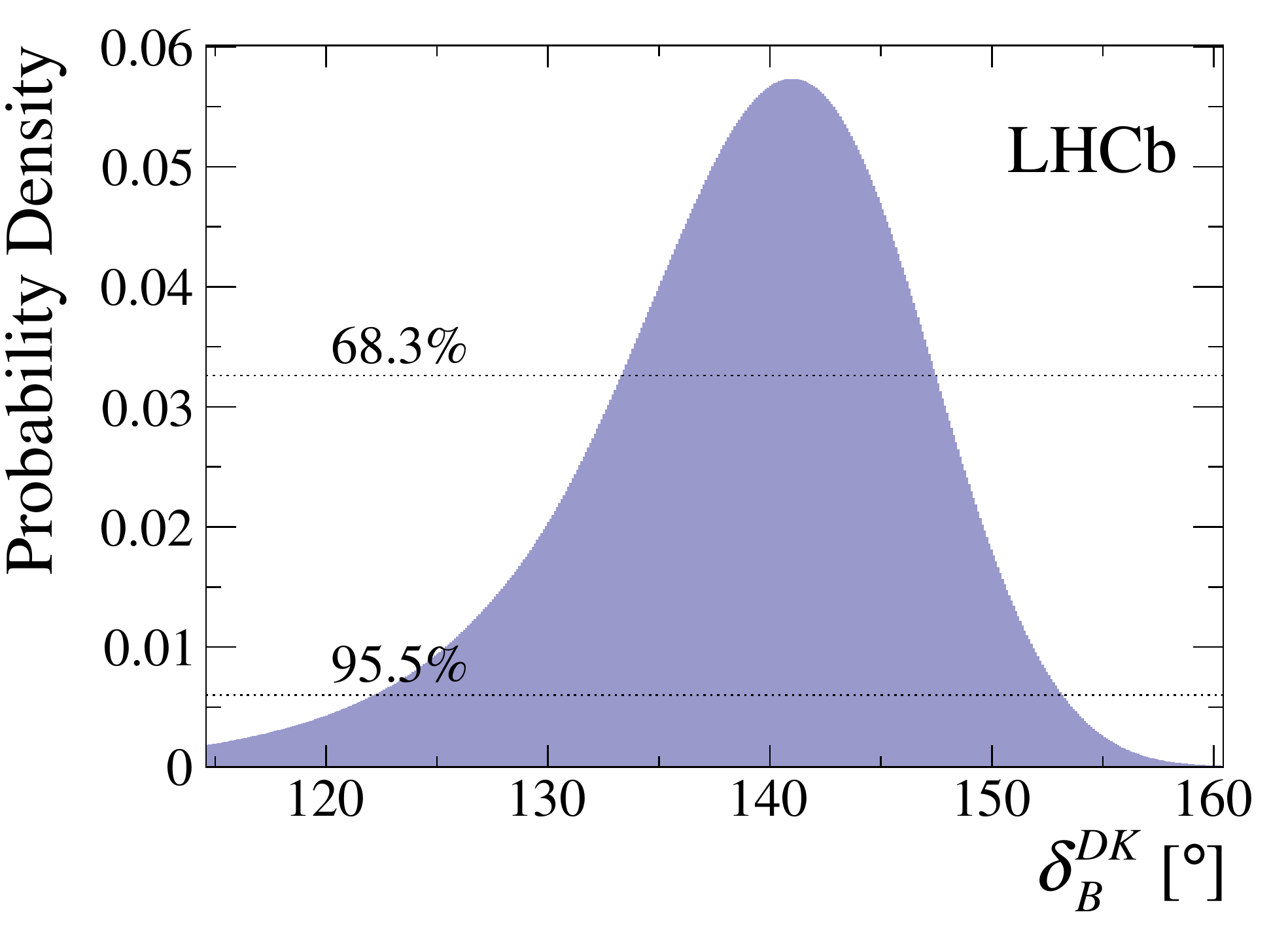}
  \includegraphics[width=.45\textwidth]{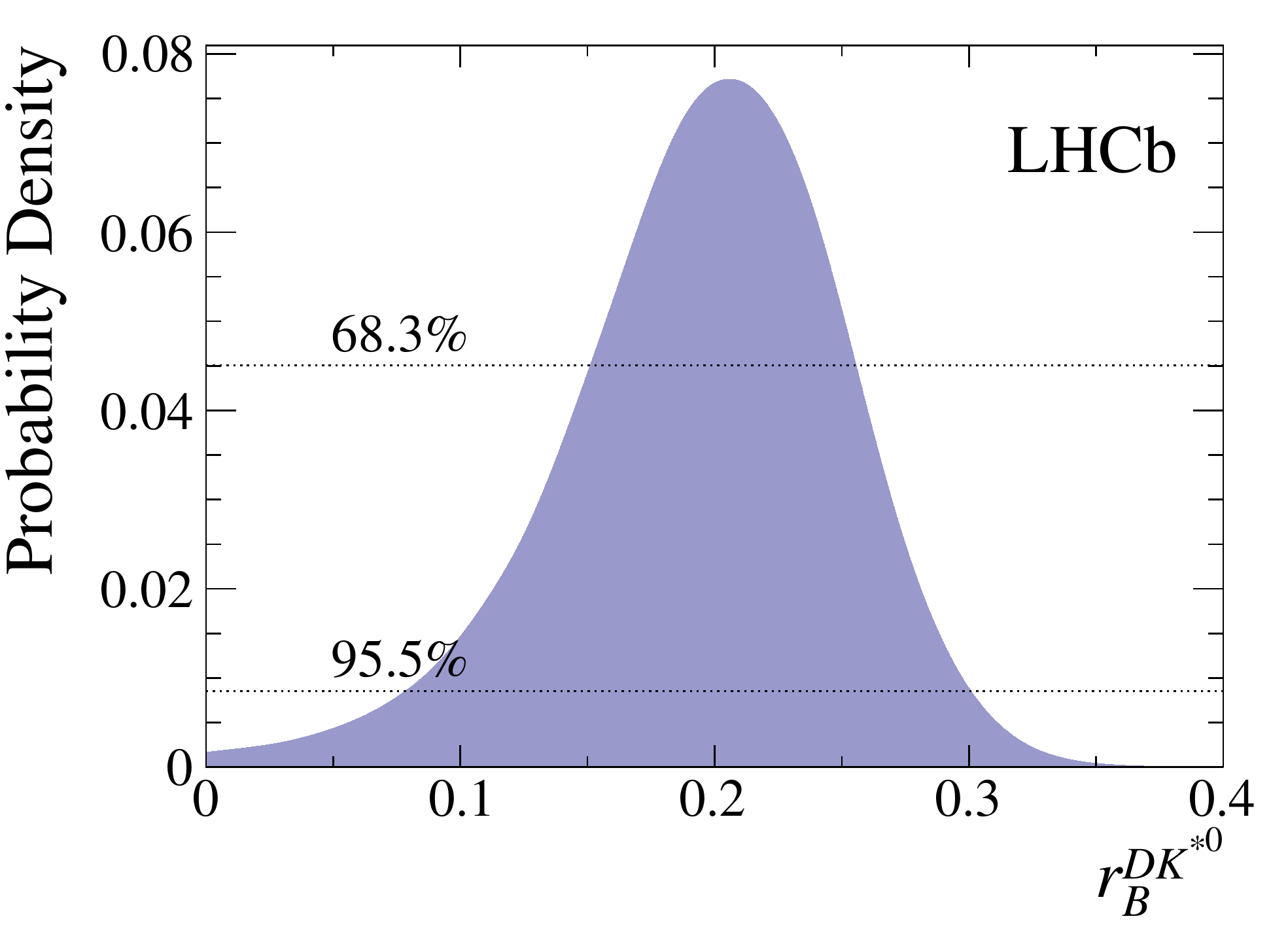}
  \includegraphics[width=.45\textwidth]{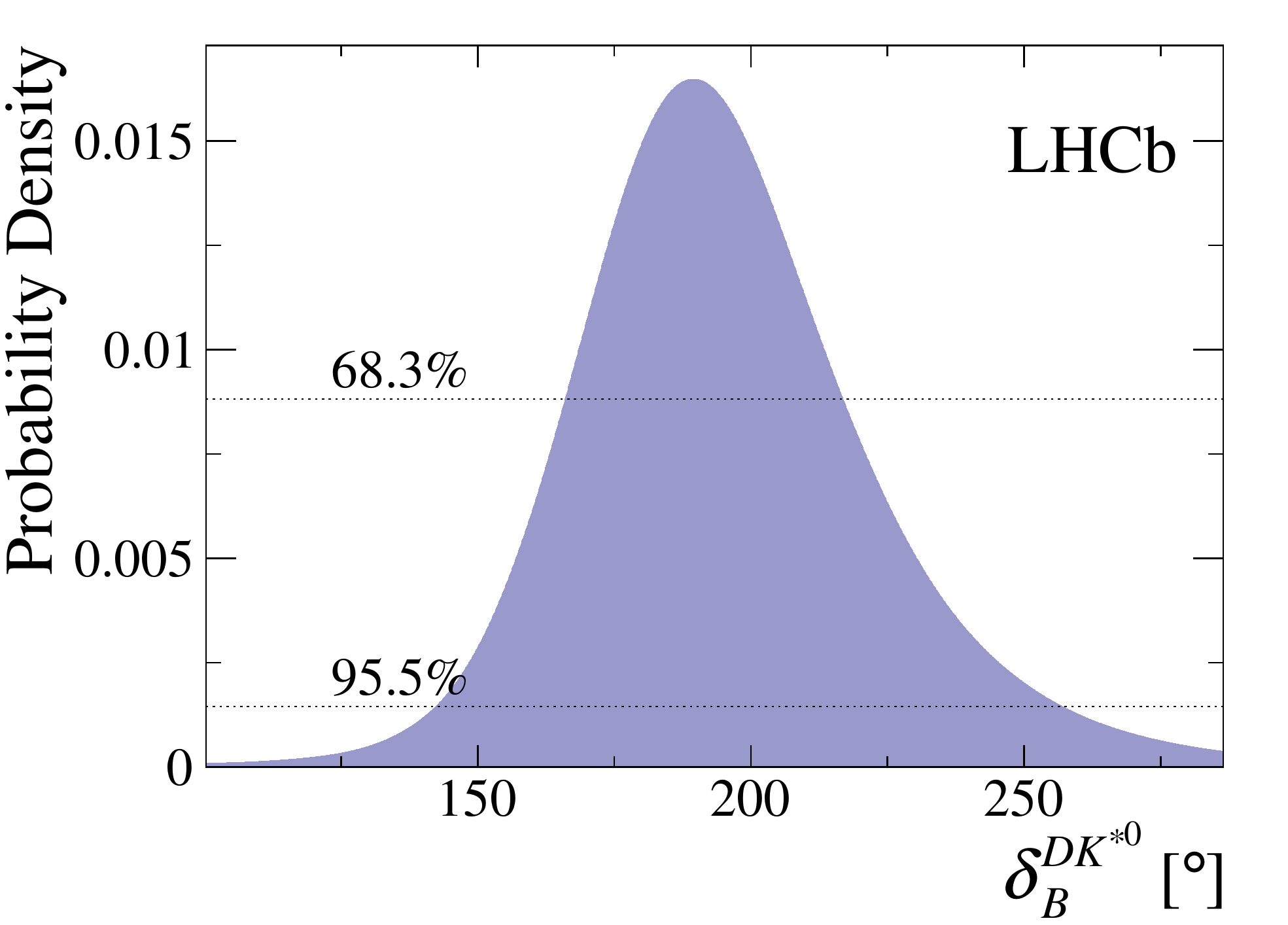}
  \includegraphics[width=.45\textwidth]{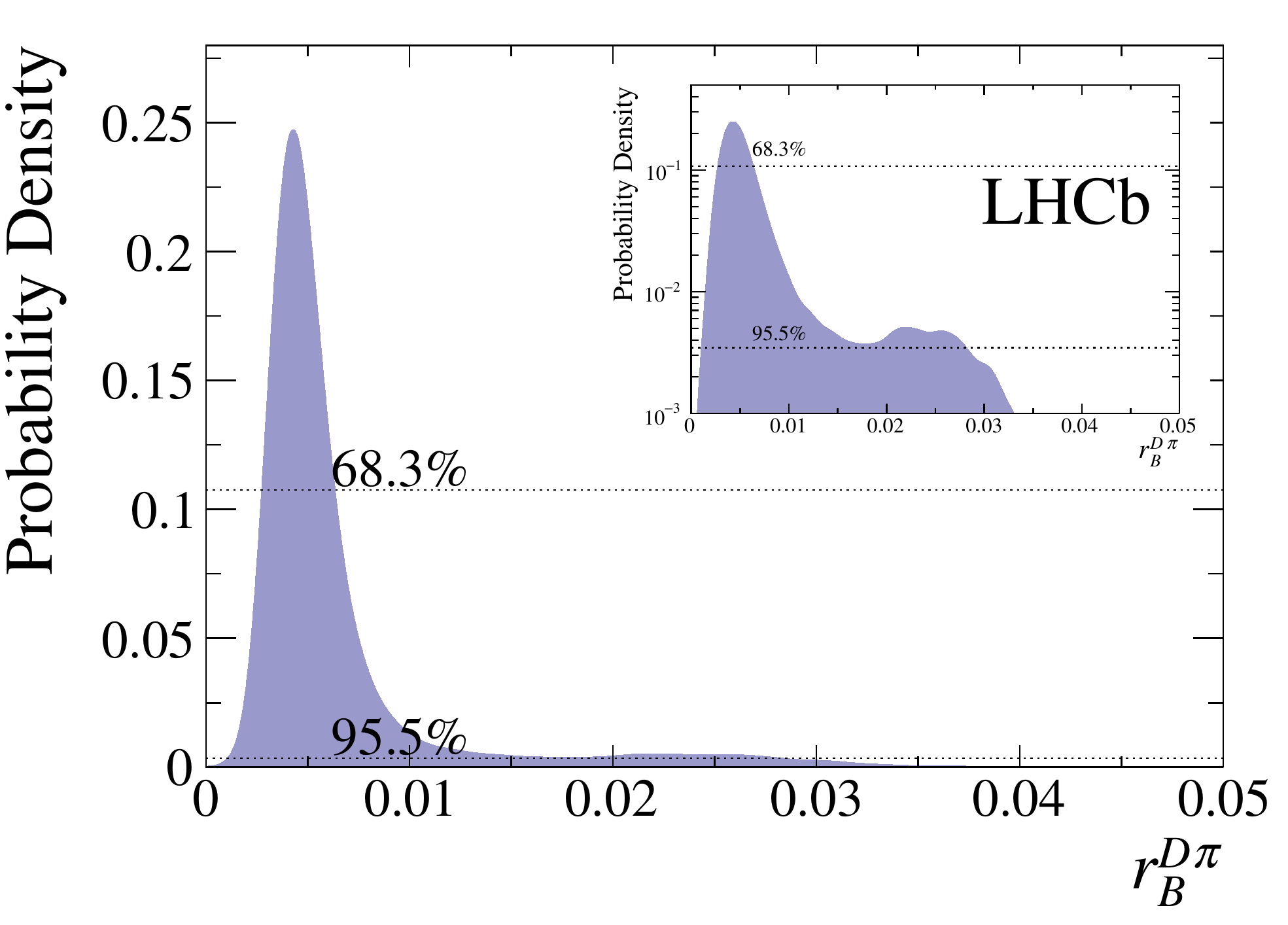}
  \includegraphics[width=.45\textwidth]{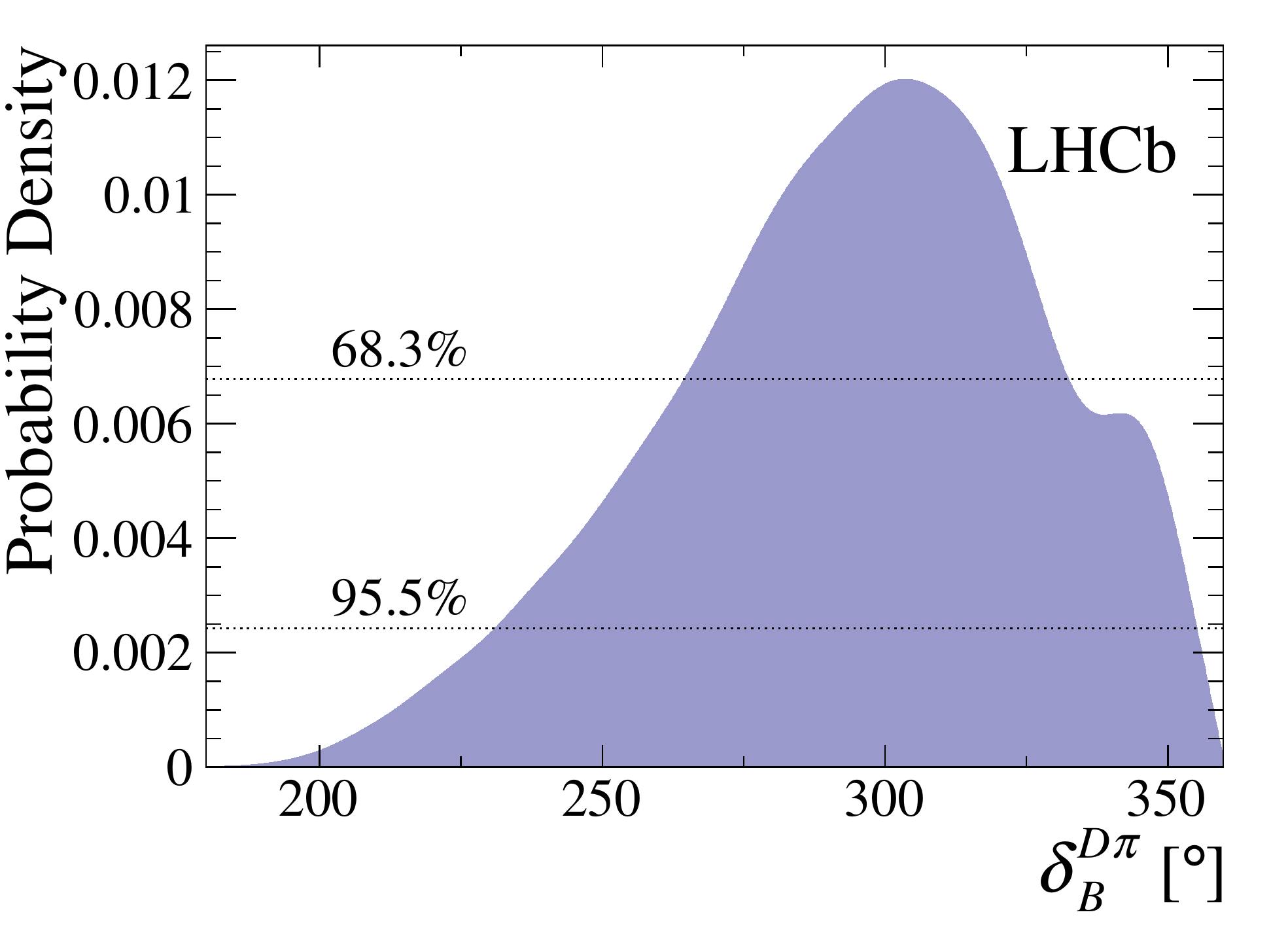}
  \includegraphics[width=.45\textwidth]{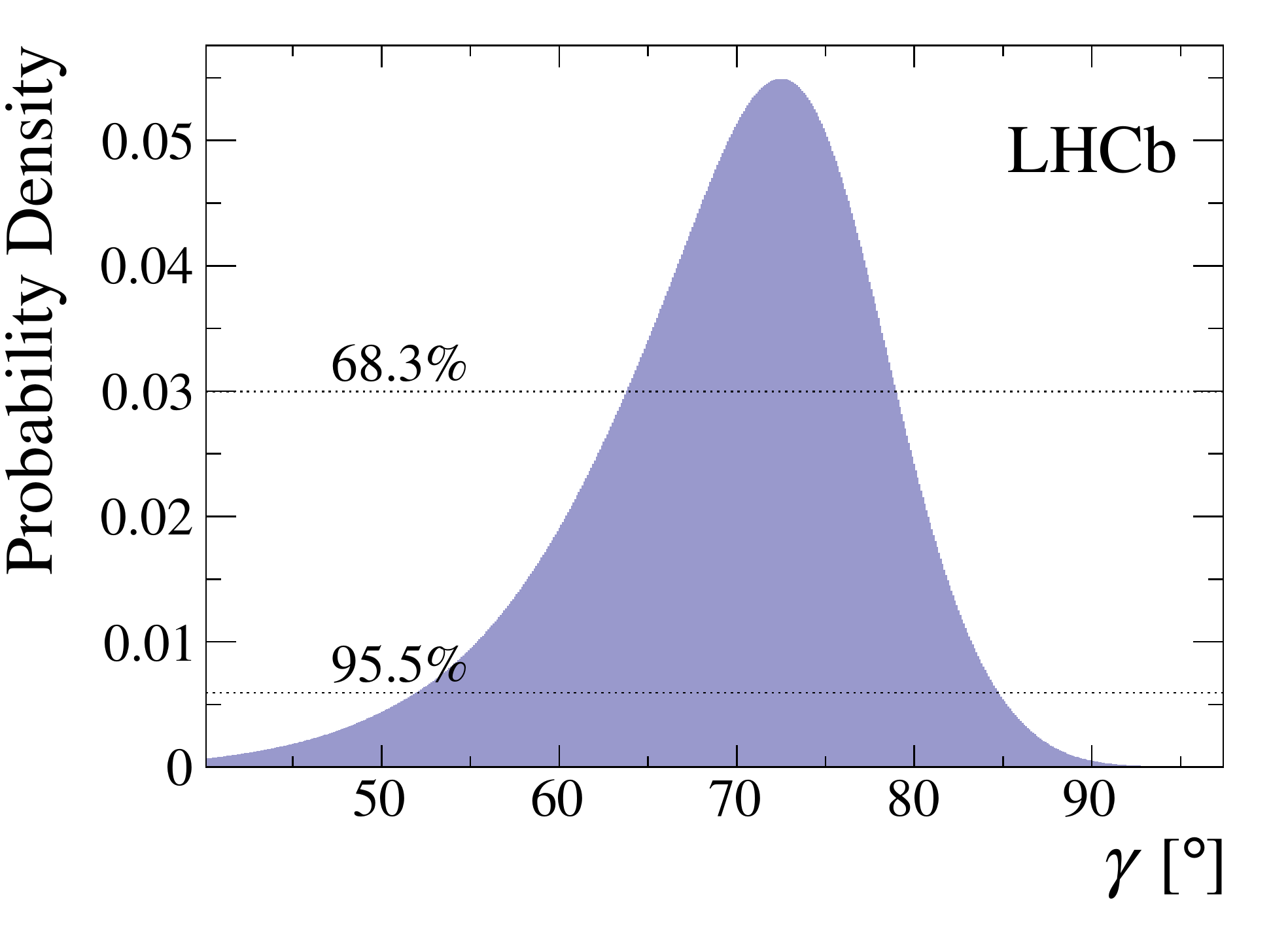}
  \caption{Posterior probability density from the Bayesian interpretation for the \Dh combination. The inset for \rbpi shows the same distribution on a logarithmic scale.}
  \label{fig:DhcombBayes}
\end{figure}

\begin{figure}
  \centering
  \includegraphics[width=.40\textwidth]{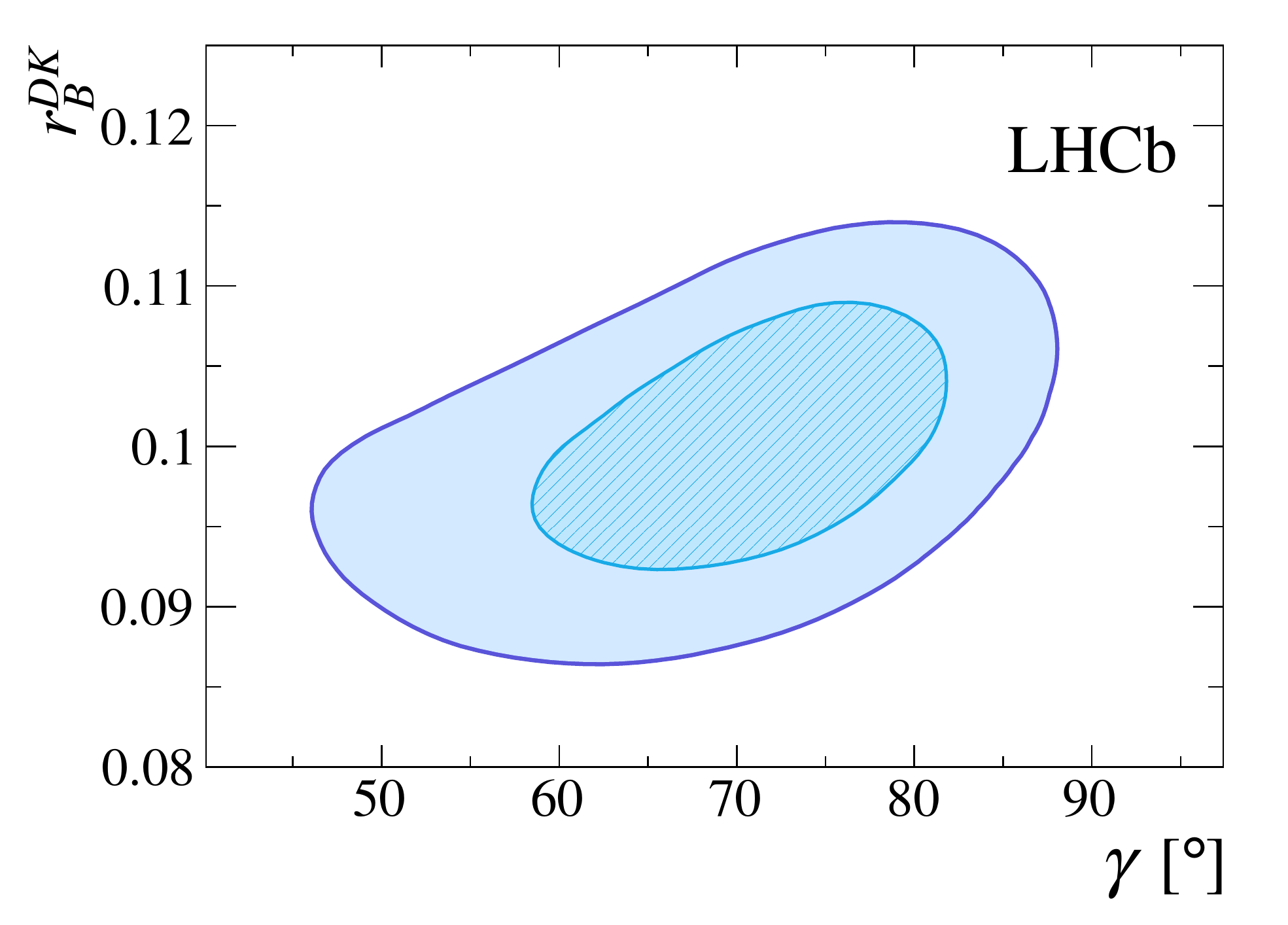}
  \includegraphics[width=.40\textwidth]{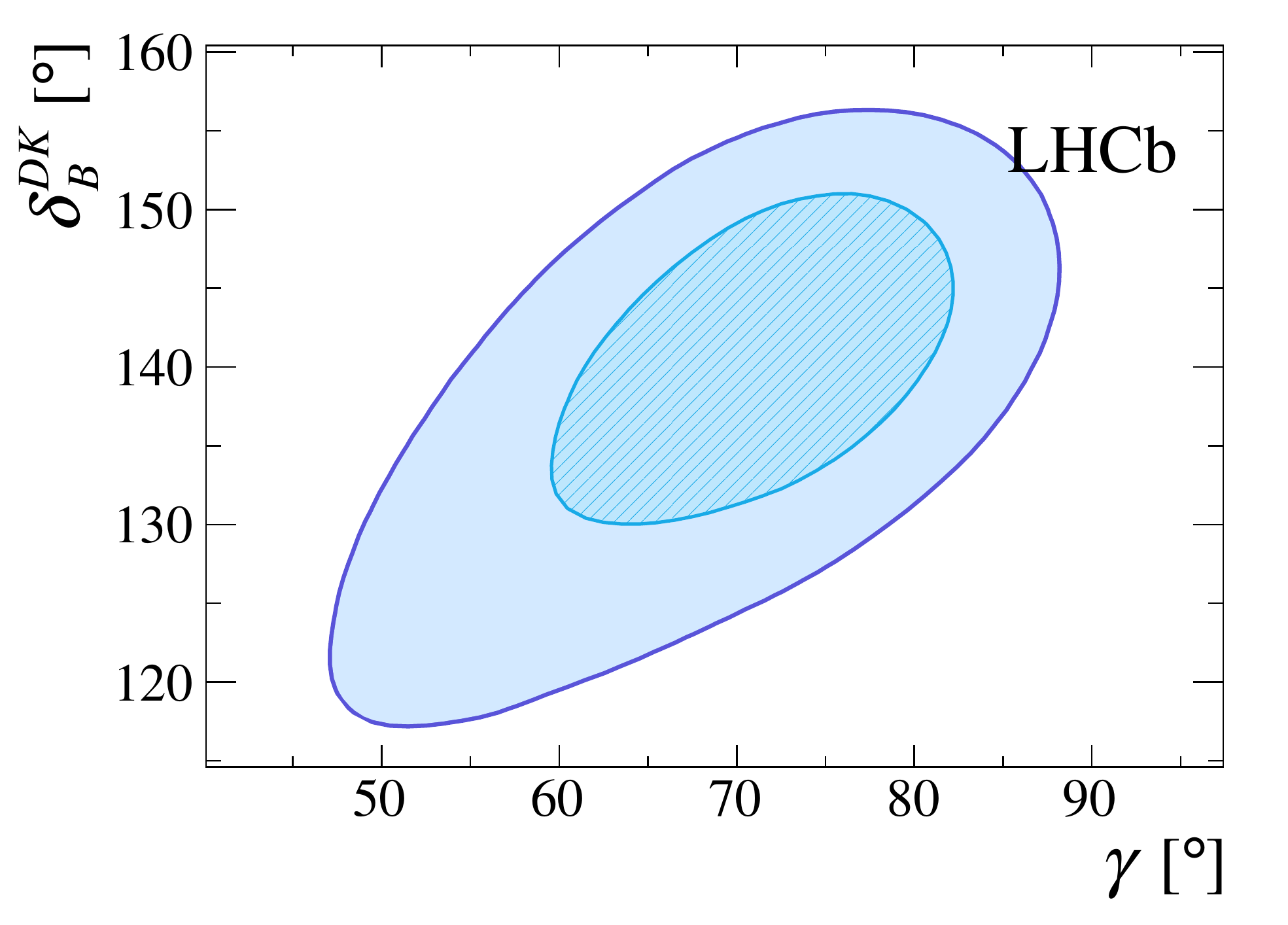}
  \includegraphics[width=.40\textwidth]{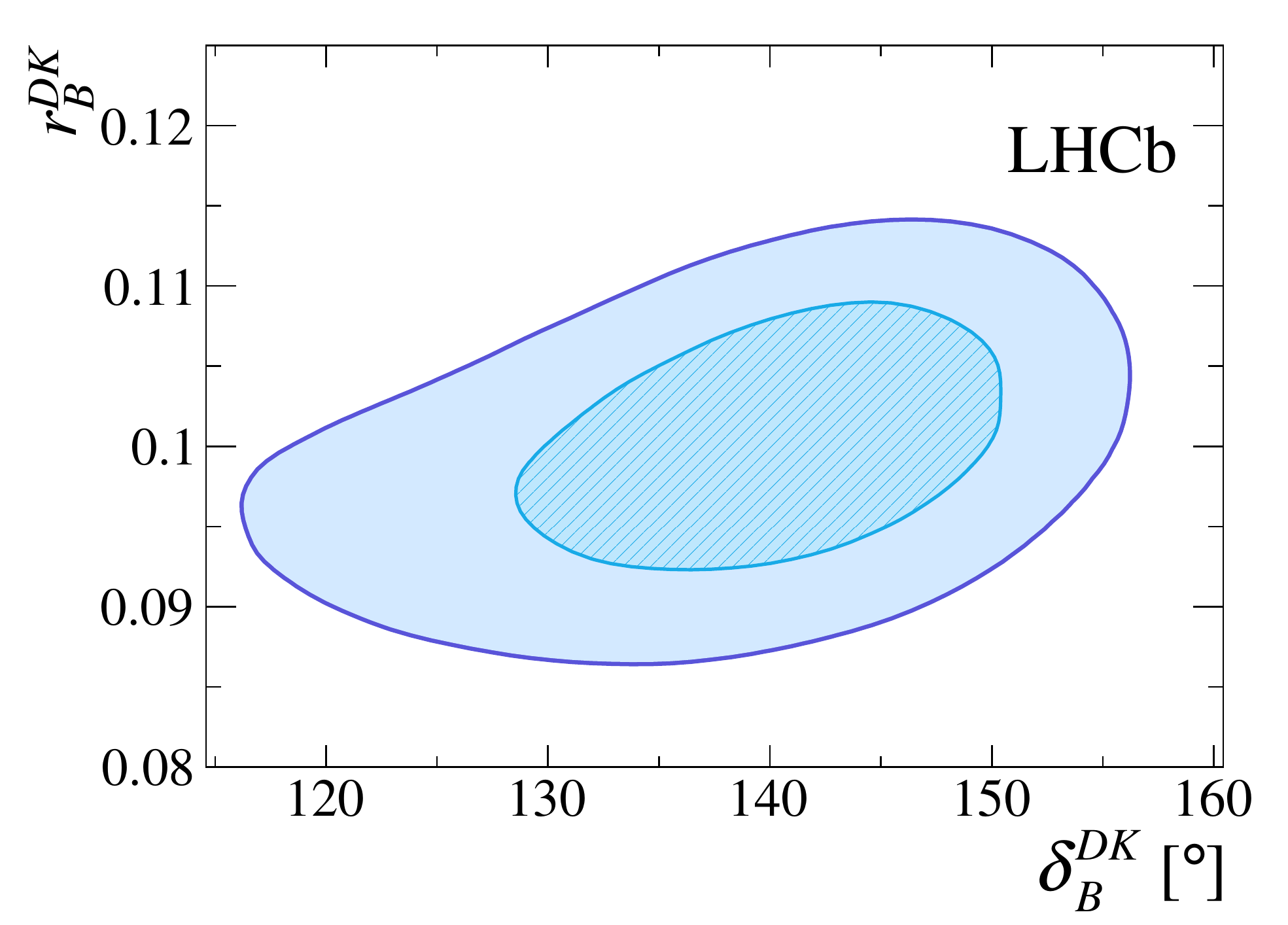}
  \includegraphics[width=.40\textwidth]{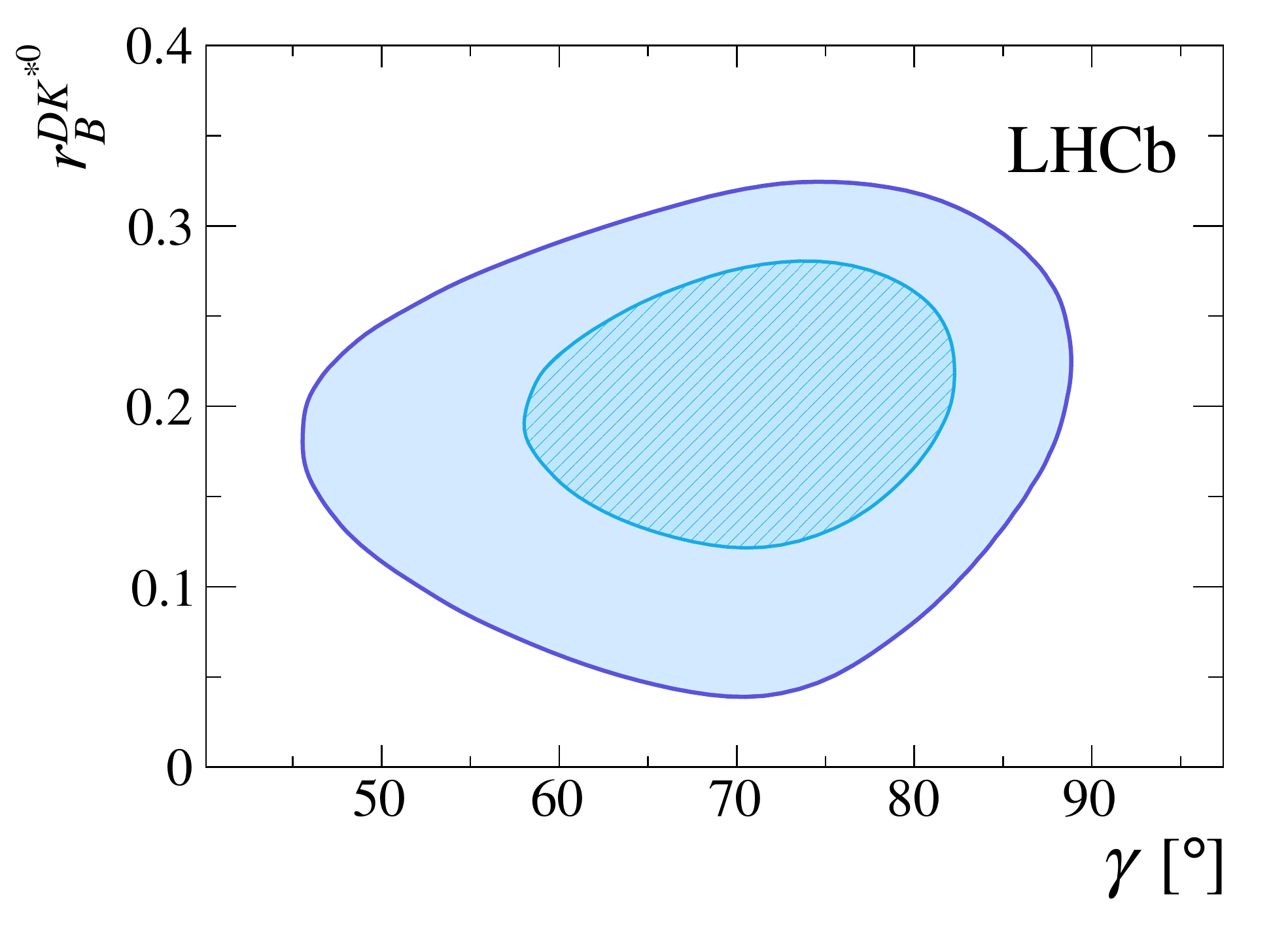}
  \includegraphics[width=.40\textwidth]{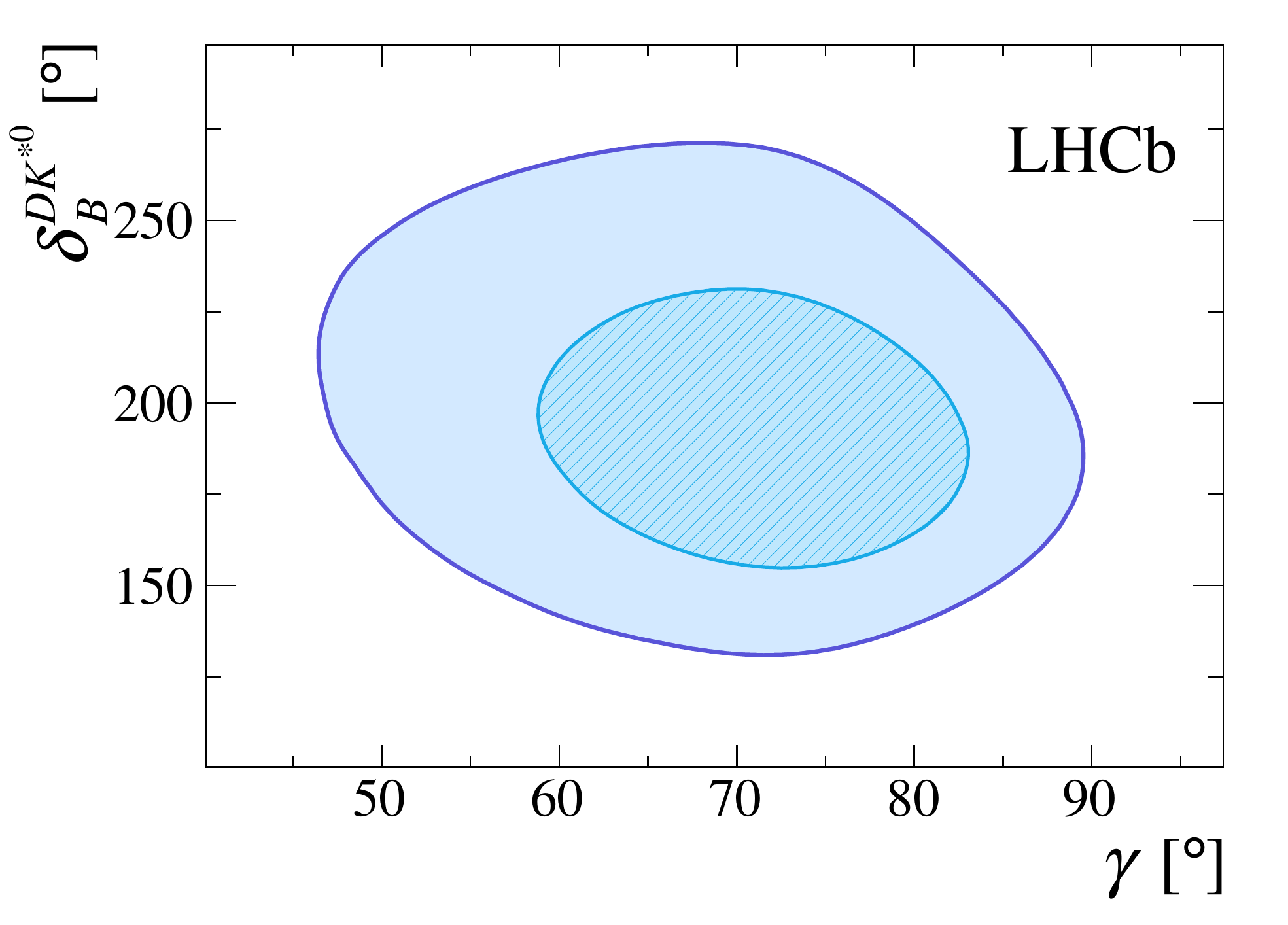}
  \includegraphics[width=.40\textwidth]{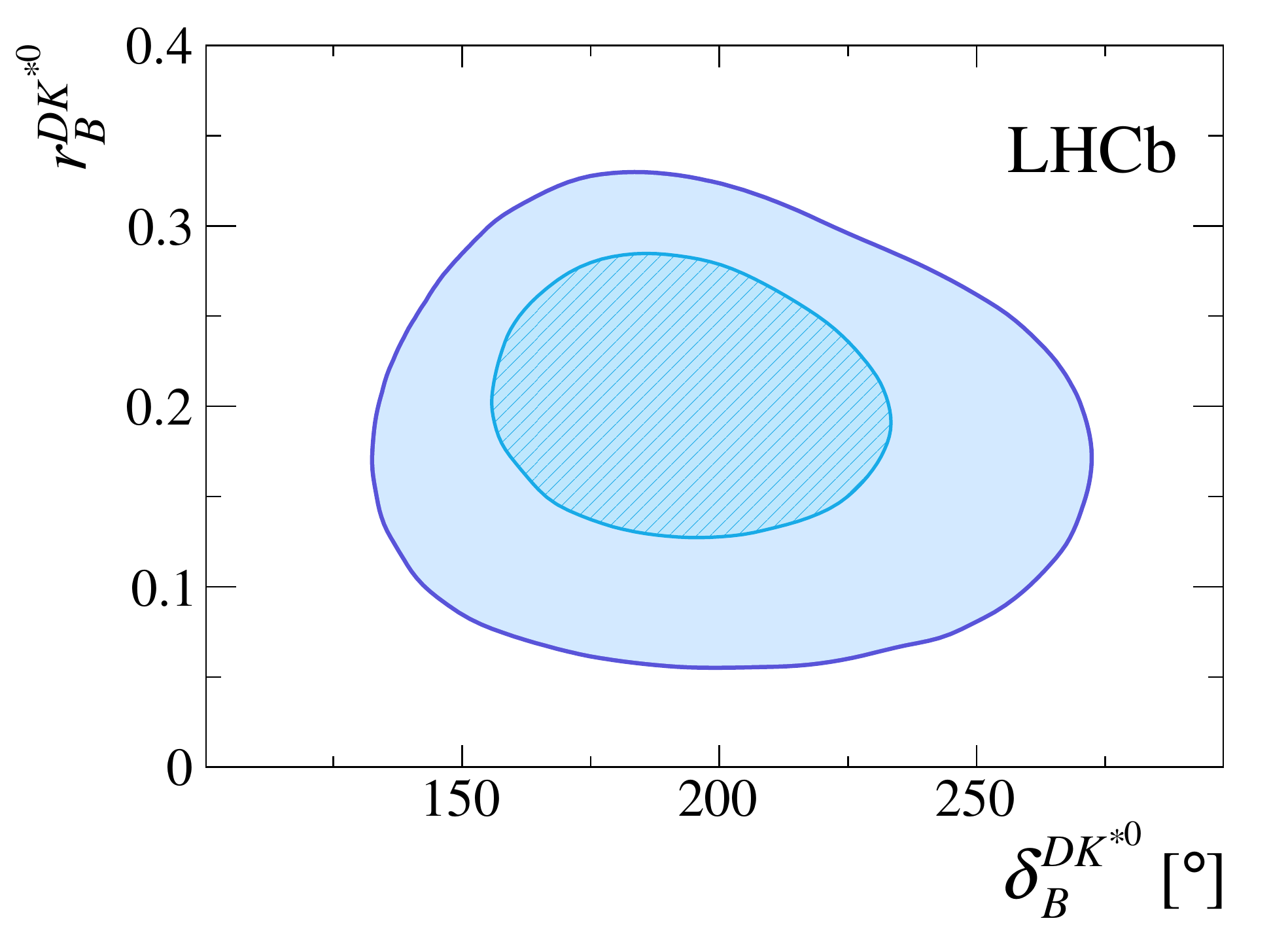}
  \includegraphics[width=.40\textwidth]{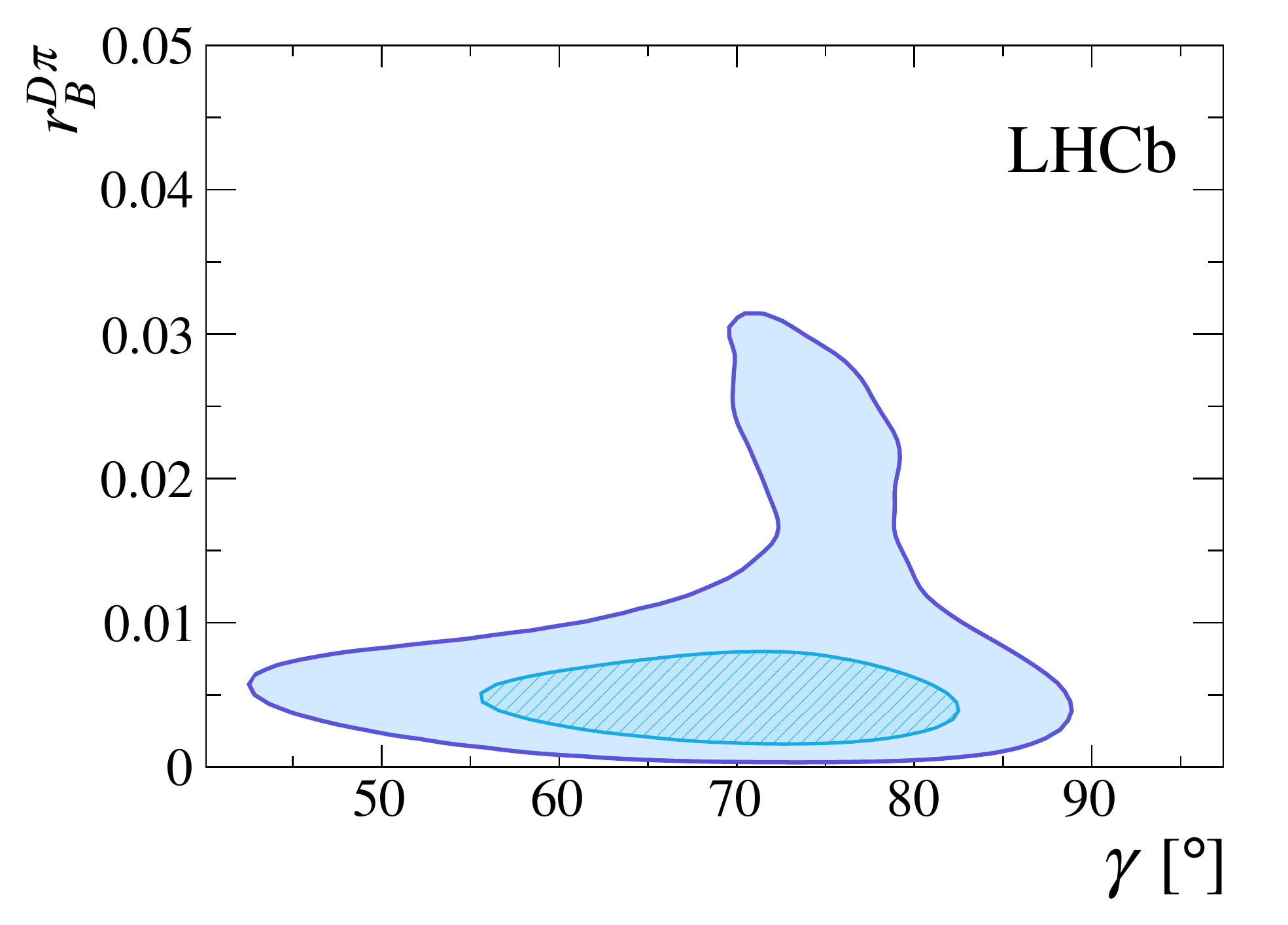}
  \includegraphics[width=.40\textwidth]{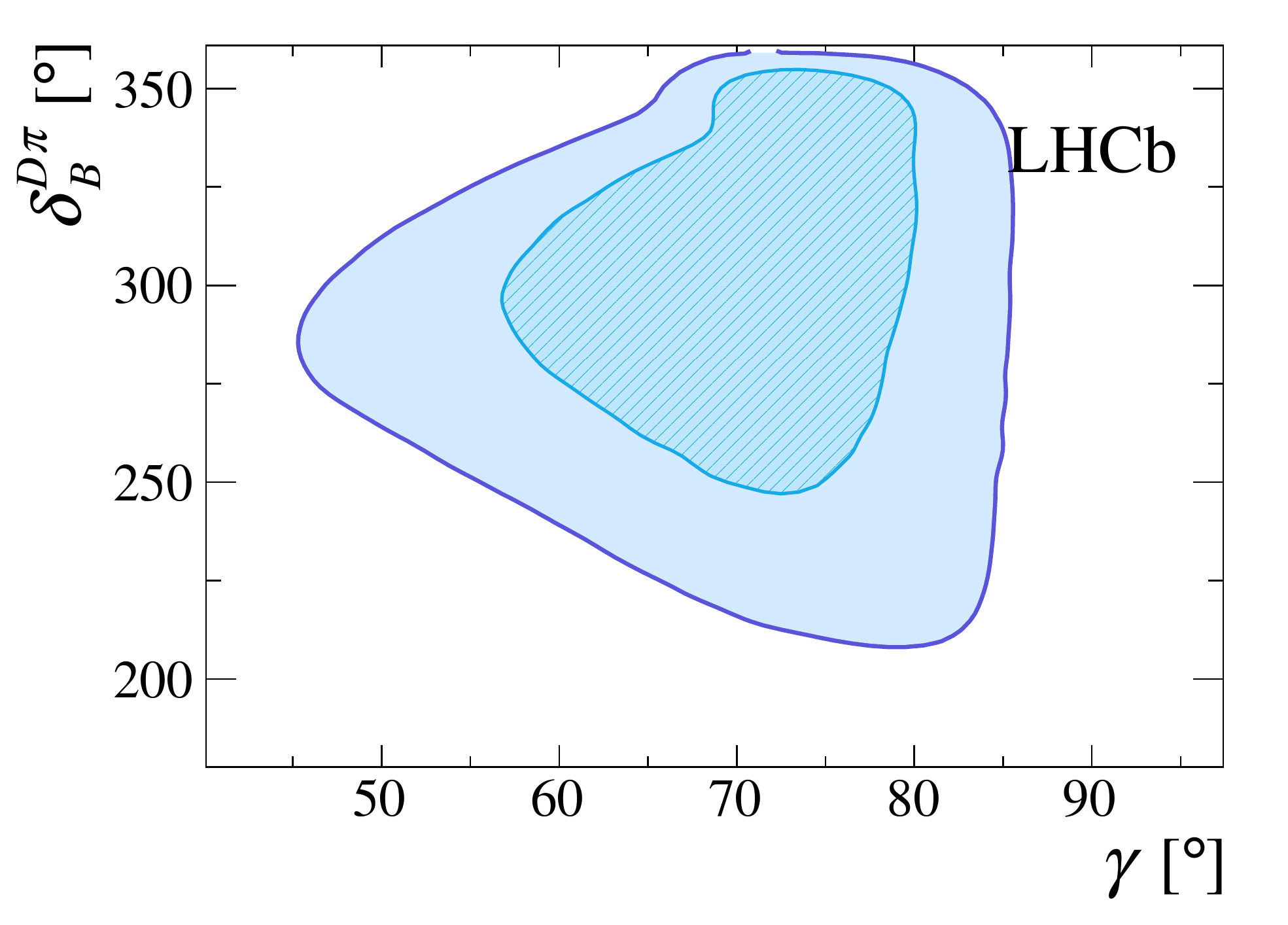}
  \includegraphics[width=.40\textwidth]{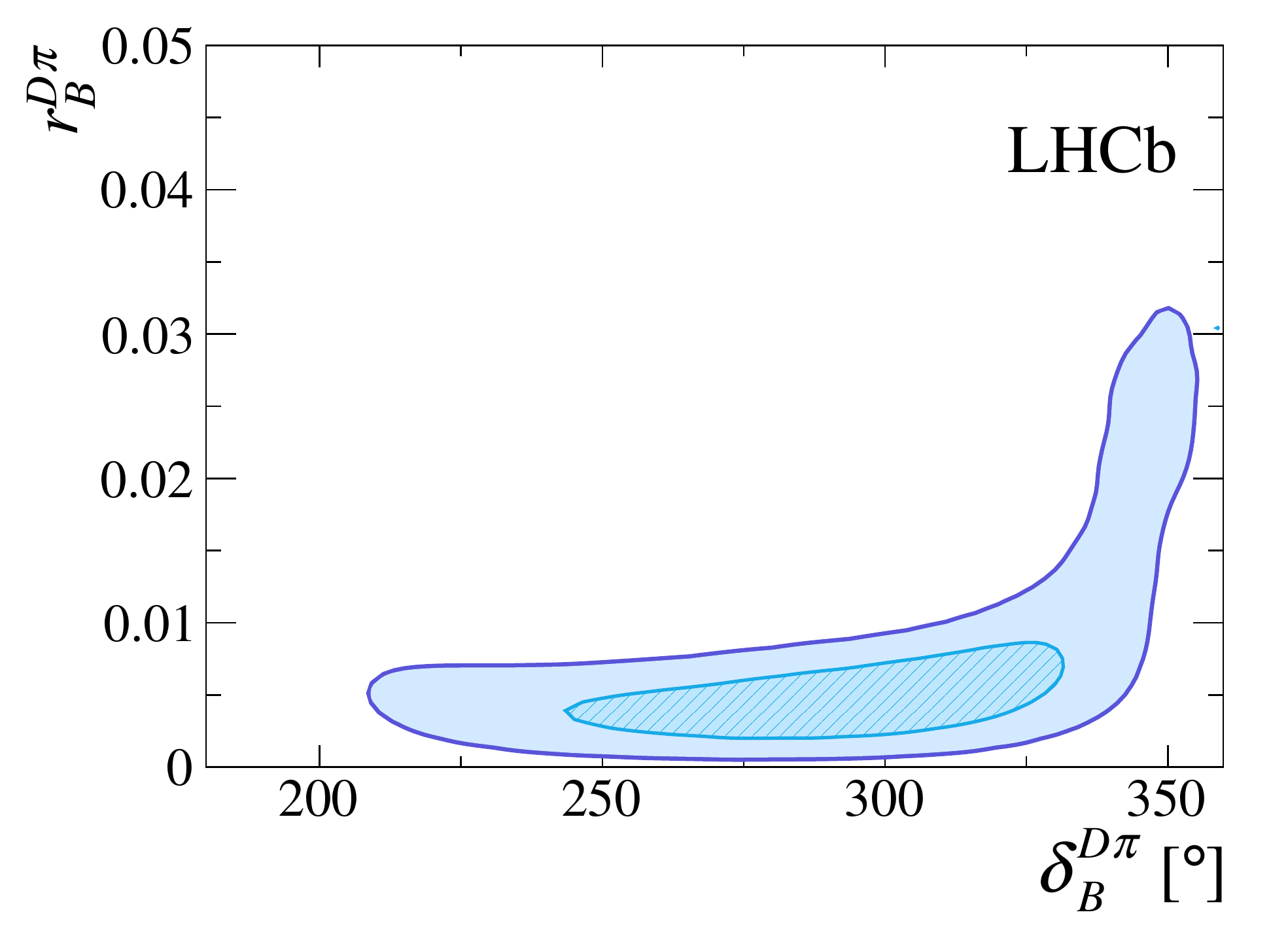}
  \caption{Two-dimensional posterior probability regions from the Bayesian interpretation for the \Dh combination. Light and dark regions show the 68.3\% and 95.5\% credible intervals respectively.}
  \label{fig:DhcombBayes2D}
\end{figure}

The results are given in Table~\ref{tab:resBayesDh} and shown in Figs.~\ref{fig:DhcombBayes} and~\ref{fig:DhcombBayes2D}.
Comparison with the frequentist treatment (Sec.~\ref{sec:results_dh}) shows that the $1\sigma$ intervals and regions differ between the two treatments, but satisfactory agreement is recovered at $2\sigma$. Such differences are not uncommon when comparing confidence and credible intervals or regions with low enough confidence level and probability, in the presence of a highly non-Gaussian likelihood function.


\begin{table}[!htb]
\caption{\label{tab:resBayesDh}Credible intervals and most probable values for
the hadronic parameters determined from the \Dh Bayesian combination.
}
\renewcommand{\arraystretch}{1.4}
\begin{center}
\resizebox{\textwidth}{!}{%
\begin{tabular}{p{2cm}cccc}
\hline
Observable & Central value & 68.3\% Interval & 95.5\% Interval & 99.7\% Interval\\
\hline
$\gamma\, (^{\circ})$	  & \gFullCentralBayes	      & \gFullOnesigBayes  	    & \gFullTwosigBayes        & \gFullThreesigBayes       \\
$\rb$		              & \rbFullCentralBayes	      & \rbFullOnesigBayes	    & \rbFullTwosigBayes       & \rbFullThreesigBayes      \\
$\db (^{\circ})$      & \dbFullCentralBayes	      & \dbFullOnesigBayes	    & \dbFullTwosigBayes       & \dbFullThreesigBayes      \\
$\rbDKstz$			      & \rbDKstzFullCentralBayes	& \rbDKstzFullOnesigBayes	& \rbDKstzFullTwosigBayes  & \rbDKstzFullThreesigBayes \\
$\dbDKstz (^{\circ})$	& \dbDKstzFullCentralBayes	& \dbDKstzFullOnesigBayes	& \dbDKstzFullTwosigBayes  & \dbDKstzFullThreesigBayes \\
\rbpi			            & \rbpiFullCentralBayes		    & \rbpiFullOnesigBayes		  & \rbpiFullTwosigBayes       & \rbpiFullThreesigBayes      \\
$\dbpi(^{\circ})$	    & \dbpiFullCentralBayes		    & \dbpiFullOnesigBayes		  & \dbpiFullTwosigBayes       & \dbpiFullThreesigBayes      \\
\hline
\end{tabular}%
}
\end{center}
\end{table}

\clearpage

\section{Conclusion}
\label{sec:conclusion}

Observables measured by LHCb that have sensitivity to the
CKM angle \g, along with auxiliary information from other experiments,
are combined to determine an improved constraint on \g.
%
%
%
Combination of all $B\to DK$-like modes results in a best fit value of $\g=\gRobustCentral^\circ$
and the
confidence intervals
\begin{align}
  \g \in \gRobustOnesig^\circ \; &\text{at} \; 68.3\%~{\rm CL}\,,\nonumber \\
  \g \in \gRobustTwosig^\circ \; &\text{at} \; 95.5\%~{\rm CL}\,\nonumber .
\end{align}

A second combination is investigated with additional inputs from $B\to D\pi$-like modes.
The frequentist and Bayesian approaches are in agreement at the $2\sigma$ level, giving intervals of $\g \in \gFullTwosig^\circ$
and $\g \in \gFullTwosigBayes^\circ \; \text{at} \; 95.5\%~{\rm CL}$, respectively.

Taking the best fit value and the $68.3\%$ CL interval of the \DK combination $\gamma$ is found to be
\begin{align}
\g = (\gQuoted)\degrees\, \nonumber,
\end{align}
where the uncertainty includes both statistical and systematic effects.
A Bayesian interpretation yields similar results, with credible intervals found to be consistent with the
corresponding confidence intervals of the frequentist treatment. The result for \g is compatible
with the world averages~\cite{CKMfitter,UTfit} and the previous LHCb average, $\g = (73^{+9}_{-10})^{\circ}$~\cite{LHCb-CONF-2014-004}.
This combination has a significantly smaller uncertainty than the previous one
and replaces it as the most precise determination of $\g$ from a single experiment to date.

Additional inputs to the combinations in the future will add extra sensitivity, this includes use of new decay modes (such as $\Bp\to\D\Kstarp$),
updates of current measurements to the full Run I data sample (such as \BsDsK) and inclusion of the Run II data sample.
Exploiting the full LHCb Run II data sample over the coming years is expected to reduce the uncertainty on \g to approximately $4\degrees$.


\section*{Acknowledgements}

\noindent We would like to acknowledge the significant efforts of our
late friend and colleague Moritz Karbach who invested considerable time and hard
work into the studies presented in this paper.
We express our gratitude to our colleagues in the CERN
accelerator departments for the excellent performance of the LHC. We
thank the technical and administrative staff at the LHCb
institutes. We acknowledge support from CERN and from the national
agencies: CAPES, CNPq, FAPERJ and FINEP (Brazil); NSFC (China);
CNRS/IN2P3 (France); BMBF, DFG and MPG (Germany); INFN (Italy);
FOM and NWO (The Netherlands); MNiSW and NCN (Poland); MEN/IFA (Romania);
MinES and FASO (Russia); MinECo (Spain); SNSF and SER (Switzerland);
NASU (Ukraine); STFC (United Kingdom); NSF (USA).
We acknowledge the computing resources that are provided by CERN, IN2P3 (France), KIT and DESY (Germany), INFN (Italy), SURF (The Netherlands), PIC (Spain), GridPP (United Kingdom), RRCKI and Yandex LLC (Russia), CSCS (Switzerland), IFIN-HH (Romania), CBPF (Brazil), PL-GRID (Poland) and OSC (USA). We are indebted to the communities behind the multiple open
source software packages on which we depend.
Individual groups or members have received support from AvH Foundation (Germany),
EPLANET, Marie Sk\l{}odowska-Curie Actions and ERC (European Union),
Conseil G\'{e}n\'{e}ral de Haute-Savoie, Labex ENIGMASS and OCEVU,
R\'{e}gion Auvergne (France), RFBR and Yandex LLC (Russia), GVA, XuntaGal and GENCAT (Spain), Herchel Smith Fund, The Royal Society, Royal Commission for the Exhibition of 1851 and the Leverhulme Trust (United Kingdom).

\newpage
\addcontentsline{toc}{section}{Appendices}
{\noindent\bf\Large Appendices}
\begin{appendix}
\section{Relationships between parameters and observables}
\label{sec:relations}

The equations given in this section reflect the relationship between the experimental observables and the parameters of interest.
For simplicity, the equations are given in the absence of \Dz--\Dzb mixing. In order to include the small ($<0.5\degrees$) effects
from \Dz--\Dzb mixing, the equations should be modified following the recommendation in Ref.~\cite{Rama:2013voa}, making use of
the \Dz decay time acceptance coefficients, $M_{xy}$, given in Table~\ref{tab:mxy}.

\begin{table}[h!]
  \begin{center}
    \caption{\Dz decay time acceptance coefficients (see Ref.~\cite{Rama:2013voa}) for each analysis.}
    \renewcommand{\arraystretch}{1.4}
    \begin{tabular}{ll}
      \hline
      Analysis          &  $M_{xy}$ \\
      \hline
      \Dhh GLW/ADS      & 0.594 \\
      \Dhhpiz GLW/ADS   & 0.592 \\
      \Dhpipipi GLW/ADS & 0.570 \\
      \BuDhpipi GLW/ADS & 0.6 \\
      \DKSKpi GLS       & 0.6 \\
      \BdDzKstz ADS     & 0.6 \\
      \hline
    \end{tabular}
  \label{tab:mxy}
  \end{center}
\end{table}

\addtocontents{toc}{\setcounter{tocdepth}{1}}
\subsection{\boldmath\BuDh, \Dhh observables}
\label{sec:relations_glwads}

\begin{alignat}{4}
& \AadsDK       & \;=&\;  \frac{2\rb\rdKpi\sin\left(\db+\ddKpi\right)\sin\gamma}{\left(\rb\right)^{2} + \left(\rdKpi\right)^{2} + 2\rb\rdKpi\cos\left(\db+\ddKpi\right)\cos\gamma}  \nonumber \\ & & & \nonumber \\
& \AadsDPi      & \;=&\;  \frac{2\rbpi\rdKpi\sin\left(\dbpi+\ddKpi\right)\sin\gamma}{\left(\rbpi\right)^{2} + \left(\rdKpi\right)^{2} + 2\rbpi\rdKpi\cos\left(\dbpi+\ddKpi\right)\cos\gamma}  \nonumber \\ & & & \nonumber \\
& \AcpDkKK      & \;=&\;  \frac{2\rb\sin\db\sin\gamma}{1 + \left(\rb\right)^{2} + 2\rb\cos\db\cos\gamma} + \DAcpKK \nonumber \\ & & & \nonumber \\
& \AcpDkPipi    & \;=&\;  \frac{2\rb\sin\db\sin\gamma}{1 + \left(\rb\right)^{2} + 2\rb\cos\db\cos\gamma} + \DAcpPipi \nonumber \\ & & & \nonumber \\
& \AcpDpiKK     & \;=&\;  \frac{2\rbpi\sin\dbpi\sin\gamma}{1 + \left(\rbpi\right)^{2} + 2\rbpi\cos\dbpi\cos\gamma} + \DAcpKK  \nonumber \\ & & & \nonumber \\
& \AcpDpiPipi   & \;=&\;  \frac{2\rbpi\sin\dbpi\sin\gamma}{1 + \left(\rbpi\right)^{2} + 2\rbpi\cos\dbpi\cos\gamma} + \DAcpPipi  \nonumber \\ & & & \nonumber \\
& \AfavDkKpi    & \;=&\;  \frac{2\rb\rdKpi\sin\left(\db-\ddKpi\right)\sin\gamma}{1 + \left(\rb\rdKpi\right)^{2} + 2\rb\rdKpi\cos\left(\db-\ddKpi\right)\cos\gamma}  \nonumber \\ & & & \nonumber \\
& \RadsDK       & \;=&\;  \frac{\left(\rb\right)^{2} + \left(\rdKpi\right)^{2} + 2\rb\rdKpi\cos\left(\db+\ddKpi\right)\cos\gamma}{1 + \left(\rb\rdKpi\right)^{2} + 2\rb\rdKpi\cos\left(\db-\ddKpi\right)\cos\gamma}  \nonumber \\ & & & \nonumber \\
& \RadsDPi      & \;=&\;  \frac{\left(\rbpi\right)^{2} + \left(\rdKpi\right)^{2} + 2\rbpi\rdKpi\cos\left(\dbpi+\ddKpi\right)\cos\gamma}{1 + \left(\rbpi\rdKpi\right)^{2} + 2\rbpi\rdKpi\cos\left(\dbpi-\ddKpi\right)\cos\gamma}   \nonumber \\ & & & \nonumber \\
& \RkpKpi       & \;=&\;   R \frac{1 + \left(\rb\rdKpi\right)^{2} + 2\rb\rdKpi\cos\left(\db-\ddKpi\right)\cos\gamma}{1 + \left(\rbpi\rdKpi\right)^{2} + 2\rbpi\rdKpi\cos\left(\dbpi-\ddKpi\right)\cos\gamma}  \nonumber \\ & & & \nonumber \\
& \RCPKK        & \;=&\; \frac{\RkpKK}{\RkpKpi}  = R \frac{1 + \rbsq + 2\rb\cos(\db)\cos(\g)}{1 + \rbpisq + 2\rbpi\cos(\dbpi)\cos(\g)} \nonumber \\ & & & \nonumber \\
& \RCPPiPi      & \;=&\; \frac{\RkpPipi}{\RkpKpi} = R \frac{1 + \rbsq + 2\rb\cos(\db)\cos(\g)}{1 + \rbpisq + 2\rbpi\cos(\dbpi)\cos(\g)}  \nonumber
\end{alignat}


\subsection{\boldmath\BuDh, \Dhpipipi observables}
\label{sec:relations_adsk3pi}

\begin{alignat}{4}
& \AadsDKKppp        & \;=&\; \frac{2\kdKppp\rb\rdKppp\sin\left(\db+\ddKppp\right)\sin\gamma}{\left(\rb\right)^{2} + \left(\rdKppp\right)^{2} + 2\kdKppp\rb\rdKppp\cos\left(\db+\ddKppp\right)\cos\gamma}  \nonumber \\ & & & \nonumber \\
& \AadsDPiKppp       & \;=&\; \frac{2\kdKppp\rbpi\rdKppp\sin\left(\dbpi+\ddKppp\right)\sin\gamma}{\left(\rbpi\right)^{2} + \left(\rdKppp\right)^{2} + 2\kdKppp\rbpi\rdKppp\cos\left(\dbpi+\ddKppp\right)\cos\gamma}  \nonumber \\ & & & \nonumber \\
& \AcpDKpppp         & \;=&\; \frac{2(2\Fpppp-1)\rb\sin\db\sin\gamma}{1 + \left(\rb\right)^{2} + 2(2\Fpppp-1)\rb\cos\db\cos\gamma}  \nonumber \\ & & & \nonumber \\
& \AcpDpipppp        & \;=&\; \frac{2(2\Fpppp-1)\rbpi\sin\dbpi\sin\gamma}{1 + \left(\rbpi\right)^{2} + 2(2\Fpppp-1)\rbpi\cos\dbpi\cos\gamma}  \nonumber \\ & & & \nonumber \\
& \AfavDkKppp        & \;=&\; \frac{2\kdKppp\rb\rdKppp\sin\left(\db-\ddKppp\right)\sin\gamma}{1 + \left(\rb\rdKppp\right)^{2} + 2\kdKppp\rb\rdKppp\cos\left(\db-\ddKppp\right)\cos\gamma}  \nonumber \\ & & & \nonumber \\
& \RadsDKKppp        & \;=&\; \frac{\left(\rb\right)^{2} + \left(\rdKppp\right)^{2} + 2\kdKppp\rb\rdKppp\cos\left(\db+\ddKppp\right)\cos\gamma}{1 + \left(\rb\rdKppp\right)^{2} + 2\kdKppp\rb\rdKppp\cos\left(\db-\ddKppp\right)\cos\gamma}  \nonumber \\ & & & \nonumber \\
& \RadsDPiKppp       & \;=&\; \frac{\left(\rbpi\right)^{2} + \left(\rdKppp\right)^{2} + 2\kdKppp\rbpi\rdKppp\cos\left(\dbpi+\ddKppp\right)\cos\gamma}{1 + \left(\rbpi\rdKppp\right)^{2} + 2\kdKppp\rbpi\rdKppp\cos\left(\dbpi-\ddKppp\right)\cos\gamma}   \nonumber \\ & & & \nonumber \\
& \RkpKppp           & \;=&\;  R \frac{1 + \left(\rb\rdKppp\right)^{2} + 2\kdKppp\rb\rdKppp\cos\left(\db-\ddKppp\right)\cos\gamma}{1 + \left(\rbpi\rdKppp\right)^{2} + 2\kdKppp\rbpi\rdKppp\cos\left(\dbpi-\ddKppp\right)\cos\gamma}  \nonumber \\ & & & \nonumber \\
& \RCPpppp           & \;=&\; \frac{\Rkppppp}{\RkpKppp} = R \frac{1 + \rbsq + 2\rb(2\Fpppp-1)\cos(\db)\cos(\g)}{1 + \rbpisq + 2\rbpi(2\Fpppp-1)\cos(\dbpi)\cos(\g)}   \nonumber
\end{alignat}


\subsection{\boldmath\BuDh, \Dhhpiz observables}
\label{sec:relations_adshpipiz}

\begin{alignat}{4}
& \AadsDKkpp     & \;=&\; \frac{2\kdKpp\rb\rdKpp\sin\left(\db+\ddKpp\right)\sin\gamma}{\left(\rb\right)^{2} + \left(\rdKpp\right)^{2} + 2\kdKpp\rb\rdKpp\cos\left(\db+\ddKpp\right)\cos\gamma}  \nonumber \\ & & & \nonumber \\
& \AadsDPikpp    & \;=&\; \frac{2\kdKpp\rbpi\rdKpp\sin\left(\dbpi+\ddKpp\right)\sin\gamma}{\left(\rbpi\right)^{2} + \left(\rdKpp\right)^{2} + 2\kdKpp\rbpi\rdKpp\cos\left(\dbpi+\ddKpp\right)\cos\gamma}  \nonumber \\ & & & \nonumber \\
& \AcpDkKKPiz    & \;=&\; \frac{2(2\FKKp-1)\rb\sin\db\sin\gamma}{1 + \left(\rb\right)^{2} + 2(2\FKKp-1)\rb\cos\db\cos\gamma}  \nonumber \\ & & & \nonumber \\
& \AcpDkPiPiPiz  & \;=&\; \frac{2(2\Fppp-1)\rb\sin\db\sin\gamma}{1 + \left(\rb\right)^{2} + 2(2\Fppp-1)\rb\cos\db\cos\gamma}  \nonumber \\ & & & \nonumber \\
& \AcpDpiKKPiz   & \;=&\; \frac{2(2\FKKp-1)\rbpi\sin\dbpi\sin\gamma}{1 + \left(\rbpi\right)^{2} + 2(2\FKKp-1)\rbpi\cos\dbpi\cos\gamma}  \nonumber \\ & & & \nonumber \\
& \AcpDpiPiPiPiz & \;=&\; \frac{2(2\Fppp-1)\rbpi\sin\dbpi\sin\gamma}{1 + \left(\rbpi\right)^{2} + 2(2\Fppp-1)\rbpi\cos\dbpi\cos\gamma}  \nonumber \\ & & & \nonumber \\
& \AfavDkKPiPiz  & \;=&\; \frac{2\kdKpp\rb\rdKpp\sin\left(\db-\ddKpp\right)\sin\gamma}{1 + \left(\rb\rdKpp\right)^{2} + 2\kdKpp\rb\rdKpp\cos\left(\db-\ddKpp\right)\cos\gamma}  \nonumber \\ & & & \nonumber \\
& \RadsDKkpp     & \;=&\; \frac{\left(\rb\right)^{2} + \left(\rdKpp\right)^{2} + 2\kdKpp\rb\rdKpp\cos\left(\db+\ddKpp\right)\cos\gamma}{1 + \left(\rb\rdKpp\right)^{2} + 2\kdKpp\rb\rdKpp\cos\left(\db-\ddKpp\right)\cos\gamma}  \nonumber \\ & & & \nonumber \\
& \RadsDPikpp    & \;=&\; \frac{\left(\rbpi\right)^{2} + \left(\rdKpp\right)^{2} + 2\kdKpp\rbpi\rdKpp\cos\left(\dbpi+\ddKpp\right)\cos\gamma}{1 + \left(\rbpi\rdKpp\right)^{2} + 2\kdKpp\rbpi\rdKpp\cos\left(\dbpi-\ddKpp\right)\cos\gamma}   \nonumber \\ & & & \nonumber \\
& \RCPKKPiz      & \;=&\; \frac{\RkpKKp}{\RkpKpp} = R \frac{1 + \rbsq + 2\rb(2\FKKp-1)\cos(\db)\cos(\g)}{1 + \rbpisq + 2\rbpi(2\FKKp-1)\cos(\dbpi)\cos(\g)}  \nonumber \\ & & & \nonumber \\
& \RCPPiPiPiz    & \;=&\; \frac{\Rkpppp}{\RkpKpp} = R \frac{1 + \rbsq + 2\rb(2\Fppp-1)\cos(\db)\cos(\g)}{1 + \rbpisq + 2\rbpi(2\Fppp-1)\cos(\dbpi)\cos(\g)} \nonumber
\end{alignat}


\subsection{\boldmath\BuDK, \DKShh observables}
\label{sec:relations_ggsz}

\begin{alignat}{4}
& \xm & \;=&\; \rb \cos(\db -\gamma)~ \nonumber \\
& \ym & \;=&\; \rb \sin(\db -\gamma)~ \nonumber \\
& \xp & \;=&\; \rb \cos(\db +\gamma)~ \nonumber \\
& \yp & \;=&\; \rb \sin(\db +\gamma)~ \nonumber
\end{alignat}


\subsection{\boldmath\BuDh, \DKSKpi observables}
\label{sec:relations_glskskpi}

\begin{alignat}{4}
& \RfavsupDkKskpi  &=\frac{1+\rbsq \rdKskpisq + 2\kdKskpi \rb \rdKskpi \cos(\db -\ddKskpi )\cos\gamma}{\rbsq + \rdKskpisq + 2\kdKskpi \rb \rdKskpi \cos(\db +\ddKskpi )\cos\gamma}\, \nonumber \\ & & & \nonumber \\
& \AfavDkKskpi     &=\frac{2\kdKppp \rb \rdKppp \sin(\db -\ddKppp )\sin\gamma}{1+\rbsq \rdKpppsq + 2\kdKppp \rb \rdKppp \cos(\db -\ddKppp )\cos\gamma}\, \nonumber \\ & & & \nonumber \\
& \AsupDkKskpi     &=\frac{2\kdKppp \rb \rdKppp \sin(\db +\ddKppp )\sin\gamma}{\rbsq + \rdKskpisq + 2\kdKskpi \rb \rdKskpi \cos(\db +\ddKskpi )\cos\gamma}\, \nonumber
\end{alignat}


\subsection{\boldmath\BuDhpipi, \Dhh observables}
\label{sec:relations_glwadsdkpipi}

\begin{alignat}{4}
& \RcpDkpipi     & \;=&\; 1 + \rbDkppsq + 2\kbDkpp\rbDkpp\cos\dbDkpp \cos\g\,, \nonumber \\ & & & \nonumber \\
& \AfavDkpipiKpi & \;=&\; \frac{2\kbDkpp\rbDkpp\rdKpi  \sin(\dbDkpp-\ddKpi)\sin\gamma} {1 + \rbDkppsq\rdKpisq + 2\kbDkpp\rbDkpp\rdKpi \cos(\dbDkpp-\ddKpi)\cos\gamma}\, \nonumber \\ & & & \nonumber \\
& \AfavDpipipiKpi& \;=&\; \frac{2\kbDppp\rbDppp\rdKpi  \sin(\dbDppp-\ddKpi)\sin\gamma} {1 + \rbDpppsq\rdKpisq + 2\kbDppp\rbDppp\rdKpi \cos(\dbDppp-\ddKpi)\cos\gamma}\, \nonumber \\ & & & \nonumber \\
& \AcpDkpipiKK   & \;=&\; \frac{2\kbDkpp\rbDkpp  \sin\dbDkpp\sin\gamma} {1 + \rbDkppsq + 2\kbDkpp\rbDkpp \cos\dbDkpp\cos\gamma} + \DAcpKK\, \nonumber \\ & & & \nonumber \\
& \AcpDkpipiPipi & \;=&\; \frac{2\kbDkpp\rbDkpp  \sin\dbDkpp\sin\gamma} {1 + \rbDkppsq + 2\kbDkpp\rbDkpp \cos\dbDkpp\cos\gamma} + \DAcpPipi\, \nonumber \\ & & & \nonumber \\
& \AcpDpipipiKK  & \;=&\; \frac{2\kbDppp\rbDppp  \sin\dbDppp\sin\gamma} {1 + \rbDpppsq + 2\kbDppp\rbDppp \cos\dbDppp\cos\gamma} + \DAcpKK\, \nonumber \\ & & & \nonumber \\
& \AcpDpipipiPipi& \;=&\; \frac{2\kbDkpp\rbDkpp  \sin\dbDkpp\sin\gamma} {1 + \rbDkppsq + 2\kbDkpp\rbDkpp \cos\dbDkpp\cos\gamma} + \DAcpPipi\, \nonumber \\ & & & \nonumber \\
& \RpDkpipiKpi   & \;=&\; \frac{\rbDkppsq  + \rdKpisq + 2\kbDkpp\rbDkpp\rdKpi \cos(\dbDkpp +\ddKpi +\gamma)}{1 + \rbDkppsq\rdKpisq + 2\kbDkpp\rbDkpp\rdKpi \cos(\dbDkpp -\ddKpi +\gamma)}\, \nonumber \\ & & & \nonumber \\
& \RmDkpipiKpi   & \;=&\; \frac{\rbDkppsq  + \rdKpisq + 2\kbDkpp\rbDkpp\rdKpi \cos(\dbDkpp +\ddKpi -\gamma)}{1 + \rbDkppsq\rdKpisq + 2\kbDkpp\rbDkpp\rdKpi \cos(\dbDkpp -\ddKpi -\gamma)}\, \nonumber \\ & & & \nonumber \\
& \RpDpipipiKpi  & \;=&\; \frac{\rbDpppsq  + \rdKpisq + 2\kbDppp\rbDppp\rdKpi \cos(\dbDppp +\ddKpi +\gamma)}{1 + \rbDpppsq\rdKpisq + 2\kbDppp\rbDppp\rdKpi \cos(\dbDppp -\ddKpi +\gamma)}\, \nonumber \\ & & & \nonumber \\
& \RmDpipipiKpi  & \;=&\; \frac{\rbDkppsq  + \rdKpisq + 2\kbDkpp\rbDkpp\rdKpi \cos(\dbDkpp +\ddKpi -\gamma)}{1 + \rbDkppsq\rdKpisq + 2\kbDkpp\rbDkpp\rdKpi \cos(\dbDkpp -\ddKpi -\gamma)}\, \nonumber
\end{alignat}


\subsection{\boldmath\BdDzKstz, \DKpi observables}
\label{sec:relations_glwadsdkst}

\small{
\begin{flalign}
  \noindent
  & \AfavDkstKpi \;= \;  \nonumber \\
  & \;\; \frac{2\kbDKstz\RbDKstz\rbDKstz\rdKpi  \sin(\dbDKstz + \DbDKstz   -\ddKpi)\sin\gamma} {1 + (\RbDKstz\rbDKstz)^2\rdKpisq + 2\kbDKstz\RbDKstz\rbDKstz\rdKpi \cos(\dbDKstz+\DbDKstz-\ddKpi)\cos\gamma}\,  \nonumber \\ & \nonumber \\
  & \RpDkstKpi   \;= \; \nonumber \\
  & \;\; \frac{(\RbDKstz\rbDKstz)^2  + \rdKpisq + 2\kbDKstz\RbDKstz\rbDKstz\rdKpi \cos(\dbDKstz +\DbDKstz +\ddKpi +\gamma)}{1 + (\RbDKstz\rbDKstz)^2\rdKpisq + 2\kbDKstz\RbDKstz\rbDKstz\rdKpi \cos(\dbDKstz + \DbDKstz -\ddKpi +\gamma)}\,  \nonumber \\ & \nonumber \\
  & \RmDkstKpi  \; = \; \nonumber  \nonumber \\
  & \;\; \frac{(\RbDKstz\rbDKstz)^2  + \rdKpisq + 2\kbDKstz\RBDKstz\rbDKstz\rdKpi \cos(\dbDKstz +\DbDKstz +\ddKpi -\gamma)}{1 + (\RbDKstz\rbDKstz)^2\rdKpisq + 2\kbDKstz\RbDKstz\rbDKstz\rdKpi \cos(\dbDKstz + \DbDKstz -\ddKpi -\gamma)}\,  \nonumber %
\end{flalign}
} %


\subsection{\boldmath\BdDKpi, \Dhh observables}
\label{sec:relations_dalitzdkst}

\begin{alignat}{4}
& \xmdkpi & \;=&\; \rbDKstz \cos(\dbDKstz -\gamma)~ \nonumber \\
& \ymdkpi & \;=&\; \rbDKstz \sin(\dbDKstz -\gamma)~ \nonumber \\
& \xpdkpi & \;=&\; \rbDKstz \cos(\dbDKstz +\gamma)~ \nonumber \\
& \ypdkpi & \;=&\; \rbDKstz \sin(\dbDKstz +\gamma)~ \nonumber
\end{alignat}


\subsection{\boldmath\BdDzKstz, \DKSpipi observables}
\label{sec:relations_ggszdkst}

\begin{alignat}{4}
& \xmdkst & \;=&\; \RbDKstz\rbDKstz \cos(\dbDKstz + \DbDKstz -\gamma)~ \nonumber \\
& \ymdkst & \;=&\; \RbDKstz\rbDKstz \sin(\dbDKstz + \DbDKstz -\gamma)~ \nonumber \\
& \xpdkst & \;=&\; \RbDKstz\rbDKstz \cos(\dbDKstz + \DbDKstz +\gamma)~ \nonumber \\
& \ypdkst & \;=&\; \RbDKstz\rbDKstz \sin(\dbDKstz + \DbDKstz +\gamma)~ \nonumber
\end{alignat}


\subsection{\boldmath\BsDsK observables}
\label{sec:relations_dsk}

\begin{alignat}{4}
& \Cpar & \;=&\; \frac{1-\rdsksq}{1+\rdsksq}\,, \nonumber \\ & & & \nonumber \\
& \Dpar & \;=&\; \frac{2\rdsk\cos(\ddsk-(\g+\phis))}{1+\rdsksq}\, \nonumber \\ & & & \nonumber \\
& \Dbpar& \;=&\; \frac{2\rdsk\cos(\ddsk+(\g+\phis))}{1+\rdsksq}\, \nonumber \\ & & & \nonumber \\
& \Spar & \;=&\; \frac{2\rdsk\sin(\ddsk-(\g+\phis))}{1+\rdsksq}\, \nonumber \\ & & & \nonumber \\
& \Sbpar& \;=&\; \frac{2\rdsk\sin(\ddsk+(\g+\phis))}{1+\rdsksq}\, \nonumber
\end{alignat}

\section{Input observable values and uncertainties}
\label{sec:inputs_vals}

The input observable values and their statistical and systematic uncertainties are listed below.
The observables labelled $D\pi$ are only used in the \Dh combination.


\subsection{\boldmath\BuDh, \Dhh analysis}
\label{sec:input_glwads}

The values and uncertainties are taken from Ref.~\cite{LHCb-PAPER-2016-003}.
The observables are defined in analogy to Eqs.~(\ref{eq:doubleRs}-\ref{eq:af}), and the measured values are
\begin{alignat}{6}
& \AadsDK          & \;=&\;            -0.403\phantom{0} \;\pm\; & 0.056\phantom{0} \;\pm\; & 0.011\,, \nonumber \\
& \AadsDPi         & \;=&\;  \phantom{-0}0.100\phantom{0} \;\pm\; & 0.031\phantom{0} \;\pm\; & 0.009\,, \nonumber \\
& \AcpDkKK         & \;=&\;  \phantom{-0}0.087\phantom{0} \;\pm\; & 0.020\phantom{0} \;\pm\; & 0.008\,, \nonumber \\
& \AcpDkPipi       & \;=&\;  \phantom{-0}0.128\phantom{0} \;\pm\; & 0.037\phantom{0} \;\pm\; & 0.012\,, \nonumber \\
& \AcpDpiKK        & \;=&\;            -0.0145           \;\pm\; & 0.0050           \;\pm\; & 0.0017\,, \nonumber \\
& \AcpDpiPipi      & \;=&\;  \phantom{-0}0.0043           \;\pm\; & 0.0086           \;\pm\; & 0.0031\,, \nonumber \\
& \AfavDkKpi       & \;=&\;            -0.0194           \;\pm\; & 0.0072           \;\pm\; & 0.0060\,, \nonumber \\
& \RadsDK          & \;=&\;  \phantom{-0}0.0188           \;\pm\; & 0.0011           \;\pm\; & 0.0010\,, \nonumber \\
& \RadsDPi         & \;=&\;  \phantom{-0}0.0036           \;\pm\; & 0.0001           \;\pm\; & 0.0001\,, \nonumber \\
& \RCPKK           & \;=&\;  \phantom{-0}0.968\phantom{0} \;\pm\; & 0.022\phantom{0} \;\pm\; & 0.021 \;\pm\; 0.010 \,, \nonumber \\
& \RCPPiPi         & \;=&\;  \phantom{-0}1.002\phantom{0} \;\pm\; & 0.040\phantom{0} \;\pm\; & 0.026 \;\pm\; 0.010 \,, \nonumber \\
& \RkpKpi          & \;=&\;  \phantom{-0}0.0779           \;\pm\; & 0.0006           \;\pm\; & 0.0019\,, \nonumber
\end{alignat}

\noindent where the first uncertainties are statistical and the second are systematic. For \RCPKK and \RCPPiPi the third
uncertainties arise from the assumption that $\rbpi=0$ as discussed in Ref.~\cite{LHCb-PAPER-2016-003} and subsequently applies
only for the \robust combination.
Their statistical and systematic correlations are given in Tables~\ref{tab:glwadshhCor}
and~\ref{tab:glwadshhCorSyst}.
The relationships between observables and parameters are given in Appendix~\ref{sec:relations_glwads}.


\subsection{\boldmath\BuDh, \Dhpipipi analysis}
\label{sec:input_adsk3pi}

The values and uncertainties are taken from Ref.~\cite{LHCb-PAPER-2016-003}.
The observables are defined in analogy to Eqs.~(\ref{eq:doubleRs}--\ref{eq:af}),
and the measured values are
\begin{alignat}{6}
& \AadsDKKppp           & \;=&\;             -0.313\phantom{0}    \;\pm\; & 0.102\phantom{0}   \;\pm\; & 0.038   \,,  \nonumber \\
& \AadsDPiKppp          & \;=&\;  \phantom{-0}0.023\phantom{0}    \;\pm\; & 0.048\phantom{0}   \;\pm\; & 0.005   \,,  \nonumber \\
& \AcpDKpppp            & \;=&\;  \phantom{-0}0.100\phantom{0}    \;\pm\; & 0.034\phantom{0}   \;\pm\; & 0.018   \,,  \nonumber \\
& \AcpDpipppp           & \;=&\;             -0.0041              \;\pm\; & 0.0079             \;\pm\; & 0.0024  \,,  \nonumber \\
& \AfavDkKppp           & \;=&\;             -0.000\phantom{0}    \;\pm\; & 0.012\phantom{0}   \;\pm\; & 0.002   \,,  \nonumber \\
& \RadsDKKppp           & \;=&\;  \phantom{-0}0.0140              \;\pm\; & 0.0015             \;\pm\; & 0.0006  \,,  \nonumber \\
& \RadsDPiKppp          & \;=&\;  \phantom{-0}0.0038              \;\pm\; & 0.0002             \;\pm\; & 0.0001  \,,  \nonumber \\
& \RCPpppp              & \;=&\;  \phantom{-0}0.975\phantom{0}    \;\pm\; & 0.037\phantom{0}   \;\pm\; & 0.019   \;\pm\; 0.005 \,,\nonumber \\
& \RkpKppp              & \;=&\;  \phantom{-0}0.0793              \;\pm\; & 0.0010             \;\pm\; & 0.0018  \,,  \nonumber
\end{alignat}

\noindent where the first uncertainty is statistical and the second is systematic. The third uncertainty for \RCPpppp
is again from the assumption that $\rbpi=0$ and subsequently applies only for the \robust combination.
Their statistical and systematic correlations are given in Tables~\ref{tab:glwadsk3piCor} and~\ref{tab:glwadsk3piCorSyst}.
The relationships between observables and parameters are given in Appendix~\ref{sec:relations_adsk3pi}.


\subsection{\boldmath\BuDh, \Dhhpiz analysis}
\label{sec:input_adshpipiz}

The values and uncertainties are taken from Ref.~\cite{LHCb-PAPER-2015-014}.
The observables are defined in analogy to Eqs.~(\ref{eq:doubleRs}--\ref{eq:af}),
and the measured values are
\begin{alignat}{6}
& \AadsDKkpp            & \;=&\;             -0.20\phantom{000}     \;\pm\; & 0.27\phantom{000}    \;\pm\; & 0.04    \,, \nonumber \\
& \AadsDPikpp           & \;=&\;  \phantom{-0}0.44\phantom{000}     \;\pm\; & 0.19\phantom{000}    \;\pm\; & 0.01    \,, \nonumber \\
& \AcpDkKKPiz           & \;=&\;  \phantom{-0}0.30\phantom{000}     \;\pm\; & 0.20\phantom{000}    \;\pm\; & 0.02    \,, \nonumber \\
& \AcpDkPiPiPiz         & \;=&\;  \phantom{-0}0.054\phantom{00}     \;\pm\; & 0.091\phantom{00}    \;\pm\; & 0.011   \,, \nonumber \\
& \AcpDpiKKPiz          & \;=&\;             -0.030\phantom{00}     \;\pm\; & 0.040\phantom{00}    \;\pm\; & 0.005   \,, \nonumber \\
& \AcpDpiPiPiPiz        & \;=&\;             -0.016\phantom{00}     \;\pm\; & 0.020\phantom{00}    \;\pm\; & 0.004   \,, \nonumber \\
& \AfavDkKPiPiz         & \;=&\;  \phantom{-0}0.010\phantom{00}     \;\pm\; & 0.026\phantom{00}    \;\pm\; & 0.005   \,, \nonumber \\
& \RadsDKkpp            & \;=&\;  \phantom{-0}0.014\phantom{00}     \;\pm\; & 0.005\phantom{00}    \;\pm\; & 0.002   \,, \nonumber \\
& \RadsDPikpp           & \;=&\;  \phantom{-0}0.00235               \;\pm\; & 0.00049              \;\pm\; & 0.00006 \,, \nonumber \\
& \RCPKKPiz             & \;=&\;  \phantom{-0}0.95\phantom{000}     \;\pm\; & 0.22\phantom{000}    \;\pm\; & 0.05    \,, \nonumber \\
& \RCPPiPiPiz           & \;=&\;  \phantom{-0}0.98\phantom{000}     \;\pm\; & 0.11\phantom{000}    \;\pm\; & 0.05    \,, \nonumber
\end{alignat}

\noindent where the first uncertainty is statistical and the second is systematic.
The statistical and systematic correlations are given in Tables~\ref{tab:glwadshhpizCor}
and~\ref{tab:glwadshhpizCorSyst}.
The relationships between observables and parameters are given in Appendix~\ref{sec:relations_adshpipiz}.


\subsection{\boldmath\BuDK, \DKShh analysis}
\label{sec:input_ggsz}

The values and uncertainties are taken from Ref.~\cite{LHCb-PAPER-2014-041}.
The results are
\begin{alignat}{3}
& \xmdk                   & \;=\;  \phantom{-}0.025  \;\pm\; & 0.025 \;\pm\; & 0.010 \;\pm\; & 0.005\,, \nonumber \\
& \ymdk                   & \;=\;  \phantom{-}0.075  \;\pm\; & 0.029 \;\pm\; & 0.005 \;\pm\; & 0.014\,, \nonumber \\
& \xpdk                   & \;=\;            -0.077  \;\pm\; & 0.024 \;\pm\; & 0.010 \;\pm\; & 0.004\,, \nonumber \\
& \ypdk                   & \;=\;            -0.022  \;\pm\; & 0.025 \;\pm\; & 0.004 \;\pm\; & 0.010\,, \nonumber
\end{alignat}

\noindent where the first uncertainty is statistical, the second is systematic, and the third is an external uncertainty due to the information on the
strong phase variation across the \DKShh phase space. Correlations between the statistical and systematic uncertainties are given in Tables~\ref{tab:ggszCor}
and~\ref{tab:ggszCorSyst}.
The relationships between observables and parameters are given in Appendix~\ref{sec:relations_ggsz}.


\subsection{\boldmath\BuDh, \DKSKpi analysis}
\label{sec:input_glskskpi}

The values and uncertainties are taken from Ref.~\cite{LHCb-PAPER-2013-068}.
The observables are defined in analogy to Eqs.~(\ref{eq:rads}--\ref{eq:af})
and are
\begin{alignat}{3}
& \RfavsupDkKskpi       & \;=\;  \phantom{-}3.855  \;\pm\; & 0.961 \;\pm\; & 0.060\,, \nonumber \\
& \AfavDkKskpi          & \;=\;  \phantom{-}0.026  \;\pm\; & 0.109 \;\pm\; & 0.029\,, \nonumber \\
& \AsupDkKskpi          & \;=\;  \phantom{-}0.336  \;\pm\; & 0.208 \;\pm\; & 0.026\,, \nonumber
\end{alignat}

\noindent where the first uncertainty is statistical and the second is systematic.
The statistical and systematic correlations are found to be negligible and not included.
The relationships between observables and parameters are given in Appendix~\ref{sec:relations_glskskpi}.


\subsection{\boldmath\BuDhpipi, \Dhh analysis}
\label{sec:input_glwadsdkpipi}

The values and uncertainties are taken from Ref.~\cite{LHCb-PAPER-2015-020}.
The observables are defined in analogy to Eqs.~(\ref{eq:doubleRs},\ref{eq:acp},\ref{eq:af}--\ref{eq:rm}) and are
\begin{alignat}{6}
& \RcpDkpipi            & \;=&\;  \phantom{-0}1.040\phantom{0}    \;\pm\; & 0.064\phantom{0}  \phantom{\;\pm\;} &  \nonumber \\
& \AfavDkpipiKpi        & \;=&\;  \phantom{-0}0.013\phantom{0}    \;\pm\; & 0.019\phantom{0}   \;\pm\; & 0.013  \,,  \nonumber \\
& \AfavDpipipiKpi       & \;=&\;             -0.002\phantom{0}    \;\pm\; & 0.003\phantom{0}   \;\pm\; & 0.011  \,,  \nonumber \\
& \AcpDkpipiKK          & \;=&\;             -0.045\phantom{0}    \;\pm\; & 0.064\phantom{0}   \;\pm\; & 0.011  \,,  \nonumber \\
& \AcpDkpipiPipi        & \;=&\;             -0.054\phantom{0}    \;\pm\; & 0.101\phantom{0}   \;\pm\; & 0.011  \,,  \nonumber \\
& \AcpDpipipiKK         & \;=&\;             -0.019\phantom{0}    \;\pm\; & 0.011\phantom{0}   \;\pm\; & 0.010  \,,  \nonumber \\
& \AcpDpipipiPipi       & \;=&\;             -0.013\phantom{0}    \;\pm\; & 0.016\phantom{0}   \;\pm\; & 0.010  \,,  \nonumber \\
& \RpDkpipiKpi          & \;=&\;  \phantom{-0}0.0107              \;\pm\; & 0.0060             \;\pm\; & 0.0011 \,,  \nonumber \\
& \RmDkpipiKpi          & \;=&\;  \phantom{-0}0.0053              \;\pm\; & 0.0045             \;\pm\; & 0.0006 \,,  \nonumber \\
& \RpDpipipiKpi         & \;=&\;  \phantom{-0}0.0043              \;\pm\; & 0.0005             \;\pm\; & 0.0002 \,,  \nonumber \\
& \RmDpipipiKpi         & \;=&\;  \phantom{-0}0.0042              \;\pm\; & 0.0005             \;\pm\; & 0.0002 \,,  \nonumber
\end{alignat}

\noindent where the first uncertainty is statistical and the second is systematic. For \RcpDkpipi, the single uncertainty includes both statistical and systematic contributions.
The only non-negligible correlations are the statistical correlations, $\rho(\AcpDkpipiKK,\AcpDkpipiPipi)=0.20$ and $\rho(\AcpDpipipiKK,\AcpDpipipiPipi)=0.08$.
The relationships between observables and parameters are given in Appendix~\ref{sec:relations_glwadsdkpipi}.


\subsection{\boldmath\BdDzKstz, \DKpi analysis}
\label{sec:input_glwadsdkst}

The values and uncertainties are taken from Ref.~\cite{LHCb-PAPER-2014-028}.
The ADS observables are defined in analogy to Eqs.~(\ref{eq:af}--\ref{eq:rm}) and are
\begin{alignat}{3}
& \AfavDkstKpi          & \;=\;            -0.03  \;\pm\; & 0.04 \;\pm\; & 0.02\,, \nonumber \\
& \RpDkstKpi            & \;=\;  \phantom{-}0.06  \;\pm\; & 0.03 \;\pm\; & 0.01\,, \nonumber \\
& \RmDkstKpi            & \;=\;  \phantom{-}0.06  \;\pm\; & 0.03 \;\pm\; & 0.01\,, \nonumber
\end{alignat}

\noindent where the first uncertainty is statistical and the second is systematic.
The statistical correlations are given in Table~\ref{tab:glwadsdkstCor},
and the systematic correlations in Table~\ref{tab:glwadsdkstCorSyst}.
The relationships between observables and parameters are given in Appendix~\ref{sec:relations_glwadsdkst}.


\subsection{\boldmath\BdDKpi, \Dhh analysis}
\label{sec:input_dalitzdkst}

The values and uncertainties are taken from Ref.~\cite{LHCb-PAPER-2015-059}.
The results are
\begin{alignat}{3}
& \xmdkpi               & \;=\;            -0.02  \;\pm\; & 0.13 \;\pm\; & 0.14\,, \nonumber \\
& \ymdkpi               & \;=\;            -0.35  \;\pm\; & 0.26 \;\pm\; & 0.41\,, \nonumber \\
& \xpdkpi               & \;=\;  \phantom{-}0.04  \;\pm\; & 0.16 \;\pm\; & 0.11\,, \nonumber \\
& \ypdkpi               & \;=\;            -0.47  \;\pm\; & 0.28 \;\pm\; & 0.22\,, \nonumber
\end{alignat}

\noindent where the first uncertainty is statistical and the second is systematic.
The correlations are given in Tables~\ref{tab:glwkstCor} and~\ref{tab:glwkstCorSyst}.
The relationships between observables and parameters are given in Appendix~\ref{sec:relations_dalitzdkst}.


\subsection{\boldmath\BdDzKstz, \DKSpipi analysis}
\label{sec:input_ggszdkst}

The values and uncertainties are taken from Ref.~\cite{LHCb-PAPER-2016-007}.
The results are
\begin{alignat}{3}
& \xmdkst               & \;=\;            -0.15  \;\pm\; & 0.14 \;\pm\; & 0.03 \;\pm\; & 0.01\,, \nonumber \\
& \ymdkst               & \;=\;  \phantom{-}0.25  \;\pm\; & 0.15 \;\pm\; & 0.06 \;\pm\; & 0.01\,, \nonumber \\
& \xpdkst               & \;=\;  \phantom{-}0.05  \;\pm\; & 0.24 \;\pm\; & 0.04 \;\pm\; & 0.01\,, \nonumber \\
& \ypdkst               & \;=\;            -0.65  \;\pm\; & 0.24 \;\pm\; & 0.08 \;\pm\; & 0.01\,, \nonumber
\end{alignat}

\noindent where the first uncertainty is statistical, the second is systematic and the third is from the Dalitz plot fit model.
The correlations are given in Table~\ref{tab:ggszkstCor}.
The relationships between observables and parameters are given in Appendix~\ref{sec:relations_ggszdkst}.


\subsection{\boldmath\BsDsK analysis}
\label{sec:input_dsk}

The values and uncertainties are taken from Ref.~\cite{LHCb-PAPER-2014-038} (with a change in the sign convention, see Appendix~\ref{sec:relations_dsk} for the explicit definition).
The results are
\begin{alignat}{3}
& \Cpar                 & \;=\;  \phantom{-}0.53  \;\pm\; & 0.25 \;\pm\; & 0.04\,, \nonumber \\
& \Dpar                 & \;=\;            -0.37  \;\pm\; & 0.42 \;\pm\; & 0.20\,, \nonumber \\
& \Dbpar                & \;=\;            -0.20  \;\pm\; & 0.41 \;\pm\; & 0.20\,, \nonumber \\
& \Spar                 & \;=\;            -1.09  \;\pm\; & 0.33 \;\pm\; & 0.08\,, \nonumber \\
& \Sbpar                & \;=\;  \phantom{-}0.36  \;\pm\; & 0.34 \;\pm\; & 0.08\,, \nonumber
\end{alignat}

\noindent where the first uncertainty is statistical and the second is systematic.
The statistical correlations are given in Table~\ref{tab:dskCor},
and the systematic correlations in Table~\ref{tab:dskCorSyst}.
The relationships between observables and parameters are given in Appendix~\ref{sec:relations_dsk}.

\pagebreak


\section{Uncertainty correlations for the input observables}
\label{sec:inputs_corrs}

The correlation matrices of the statistical and systematic uncertainties are given below.
The observables labelled $D\pi$ are only used in the \Dh combination.


\begin{landscape}
\begin{table}[!hbtp]
\small
\centering
\caption{Correlation matrix of the statistical uncertainties for the \BuDh, \Dzhh observables~\cite{LHCb-PAPER-2016-003}.}
\label{tab:glwadshhCor}
\renewcommand{\arraystretch}{1.4}
\begin{tabular}{ l |llllllllllll}
\hline
                & \AadsDK & \AadsDPi & \AcpDkKK & \AcpDkPipi & \AcpDpiKK & \AcpDpiPipi & \AfavDkKpi & \RadsDK & \RadsDPi & \RCPKK & \RCPPiPi & \RkpKpi\\
\hline
\AadsDK         &  $\phantom{-}1$ &  $-0.047$ &  $\phantom{-}0.002$ &  $\phantom{-}0.001$ &  $\phantom{-}0.009$ &  $\phantom{-}0.005$ &   $\phantom{-}0.008$ &  $\phantom{-}0.102$ &  $-0.003$ &  $\phantom{-}0$ &  $\phantom{-}0$ &  $\phantom{-}0$  \\
\AadsDPi        &  $-0.047$ &  $\phantom{-}1$ &  $\phantom{-}0.004$ &  $\phantom{-}0.003$ &  $\phantom{-}0.017$ &  $\phantom{-}0.010$ &   $\phantom{-}0.014$ &  $\phantom{-}0.015$ &  $-0.043$ &  $\phantom{-}0$ &  $\phantom{-}0$ &  $\phantom{-}0$  \\
\AcpDkKK        &  $\phantom{-}0.002$ &  $\phantom{-}0.004$ &  $\phantom{-}1$ &  $\phantom{-}0.004$ &  $-0.007$ &  $\phantom{-}0.016$ &   $\phantom{-}0.024$ &  $\phantom{-}0$ &  $\phantom{-}0$ &  $-0.014$ &  $\phantom{-}0$ &  $-0.001$  \\
\AcpDkPipi      &  $\phantom{-}0.001$ &  $\phantom{-}0.003$ &  $\phantom{-}0.004$ &  $\phantom{-}1$ &  $\phantom{-}0.016$ &  $-0.036$ &   $\phantom{-}0.014$ &  $-0.001$ &  $\phantom{-}0$ &  $\phantom{-}0$ &  $-0.038$ &  $-0.002$  \\
\AcpDpiKK       &  $\phantom{-}0.009$ &  $\phantom{-}0.017$ &  $-0.007$ &  $\phantom{-}0.016$ &  $\phantom{-}1$ &  $\phantom{-}0.064$ &   $\phantom{-}0.092$ &  $\phantom{-}0$ &  $\phantom{-}0$ &  $-0.001$ &  $\phantom{-}0$ &  $-0.001$  \\
\AcpDpiPipi     &  $\phantom{-}0.005$ &  $\phantom{-}0.010$ &  $\phantom{-}0.016$ &  $-0.036$ &  $\phantom{-}0.064$ &  $\phantom{-}1$ &   $\phantom{-}0.053$ &  $\phantom{-}0$ &  $\phantom{-}0$ &  $\phantom{-}0$ &  $-0.003$ &  $\phantom{-}0$  \\
\AfavDkKpi      &  $\phantom{-}0.008$ &  $\phantom{-}0.014$ &  $\phantom{-}0.024$ &  $\phantom{-}0.014$ &  $\phantom{-}0.092$ &  $\phantom{-}0.053$ &   $\phantom{-}1$ &  $\phantom{-}0$ &  $\phantom{-}0$ &  $\phantom{-}0$ &  $\phantom{-}0$ &  $\phantom{-}0$  \\
\RadsDK         &  $\phantom{-}0.102$ &  $\phantom{-}0.015$ &  $\phantom{-}0$ &  $-0.001$ &  $\phantom{-}0$ &  $\phantom{-}0$ &   $\phantom{-}0$ &  $\phantom{-}1$ &  $-0.022$ &  $\phantom{-}0.040$ &  $\phantom{-}0.025$ &  $-0.114$  \\
\RadsDPi        &  $-0.003$ &  $-0.043$ &  $\phantom{-}0$ &  $\phantom{-}0$ &  $\phantom{-}0$ &  $\phantom{-}0$ &   $\phantom{-}0$ &  $-0.022$ &  $\phantom{-}1$ &  $-0.005$ &  $-0.003$ &  $\phantom{-}0.011$  \\
\RCPKK          &  $\phantom{-}0$ &  $\phantom{-}0$ &  $-0.014$ &  $\phantom{-}0$ &  $-0.001$ &  $\phantom{-}0$ &   $\phantom{-}0$ &  $\phantom{-}0.040$ &  $-0.005$ &  $\phantom{-}1$ &  $\phantom{-}0.060$ &  $-0.317$  \\
\RCPPiPi        &  $\phantom{-}0$ &  $\phantom{-}0$ &  $\phantom{-}0$ &  $-0.038$ &  $\phantom{-}0$ &  $-0.003$ &   $\phantom{-}0$ &  $\phantom{-}0.025$ &  $-0.003$ &  $\phantom{-}0.060$ &  $\phantom{-}1$ &  $-0.176$  \\
\RkpKpi         &  $\phantom{-}0$ &  $\phantom{-}0$ &  $-0.001$ &  $-0.002$ &  $-0.001$ &  $\phantom{-}0$ &   $\phantom{-}0$ &  $-0.114$ &  $\phantom{-}0.011$ &  $-0.317$ &  $-0.176$ &  $\phantom{-}1$  \\
\hline
\end{tabular}

\end{table}
\end{landscape}

\begin{landscape}
\begin{table}[!hbtp]
\small
\centering
\caption{Correlation matrix of the systematic uncertainties for the \BuDh, \Dzhh observables~\cite{LHCb-PAPER-2016-003}.}
\label{tab:glwadshhCorSyst}
\renewcommand{\arraystretch}{1.4}
  \begin{tabular}{ l |llllllllllll}
\hline
                & \AadsDK & \AadsDPi & \AcpDkKK & \AcpDkPipi & \AcpDpiKK & \AcpDpiPipi & \AfavDkKpi & \RadsDK & \RadsDPi & \RCPKK & \RCPPiPi & \RkpKpi\\
\hline
\AadsDK         &  $\phantom{-}1$ &  $\phantom{-}0.36$ &  $-0.06$ &  $\phantom{-}0.27$ &  $\phantom{-}0.30$ &  $\phantom{-}0.04$ &  $\phantom{-}0.09$ &  $\phantom{-}0.78$ &  $-0.43$ &  $-0.04$ &  $\phantom{-}0.23$ &  $-0.14$  \\
\AadsDPi        &  $\phantom{-}0.36$ &  $\phantom{-}1$ &  $-0.03$ &  $\phantom{-}0.31$ &  $\phantom{-}0.22$ &  $-0.06$ &  $-0.55$ &  $\phantom{-}0.59$ &  $-0.47$ &  $-0.01$ &  $\phantom{-}0.12$ &  $-0.04$  \\
\AcpDkKK        &  $-0.06$ &  $-0.03$ &  $\phantom{-}1$ &  $-0.02$ &  $-0.80$ &  $\phantom{-}0.09$ &  $\phantom{-}0.09$ &  $-0.10$ &  $-0.06$ &  $\phantom{-}0.03$ &  $-0.28$ &  $\phantom{-}0.07$  \\
\AcpDkPipi      &  $\phantom{-}0.27$ &  $\phantom{-}0.31$ &  $-0.02$ &  $\phantom{-}1$ &  $\phantom{-}0.19$ &  $-0.42$ &  $-0.01$ &  $\phantom{-}0.35$ &  $-0.28$ &  $-0.22$ &  $\phantom{-}0.11$ &  $-0.07$  \\
\AcpDpiKK       &  $\phantom{-}0.30$ &  $\phantom{-}0.22$ &  $-0.80$ &  $\phantom{-}0.19$ &  $\phantom{-}1$ &  $\phantom{-}0.11$ &  $\phantom{-}0.09$ &  $\phantom{-}0.37$ &  $-0.21$ &  $-0.16$ &  $\phantom{-}0.20$ &  $-0.13$  \\
\AcpDpiPipi     &  $\phantom{-}0.04$ &  $-0.06$ &  $\phantom{-}0.09$ &  $-0.42$ &  $\phantom{-}0.11$ &  $\phantom{-}1$ &  $\phantom{-}0.30$ &  $-0.03$ &  $\phantom{-}0.06$ &  $-0.08$ &  $-0.03$ &  $-0.09$  \\
\AfavDkKpi      &  $\phantom{-}0.09$ &  $-0.55$ &  $\phantom{-}0.09$ &  $-0.01$ &  $\phantom{-}0.09$ &  $\phantom{-}0.30$ &  $\phantom{-}1$ &  $-0.11$ &  $\phantom{-}0.05$ &  $-0.01$ &  $-0.02$ &  $-0.02$  \\
\RadsDK         &  $\phantom{-}0.78$ &  $\phantom{-}0.59$ &  $-0.10$ &  $\phantom{-}0.35$ &  $\phantom{-}0.37$ &  $-0.03$ &  $-0.11$ &  $\phantom{-}1$ &  $-0.57$ &  $-0.14$ &  $\phantom{-}0.33$ &  $-0.22$  \\
\RadsDPi        &  $-0.43$ &  $-0.47$ &  $-0.06$ &  $-0.28$ &  $-0.21$ &  $\phantom{-}0.06$ &  $\phantom{-}0.05$ &  $-0.57$ &  $\phantom{-}1$ &  $\phantom{-}0.19$ &  $\phantom{-}0.10$ &  $\phantom{-}0.02$  \\
\RCPKK          &  $-0.04$ &  $-0.01$ &  $\phantom{-}0.03$ &  $-0.22$ &  $-0.16$ &  $-0.08$ &  $-0.01$ &  $-0.14$ &  $\phantom{-}0.19$ &  $\phantom{-}1$ &  $\phantom{-}0.17$ &  $\phantom{-}0.21$  \\
\RCPPiPi        &  $\phantom{-}0.23$ &  $\phantom{-}0.12$ &  $-0.28$ &  $\phantom{-}0.11$ &  $\phantom{-}0.20$ &  $-0.03$ &  $-0.02$ &  $\phantom{-}0.33$ &  $\phantom{-}0.10$ &  $\phantom{-}0.17$ &  $\phantom{-}1$ &  $-0.11$  \\
\RkpKpi         &  $-0.14$ &  $-0.04$ &  $\phantom{-}0.07$ &  $-0.07$ &  $-0.13$ &  $-0.09$ &  $-0.02$ &  $-0.22$ &  $\phantom{-}0.02$ &  $\phantom{-}0.21$ &  $-0.11$ &  $\phantom{-}1$  \\
\hline
\end{tabular}

\end{table}
\end{landscape}


\begin{landscape}
\begin{table}[!hbtp]
\small
\centering
\caption{Correlation matrix of the statistical uncertainties for the \BuDh, \Dzhpipipi observables~\cite{LHCb-PAPER-2016-003}.}
\label{tab:glwadsk3piCor}
\renewcommand{\arraystretch}{1.4}
  \begin{tabular}{ l |lllllllll}
\hline
                & \AadsDKKppp & \AadsDPiKppp & \AcpDKpppp & \AcpDpipppp & \AfavDkKppp & \RadsDKKppp & \RadsDPiKppp & \RCPpppp & \RkpKppp\\
\hline
\AadsDKKppp     &  $\phantom{-}1$ &  $-0.062$ &  $\phantom{-}0.002$ &  $\phantom{-}0.009$ &   $0.006$ &  $\phantom{-}0.082$ &  $-0.004$ &  $\phantom{-}0.002$ &  $\phantom{-}0.003$  \\
\AadsDPiKppp    &  $-0.062$ &  $\phantom{-}1$ &  $\phantom{-}0.005$ &  $\phantom{-}0.020$ &   $0.017$ &  $\phantom{-}0.013$ &  $-0.022$ &  $\phantom{-}0$ &  $\phantom{-}0$  \\
\AcpDKpppp      &  $\phantom{-}0.002$ &  $\phantom{-}0.005$ &  $\phantom{-}1$ &  $-0.020$ &   $0.024$ &  $-0.001$ &  $\phantom{-}0$ &  $-0.018$ &  $-0.002$  \\
\AcpDpipppp     &  $\phantom{-}0.009$ &  $\phantom{-}0.020$ &  $-0.020$ &  $\phantom{-}1$ &   $0.097$ &  $\phantom{-}0$ &  $\phantom{-}0$ &  $-0.004$ &  $\phantom{-}0$  \\
\AfavDkKppp     &  $\phantom{-}0.006$ &  $\phantom{-}0.017$ &  $\phantom{-}0.024$ &  $\phantom{-}0.097$ &   $    1$ &  $\phantom{-}0$ &  $\phantom{-}0$ &  $\phantom{-}0$ &  $\phantom{-}0.001$  \\
\RadsDKKppp     &  $\phantom{-}0.082$ &  $\phantom{-}0.013$ &  $-0.001$ &  $\phantom{-}0$ &   $    0$ &  $\phantom{-}1$ &  $-0.046$ &  $\phantom{-}0.041$ &  $-0.099$  \\
\RadsDPiKppp    &  $-0.004$ &  $-0.022$ &  $\phantom{-}0$ &  $\phantom{-}0$ &   $    0$ &  $-0.046$ &  $\phantom{-}1$ &  $-0.004$ &  $\phantom{-}0.012$  \\
\RCPpppp        &  $\phantom{-}0.002$ &  $\phantom{-}0$ &  $-0.018$ &  $-0.004$ &   $    0$ &  $\phantom{-}0.041$ &  $-0.004$ &  $\phantom{-}1$ &  $-0.308$  \\
\RkpKppp        &  $\phantom{-}0.003$ &  $\phantom{-}0$ &  $-0.002$ &  $\phantom{-}0$ &   $0.001$ &  $-0.099$ &  $\phantom{-}0.012$ &  $-0.308$ &  $\phantom{-}1$  \\
\hline
\end{tabular}

\end{table}
\end{landscape}

\begin{landscape}
\begin{table}[!hbtp]
\small
\centering
\caption{Correlation matrix of the systematic uncertainties for the \BuDh, \Dzhpipipi observables~\cite{LHCb-PAPER-2016-003}.}
\label{tab:glwadsk3piCorSyst}
\renewcommand{\arraystretch}{1.4}
  \begin{tabular}{ l |ccccccccc}
\hline
                & \AadsDKKppp & \AadsDPiKppp & \AcpDKpppp & \AcpDpipppp & \AfavDkKppp & \RadsDKKppp & \RadsDPiKppp & \RCPpppp & \RkpKppp\\
\hline
\AadsDKKppp     &    $\phantom{-}1$  &    $-0.09$  &    $-0.04$  &    $\phantom{-}0.02$  &    $\phantom{-}0.02$  &    $\phantom{-}0.87$  &    $\phantom{-}0.05$  &    $-0.04$  &    $-0.13$  \\
\AadsDPiKppp    &    $-0.09$  &    $\phantom{-}1$  &    $-0.34$  &    $\phantom{-}0.43$  &    $\phantom{-}0.05$  &    $\phantom{-}0.10$  &    $\phantom{-}0.46$  &    $-0.04$  &    $\phantom{-}0.17$  \\
\AcpDKpppp      &    $-0.04$  &    $-0.34$  &    $\phantom{-}1$  &    $\phantom{-}0.31$  &    $\phantom{-}0.09$  &    $\phantom{-}0.03$  &    $-0.35$  &    $\phantom{-}0.07$  &    $-0.14$  \\
\AcpDpipppp     &    $\phantom{-}0.02$  &    $\phantom{-}0.43$  &    $\phantom{-}0.31$  &    $\phantom{-}1$  &    $\phantom{-}0.32$  &    $\phantom{-}0.01$  &    $\phantom{-}0.24$  &    $-0.07$  &    $\phantom{-}0.08$  \\
\AfavDkKppp     &    $\phantom{-}0.02$  &    $\phantom{-}0.05$  &    $\phantom{-}0.09$  &    $\phantom{-}0.32$  &    $\phantom{-}1$  &    $\phantom{-}0.02$  &    $-0.02$  &    $\phantom{-}0.02$  &    $\phantom{-}0$  \\
\RadsDKKppp     &    $\phantom{-}0.87$  &    $\phantom{-}0.10$  &    $\phantom{-}0.03$  &    $\phantom{-}0.01$  &    $\phantom{-}0.02$  &    $\phantom{-}1$  &    $\phantom{-}0.14$  &    $\phantom{-}0.04$  &    $-0.04$  \\
\RadsDPiKppp    &    $\phantom{-}0.05$  &    $\phantom{-}0.46$  &    $-0.35$  &    $\phantom{-}0.24$  &    $-0.02$  &    $\phantom{-}0.14$  &    $\phantom{-}1$  &    $-0.06$  &    $\phantom{-}0.13$  \\
\RCPpppp        &    $-0.04$  &    $-0.04$  &    $\phantom{-}0.07$  &    $-0.07$  &    $\phantom{-}0.02$  &    $\phantom{-}0.04$  &    $-0.06$  &    $\phantom{-}1$  &    $\phantom{-}0.11$  \\
\RkpKppp        &    $-0.13$  &    $\phantom{-}0.17$  &    $-0.14$  &    $\phantom{-}0.08$  &    $\phantom{-}0$  &    $-0.04$  &    $\phantom{-}0.13$  &    $\phantom{-}0.11$  &    $\phantom{-}1$  \\
\hline
\end{tabular}

\end{table}
\end{landscape}


\begin{landscape}
\begin{table}[!hbtp]
\small
\centering
\caption{Correlation matrix of the statistical uncertainties for the \BuDh, \Dzhhpiz observables~\cite{LHCb-PAPER-2014-038}.}
\label{tab:glwadshhpizCor}
\renewcommand{\arraystretch}{1.4}
  \begin{tabular}{ l |lllllllllll}
\hline
                & \AadsDKkpp & \AadsDPikpp & \AcpDkKKPiz & \AcpDkPiPiPiz & \AcpDpiKKPiz & \AcpDpiPiPiPiz & \AfavDkKPiPiz & \RadsDKkpp & \RadsDPikpp & \RCPKKPiz & \RCPPiPiPiz\\
\hline
\AadsDKkpp      &    $\phantom{-}1$  &    $-0.04$  &    $\phantom{-}0$  &    $\phantom{-}0$  &    $\phantom{-}0$  &    $\phantom{-}0.01$  &    $\phantom{-}0.01$  &    $\phantom{-}0.13$  &    $\phantom{-}0$  &    $\phantom{-}0$  &    $\phantom{-}0$   \\
\AadsDPikpp     &    $-0.04$  &    $\phantom{-}1$  &    $\phantom{-}0$  &    $\phantom{-}0$  &    $\phantom{-}0$  &    $\phantom{-}0.01$  &    $\phantom{-}0.01$  &    $-0.01$  &    $-0.34$  &    $\phantom{-}0$  &    $\phantom{-}0$   \\
\AcpDkKKPiz     &    $\phantom{-}0$  &    $\phantom{-}0$  &    $\phantom{-}1$  &    $\phantom{-}0$  &    $-0.04$  &    $\phantom{-}0.01$  &    $\phantom{-}0.01$  &    $\phantom{-}0$  &    $\phantom{-}0$  &    $-0.20$  &    $-0.01$   \\
\AcpDkPiPiPiz   &    $\phantom{-}0$  &    $\phantom{-}0$  &    $\phantom{-}0$  &    $\phantom{-}1$  &    $\phantom{-}0.01$  &    $-0.04$  &    $\phantom{-}0.02$  &    $\phantom{-}0$  &    $\phantom{-}0$  &    $\phantom{-}0$  &    $-0.04$   \\
\AcpDpiKKPiz    &    $\phantom{-}0$  &    $\phantom{-}0$  &    $-0.04$  &    $\phantom{-}0.01$  &    $\phantom{-}1$  &    $\phantom{-}0.04$  &    $\phantom{-}0.04$  &    $\phantom{-}0$  &    $\phantom{-}0$  &    $\phantom{-}0$  &    $\phantom{-}0$   \\
\AcpDpiPiPiPiz  &    $\phantom{-}0.01$  &    $\phantom{-}0.01$  &    $\phantom{-}0.01$  &    $-0.04$  &    $\phantom{-}0.04$  &    $\phantom{-}1$  &    $\phantom{-}0.08$  &    $\phantom{-}0$  &    $\phantom{-}0$  &    $\phantom{-}0$  &    $\phantom{-}0$   \\
\AfavDkKPiPiz   &    $\phantom{-}0.01$  &    $\phantom{-}0.01$  &    $\phantom{-}0.01$  &    $\phantom{-}0.02$  &    $\phantom{-}0.04$  &    $\phantom{-}0.08$  &    $\phantom{-}1$  &    $\phantom{-}0$  &    $\phantom{-}0$  &    $\phantom{-}0$  &    $\phantom{-}0$   \\
\RadsDKkpp      &    $\phantom{-}0.13$  &    $-0.01$  &    $\phantom{-}0$  &    $\phantom{-}0$  &    $\phantom{-}0$  &    $\phantom{-}0$  &    $\phantom{-}0$  &    $\phantom{-}1$  &    $\phantom{-}0.03$  &    $\phantom{-}0$  &    $\phantom{-}0.01$   \\
\RadsDPikpp     &    $\phantom{-}0$  &    $-0.34$  &    $\phantom{-}0$  &    $\phantom{-}0$  &    $\phantom{-}0$  &    $\phantom{-}0$  &    $\phantom{-}0$  &    $\phantom{-}0.03$  &    $\phantom{-}1$  &    $\phantom{-}0$  &    $\phantom{-}0$   \\
\RCPKKPiz       &    $\phantom{-}0$  &    $\phantom{-}0$  &    $-0.20$  &    $\phantom{-}0$  &    $\phantom{-}0$  &    $\phantom{-}0$  &    $\phantom{-}0$  &    $\phantom{-}0$  &    $\phantom{-}0$  &    $\phantom{-}1$  &    $\phantom{-}0.02$   \\
\RCPPiPiPiz     &    $\phantom{-}0$  &    $\phantom{-}0$  &    $-0.01$  &    $-0.04$  &    $\phantom{-}0$  &    $\phantom{-}0$  &    $\phantom{-}0$  &    $\phantom{-}0.01$  &    $\phantom{-}0$  &    $\phantom{-}0.02$  &    $\phantom{-}1$   \\
\hline
\end{tabular}

\end{table}
\end{landscape}

\begin{landscape}
\begin{table}[!hbtp]
\small
\centering
\caption{Correlation matrix of the systematic uncertainties for the \BuDh, \Dzhhpiz observables~\cite{LHCb-PAPER-2014-038}.}
\label{tab:glwadshhpizCorSyst}
\renewcommand{\arraystretch}{1.4}
  \begin{tabular}{ l |lllllllllll}
\hline
                & \AadsDKkpp & \AadsDPikpp & \AcpDkKKPiz & \AcpDkPiPiPiz & \AcpDpiKKPiz & \AcpDpiPiPiPiz & \AfavDkKPiPiz & \RadsDKkpp & \RadsDPikpp & \RCPKKPiz & \RCPPiPiPiz\\
\hline
\AadsDKkpp      &    $\phantom{-}1$  &    $\phantom{-}0.03$  &    $\phantom{-}0.07$  &    $\phantom{-}0.07$  &    $\phantom{-}0.18$  &    $\phantom{-}0.17$  &    $-0.16$  &    $\phantom{-}0.81$  &    $\phantom{-}0.32$  &    $\phantom{-}0.02$  &    $\phantom{-}0.13$   \\
\AadsDPikpp     &    $\phantom{-}0.03$  &    $\phantom{-}1$  &    $\phantom{-}0.28$  &    $\phantom{-}0.31$  &    $\phantom{-}0.67$  &    $\phantom{-}0.68$  &    $-0.63$  &    $-0.18$  &    $-0.49$  &    $\phantom{-}0$  &    $-0.04$   \\
\AcpDkKKPiz     &    $\phantom{-}0.07$  &    $\phantom{-}0.28$  &    $\phantom{-}1$  &    $\phantom{-}0.77$  &    $\phantom{-}0.07$  &    $\phantom{-}0.05$  &    $\phantom{-}0.05$  &    $\phantom{-}0.08$  &    $-0.08$  &    $-0.33$  &    $-0.18$   \\
\AcpDkPiPiPiz   &    $\phantom{-}0.07$  &    $\phantom{-}0.31$  &    $\phantom{-}0.77$  &    $\phantom{-}1$  &    $\phantom{-}0.05$  &    $\phantom{-}0.02$  &    $-0.06$  &    $\phantom{-}0.13$  &    $-0.11$  &    $-0.14$  &    $-0.25$   \\
\AcpDpiKKPiz    &    $\phantom{-}0.18$  &    $\phantom{-}0.67$  &    $\phantom{-}0.07$  &    $\phantom{-}0.05$  &    $\phantom{-}1$  &    $\phantom{-}0.88$  &    $-0.82$  &    $-0.04$  &    $\phantom{-}0.02$  &    $-0.04$  &    $\phantom{-}0.02$   \\
\AcpDpiPiPiPiz  &    $\phantom{-}0.17$  &    $\phantom{-}0.68$  &    $\phantom{-}0.05$  &    $\phantom{-}0.02$  &    $\phantom{-}0.88$  &    $\phantom{-}1$  &    $-0.87$  &    $-0.03$  &    $\phantom{-}0$  &    $\phantom{-}0$  &    $\phantom{-}0.01$   \\
\AfavDkKPiPiz   &    $-0.16$  &    $-0.63$  &    $\phantom{-}0.05$  &    $-0.06$  &    $-0.82$  &    $-0.87$  &    $\phantom{-}1$  &    $-0.05$  &    $\phantom{-}0.06$  &    $\phantom{-}0.04$  &    $\phantom{-}0$   \\
\RadsDKkpp      &    $\phantom{-}0.81$  &    $-0.18$  &    $\phantom{-}0.08$  &    $\phantom{-}0.13$  &    $-0.04$  &    $-0.03$  &    $-0.05$  &    $\phantom{-}1$  &    $\phantom{-}0.33$  &    $-0.03$  &    $-0.02$   \\
\RadsDPikpp     &    $\phantom{-}0.32$  &    $-0.49$  &    $-0.08$  &    $-0.11$  &    $\phantom{-}0.02$  &    $\phantom{-}0$  &    $\phantom{-}0.06$  &    $\phantom{-}0.33$  &    $\phantom{-}1$  &    $\phantom{-}0.02$  &    $-0.02$   \\
\RCPKKPiz       &    $\phantom{-}0.02$  &    $\phantom{-}0$  &    $-0.33$  &    $-0.14$  &    $-0.04$  &    $\phantom{-}0$  &    $\phantom{-}0.04$  &    $-0.03$  &    $\phantom{-}0.02$  &    $\phantom{-}1$  &    $\phantom{-}0.38$   \\
\RCPPiPiPiz     &    $\phantom{-}0.13$  &    $-0.04$  &    $-0.18$  &    $-0.25$  &    $\phantom{-}0.02$  &    $\phantom{-}0.01$  &    $\phantom{-}0$  &    $-0.02$  &    $-0.02$  &    $\phantom{-}0.38$  &    $\phantom{-}1$   \\
\hline
\end{tabular}

\end{table}
\end{landscape}


\begin{table}[!hbtp]
\small
\centering
\caption{Correlation matrix of the statistical uncertainties for the \BuDK, \DzKShh observables~\cite{LHCb-PAPER-2014-041}.}
\label{tab:ggszCor}
\renewcommand{\arraystretch}{1.4}
  \begin{tabular}{ l |llll}
\hline
                  & \xmdk & \ymdk & \xpdk & \ypdk\\
\hline
\xmdk             &  $\phantom{-}1$ &  $-0.247$ &  $\phantom{-}0.038$ &  $-0.003$  \\
\ymdk             &  $-0.247$ &  $\phantom{-}1$ &  $-0.011$ &  $\phantom{-}0.012$  \\
\xpdk             &  $\phantom{-}0.038$ &  $-0.011$ &  $\phantom{-}1$ &  $\phantom{-}0.002$  \\
\ypdk             &  $-0.003$ &  $\phantom{-}0.012$ &  $\phantom{-}0.002$ &  $\phantom{-}1$  \\
\hline
\end{tabular}

\end{table}

\begin{table}[!hbtp]
\small
\centering
\caption{Correlation matrix of the systematic uncertainties for the \BuDK, \DzKShh observables~\cite{LHCb-PAPER-2014-041}.}
\label{tab:ggszCorSyst}
\renewcommand{\arraystretch}{1.4}
  \begin{tabular}{ l |llll}
\hline
                  & \xmdk & \ymdk & \xpdk & \ypdk\\
\hline
\xmdk             &  $\phantom{-}1$ &  $\phantom{-}0.005$ &  $-0.025$ &  $\phantom{-}0.070$  \\
\ymdk             &  $\phantom{-}0.005$ &  $\phantom{-}1$ &  $\phantom{-}0.009$ &  $-0.141$  \\
\xpdk             &  $-0.025$ &  $\phantom{-}0.009$ &  $\phantom{-}1$ &  $\phantom{-}0.008$  \\
\ypdk             &  $\phantom{-}0.070$ &  $-0.141$ &  $\phantom{-}0.008$ &  $\phantom{-}1$  \\
\hline
\end{tabular}

\end{table}









\begin{table}[!hbtp]
\small
\centering
\caption{Correlation matrix of the statistical uncertainties for the \BdDzKstz, \DKpi observables~\cite{LHCb-PAPER-2014-028}.}
\label{tab:glwadsdkstCor}
\renewcommand{\arraystretch}{1.4}
  \begin{tabular}{ l |lll}
\hline
                & \AfavDkstKpi & \RpDkstKpi & \RmDkstKpi\\
\hline
\AfavDkstKpi    & $\phantom{-}1$ &  $\phantom{-}0.091$ &  $\phantom{-}0.083$  \\
\RpDkstKpi      & $\phantom{-}0.091$ &  $\phantom{-}1$ &  $-0.081$  \\
\RmDkstKpi      & $-0.083$ &  $-0.081$ &  $\phantom{-}1$  \\
\hline
\end{tabular}

\end{table}

\begin{table}[!hbtp]
\small
\centering
\caption{Correlation matrix of the systematic uncertainties for the \BdDzKstz, \DKpi observables~\cite{LHCb-PAPER-2014-028}.}
\label{tab:glwadsdkstCorSyst}
\renewcommand{\arraystretch}{1.4}
  \begin{tabular}{ l |lll}
\hline
                & \AfavDkstKpi & \RpDkstKpi & \RmDkstKpi\\
\hline
\AfavDkstKpi    &   $    1$ &   $0.008$ &   $0.008$  \\
\RpDkstKpi      &   $0.008$ &   $    1$ &   $0.997$  \\
\RmDkstKpi      &   $0.008$ &   $0.997$ &   $    1$  \\
\hline
\end{tabular}

\end{table}

\clearpage

\begin{table}[!hbtp]
\small
\centering
\caption{Correlation matrix of the statistical uncertainties for \BdDzKpi, \Dhh observables~\cite{LHCb-PAPER-2015-059}.}
\label{tab:glwkstCor}
\renewcommand{\arraystretch}{1.4}
  \begin{tabular}{ l |llll}
\hline
                & \xmdkpi & \ymdkpi & \xpdkpi & \ypdkpi\\
\hline
\xmdkpi         &   $    1$ &   $0.341$ &   $0.104$ &   $0.130$  \\
\ymdkpi         &   $0.341$ &   $    1$ &   $0.054$ &   $0.154$  \\
\xpdkpi         &   $0.104$ &   $0.054$ &   $    1$ &   $0.501$  \\
\ypdkpi         &   $0.130$ &   $0.154$ &   $0.501$ &   $    1$  \\
\hline
\end{tabular}

\end{table}

\begin{table}[!hbtp]
\small
\centering
\caption{Correlation matrix of the systematic uncertainties for \BdDzKpi, \Dhh observables~\cite{LHCb-PAPER-2015-059}.}
\label{tab:glwkstCorSyst}
\renewcommand{\arraystretch}{1.4}
  \begin{tabular}{ l |llll}
\hline
                & \xmdkpi & \ymdkpi & \xpdkpi & \ypdkpi\\
\hline
\xmdkpi         &   $    1$ &   $0.872$ &   $0.253$ &   $0.368$  \\
\ymdkpi         &   $0.872$ &   $    1$ &   $0.293$ &   $0.414$  \\
\xpdkpi         &   $0.253$ &   $0.293$ &   $    1$ &   $0.731$  \\
\ypdkpi         &   $0.368$ &   $0.414$ &   $0.731$ &   $    1$  \\
\hline
\end{tabular}

\end{table}


\begin{table}[!hbtp]
\small
\centering
\caption{Correlation matrix of the statistical uncertainties for the \BdDzKstz, \DKSpipi observables~\cite{LHCb-PAPER-2016-007}.}
\label{tab:ggszkstCor}
\renewcommand{\arraystretch}{1.4}
  \begin{tabular}{ l |llll}
\hline
                & \xmdkst & \ymdkst & \xpdkst & \ypdkst\\
\hline
\xmdkst         &       1 &   0.143 &       0 &       0  \\
\ymdkst         &   0.143 &       1 &       0 &       0  \\
\xpdkst         &       0 &       0 &       1 &   0.143  \\
\ypdkst         &       0 &       0 &   0.143 &       1  \\
\hline
\end{tabular}

\end{table}


\begin{table}[!hbtp]
\small
\centering
\caption{Correlation matrix of the statistical uncertainties for the \BsDsK observables~\cite{LHCb-PAPER-2014-038}.}
\label{tab:dskCor}
\renewcommand{\arraystretch}{1.4}
  \begin{tabular}{ l |lllll}
\hline
                & \Cpar & \Dpar & \Dbpar & \Spar & \Sbpar \\
\hline
\Cpar           &  $\phantom{-}1$ &  $-0.084$ &  $-0.103$ &  $-0.008$ &  $\phantom{-}0.045$  \\
\Dpar           &  $-0.084$ &  $\phantom{-}1$ &  $\phantom{-}0.544$ &  $\phantom{-}0.117$ &  $-0.022$  \\
\Dbpar          &  $-0.103$ &  $\phantom{-}0.544$ &  $\phantom{-}1$ &  $\phantom{-}0.067$ &  $-0.032$  \\
\Spar           &  $-0.008$ &  $\phantom{-}0.117$ &  $\phantom{-}0.067$ &  $\phantom{-}1$ &  $-0.002$  \\
\Sbpar          &  $\phantom{-}0.045$ &  $-0.022$ &  $-0.032$ &  $-0.002$ &  $\phantom{-}1$  \\
\hline
\end{tabular}

\end{table}

\begin{table}[!hbtp]
\small
\centering
\caption{Correlation matrix of the systematic uncertainties for the \BsDsK observables~\cite{LHCb-PAPER-2014-038}.}
\label{tab:dskCorSyst}
\renewcommand{\arraystretch}{1.4}
  \begin{tabular}{ l |lllll}
\hline
                & \Cpar & \Dpar & \Dbpar & \Spar & \Sbpar \\
\hline
\Cpar           &  $\phantom{-}1$ &  $-0.22$ &  $-0.22$ &  $-0.04$ &  $\phantom{-}0.03$  \\
\Dpar           &  $-0.22$ &  $\phantom{-}1$ &  $\phantom{-}0.96$ &  $\phantom{-}0.17$ &  $-0.14$  \\
\Dbpar          &  $-0.22$ &  $\phantom{-}0.96$ &  $\phantom{-}1$ &  $\phantom{-}0.17$ &  $-0.14$  \\
\Spar           &  $-0.04$ &  $\phantom{-}0.17$ &  $\phantom{-}0.17$ &  $\phantom{-}1$ &  $-0.09$  \\
\Sbpar          &  $\phantom{-}0.03$ &  $-0.14$ &  $-0.14$ &  $-0.09$ &  $\phantom{-}1$  \\
\hline
\end{tabular}

\end{table}

\section{External constraint values and uncertainties}
\label{sec:inputs_aux_vals}


\subsection{Input from global fit to charm data}
\label{sec:input_hfag}

The values and uncertainties are taken from Ref.~\cite{Amhis:2014hma}.
The observables are
\begin{alignat}{3}
& \xd                   & \;=\;  \phantom{-0}0.0037  \;\pm\; & \phantom{0}0.0016\,, \nonumber \\
& \yd                   & \;=\;  \phantom{-0}0.0066  \;\pm\; & \phantom{0}0.0009\,, \nonumber \\
& \ddKpi                & \;=\;  \phantom{-000}3.35  \;\pm\; & \phantom{000}0.21\, \rad, \nonumber \\
& \RdKpi                & \;=\;  \phantom{-}0.00349  \;\pm\; & 0.00004\,, \nonumber \\
& \DAcpPipi             & \;=\;  \phantom{-0}0.0010  \;\pm\; & \phantom{0}0.0015\,, \nonumber \\
& \DAcpKK               & \;=\;  \phantom{0}-0.0015  \;\pm\; & \phantom{0}0.0014\,. \nonumber
\end{alignat}

\noindent Here the value of \ddKpi has been shifted by $\pi$ to comply with the phase convention used in the combination.
The correlations of the charm parameters are given in Table~\ref{tab:hfagCor}.


\subsection{\boldmath Input for \DzKpipipi and \DzKpipiz decays}
\label{sec:input_cleo}

The values and uncertainties are taken from Ref.~\cite{Evans:2016tlp}.
The values used are
\begin{alignat}{3}
& \kdKppp               & \;=\;  \phantom{-00}0.43  \;\pm\; & \phantom{00}0.17\,, \nonumber \\
& \ddKppp               & \;=\;  \phantom{-00}2.23  \;\pm\; & \phantom{00}0.49\,\rad, \nonumber \\
& \kdKpp                & \;=\;  \phantom{-00}0.81  \;\pm\; & \phantom{00}0.06\,, \nonumber \\
& \ddKpp                & \;=\;  \phantom{-00}3.46  \;\pm\; & \phantom{00}0.26\,\rad, \nonumber \\
& \rdKppp               & \;=\;  \phantom{-}0.0549  \;\pm\; & 0.0006\,, \nonumber \\
& \rdKpp                & \;=\;  \phantom{-}0.0447  \;\pm\; & 0.0012\,. \nonumber
\end{alignat}

\noindent The correlation matrix is given in Table~\ref{tab:cleoCor}.


\subsection{\boldmath $CP$ content of \Dhhpiz and \Dpipipipi decays}
\label{sec:input_fcp}

The values and uncertainties are taken from Ref.~\cite{Malde:2015mha}.
The values used are
\begin{alignat}{3}
& \Fppp                 & \;=\;  \phantom{-}0.973  \;\pm\; & 0.017\,, \nonumber \\
& \FKKp                 & \;=\;  \phantom{-}0.732  \;\pm\; & 0.055\,, \nonumber \\
& \Fpppp                & \;=\;  \phantom{-}0.737  \;\pm\; & 0.032\,. \nonumber
\end{alignat}

%


\subsection{\boldmath Input for \DKSKpi parameters}
\label{sec:input_cleo_glskskpi}

The following constraints from Ref.~\cite{Insler:2012pm} are used:
\begin{align}
  \RdKskpi         &=  \phantom{-}0.356 \pm 0.034 \pm 0.007\,,\nonumber \\
  \ddKskpi         &=  -0.29 \pm 0.32\,\rad,\nonumber \\
  \kdKskpi         &=  \phantom{-}0.94 \pm 0.16\,.\nonumber
\end{align}
In addition the following contraint from Ref.~\cite{LHCb-PAPER-2015-026} is used
\begin{align}
  \RdKskpi &= 0.370 \pm 0.003 \pm 0.012\,.\nonumber
\end{align}
The correlation between \ddKskpi and \kdKskpi is determined from the experimental likelihood
to be $\rho(\ddKskpi,\kdKskpi)=-0.60$.


\subsection{\boldmath Constraints on the \BdDKstz hadronic parameters}
\label{sec:input_glwadsdkst_coherence}

The values and uncertainties are taken from Ref.~\cite{LHCb-PAPER-2015-059}.
The values used are
\begin{align}
\kbDKstz     &=  0.958 \pm 0.008 \pm 0.024, \nonumber\\
\RBDKstz     &=  1.020 \pm 0.020 \pm 0.060, \nonumber \\
\DBDKstz     &=  0.020 \pm 0.025 \pm 0.110\, \rad, \nonumber
\end{align}
where the first uncertainty is statistical and the second systematic.
These are taken to be uncorrelated.


\subsection{\boldmath Constraint on \phis}
\label{sec:input_phis}

The value used is taken from Ref.~\cite{LHCb-PAPER-2014-059} as
\begin{align}
\phis = -0.010 \pm 0.039 \rad\,. \nonumber
\end{align}

\clearpage
\section{Uncertainty correlations for the external constraints}
\label{sec:inputs_aux_corrs}


\begin{table}[!hbtp]
\small
\centering
\caption{Correlations of the HFAG charm parameters (CHARM 2015, ``Fit 3'', \CP violation allowed)~\cite{Amhis:2014hma}.}
\label{tab:hfagCor}
\renewcommand{\arraystretch}{1.4}
  \begin{tabular}{ l |llllll}
\hline
                & \xd & \yd & \ddKpi & \RdKpi & \DAcpPipi & \DAcpKK\\
\hline
\xd             & $\phantom{-}1$ &  $-0.361$ &  $-0.332$ &  $\phantom{-}0.234$ &  $\phantom{-}0.117$ &  $\phantom{-}0.146$  \\
\yd             & $ -0.361$ &  $\phantom{-}1$ &  $\phantom{-}0.941$ &  $\phantom{-}0.234$ &  $-0.180$ &  $-0.221$  \\
\ddKpi          & $ -0.332$ &  $\phantom{-}0.941$ &  $\phantom{-}1$ &  $\phantom{-}0.439$ &  $-0.200$ &  $-0.237$  \\
\RdKpi          & $  0.234$ &  $\phantom{-}0.234$ &  $\phantom{-}0.439$ &  $\phantom{-}1$ &  $-0.078$ &  $-0.067$  \\
\DAcpPipi       & $  0.117$ &  $-0.180$ &  $-0.200$ &  $-0.078$ &  $\phantom{-}1$ &  $\phantom{-}0.726$  \\
\DAcpKK         & $  0.146$ &  $-0.221$ &  $-0.237$ &  $-0.067$ &  $\phantom{-}0.726$ &  $\phantom{-}1$  \\
\hline
\end{tabular}

\end{table}

\begin{table}[!hbtp]
\small
\centering
\caption{Correlations of the \DzKpipipi and \DzKpipiz parameters from CLEO and LHCb~\cite{Evans:2016tlp}.}
\label{tab:cleoCor}
\renewcommand{\arraystretch}{1.4}
  \begin{tabular}{ l |llllll}
\hline
                & \kdKppp & \ddKppp & \kdKpp & \ddKpp & \rdKppp & \rdKpp\\
\hline
\kdKppp         &  $\phantom{-}1$ &  $-0.67$ &  $\phantom{-}0.04$ &  $-0.05$ &  $-0.48$ &  $-0.04$  \\
\ddKppp         &  $-0.67$ &  $\phantom{-}1$ &  $\phantom{-}0.02$ &  $\phantom{-}0.15$ &  $\phantom{-}0.12$ &  $\phantom{-}0.08$  \\
\kdKpp          &  $\phantom{-}0.04$ &  $\phantom{-}0.02$ &  $\phantom{-}1$ &  $\phantom{-}0.23$ &  $-0.04$ &  $-0.04$  \\
\ddKpp          &  $-0.05$ &  $\phantom{-}0.15$ &  $\phantom{-}0.23$ &  $\phantom{-}1$ &  $-0.02$ &  $\phantom{-}0.36$  \\
\rdKppp         &  $-0.48$ &  $\phantom{-}0.12$ &  $-0.04$ &  $-0.02$ &  $\phantom{-}1$ &  $-0.03$  \\
\rdKpp          &  $-0.04$ &  $\phantom{-}0.08$ &  $-0.04$ &  $\phantom{-}0.36$ &  $-0.03$ &  $\phantom{-}1$  \\
\hline
\end{tabular}

\end{table}

\clearpage

\section{Fit parameter correlations}
\label{sec:fitout}

\subsection{\DK combination}
\label{sec:app_dk_corr}

\begin{table}[h!]
  \centering
  \caption{Fit parameter correlations for the \DK combination. The fit results are given in Table~\ref{tab:resultrobust}}
  \label{tab:dk_fit_corr}
  \begin{tabular}{ l | l l l l l }
   \hline
              & \g              & \rb             & \db             & \rbDKstz        & \dbDKstz  \\
   \hline
    \g        & \phantom{-}1    & \phantom{-}0.54 & \phantom{-}0.44 & \phantom{-}0.21 & -0.15     \\
    \rb       & \phantom{-}0.54 & \phantom{-}1    & \phantom{-}0.39 & \phantom{-}0.11 & -0.08     \\
    \db       & \phantom{-}0.44 & \phantom{-}0.39 & \phantom{-}1    & \phantom{-}0.08 & -0.05     \\
    \rbDKstz  & \phantom{-}0.21 & \phantom{-}0.11 & \phantom{-}0.08 & \phantom{-}1    & -0.13     \\
    \dbDKstz  & -0.15           & -0.08           & -0.05           & -0.13           & 1         \\
   \hline
  \end{tabular}
\end{table}

\subsection{\Dh combination}
\label{sec:app_dh_corr}

\begin{table}[h!]
  \centering
  \caption{Fit parameter correlations for the \Dh combination solution 1. The fit results are given in Table~\ref{tab:resultdh}}
  \label{tab:dk_fit_corr}
  \begin{tabular}{ l | l l l l l l l}
   \hline
              & \g              & \rb             & \db             & \rbDKstz        & \dbDKstz        & \rbpi           & \dbpi           \\
   \hline
    \g        & \phantom{-}1    & \phantom{-}0.19 & \phantom{-}0.23 & \phantom{-}0.10 & -0.07           & -0.59           & -0.22           \\
    \rb       & \phantom{-}0.19 & \phantom{-}1    & \phantom{-}0.23 & \phantom{-}0.02 & \phantom{-}0    & -0.20           & \phantom{-}0.02 \\
    \db       & \phantom{-}0.23 & \phantom{-}0.23 & \phantom{-}1    & \phantom{-}0.02 & \phantom{-}0    & -0.09           & \phantom{-}0.42 \\
    \rbDKstz  & \phantom{-}0.10 & \phantom{-}0.02 & \phantom{-}0.02 & \phantom{-}1    & -0.10           & -0.06           & -0.03           \\
    \dbDKstz  & -0.07           & \phantom{-}0    & \phantom{-}0    & -0.10           & 1               & \phantom{-}0.04 & \phantom{-}0.03 \\
    \rbpi     & -0.59           & -0.20           & -0.09           & -0.06           & \phantom{-}0.04 & 1               & \phantom{-}0.45 \\
    \dbpi     & -0.22           & \phantom{-}0.02 & \phantom{-}0.42 & -0.03           & \phantom{-}0.03 & \phantom{-}0.45 & 1               \\
   \hline
  \end{tabular}
\end{table}

\begin{table}[h!]
  \centering
  \caption{Fit parameter correlations for the \Dh combination solution 2. The fit results are given in Table~\ref{tab:resultdh}}
  \label{tab:dk_fit_corr}
  \begin{tabular}{ l | l l l l l l l}
   \hline
              & \g              & \rb             & \db             & \rbDKstz        & \dbDKstz        & \rbpi           & \dbpi           \\
   \hline
    \g        & \phantom{-}1    & \phantom{-}0.52 & \phantom{-}0.51 & \phantom{-}0.22 & -0.16           & -0.12           & \phantom{-}0.01 \\
    \rb       & \phantom{-}0.52 & \phantom{-}1    & \phantom{-}0.41 & \phantom{-}0.11 & -0.08           & \phantom{-}0.03 & \phantom{-}0.10 \\
    \db       & \phantom{-}0.51 & \phantom{-}0.41 & \phantom{-}1    & \phantom{-}0.11 & -0.06           & -0.19           & -0.01           \\
    \rbDKstz  & \phantom{-}0.22 & \phantom{-}0.11 & \phantom{-}0.11 & \phantom{-}1    & -0.13           & -0.02           & \phantom{-}0    \\
    \dbDKstz  & -0.16           & -0.08           & -0.06           & -0.13           & 1               & \phantom{-}0.01 & \phantom{-}0    \\
    \rbpi     & -0.12           & \phantom{-}0.03 & -0.19           & -0.02           & \phantom{-}0.01 & 1               & \phantom{-}0.83 \\
   \dbpi     & \phantom{-}0.01  & \phantom{-}0.10 & -0.01           & \phantom{-}0    & \phantom{-}0    & \phantom{-}0.83 & 1               \\
   \hline
  \end{tabular}
\end{table}

\end{appendix}

\newpage
\newpage

\addcontentsline{toc}{section}{References}
\setboolean{inbibliography}{true}
\bibliographystyle{LHCb}
\bibliography{LHCb-PAPER,LHCb-CONF,references,bayesian}

\newpage



\centerline{\large\bf LHCb collaboration}
\begin{flushleft}
\small
R.~Aaij$^{40}$,
B.~Adeva$^{39}$,
M.~Adinolfi$^{48}$,
Z.~Ajaltouni$^{5}$,
S.~Akar$^{6}$,
J.~Albrecht$^{10}$,
F.~Alessio$^{40}$,
M.~Alexander$^{53}$,
S.~Ali$^{43}$,
G.~Alkhazov$^{31}$
% ...

{\footnotesize \it
% ...
$ ^{5}$Clermont Universit{\'e}, Universit{\'e} Blaise Pascal, CNRS/IN2P3, LPC, Clermont-Ferrand, France\\
$ ^{6}$CPPM, Aix-Marseille Universit{\'e}, CNRS/IN2P3, Marseille, France\\
% ...
$ ^{10}$Fakult{\"a}t Physik, Technische Universit{\"a}t Dortmund, Dortmund, Germany\\
% ...
$ ^{31}$Petersburg Nuclear Physics Institute (PNPI), Gatchina, Russia\\
% ...
$ ^{39}$Universidad de Santiago de Compostela, Santiago de Compostela, Spain\\
$ ^{40}$European Organization for Nuclear Research (CERN), Geneva, Switzerland\\
% ...
$ ^{43}$Nikhef National Institute for Subatomic Physics, Amsterdam, The Netherlands\\
% ...
$ ^{48}$H.H. Wills Physics Laboratory, University of Bristol, Bristol, United Kingdom\\
% ...
$ ^{53}$School of Physics and Astronomy, University of Glasgow, Glasgow, United Kingdom\\
% ...
}
\end{flushleft}


\end{document}